\documentclass[10pt]{article}
\usepackage{amsmath}
\usepackage{amsfonts}
\usepackage{latexsym,amssymb,enumerate,ulem,pifont,calc}

\makeatletter

\@addtoreset{equation}{section}
\makeatother

\begin{document}

\begin{titlepage}

\vskip 2.0 cm
\begin{center}  {\huge{\bf BPS Black Hole Entropy and Attractors\\\vskip 0.2 cm in Very Special Geometry}} \\\vskip 0.5 cm {\Large{\bf Cubic Forms, Gradient Maps and their Inversion}}

\vskip 2.5 cm

{\large{\bf Bert van Geemen$^1$}, {\bf Alessio Marrani$^{2,3}$}, and {\bf Francesco Russo$^{4}$}}

\vskip 1.0 cm

$^1${\sl
Dipartimento di Matematica, Universit\`a di Milano,\\Via Saldini 50, I-20133
Milano, Italy\\
\texttt{lambertus.vangeemen@unimi.it}}\\

\vskip 0.5
cm

$^2${\sl Centro Studi e Ricerche Enrico Fermi,\\Via Panisperna 89A,
I-00184, Roma, Italy}\\

\vskip 0.5 cm

$^3${\sl Dipartimento di Fisica
e Astronomia Galileo Galilei, Universit\`a di Padova,\\and INFN, sezione
di
Padova, Via Marzolo 8, I-35131 Padova, Italy\\
\texttt{jazzphyzz@gmail.com}}\\

\vskip 0.5 cm

$^4${\sl Dipartimento di
Matematica e Informatica, Universit\`a di Catania,\\ Viale A. Doria 5,

I-95125 Catania, Italy\\ \texttt{frusso@dmi.unict.it}}

\vskip 0.5
cm


 \end{center}

 \vskip 2.0 cm

\begin{%
abstract}

We consider Bekenstein-Hawking entropy and attractors in
extremal BPS black
holes of $\mathcal{N}=2$, $D=4$ ungauged supergravity
obtained as reduction
of minimal, matter-coupled $D=5$ supergravity. They
are generally expressed
in terms of solutions to an inhomogeneous system of
coupled quadratic
equations, named \textit{BPS system}, depending on the
cubic prepotential as
well as on the electric-magnetic fluxes in the
extremal black hole
background. Focussing on homogeneous \textit{%
non-symmetric} scalar manifolds
(whose classification is known in terms of $%
L(q,P,\dot{P})$ models), under
certain assumptions on the Clifford matrices
pertaining to the related cubic
prepotential, we formulate and prove an
\textit{invertibility condition} for
the gradient map of the corresponding
cubic form (to have a birational inverse map which is
given by homogeneous
polynomials of degree four), and therefore for the solutions
to the BPS
system to be explicitly determined, in turn providing novel,
explicit expressions for the BPS black hole entropy and the related attractors as
solution of the BPS attractor equations.
After a general treatment, we present a number of explicit examples with $%
\dot{P}%
=0$, such as $L(q,P)$, $1\leqslant q\leqslant 3$ and $P\geqslant 1$,
or $%
L(q,1)$, $4\leqslant q\leqslant 9$, and one model with $\dot{P}=1$,
namely $%
L(4,1,1)$. We also briefly comment on Kleinian signatures and split
algebras. In particular, we provide, for the first time, the explicit form of the BPS black hole entropy and of the related BPS attractors for the infinite class of $L(1,P)$ $P\geqslant 2$ non-symmetric models of $\mathcal{N}=2$, $D=4$ supergravity.

 \end{abstract}
\vspace{24pt} \end{titlepage}


\newpage \tableofcontents \newpage


\section{Introduction}

In the last decades, the theoretical and phenomenological implications of
the physics of black holes \cite{wald} had a profound and fertile impact on
many branches of science, from astrophysics, cosmology, particle physics, to
mathematical physics \cite{moore}, quantum information theory \cite{Duff},
and, recently, number theory \cite{Bhargava}. Remarkably, the singularity
theorems proved by Penrose and Hawking \cite{ph} imply that the black holes
are an unavoidable consequence of Einstein's theory of General Relativity,
as well as of its modern generalizations such as supergravity \cite{black,
Sen, Cvetic}, superstrings and M-theory \cite{review}. Classically, the
gravitational force inside the event horizon of a black hole is so strong
that nothing, not even light, can escape. However, in the 70s Hawking showed
that quantum effects cause black holes to thermally radiate, and eventually
evaporate \cite{Hawk}.

While the frontiers of physics are progressing also the 21st century, it
should not be forgotten that the physics of the 20th century is conceptually
founded on two theories which are mutually incompatible. On one side,
Quantum Mechanics governs the microscopic world of the basic constituents of
matter, such as molecules, atoms, nuclei and beyond. On the other side,
General Relativity describes gravity and the macroscopic, large-scale
structures, ranging from planetary orbits to the Universe in its entirety.
As the energy increases, Quantum Mechanics and General Relativity inevitably
meet, giving rise to startling, even paradoxical, consequences.

A tantalizing aspect of the physics of a black hole is that its
thermodynamical features seem to encode fundamental insights of a not yet
formulated theory of Quantum Gravity, which should necessarily arise from
the reconciliation of the two aforementioned apparently contradictory
physical theories. In this framework, a crucial relevance owes to the
Bekenstein-Hawking entropy-area formula \cite{entrop}:
\begin{equation}
S=\frac{k_{B}}{\ell _{P}^{2}}\,\frac{A_{H}}{4}\,\,,  \label{bekhaw}
\end{equation}%
where\footnote{%
We will use the so-called natural units henceforth: $\hbar =c=G=k_{B}=1$.} $%
k_{B}$ is the Boltzmann constant, $\ell _{P}^{2}=G\hbar /c^{3}$ is the
squared Planck length, whereas $A_{H}$ denotes the area of the event horizon
of the black hole itself. This formula relates a thermodynamical quantity
(the entropy $S$) to a geometric quantity (the area $A_{H}$), and after much
theoretical work it still puzzles the scientific community. In fact, a
crucial issue that Quantum Gravity must necessarily address concerns on the
origin of $S$ at a fundamental level. At the classical level, (\ref{bekhaw})
yields that black hole entropy is determined by the area of the event
horizon, which is a macroscopic and geometric quantity; however, black hole
entropy must also enjoy a microscopic, statistical derivation, accounting
for fundamental microscopic degrees of freedom.

Since superstring theory and M-theory are the most serious candidates for a
theory of Quantum Gravity, they are expected to provide a microscopic,
statistical explanation of the entropy-area law (\ref{bekhaw}) \cite%
{blackmicro}. Black holes are typical non-perturbative objects, since they
describe a physical regime in which the gravitational field is very strong;
thus, only a non-perturbative approach can successfully deal with them.
Progress in this direction came after 1995 \cite{witten}, through the
recognition of the role of string dualities, which allow one to relate the
strong coupling regime of one superstring model to the weak coupling regime
of another. Remarkably, there is evidence that the string dualities are all
encoded into the global symmetry group (the $U$-duality, also named
electric-magnetic, group) of the low energy supergravity effective action
\cite{HT-1}.

Black holes, and in particular their extremal configurations \cite%
{Extreme-BH}, are embedded in a natural way in supergravity theories, which,
being invariant under local super-Poincar\'{e} transformations, include
General Relativity, providing a consistent description of the graviton
coupled to other fields in a supersymmetric framework. Extremal black holes
have become objects of crucial relevance in the context of superstrings
after 1995 \cite{blackmicro,review,black,ortino} : the classical solutions
of supergravity that preserve a fraction of the original supersymmetries can
be interpreted as non-perturbative states, necessary to complete the
perturbative string spectrum and make it invariant under the many
conjectured duality symmetries \cite{schw,sesch,HT-1,gmv,vasch}. In such a
framework, extremal black holes, as well as their parent $p$-branes in
higher dimensions, are conceived as additional particle-like states that
compose the spectrum of a fundamental quantum theory. Similar to monopoles
in gauge theories, these non-perturbative quantum states originate from
regular solutions of the classical field equations, i.e. the very same
Einstein equations on which General Relativity relies. The crucial new
ingredient, in this respect, is Supersymmetry, which requires a precise
balance between vector fields and scalar fields in the bosonic spectrum of
the theory. As such, the general framework we are going to deal with is
provided by the so-called Einstein--Maxwell-scalar theories, whose global
mathematical thorough treatment has been recently given in \cite{Shahbazi}
(see also \cite{Alek}).

Supergravity theories provide a low-energy effective theory description of
superstring and M-theory, holding at the lowest order in the string loop
expansion, when the space-time curvature is much smaller than the typical
string scale (string tension). Consequently, the supergravity description of
extremal black holes can be trusted only when the radius of the event
horizon is much larger than the string scale, corresponding to the regime of
large charges. We will not be dealing with further corrections, introduced
by string theory, which give rise to higher derivative terms in the low
energy effective action, such that the black hole entropy is expected to be
corrected by subleading terms in the limit of small curvature :it is well
known that these corrections determine a deviation from the area law for the
entropy \cite{w,cdm}.

The cosmic censorship conjecture \cite{CCC} is naturally realized by
conceiving extremal black holes as solitonic solutions of $\mathcal{N}$%
-extended locally supersymmetric theory of Einstein gravity : in fact,
denoting with $\mathcal{N}$ the number of spinor supercharges in $3+1$
space-time dimensions, when $\mathcal{N}\geqslant 2$ the so-called \textit{%
BPS} (Bogomol'nyi-Prasad-Sommerfeld) \textit{bound }\cite{BPS bound},%
\begin{equation}
M\geqslant |Q|,  \label{BPS1}
\end{equation}%
where $M$ and $Q$ respectively are the mass and the magnetic (or electric)
charge of the black hole, is just a consequence of the supersymmetry
algebra, implying that no naked singularities can occur.

When the black hole solution is embedded into a $\mathcal{N}$-extended
supergravity theory, the model is characterized by a certain $\mathcal{N}$%
-dependent number of scalar fields, collectively denoted by $\phi $. In this
framework, the charge $Q$ is to be replaced by the maximum eigenvalue of the
$\mathcal{N}\times \mathcal{N}$ central charge matrix appearing in the
r.h.s. of the supersymmetry algebra (depending on the expectation value $%
\phi _{H}\left( p,q\right) $ of the scalar fields on the event horizon,
where $p$'s and $q$'s respectively denote magnetic and electric charges of
the black hole) :
\begin{equation}
M=M(p,q)\geqslant |Z_{\max }(\phi _{H}\left( p,q\right) ,p,q)|  \label{BPS2}
\end{equation}%
In the present paper, we will be dealing only with extremal black holes, in
which the BPS bound (\ref{BPS1})-(\ref{BPS2}) is saturated.

Extremal black holes enjoy the following peculiar and crucial feature:
despite the fact that the dynamics depends on scalar fields, the event
horizon of the black hole loses all information about the scalars, and this
holds regardless of the supersymmetry-preserving features of the solution.
This phenomenon is described by the so-called \textit{attractor mechanism}
\cite{AM,book}: independently of their boundary conditions at spatial
infinity, scalar fields flow to a fixed point given by a certain ratio of
electric and magnetic charges, when approaching the event horizon. In this
framework, the scalar fields are \textit{moduli}, i.e. they are continuous
parameters which can be freely specified at infinity, raising the dangerous
possibility that the black hole entropy might depend on their values.
Indeed, such a dependence presumably would lead to a violation of the second
law of thermodynamics, since it would allow one to quasi-statically decrease
the entropy by varying the moduli. Instead, the black hole entropy turns out
to depend only on the values acquired by the scalar fields at the event
horizon, which in turn only depend on the conserved charges ($p$ and $q$)
associated to gauge invariance of the black hole solution itself: in this
sense, the entropy of extremal black holes is a topological quantity,
because it is fixed in terms of the quantized electric and magnetic charges,
while it does not depend on moduli.

For extremal black holes in Maxwell-Einstein supergravity theories with $%
\mathcal{N}=2$ local supersymmetry in $3+1$ space-time dimensions \cite%
{rev-Ferrara}, the saturation of the BPS bound (\ref{BPS2}),%
\begin{equation}
M\left( p,q\right) =|Z(\phi _{H}\left( p,q\right) ,p,q)|,  \label{extr1}
\end{equation}%
yields, for the black hole entropy,
\begin{equation}
S\left( p,q\right) =\frac{A_{H}(p,q)}{4}=\pi |Z(\phi _{H}\left( p,q\right)
,p,q)|^{2}\,,  \label{bert}
\end{equation}%
where $Z$ is the $\mathcal{N}=2$ central charge function \cite{CDFp}. The
attractor values of the scalar fields at the event horizon, here
collectively denoted by $\phi _{H}\left( p,q\right) $, arise as solutions to
the so-called \textit{BPS attractor equations} :%
\begin{equation}
\left. \mathcal{D}_{\phi }Z(\phi ,p,q)\right\vert _{\phi =\phi _{H}(p,q)}=0,
\end{equation}%
where $\mathcal{D}_{\phi }$ denotes the K\"{a}hler-covariant differential
operator acting on the scalar manifold (target space of moduli fields). The
entropy $S$ generally enjoys a $U$-duality-invariant expression (homogeneous
of degree two) in terms of electric and magnetic charges, only depending on
the nature of the $U$-duality groups and on the appropriate representations
of electric and magnetic charges \cite{uinvar}.

Through the years, the attractor mechanism has been discovered to have a
broader application \cite{ortin, fegika, non-BPS, bfm1,bfgm} beyond the BPS
cases, being a peculiarity of all extremal black-holes, BPS or not. Even for
these more general cases, because of the topological nature of the
extremality condition, the entropy formula turns out to be still given by a $%
U$-duality invariant expression built out of electric and magnetic
charges.\bigskip

The present paper is devoted to the determination of the explicit expression
of two purely charge-dependent quantities, characterizing the physics of BPS
extremal black holes : the Bekenstein-Hawking entropy $S\left( p,q\right) $
and the attractor values, collectively denoted by $\phi _{H}\left(
p,q\right) $, acquired by the scalar fields when approaching the (unique)
event horizon (regardless of the boundary conditions of their evolution
dynamics). We will consider ungauged Maxwell-Einstein supergravity theories
with $\mathcal{N}=2$ extended local supersymmetry in $3+1$ space-time
dimensions, in the case in which the special K\"{a}hler geometry of the
vector multiplets' scalar manifold is determined by a cubic holomorphic
prepotential (very special geometry). In fact, in such a framework only the
models in which the scalar manifold is a symmetric coset have been
thoroughly investigated : exploiting the relation to the theory of cubic
(simple and semisimple) Jordan algebras and related Freudenthal triple
systems, which hold in all cases but the so-called Luciani models (with
quadratic prepotential), the explicit expressions of $S\left( p,q\right) $
and of $\phi _{H}\left( p,q\right) $ have been explicitly computed for
extremal, both BPS and non-BPS, black holes (cfr. e.g. \cite{FGimonK} and
refs. therein).

On the other hand, very little is known in the case in which the (vector
multiplets') scalar manifold is not symmetric. In very special geometry, the
BPS entropy and the BPS attractor values of the scalar fields have been
computed by Shmakova \cite{Shmakova}, up to the solution of an inhomogeneous
system of quadratic algebraic equations, named \textit{BPS system}. A
noteworthy, countably infinite class of cubic non-symmetric models is
provided by homogeneous non-symmetric models, which have been classified in
\cite{dWVP} (in a mathematical context, see also the subsequent
classification in \cite{Cortes}). These models are quite interesting from a
physical point of view, because some of them naturally occur in the four
dimensional effective supergravity description of brane dynamics, when their
brane and bulk degrees of freedom get unified. To the best of our knowledge,
only \cite{DFT-Hom-07} dealt with such a class of models, but did not
investigate the explicit determination of $S\left( p,q\right) $ and $\phi
_{H}\left( p,q\right) $. In the present work, we will rely on the existing
classification of homogeneous non-symmetric special manifolds, and we will
formulate a (sufficient but not necessary) condition for the BPS system to
be explicitly solved. Thus, within the validity of such a condition (which
we will prove to actually hold for an infinite, countable number of models),
we will explicitly determine the expression of the Bekenstein-Hawking
(semi)classical black hole entropy $S\left( p,q\right) $ as well as of the
purely charge-dependent attractor values $\phi _{H}\left( p,q\right) $
acquired by the scalar fields at the event horizon of asymptotically flat,
spherically symmetric, static, dyonic, extremal BPS black holes.\bigskip

The paper is organized as follows.

In Secs.\ \ref{sec-BPS}-\ref{sec-BPS-2} we introduce the Bekenstein-Hawking
BPS black hole entropy and the BPS attractor values of scalar fields in $%
\mathcal{N}=2$ very special geometry, relating their explicit expressions to
the solution of the corresponding BPS system, or equivalently to the
inversion of the gradient map of the cubic form defining the corresponding
cubic holomorphic prepotential. Then, in Sec.\ \ref{HVSG} we specialize the
treatment to homogeneous very special geometry, briefly recalling some basic
facts on symmetric and non-symmetric spaces in Secs. \ref{symmd} resp. \ref%
{nsymmd}. In Sec.\ \ref{HNS} we review the classification of homogeneous
special $d$-spaces \cite{dWVP}, and in Sec.\ \ref{Lq1} we briefly consider
the class $L(q,1)$. Sec.\ \ref{clifmat}, which is in turn split into eight
Subsecs., is then devoted to the introduction and review of another
important ingredient of our treatment: Euclidean Clifford algebras. Next,
Secs.\ \ref{invertgamma} and \ref{checkinvmap} contain the main results of
the present paper: after enouncing an invertibility condition for the
gradient map of the cubic form in Sec.\ \ref{main} (then proved in Sec.\ \ref%
{checkinvmap}), the BPS system is explicitly solved in Secs. \ref{explin}-%
\ref{Sol-BPS}, and the explicit expressions of the BPS entropy and of the
BPS attractor values of scalar fields at the horizon are computed in Sec.\ %
\ref{BPS-S-gen}. Then, Sec.\ \ref{complete} introduces the so-called
complete models, whose known examples coincide with the symmetric $d$%
-spaces, recalled in Sec.\ \ref{symmetric}. Secs.\ \ref{explmodels}, \ref%
{L420} and \ref{LqP0} present a threefold wealth of models in which the
invertibility condition of Sec.\ \ref{main} holds true, and in some cases,
such as the models $L(1,2)$ and $L(1,3)$ in Secs.\ \ref{L120} resp. \ref%
{L130} (then generalized\footnote{%
In Sec.\ \ref{geom-fact} we will also briefly present a geometric point of
view on the factorization of the inverse map of the gradient map, whose
detailed investigation goes beyond the scope of this paper.} as $L(1,P)$
with $P\geqslant 2$ in Sec.\ \ref{L1P0}), the corresponding BPS system is
explicitly solved (with details given in Appendices \ref{App-L(1,2)} and \ref%
{App-L(1,3)}), thus providing the fully fledged expressions of the BPS
entropy and attractors. The non-uniqueness of the matrices $\Omega _{K}$'s
occurring in the invertibility condition of Sec.\ \ref{main} is discussed in
Sec.\ \ref{L1PL81}, and in Sec.\ \ref{Des-Sub} the difference between
descendant models and submodels is highlighted (with details given in App.\ %
\ref{App-Lie}). Moreover, Sec.\ \ref{L411} discusses the unique model of the
present paper which has a non-vanishing $\dot{P}$ : namely, the model $%
L(4,1,1)$. A brief discussion of cubic models determined by Kleinian (rather
than Lorentzian) quadratic polynomials (thus giving rise to a non-special
geometry) is provided in Sec.\ \ref{55}. Finally, Sec.\ \ref{L910} deals
with the model $L(9,1)$, as an example of model in which the invertibility
criterion of Sec.\ \ref{main} does not seem to be applicable, and thus other
approaches are needed to prove or disprove the invertibility of the BPS
system. Finally, some outlook and hints for further developments are
provided in the concluding Sec.\ \ref{Conclusion}.


\section{BPS Black hole entropy and attractors in very special geometry\label%
{sec-BPS}}

A large class of four-dimensional Maxwell-Einstein gravity theories with
local $\mathcal{N}=2$ supersymmetry can be obtained by an $S^{1}$%
-compactification of five-dimensional minimal supergravity. In such a case,
the K\"{a}hler-Hodge geometry of the vector multiplets' scalar manifolds in $%
D=4$ is named \textit{very special }\cite%
{spegeo,Strominger-SKG,N=2-Big,Freed}, and it is determined by an
holomorphic prepotential of the type\footnote{%
Einstein summation convention on repeated indices is understood throughout.
Lowercase Latin indices run $1,...,N$ throughout; $N$ denotes the number of
vector multiplets. The index $0$ pertains to the ($D=4$) graviphotonic
sector.}%
\begin{equation}
F\left( X\right) :=\frac{1}{3!}d_{ijk}\frac{X^{i}X^{j}X^{k}}{X^{0}},
\label{F(X)}
\end{equation}%
where $d_{ijk}$ is a completely symmetric real tensor, and the $X^{\Lambda }$%
's, $\Lambda =0,i$, are the contravariant symplectic sections of the K\"{a}%
hler-Hodge target space of scalar fields. The symplectic frame in which $%
F\left( X\right) $ (\ref{F(X)}) is specified is the one of the so-called
\textquotedblleft $4D/5D$ special coordinates" (see for instance \cite%
{N=2-Big, CFM}) : the manifest symmetry is the electric-magnetic ($U$-)
duality\footnote{%
In this paper, $U$-duality is referred to as the \textquotedblleft
continuous\textquotedblright\ symmetries of \cite{CJ-1}. Their discrete
versions are the $U$-duality non-perturbative string theory symmetries
introduced by Hull and Townsend \cite{HT-1}.} group of the parent theory in $%
D=5$ (which leaves $d_{ijk}$ invariant).

From the treatment given in \cite{Shmakova}, in the general case in which
all electric and magnetic charges associated to the black hole solution are
non-vanishing, the Bekenstein-Hawking entropy of static, spherically
symmetric, BPS extremal dyonic black holes in ungauged $\mathcal{N}=2$, $D=4$
Maxwell-Einstein supergravity whose vector multiplets' scalar manifold
displays a \textit{very special} K\"{a}hler geometry, reads\footnote{%
This formula fixes a typo in Eq. (12) of \cite{Shmakova}. For further,
recent insight on the BPS\ entropy in very special geometry, see \cite{Alek}.%
}%
\begin{equation}
\frac{S}{\pi }=\frac{1}{3\left\vert p^{0}\right\vert }\sqrt{\frac{4}{3}%
\left( \Delta _{i}x^{i}\right) ^{2}-9\left[ p^{0}\left( p\cdot q\right)
-2I_{3}(p)\right] ^{2}},  \label{S-BPS}
\end{equation}%
where $p^{0}$, $p^{i}$, $q_{0}$ and $q_{i}$ are the magnetic resp. electric
black hole charges, and%
\begin{eqnarray}
\Delta _{i} &:&=\frac{1}{2}d_{ijk}p^{j}p^{k}-p^{0}q_{i}=\frac{\partial
I_{3}(p)}{\partial p^{i}}-p^{0}q_{i};  \label{rec0} \\
p\cdot q &:&=p^{0}q_{0}+p^{i}q_{i};  \label{rec1} \\
I_{3}(p) &:&=\frac{1}{3!}d_{ijk}p^{i}p^{j}p^{k}.  \label{rec2}
\end{eqnarray}%
Note that, as it must be, $S$ (\ref{S-BPS}) is homogeneous of degree $2$ in
the black hole charges. Furthermore, $q_{0}$ enters the expression (\ref%
{S-BPS}) only through the quantity $p\cdot q$ (\ref{rec1}).

The $x^{i}$'s appearing in (\ref{S-BPS}) are the solutions $x^{i}\left(
\Delta _{j};d_{klm}\right) $ of the system of algebraic quadratic equations%
\begin{equation}
\frac{1}{3!}d_{ijk}x^{j}x^{k}=\Delta _{i},  \label{BPS}
\end{equation}%
which we will henceforth name the \textit{BPS system}\footnote{%
Let us recall that the explicit solution of the BPS system is also relevant
for the solution of the attractor equations in asymptotically $AdS$, dyonic,
extremal $\frac{1}{4}$-BPS black holes of $U(1)$ Fayet-Iliopoulos gauged
Maxwell-Einstein $\mathcal{N}=2$ supergravity in four space-time dimensions
\cite{Halmagyi1, Halmagyi2}.} pertaining to the model of $\mathcal{N}=2$, $%
D=4$ ungauged supergravity (coupled to vector multiplets\footnote{%
The coupling to hypermultiplets can be disregarded, because their equations
of motion decouple completely in the ungauged case, and so they do not
contribute at all to the extremal black hole entropy.}) under consideration.
Since the $\Delta _{i}$'s are homogeneous of degree two in the black hole
charges $p^{i},p^{0}$ and $q_{i}$, (\ref{BPS}) implies that the $x^{i}$'s
are homogeneous of degree one in the same variables:%
\begin{equation}
\frac{\partial x^{i}}{\partial p^{j}}p^{j}+\frac{\partial x^{i}}{\partial
p^{0}}p^{0}+\frac{\partial x^{i}}{\partial q_{j}}q_{j}=x^{i}.  \label{homm}
\end{equation}%
Furthermore, the $x^{i}$'s contribute to the black hole entropy only through
the square of the quantity\footnote{%
Recalling (\ref{rec2}), such a definition implies that $I_{3}(p)=\mathcal{V}%
(p)$.}%
\begin{equation}
\Delta _{i}x^{i}=\frac{1}{3!}d_{ijk}x^{i}x^{j}x^{k}=:\mathcal{V}(x).
\label{expr1}
\end{equation}%
The number of equations in the system (\ref{BPS}) is equal to the number $N$
of complex scalar fields, which in the large volume limit of Calabi-Yau
compactifications of type II superstrings correspond to the Calabi-Yau
\textit{moduli} fields. The l.h.s. of (\ref{BPS}) is given by quadratic
forms with coefficients $d_{(i)jk}$, while the r.h.s. is arbitrary and
depends on the values of electric and magnetic charges of the extremal BPS
black hole. In other words, it defines $N$ quadratic hypersurfaces (each of
dimension $N-1$) in an $N$-dimensional space, and the intersection of these
hypersurfaces is the set of solutions of the system (\ref{BPS}). Therefore,
the BPS system may also not admit any analytical (or in closed form) real
solution at all.

Note that the condition%
\begin{equation}
\frac{4}{3}\left( \Delta _{i}x^{i}\right) ^{2}-9\left[ p^{0}\left( p\cdot
q\right) -2I_{3}(p)\right] ^{2}>0
\end{equation}%
is a consistency condition for the BPS entropy (\ref{S-BPS}) to be well
defined. We will see below what is the general (set of) BPS condition(s) in $%
\mathcal{N}=2$ ungauged supergravity with cubic prepotential (cfr. (\ref%
{BPS-cond-gen}) below).

By switching the notation of scalar fields (at the horizon) from $\phi _{H}$
to $z_{H}^{i}$, and denoting with $\mathbf{i}$ the imaginary unit of $%
\mathbb{C}$, the explicit solutions to the BPS Attractor Equations read \cite%
{Shmakova}%
\begin{eqnarray}
z_{H}^{i}\left( p^{0},p^{k},q_{0},q_{k}\right) &=&\frac{3}{2}\frac{x^{i}}{%
p^{0}\Delta _{j}x^{j}}\left[ p^{0}\left( p\cdot q\right) -2I_{3}(p)\right] +%
\frac{p^{i}}{p^{0}}-\mathbf{i}\frac{3}{2}\frac{x^{i}}{\left\vert \Delta
_{j}x^{j}\right\vert }\frac{S}{\pi }  \notag \\
&=&\frac{3}{2}\frac{x^{i}}{\Delta _{j}x^{j}}\left[ \frac{p^{0}\left( p\cdot
q\right) -2I_{3}(p)}{p^{0}}-\mathbf{i}\frac{S}{\pi }\text{sgn}\left( \Delta
_{j}x^{j}\right) \right] +\frac{p^{i}}{p^{0}},  \label{BPS-crit-points}
\end{eqnarray}%
where $S$ is given by (\ref{S-BPS}), and the subscript \textquotedblleft $H$%
" denotes the evaluation at the (unique) event horizon of the extremal BPS
black hole.

Finally, it is here worth recalling that in very special geometry, by
construction, the quantity\linebreak\ $d_{ijk}$Im$\left( z_{H}^{i}\right) $Im%
$\left( z_{H}^{j}\right) $Im$\left( z_{H}^{k}\right) $ must have a definite
sign, say negative, and this imposes a further constraint on the sign of $%
x^{i}$'s :%
\begin{equation}
d_{ijk}\text{Im}\left( z_{H}^{i}\right) \text{Im}\left( z_{H}^{j}\right)
\text{Im}\left( z_{H}^{k}\right) <0\Leftrightarrow -\frac{27}{8}\frac{S^{3}}{%
\pi ^{3}}\frac{d_{ijk}x^{i}x^{j}x^{k}}{\left( \Delta _{l}x^{l}\right) ^{3}}%
\text{sgn}\left( \Delta _{m}x^{m}\right) =-\frac{27}{8}\frac{S^{3}}{\pi ^{3}}%
\frac{d_{ijk}x^{i}x^{j}x^{k}}{\left\vert \Delta _{l}x^{l}\right\vert ^{3}}<0.
\label{d<0}
\end{equation}%
By exploiting (\ref{BPS}), condition (\ref{d<0}) can be rewritten as%
\begin{equation}
-\frac{27\cdot 6}{8}\frac{S^{3}}{\pi ^{3}}\frac{\text{sgn}\left( \Delta
_{j}x^{j}\right) }{\left( \Delta _{k}x^{k}\right) ^{2}}<0\Leftrightarrow
\Delta _{j}x^{j}>0\Leftrightarrow \mathcal{V}(x)>0,  \label{d<0-2}
\end{equation}%
and thus (\ref{BPS-crit-points}) gets simplified to\footnote{%
This formula matches Eq. (24) of \cite{Shmakova}.}%
\begin{equation}
z_{H}^{i}\left( p^{0},p^{k},q_{0},q_{k}\right) =\frac{3}{2}\frac{x^{i}}{%
\Delta _{j}x^{j}}\left[ \frac{p^{0}\left( p\cdot q\right) -2I_{3}(p)}{p^{0}}-%
\mathbf{i}\frac{S}{\pi }\right] +\frac{p^{i}}{p^{0}},
\label{BPS-crit-points-2}
\end{equation}%
with $S$ given by (\ref{S-BPS}). The expression (\ref{BPS-crit-points-2})
highlights the relation between the real and imaginary parts of $z_{H}^{i}$
(i.e., the attractor configurations of axions resp. dilatons) and the $x^{i}$%
's themselves.

All in all, the BPS formul\ae\ (\ref{S-BPS}) and (\ref{BPS-crit-points-2})
are well defined for%
\begin{equation}
\text{BPS}:\left\{
\begin{array}{l}
\mathcal{V}\left( x\right) >0\Leftrightarrow \Delta _{i}x^{i}>0; \\
\\
4\left( \Delta _{i}x^{i}\right) ^{2}-27\left[ p^{0}\left( p\cdot q\right)
-2I_{3}(p)\right] ^{2}>0.%
\end{array}%
\right.  \label{BPS-cond-gen}
\end{equation}

\section{\label{sec-BPS-2}BPS systems and the gradient map $\protect\nabla _{%
\mathcal{V}}$}

Given a rank-3 completely symmetric tensor $d_{ijk}=d_{(ijk)}$ in $N$
dimensions ($i,j,k=1,...,N$), we recall the cubic form defined in (\ref%
{expr1}) :%
\begin{equation}
\mathcal{V}(x):=\frac{1}{3!}d_{ijk}x^{i}x^{j}x^{k}.  \label{V}
\end{equation}%
From the Euler formula, since $\mathcal{V}$ (\ref{V}) is a homogeneous
polynomial of degree $3$ in the $x$'s, it follows that%
\begin{equation}
\frac{\partial \mathcal{V}\left( x\right) }{\partial x^{i}}x^{i}=3\mathcal{V}%
\left( x\right) .  \label{Euler}
\end{equation}%
More subtly, since the $x^{i}$'s are themselves homogeneous of degree 1 in
the black hole charges $p^{i}$, $p^{0}$ and $q_{i}$, it holds that%
\begin{eqnarray}
&&\frac{\partial \mathcal{V}\left( x\left( p^{j},p^{0},q_{j}\right) \right)
}{\partial p^{i}}p^{i}+\frac{\partial \mathcal{V}\left( x\left(
p^{j},p^{0},q_{j}\right) \right) }{\partial p^{0}}p^{0}+\frac{\partial
\mathcal{V}\left( x\left( p^{j},p^{0},q_{j}\right) \right) }{\partial q_{i}}%
q_{i}  \notag \\
&=&\frac{\partial \mathcal{V}\left( x\left( p^{j},p^{0},q_{j}\right) \right)
}{\partial x^{k}}\frac{\partial x^{k}}{\partial p^{i}}p^{i}+\frac{\partial
\mathcal{V}\left( x\left( p^{j},p^{0},q_{j}\right) \right) }{\partial x^{k}}%
\frac{\partial x^{k}}{\partial p^{0}}p^{0}+\frac{\partial \mathcal{V}\left(
x\left( p^{j},p^{0},q_{j}\right) \right) }{\partial x^{k}}\frac{\partial
x^{k}}{\partial q_{i}}q_{i}  \notag \\
&=&\frac{1}{2}d_{klm}x^{l}x^{m}\left( \frac{\partial x^{k}}{\partial p^{i}}%
p^{i}+\frac{\partial x^{k}}{\partial p^{0}}p^{0}+\frac{\partial x^{k}}{%
\partial q_{i}}q_{i}\right) =\frac{1}{2}d_{klm}x^{k}x^{l}x^{m}=3\mathcal{V}%
\left( x\right) ,
\end{eqnarray}%
where we used (\ref{homm}).

The BPS system (\ref{BPS}) can then be defined as the non-homogeneous system
of $N$ quadratic equations in $N$ unknowns $x^{i}$'s ($i=1,...,N$)
\begin{equation}
\frac{\partial \mathcal{V}\left( x\right) }{\partial x^{i}}=3\frac{\partial
\Delta \left( p,q\right) }{\partial p^{i}},  \label{1}
\end{equation}%
where%
\begin{equation}
\Delta \left( p,q\right) :=\frac{1}{3!}%
d_{ijk}p^{i}p^{j}p^{k}-p^{0}p^{i}q_{i}=I_{3}(p)-p^{0}p^{i}q_{i},
\label{Delta}
\end{equation}%
is a real quantity depending, for a given $d_{ijk}$, on some real given
(background) constants, namely the magnetic and electric charges $p^{i}$, $%
p^{0}$ and $q_{i}$ of the extremal black hole. By (\ref{Delta}), (\ref{1})
can be rewritten as%
\begin{equation}
\frac{1}{2}d_{ijk}x^{j}x^{k}=3\Delta _{i}\left( p,q\right) ,  \label{1-bis}
\end{equation}%
where%
\begin{equation}
\Delta _{i}\left( p,q\right) :=\frac{\partial \Delta \left( p,q\right) }{%
\partial p^{i}}=\frac{1}{2}d_{ijk}p^{j}p^{k}-p^{0}q_{i}.  \label{Delta!}
\end{equation}%
Thus, by introducing the gradient operators $\nabla _{x}:=\left\{ \partial
_{x^{i}}\right\} _{i=1,...,N}$ and $\nabla _{p}:=\left\{ \partial
_{p^{i}}\right\} _{i=1,...,N}$, the BPS system (\ref{BPS}) (or, equivalently
(\ref{1})) can be cast as follows\footnote{%
We recall that the symplectic bundle of special geometry is \textit{flat}
\cite{Strominger-SKG, Andrianopoli, d-geometries}.} :%
\begin{gather}
\nabla _{x}\mathcal{V}\left( x\right) =3\nabla _{p}\Delta \left( p,q\right) ;
\label{1-tris} \\
\Updownarrow  \notag \\
\partial _{x^{i}}\mathcal{V}\left( x\right) =3\Delta _{i}\left( p,q\right)
,~\forall i.  \label{1-quater}
\end{gather}

All quantities involved in these formul\ae\ are real, and, for $N$,\ $%
d_{ijk} $\ and $\Delta _{i}\left( p,q\right) $\ given in input, real
solutions $x^{i}=x^{i}\left( \Delta _{j};d_{klm}\right) $ to the \ system (%
\ref{1-quater}) are searched \textit{in closed form}, in such a way that,
when plugged into (\ref{S-BPS}), one can obtain a closed form expression for
the Bekenstein-Hawking entropy (\ref{S-BPS}) as well as for the attractor
values of scalar fields (\ref{BPS-crit-points-2}) of extremal BPS black
holes also in homogeneous non-symmetric very special geometry. Of course, as
already mentioned, the system (\ref{1-quater}) might have no analytical real
solution in the general case : it describes $N$ quadratic hypersurfaces
(each of dimension $N-1$) in $\mathbb{R}^{N}$, and the intersection of these
hypersurfaces is a solution of the system.

An important quantity is the Hessian matrix of the cubic form $\mathcal{V}%
\left( x\right) $ (\ref{V}) :%
\begin{equation}
H_{ij}\left( x\right) :=\frac{\partial ^{2}\mathcal{V}(x)}{\partial
x^{i}\partial x^{j}}=d_{ijk}x^{k}=H_{\left( ij\right) }\left( x\right) ,
\label{Hessian}
\end{equation}%
which can be regarded as the Jacobian matrix of the quadratic map
represented by the gradient map $\nabla _{\mathcal{V}}$ of $\mathcal{V}$
itself\footnote{%
Note that each component of the map $\nabla _{\mathcal{V}}$ is an
homogeneous polynomial of degree 2 in the $x$'s, thus $\left( \nabla _{%
\mathcal{V}}\right) _{i}\left( x\right) =\left( \nabla _{\mathcal{V}}\right)
_{i}\left( -x\right) $.}%
\begin{eqnarray}
\nabla _{\mathcal{V}} &:&\mathbb{R}^{N}\rightarrow \left( \mathbb{R}%
^{N}\right) ^{\ast }\simeq \mathbb{R}^{N};  \label{Q-1} \\
\left( \nabla _{\mathcal{V}}\right) _{i}\left( x\right) &:&=\frac{\partial
\mathcal{V}\left( x\right) }{\partial x^{i}}=\frac{1}{2}d_{ijk}x^{j}x^{k}.
\label{Q-2}
\end{eqnarray}%
Since each coefficient $H_{ij}\left( x\right) $ of the symmetric matrix $H$
is a linear homogeneous polynomial in the $x$'s, the determinant of the $%
N\times N$ matrix $H(x)$ is a homogeneous polynomial of degree $N$ in the $x$%
's, as evident from (\ref{Hessian}). Even if we will not exploit it in the
subsequent treatment, the condition of non-vanishing $\det H\left( x\right) $
(i.e. $H(x)$, and thus\ $\nabla _{\mathcal{V}}$, of maximal rank) can be
used to establish whether the gradient map $\nabla _{\mathcal{V}}$ can be
inverted; in fact, Dini's Theorem ensures that if $\det H\left( x\right)
\neq 0$ then locally the map $\nabla _{\mathcal{V}}$ is a diffeomorphism,
and therefore $x$ is an isolated point in the fiber $\left( \nabla _{%
\mathcal{V}}\right) ^{-1}\nabla _{\mathcal{V}}\left( x\right) $ of $\left(
\nabla _{\mathcal{V}}\right) ^{-1}$ over $\nabla _{\mathcal{V}}\left(
x\right) $.

\section{\label{HVSG}Homogeneous very special geometry}

\subsection{\label{symmd}Symmetric $d$-manifolds}

The tensors $d_{ijk}$ as in (\ref{F(X)}) giving rise to homogeneous very
special K\"{a}hler spaces (which, for this reason, have been named $d$%
\textit{-manifolds}) have been classified in \cite{dWVP} (see also \cite%
{dWVVP,dWVP2}). A noteworthy subclass is represented by the \textit{symmetric%
} $d$-manifolds \cite{dWVVP}, whose $d_{ijk}$'s have been reconsidered also
e.g. in \cite{FGimonK, Raju-2, Magic-Wissanji}. Symmetric $d$-manifolds are
characterized by a purely numerical (constant) contravariant tensor $d^{ijk}$
such that the so-called \textquotedblleft adjoint identity" of cubic Jordan
algebras holds \cite{GST2,CVP}:%
\begin{equation}
d_{(ij|k}d_{l|mn)}d^{klp}=\frac{4}{3}\delta _{(i}^{p}d_{jmn)}.
\label{adj-id}
\end{equation}%
Thus, when
\begin{equation}
d^{rst}\Delta _{r}\Delta _{s}\Delta _{t}>0,  \label{ddd}
\end{equation}%
a solution to the BPS system (\ref{BPS}) is given by (see Sec.\ 3.3.1 of
\cite{Halmagyi1})%
\begin{equation}
x^{i}=\pm \frac{3}{\sqrt{2}}\frac{d^{ijk}\Delta _{j}\Delta _{k}}{\sqrt{%
d^{lmn}\Delta _{l}\Delta _{m}\Delta _{n}}}.  \label{BPS-symm}
\end{equation}

Indeed, by exploiting the adjoint identity (\ref{adj-id}), (\ref{BPS-symm})
yields to%
\begin{equation}
\frac{1}{3!}d_{ijk}x^{j}x^{k}=\frac{1}{3!}\frac{9}{2}\frac{%
d_{ijk}d^{jlm}d^{knp}\Delta _{l}\Delta _{m}\Delta _{n}\Delta _{p}}{%
d^{rst}\Delta _{r}\Delta _{s}\Delta _{t}}\overset{\text{(\ref{adj-id})}}{=}%
\frac{\delta _{i}^{(l}d^{mnp)}\Delta _{l}\Delta _{m}\Delta _{n}\Delta _{p}}{%
d^{rst}\Delta _{r}\Delta _{s}\Delta _{t}}=\Delta _{i},
\end{equation}%
thus obtaining (\ref{BPS}).

Furthermore, by using (\ref{adj-id}), one computes that%
\begin{eqnarray}
d^{rst}\Delta _{r}\Delta _{s}\Delta _{t} &=&d^{rst}\left( \frac{1}{2}%
d_{ruv}p^{u}p^{v}-p^{0}q_{r}\right) \left( \frac{1}{2}%
d_{smn}p^{m}p^{n}-p^{0}q_{s}\right) \left( \frac{1}{2}%
d_{tpq}p^{p}p^{q}-p^{0}q_{t}\right)  \notag \\
&=&\frac{1}{8}d^{rst}d_{ruv}d_{smn}d_{tpq}p^{u}p^{v}p^{m}p^{n}p^{p}p^{q}-%
\frac{3}{4}p^{0}d^{rst}d_{ruv}d_{smn}p^{u}p^{v}p^{m}p^{n}q_{t}  \notag \\
&&+\frac{3}{2}\left( p^{0}\right)
^{2}d^{rst}d_{ruv}p^{u}p^{v}q_{s}q_{t}-\left( p^{0}\right)
^{3}d^{rst}q_{r}q_{s}q_{t}  \notag \\
&=&\frac{1}{6}\left( d_{smn}p^{s}p^{m}p^{n}\right)
^{2}-p^{0}p^{j}q_{j}d_{smn}p^{s}p^{m}p^{n}  \notag \\
&&+\frac{3}{2}\left( p^{0}\right)
^{2}d^{rst}d_{ruv}p^{u}p^{v}q_{s}q_{t}-\left( p^{0}\right)
^{3}d^{smn}q_{s}q_{m}q_{n}  \notag \\
&=&6\left[ I_{3}^{2}(p)-p^{0}p^{j}q_{j}I_{3}(p)+\left( p^{0}\right)
^{2}\left\{ I_{3}(p),I_{3}(q)\right\} -\left( p^{0}\right) ^{3}I_{3}(q)%
\right] .
\end{eqnarray}%
Thus, the consistency condition (\ref{ddd}) can be recast as follows :%
\begin{equation}
I_{3}^{2}(p)-p^{0}p^{j}q_{j}I_{3}(p)+\left( p^{0}\right) ^{2}\left\{
I_{3}(p),I_{3}(q)\right\} -\left( p^{0}\right) ^{3}I_{3}(q)>0.  \label{ddd-2}
\end{equation}%
Moreover, the condition (\ref{d<0-2}) can be rewritten as%
\begin{equation}
\Delta _{j}x^{j}>0\Leftrightarrow \pm \frac{3}{\sqrt{2}}\frac{d^{ijk}\Delta
_{i}\Delta _{j}\Delta _{k}}{\sqrt{d^{lmn}\Delta _{l}\Delta _{m}\Delta _{n}}}%
=\pm \frac{3}{\sqrt{2}}\sqrt{d^{lmn}\Delta _{l}\Delta _{m}\Delta _{n}}>0,
\end{equation}%
implying that only the branch \textquotedblleft $+$" of (\ref{BPS-symm}) is
consistent; thus selecting it (for $d^{lmn}\Delta _{l}\Delta _{m}\Delta
_{n}>0$),
\begin{equation}
x^{i}=\frac{3}{\sqrt{2}}\frac{d^{ijk}\Delta _{j}\Delta _{k}}{\sqrt{%
d^{lmn}\Delta _{l}\Delta _{m}\Delta _{n}}},  \label{BPS-symm-2}
\end{equation}%
one obtains that the explicit solutions (\ref{BPS-crit-points}) to the BPS
Attractor Equations read
\begin{equation}
z_{H}^{i}\left( p^{0},p^{k},q_{0},q_{k}\right) =\frac{3}{2}\frac{%
d^{ijk}\Delta _{j}\Delta _{k}}{d^{lmn}\Delta _{l}\Delta _{m}\Delta _{n}}%
\left[ \frac{p^{0}\left( p\cdot q\right) -2I_{3}(p)}{p^{0}}-\mathbf{i}\frac{S%
}{\pi }\right] +\frac{p^{i}}{p^{0}}.  \label{symm-BPS-crit-points}
\end{equation}

By substituting (\ref{BPS-symm}) into (\ref{S-BPS}), one obtains the BPS
entropy $S$ in terms of the quartic invariant polynomial $I_{4}$ (cfr. e.g.
\cite{CFM} and Refs. therein), which can be proved to (finitely) generate
the ring of invariant polynomials of the non-transitive action of the $4D$
U-duality group over the black hole charges' representation space \cite%
{Kac-80}. By defining%
\begin{eqnarray}
I_{3}(q) &:&=\frac{1}{3!}d^{ijk}q_{i}q_{j}q_{k}; \\
\left\{ I_{3}(p),I_{3}(q)\right\} &:&=\frac{\partial I_{3}(p)}{\partial p^{i}%
}\frac{\partial I_{3}(q)}{\partial q_{i}}=\frac{1}{4}%
d^{ijk}d_{ilm}p^{l}p^{m}q_{j}q_{k},
\end{eqnarray}%
and recalling (\ref{rec1}) and (\ref{rec2}), the expression (\ref{S-BPS})
the BPS entropy can be remarkably simplified into the following formula :%
\begin{equation}
\frac{S}{\pi }=\sqrt{-(p\cdot q)^{2}+4q_{0}I_{3}(p)-4p^{0}I_{3}(q)+4\left\{
I_{3}(p),I_{3}(q)\right\} }=:\sqrt{I_{4}},  \label{S-BPS-symm}
\end{equation}%
and therefore (\ref{symm-BPS-crit-points}) further simplifies to%
\begin{equation}
z_{H}^{i}\left( p^{0},p^{k},q_{0},q_{k}\right) =\frac{3}{2}\frac{%
d^{ijk}\Delta _{j}\Delta _{k}}{d^{lmn}\Delta _{l}\Delta _{m}\Delta _{n}}%
\left[ \frac{p^{0}\left( p\cdot q\right) -2I_{3}(p)}{p^{0}}-\mathbf{i}\sqrt{%
I_{4}}\right] +\frac{p^{i}}{p^{0}}.  \label{symm-BPS-crit-points-2}
\end{equation}%
It should be remarked that the condition $I_{4}>0$ may be weaker than the
actual BPS condition; in fact, it satisfied by both BPS and non-BPS
attractors (these latter with vanishing central charge). Indeed, in the
symmetric $d$-spaces the strictly BPS conditions (\ref{BPS-cond-gen}) can be
specified as the following system of inequalities :%
\begin{eqnarray}
&&\left\{
\begin{array}{l}
I_{4}>0\Leftrightarrow -(p\cdot
q)^{2}+4q_{0}I_{3}(p)-4p^{0}I_{3}(q)+4\left\{ I_{3}(p),I_{3}(q)\right\} >0;
\\
\\
d^{ijk}\Delta _{i}\Delta _{j}\Delta _{k}>0\Leftrightarrow
I_{3}^{2}(p)-p^{0}p^{j}q_{j}I_{3}(p)+\left( p^{0}\right) ^{2}\left\{
I_{3}(p),I_{3}(q)\right\} -\left( p^{0}\right) ^{3}I_{3}(q)>0.%
\end{array}%
\right.  \notag \\
&&  \label{BPS-cond}
\end{eqnarray}

\subsection{\label{nsymmd}Non-symmetric $d$-manifolds}

In \textit{non-symmetric} $d$-manifolds, the tensor $d^{ijk}$ still exists%
\footnote{%
As mentioned above, for \textit{symmetric} $d$-spaces, it should hold that $%
\frac{\partial d^{ijk}}{\partial \hat{\lambda}^{l}}=0$, and this can be
checked by exploiting the explicit expression of $d_{ijk}$ in such cases
\cite{dWVP, dWVVP,dWVP2,FGimonK}.}, but it generally depends on the rescaled
imaginary parts (denoted below by $\hat{\lambda}^{i}$) of the scalar fields $%
z^{i}$; indeed, within the conventions of \cite{CFM}, the \textquotedblleft
dual" $d^{ijk}$ cubic tensor is generally defined as%
\begin{eqnarray}
d^{ijk} &:&=a^{il}a^{jm}a^{kn}d_{lmn};  \label{d^ijk} \\
a^{ij} &:&=\frac{1}{2}\left( \hat{\lambda}^{i}\hat{\lambda}^{j}-2\hat{\kappa}%
^{ij}\right) ,
\end{eqnarray}%
where the scalar fields read%
\begin{eqnarray}
z^{i} &=&:x^{i}-i\mathcal{V}^{1/3}\hat{\lambda}^{i}; \\
\frac{1}{3!}d_{ijk}\hat{\lambda}^{i}\hat{\lambda}^{j}\hat{\lambda}^{k} &=&:1,
\end{eqnarray}%
and $\hat{\kappa}^{ij}$ is the inverse matrix of
\begin{equation}
\hat{\kappa}_{jk}:=d_{jkl}\hat{\lambda}^{l},~~\hat{\kappa}^{ij}\hat{\kappa}%
_{jk}=:\delta _{k}^{i}.  \label{jjjj}
\end{equation}%
Thus, it generally holds that%
\begin{equation}
\frac{\partial d^{ijk}}{\partial \hat{\lambda}^{l}}\neq 0.
\end{equation}%
This can also be obtained by the generalization of the adjoint identity (\ref%
{adj-id}) in non-symmetric very special geometry\footnote{\textit{At least}
in the homogeneous non-symmetric case, such a generalization can be regarded
as the \textquotedblleft generalized adjoint identity" holding for the
Hermitian part of the rank-3 Vinberg's T-algebras \cite{Vinberg}.}, which
reads \cite{Raju-2}:%
\begin{equation}
d_{(ij|k}d_{l|mn)}d^{klp}=\frac{4}{3}\delta _{(i}^{p}d_{jmn)}+E_{~ijmn}^{p},
\label{gen-adj-id}
\end{equation}%
where the so-called \textquotedblleft $E$-tensor" is defined as \cite{Raju-1}%
\begin{eqnarray}
E^{p}~_{ijmn} &=&a^{pk}E_{p|ijmn}; \\
E_{p|ijmn} &=&E_{p|(ijmn)}:=-\frac{1}{12}\left[
\begin{array}{c}
\left( 4\hat{\kappa}_{(i}d_{jmn)}-3\hat{\kappa}_{(ij}\hat{\kappa}%
_{mn)}\right) \hat{\kappa}_{p}+12d_{p(ij}\hat{\kappa}_{mn)} \\
\\
-16\hat{\kappa}_{p(i}d_{jmn)}-12d_{q(ij}d_{mn)r}d_{pst}\hat{\kappa}^{qs}\hat{%
\kappa}^{rt}%
\end{array}%
\right] ,
\end{eqnarray}%
where%
\begin{equation}
\hat{\kappa}_{i}:=\hat{\kappa}_{ij}\hat{\lambda}^{j}=d_{ijk}\hat{\lambda}^{j}%
\hat{\lambda}^{k}.  \label{k-hat}
\end{equation}

\subsection{\label{HNS}Classification of homogeneous $d$-manifolds : $L(q,P,%
\dot{P})$}

In the present investigation, we will focus on the solution of the BPS
system (\ref{BPS}) in the case in which the $d$-manifolds (coupled to $%
\mathcal{N}=2$, $D=4$ supergravity) are \textit{homogeneous} (and thus the $%
U $-duality group has a non-linear but transitive action on the target space
of scalar fields) but\footnote{%
Notice that all (known and classified) homogeneous non-symmetric manifolds
are of $d$-type, even if, as far as we know, there is no proof that
homogeneous non-symmetricity implies $d$-type.} \textit{non-symmetric}%
\footnote{%
Within the black hole effective potential formalism, a discussion of the
various classes of attractors and related black hole entropies in
homogeneous scalar manifolds of $\mathcal{N}=2$, $D=4$ supergravity has been
given in \cite{DFT-Hom-07}. Therein, in (2.23)-(2.25) the explicit
expression of the $E$-tensor (which is non-vanishing for homogeneous
non-symmetric cases) has been computed.}. To this aim, we now recall some
basic facts on the classification of homogeneous $d$-spaces.

In \cite{dWVP,dWVP2,dWVVP}, \textit{homogeneous} very special K\"{a}hler
spaces arising as non-compact Riemannian scalar manifolds of vector
multiplets in $\mathcal{N}=2$, $D=4$ supergravity have been classified%
\footnote{%
In \cite{Cortes} this classification has been rephrased in terms of normal $%
J $-algebras; for a recent survey, see also \cite{Alek}.}. They are denoted
as \textquotedblleft $L$-spaces" (or \textquotedblleft $L$-models"),
specified by three sets of integer parameters : $L(q,P,\dot{P})$. In such a
classification, the index $i=1,...,N$ is partitioned as\footnote{%
In the present paper, attention should be paid to the three different uses
of `$\cdot $' : it indicates a scalar product involving naught and $i$%
-indices (as in (\ref{rec1})), or an algebraic multiplication (as in the
third row of (\ref{split-index})), or a scalar product involving only $i$%
-indices (split into $s$, and $x$- and $y$- indices), as in (\ref{jjj1})-(%
\ref{jjj4}). We hope that such different meanings are easily inferred from
the context.}
\begin{eqnarray}
\left\{ i\right\} &=&s,\left\{ I\right\} ,\left\{ \alpha \right\} ;  \notag
\\
I &=&0,1,...,q+1;  \notag \\
\alpha &=&1,...,\mathcal{D}_{q+1}\cdot \left( P+\dot{P}\right) ;  \notag \\
q &\in &\mathbb{N}\cup \left\{ 0,-1\right\} ,\text{~}P,\dot{P}\in \mathbb{N}%
\cup \left\{ 0\right\} ,  \label{split-index}
\end{eqnarray}%
where $\mathcal{D}_{q+1}$ is a certain function of $q$ valued in $\mathbb{N}$
(see e.g. Table 1 of \cite{dWVP}) :%
\begin{equation}
\begin{array}{cc}
q & \mathcal{D}_{q+1} \\
-1 & 1 \\
0 & 1 \\
1 & 2 \\
2 & 4 \\
3 & 8 \\
4 & 8 \\
5 & 16 \\
6 & 16 \\
7 & 16 \\
\mathbf{n}+7 & 16\cdot \mathcal{D}_{\mathbf{n}}%
\end{array}
\label{table D}
\end{equation}%
Note that we set%
\begin{equation}
N=q+3+\mathcal{D}_{q+1}\cdot \left( P+\dot{P}\right) .  \label{exprr}
\end{equation}

The electric-magnetic ($U$-)duality group $\mathcal{G}$ of the $L(q,P,\dot{P}%
)$ models has the following graded structure \cite{dWVP, dWVP2}:%
\begin{eqnarray}
\mathcal{G} &=&\mathcal{G}_{0}\ltimes \left( \mathcal{G}_{1}\times \mathcal{G%
}_{2}\right) ,  \label{G} \\
\mathcal{G}_{0} &=&\left( SO\left( q+2,2\right) \times S_{q}\left( P,\dot{P}%
\right) \times SO(1,1)\right) _{0};  \label{G0} \\
\mathcal{G}_{1} &=&\left( \mathbf{\psi }_{q+2,2},\mathbf{F}_{q,P,\dot{P}%
}\right) _{1};  \label{G1} \\
\mathcal{G}_{2} &=&\left( \mathbf{1},\mathbf{1}\right) _{2},  \label{G2}
\end{eqnarray}%
where the subscripts (outside round brackets) denote the weights w.r.t. the $%
SO(1,1)$ factor in $\mathcal{G}_{0}$. Moreover, $S_{q}\left( P,\dot{P}%
\right) $ is the compact, metric-preserving group in the centralizer of the
real Euclidean Clifford algebra $Cl(q+1,0)$ in the $\mathcal{D}_{q+1}\left(
P+\dot{P}\right) $-dimensional representation (see e.g. in Table 3 of \cite%
{dWVVP}),%
\begin{equation}
\begin{array}{cc}
q & S_{q}(P,\dot{P}) \\
-1 & SO(P) \\
0 & SO(P)\otimes SO(\dot{P}) \\
1 & SO(P) \\
2 & U(P) \\
3 & USp(2P) \\
4 & USp(2P)\otimes USp(2\dot{P}) \\
5 & USp(2P) \\
6 & U(P) \\
7 & SO(P) \\
\mathbf{n}+7 & \text{as~for~}q+1=\mathbf{n}.%
\end{array}
\label{table DD}
\end{equation}%
Also, $\mathbf{\psi }_{q+2,2}$ denotes the (semi)spinor representation of $%
Spin(q+2,2)$, and $\mathbf{F}_{q,P,\dot{P}}$ stands for the fundamental
representation of $S_{q}\left( P,\dot{P}\right) $ itself.

The vector multiplets' scalar manifold is a non-symmetric projective special
K\"{a}hler, non-compact Riemannian coset of the form%
\begin{equation}
\mathcal{M}_{4}:=\mathcal{G}/H,
\end{equation}%
where $H$ is the maximal compact subgroup of $\mathcal{G}_{0}$ itself,%
\begin{equation}
H=mcs\left( \mathcal{G}_{0}\right) =SO\left( q+2\right) \times S_{q}\left( P,%
\dot{P}\right) \times SO(2).  \label{H}
\end{equation}%
By respectively denoting with $\mathfrak{g}$ and $\mathfrak{h}$ the Lie
algebras of the groups $G$ (\ref{G})-(\ref{G2}) and $H$ (\ref{H}), the
scalar fields coordinatizing $\mathcal{M}_{4}$ group into $H$%
-representations, such that the Lie algebra of the coset $\mathcal{M}_{4}$
enjoys the following 5-graded structure\footnote{%
Note that the presence of the upperscript \textquotedblleft $(^{\prime })$"
means that (\ref{Lie-coset}) and (\ref{R4}) hold up to possible spinor
conjugation of (semi)spinors $\mathbf{\psi }$.},
\begin{eqnarray}
\text{Lie}\left( \mathcal{M}_{4}\right) &=&\mathfrak{g}\ominus \mathfrak{h}
\notag \\
&=&\left( \mathbf{q+2},\mathbf{1}\right) _{0,-2}\oplus \left( \mathbf{\psi }%
_{q+2}^{(\prime )},\mathbf{F}_{q,P,\dot{P}}\right) _{1,-1}\oplus \left(
\mathbf{1},\mathbf{1}\right) _{0,0}\oplus \left( \mathbf{1},\mathbf{1}%
\right) _{2,0}\oplus \left( \mathbf{\psi }_{q+2},\mathbf{F}_{q,P,\dot{P}%
}\right) _{1,1}\oplus \left( \mathbf{q+2},\mathbf{1}\right) _{0,2},  \notag
\\
&&  \label{Lie-coset}
\end{eqnarray}%
where the first subscript reports (as above) the weight w.r.t. the $SO(1,1)$
factor in $\mathcal{G}_{0}$, whereas the second subscript denotes the charge
w.r.t. the $SO(2)\simeq U(1)$ factor in $H$ (\ref{H}). The total (real)
dimension of $\mathcal{M}_{4}$ is then%
\begin{equation}
\dim \mathcal{M}_{4}=\dim \text{Lie}\left( \mathcal{M}_{4}\right) =2+2\left(
q+2\right) +2\mathcal{D}_{q+1}\left( P+\dot{P}\right) =2\dim \mathcal{M}%
_{5}+2,
\end{equation}%
where%
\begin{equation}
\dim \mathcal{M}_{5}=q+2+\mathcal{D}_{q+1}\left( P+\dot{P}\right)
\end{equation}%
is the (real) dimension of the vector multiplets' real scalar manifold of
the parent theory in 5 space-time (Lorentzian) dimensions \cite{dWVP, dWVP2}.

The 2-form field strengths and their duals fit into the (generally
reducible) $\mathcal{G}_{0}$-module%
\begin{equation}
\mathbf{R}_{4}:=\left( \mathbf{q+4},\mathbf{1}\right) _{-1}\oplus \left(
\mathbf{\psi }_{q+2,2}^{(\prime )},\mathbf{F}_{q,P,\dot{P}}\right)
_{0}\oplus \left( \mathbf{q+4},\mathbf{1}\right) _{1},  \label{R4}
\end{equation}%
which has thus a 3-graded structure w.r.t. the $SO(1,1)$ factor in $\mathcal{%
G}_{0}$. The $\mathcal{G}_{0}$-module (\ref{R4}), of real dimension%
\begin{equation}
\dim \mathbf{R}_{4}=2\left( q+4\right) +2\mathcal{D}_{q+1}\left( P+\dot{P}%
\right) ,  \label{dimR4}
\end{equation}
is anti-self-conjugate (\textit{i.e.}, it is symplectic), and the
corresponding ring of $\mathcal{G}$-invariant homogeneous polynomials is
one-dimensional (\textit{cfr.} the discussion in \cite{Alek}). As far as we
know, the expression (\ref{R4}) has never been presented in the literature
thus far.

The non-vanishing components of $d_{ijk}=d_{(ijk)}$ are%
\begin{equation}
\frac{1}{3!}d_{ijk}:\left\{
\begin{array}{l}
\frac{1}{3!}d_{sIJ}:=\eta _{IJ}=\text{diag}\left( -1,\underset{q+1}{%
\underbrace{1,...,1}}\right) ; \\
\\
\frac{1}{3!}d_{I\alpha \beta }:=\left( \Gamma _{I}\right) _{\alpha \beta },%
\end{array}%
\right.  \label{d_ijk}
\end{equation}%
where, for $I=0,1,...,q+1$, the square matrices $\Gamma _{I}$, of size $%
\mathcal{D}_{q+1}\cdot \left( P+\dot{P}\right) $ and components $\left(
\Gamma _{I}\right) _{\alpha \beta }$ are defined as follows : $\Gamma _{0}:=%
\mathbb{I}_{\mathcal{D}_{q+1}\cdot \left( P+\dot{P}\right) }$, and $\left\{
\Gamma _{I}\right\} _{I\neq 0}$ are $\Gamma $-matrices that provide a
\textit{real} representation of the Euclidean Clifford algebra $Cl\left(
q+1,0\right) $; see Sec.\ \ref{clifmat}. All the other, unwritten,
components of $d_{ijk}$ vanish. We should also recall that non-vanishing
(and independent) values of $\dot{P}$ are possible only when $q=4m$, with $%
m\in \mathbb{N}\cup \left\{ 0\right\} $. Indeed, for $q=0$ mod $4$ the
representations $\Gamma _{I}$ and $-\Gamma _{I}$ are not equivalent, and a
reducible representation is given by $\Gamma =\eta \otimes \Gamma _{I}$
(with $\eta =$diag$\left( \underset{P~\text{times}}{\underbrace{1,...,1}},%
\underset{\dot{P}~\text{times}}{\underbrace{-1,...,-1}}\right) $) and thus
characterized by the multiplicity of each of these representations, namely $%
P $ and $\dot{P}$; of course, an overall sign change of all the gamma
matrices can always be re-absorbed, and this is the reason why $L(4m,P,\dot{P%
})=L(4m,\dot{P},P)$. If the representation consists of copies of only one
version of the irreducible representations, then we denote it by $L(4m,P)$.

In correspondence with the splitting (\ref{split-index}) of the index set,
we introduce new variables $s$, $x^{I}$ and $y^{\alpha }$, and define%
\begin{equation}
\xi :=~^{T}\left( s,x^{I},y^{\alpha }\right) .  \label{csii}
\end{equation}%
Then, (\ref{V}) and (\ref{Delta}) can be written as%
\begin{eqnarray}
\mathcal{V}(x) &=&s\eta _{IJ}x^{I}x^{J}+x^{I}\left( \Gamma _{I}\right)
_{\alpha \beta }y^{\alpha }y^{\beta }=sq(x)+x^{I}Q_{I}\left( y\right) ;
\label{VV} \\
\Delta \left( p,q\right) &=&\mathcal{V}(p)-p^{0}\left(
p^{s}q_{s}+p^{I}q_{I}+p^{\alpha }q_{\alpha }\right)  \notag \\
&=&p^{s}\eta _{IJ}p^{I}p^{J}+p^{I}\left( \Gamma _{I}\right) _{\alpha \beta
}p^{\alpha }p^{\beta }-p^{0}\left( p^{s}q_{s}+p^{I}q_{I}+p^{\alpha
}q_{\alpha }\right) ,
\end{eqnarray}%
where\footnote{%
The parameter $q\in \mathbb{N}\cup \left\{ 0,-1\right\} $ occurring in de
Wit-Van Proeyen's classification \cite{dWVP} should not be confused with the
quadratic form $q(x)$ (\ref{q(x)}).}%
\begin{eqnarray}
q(x) &:&=\eta _{IJ}x^{I}x^{J};  \label{q(x)} \\
Q_{I}\left( y\right) &:&=\left( \Gamma _{I}\right) _{\alpha \beta }y^{\alpha
}y^{\beta }.  \label{Q_I(y)}
\end{eqnarray}%
The BPS system (\ref{BPS}) (or, equivalently, (\ref{1-quater})) is an\
inhomogeneous system of $N$ quadratic equations in $N$ unknowns ($%
s,x^{I},y^{\alpha }$), and it acquires the following form :%
\begin{equation}
\left\{
\begin{array}{l}
q(x)=3\Delta _{s}; \\
\\
2s\eta _{IJ}x^{J}+Q_{I}\left( y\right) =3\Delta _{I}; \\
\\
2x^{I}\left( \Gamma _{I}\right) _{\alpha \beta }y^{\beta }=3\Delta _{\alpha
}.%
\end{array}%
\right.  \label{2}
\end{equation}

The open orbit of electric and magnetic charges supporting $\frac{1}{2}$-BPS
\textquotedblleft large" extremal black holes is an homogeneous
non-symmetric manifold,%
\begin{equation}
\mathcal{O}_{BPS}=\frac{\mathcal{G}}{H_{BPS}},~~\text{with~}H_{BPS}=SO\left(
q+2\right) \times S_{q}\left( P,\dot{P}\right) .  \label{HBPS}
\end{equation}%
of real dimension%
\begin{equation}
\dim \mathcal{O}_{BPS}=2q+7+2\mathcal{D}_{q+1}\left( P+\dot{P}\right)
\overset{\text{(\ref{dimR4})}}{=}\dim \mathbf{R}_{4}-1.  \label{dimOBPS}
\end{equation}%
Since $\mathcal{O}_{BPS}$ is a generic, open orbit of the non-transitive
action of the Lie group $\mathcal{G}$ (\ref{G})-(\ref{G2}) on the
representation space $\mathbf{R}_{4}$, the result (\ref{dimOBPS}) is
consistent with the general counting formula :%
\begin{equation}
\dim \mathbf{R}_{4}=\dim \mathcal{O}_{BPS}+1,
\end{equation}%
because, as mentioned above, the ring of $\mathcal{G}$-invariant homogeneous
polynomials on $\mathbf{R}_{4}$ is one-dimensional \cite{Alek}. Moreover,
the fact that the stabilizer $H_{BPS}$ (\ref{HBPS}) of $\mathcal{O}_{BPS}$
is compact implies that no \textit{moduli space} of BPS attractors exists at
all, as pointed out in \cite{FM-moduli} and firstly noticed in \cite{fegika}.

\subsubsection{$L(q,1)$ and \textquotedblleft magic\textquotedblright\
enhancements \label{Lq1}}

A noteworthy subclass of homogeneous $d$-manifolds is provided by the models
$L(q,1,0)\equiv L(q,1)$ \cite{dWVP}, which have $P=1$ and $\dot{P}=0$. From (%
\ref{G})-(\ref{G2}), the electric-magnetic ($U$-)duality group $\mathcal{G}$
of the $L(q,1)$ models has the following graded structure :%
\begin{eqnarray}
\mathcal{G} &=&\mathcal{G}_{0}\ltimes \left( \mathcal{G}_{1}\times \mathcal{G%
}_{2}\right) ,  \label{GG} \\
\mathcal{G}_{0} &=&\left( SO\left( q+2,2\right) \times S_{q}\left(
1,0\right) \times SO(1,1)\right) _{0}; \\
\mathcal{G}_{1} &=&\left( \mathbf{\psi }_{q+2,2},\mathbf{F}_{q,1,0}\right)
_{1}; \\
\mathcal{G}_{2} &=&\left( \mathbf{1},\mathbf{1}\right) _{2},  \label{GG2}
\end{eqnarray}%
where the subscripts (outside round brackets) denote the weights w.r.t. the $%
SO(1,1)$ factor in $\mathcal{G}_{0}$. Moreover, $S_{q}\left( 1,0\right) $ is
the compact, metric-preserving group in the centralizer of the real
Euclidean Clifford algebra $Cl(q+1,0)$ in the $\mathcal{D}_{q+1}$%
-dimensional representation,%
\begin{equation}
\begin{array}{cc}
q & S_{q}(1,0) \\
-1 & \mathbb{I} \\
0 & \mathbb{I} \\
1 & \mathbb{I} \\
2 & U(1) \\
3 & USp(2) \\
4 & USp(2) \\
5 & USp(2) \\
6 & U(1) \\
7 & \mathbb{I} \\
\mathbf{n}+7 & \text{as~for~}q+1=\mathbf{n},%
\end{array}%
\end{equation}%
where $\mathbb{I}$ denotes the identity. It is interesting to note that for $%
q=1,2,4,8$, the Lie algebra of $S_{q}(1,0)=\mathbb{I},U(1),USp(2),\mathbb{I}$
can be regarded as%
\begin{equation}
\text{Lie}\left( S_{q}(1,0)\right) =\varnothing ,\mathfrak{u}_{1},\mathfrak{%
usp}_{2},\varnothing =\mathfrak{tri}\left( \mathbb{A}\right) \ominus
\mathfrak{so}(\mathbb{A}),~\text{with~}\mathbb{A}=\mathbb{R},\mathbb{C},%
\mathbb{H},\mathbb{O},
\end{equation}%
respectively. \textquotedblleft $\mathfrak{tri}$" and \textquotedblleft $%
\mathfrak{so}$" respectively denote the triality and norm-preserving Lie
algebras of the corresponding division algebra $\mathbb{A}$ (see e.g. \cite%
{triality-refs}).

Furthermore, the underlying real vector space $\mathbf{V}$ has the following
structure :
\begin{equation}
\mathbf{V}={\mathbb{R}}\oplus V\oplus \mathcal{S},\qquad \dim V\,=q+2,\quad
\dim \mathcal{S}\,=\,\mathcal{D}_{q+1},  \label{Vbold}
\end{equation}%
where $\mathcal{D}_{q+1}$ is the \textit{minimal} dimension for which there
are Clifford matrices $\Gamma _{1},\ldots ,\Gamma _{q+1}\in M_{\mathcal{D}%
_{q+1}}(\mathbb{R})$, and it is given in (\ref{table D}) (cfr. also Table 1
of \cite{dWVVP} or Table 3 of \cite{dWVVP}). The total dimensions of $%
\mathbf{V}$ is thus $1+(q+2)+\mathcal{D}_{q+1}$. By defining the standard
quadratic form in ${\mathbb{R}}^{h}$ as
\begin{equation}
q_{h}\left( x\right) :=\left( x^{1}\right) ^{2}\,+\,\ldots \,+\,\left(
x^{h}\right) ^{2}=\sum_{I=1}^{h}\left( x^{I}\right) ^{2},  \label{q_n(x)}
\end{equation}%
there exists a Lorentzian (mostly plus) quadratic form on $V$ :
\begin{equation}
q(x^{0},x^{1},\ldots ,x^{q+1})\,:=-\left( x^{0}\right)
^{2}+q_{q+1}(x):=\,-\left( x^{0}\right) ^{2}\,+\left( x^{1}\right)
^{2}+\,\ldots \,+\,\left( x^{q+1}\right) ^{2}~.  \label{qqqq}
\end{equation}%
On the other hand, the $\Gamma $-matrices\footnote{%
Notice that one can take \textquotedblleft diagonal
blocks\textquotedblright\ of $\Gamma $-matrices to produce new ones.} of the
Euclidean Clifford algebra $Cl(q+1,0)$ will enter (only) as the matrices
defining quadratic forms $Q_{I}$ on $\mathcal{S}$, and they will all be
symmetric matrices \cite{dWVP}. In a different context, these models have
been treated in \cite{H2-1}.

In $L(q,1)$ models, the 2-form field strengths and their duals fit into the
(generally reducible) 3-graded $\left( SO\left( q+2,2\right) \times
S_{q}\left( 1,0\right) \times SO(1,1)\right) $-module%
\begin{equation}
\mathbf{R}_{4}:=\left( \mathbf{q+4},\mathbf{1}\right) _{-1}\oplus \left(
\mathbf{\psi }_{q+2,2}^{(\prime )},\mathbf{F}_{q,1,0}\right) _{0}\oplus
\left( \mathbf{q+4},\mathbf{1}\right) _{1}.  \label{R42}
\end{equation}%
The open orbit of electric and magnetic charges supporting $\frac{1}{2}$-BPS
\textquotedblleft large" extremal black holes reads%
\begin{equation}
\mathcal{O}_{BPS}=\frac{\mathcal{G}}{H_{BPS}},~~\text{with~}H_{BPS}=SO\left(
q+2\right) \times S_{q}\left( 1,0\right) ,  \label{HBPS2}
\end{equation}%
with real dimension%
\begin{equation}
\dim \mathcal{O}_{BPS}=2q+7+2\mathcal{D}_{q+1}.  \label{dimOBPS2}
\end{equation}%
\bigskip

Within the class $L(q,1)$, there are special models corresponding to
homogeneous \textit{symmetric} $d$-manifolds : they occur when, for $%
\mathcal{D}_{q+1}=2^{g}=2,4,8,16$, with $g=1,2,3,4$, the \textit{maximal}
corresponding $q$, namely $q=2^{g-1}=1,2,4,8$ such that $\mathcal{D}%
_{q+1}=2q $, is taken; the corresponding models, named \textquotedblleft
magic" \cite{GST1, GST2, GST3}, all pertain to $\mathcal{N}=2$
Maxwell-Einstein supergravity :
\begin{eqnarray}
&&%
\begin{array}{c}
L(1,1), \\
\dim \mathbf{V}=1+3+2=6, \\
Cl\left( 2,0\right) , \\
J_{3}^{\mathbb{R}};%
\end{array}%
\quad
\begin{array}{c}
L(2,1), \\
\dim \mathbf{V}=1+4+4=9, \\
Cl\left( 3,0\right) , \\
J_{3}^{\mathbb{C}};%
\end{array}%
\quad  \notag \\
&&~  \notag \\
&&~  \label{form-Eucl} \\
&&%
\begin{array}{c}
L(4,1), \\
\dim \mathbf{V}=1+6+8=15, \\
Cl\left( 5,0\right) , \\
J_{3}^{\mathbb{H}};%
\end{array}%
\quad
\begin{array}{c}
L(8,1), \\
\dim \mathbf{V}=1+10+16=27, \\
Cl\left( 9,0\right) , \\
J_{3}^{\mathbb{O}},%
\end{array}
\notag
\end{eqnarray}%
respectively. In these models, $\mathbf{V}$ has dimension $3q+3$, and it can
be realized as the simple cubic Jordan algebra $J_{3}^{\mathbb{A}}$ of
Hermitian $3\times 3$ matrices over the division algebra division algebra $%
\mathbb{A}=\mathbb{R},\mathbb{C},\mathbb{H},\mathbb{O}$, with $\mathbb{O}$
denoting the algebra of octonions \cite{JVNW}. As we have already mentioned
(cfr. Sec.\ \ref{symmd}), the entropy of extremal black holes in these
models is explicitly known, and it is given by the (unique) quartic
invariant polynomial, $I_{4}$. The symmetries of \textquotedblleft magic"
Maxwell-Einstein supergravity theories in $D=3$, $4$ and $5$ space-time
dimensions have been discussed along the years from various perspectives,
see e.g. \cite{GP1,GP2,GP3,GP4} (and Refs. therein).

It is interesting to remark that the $J_{3}^{\mathbb{A}}$-based
\textquotedblleft magic" models $L(q,1)$ with $q=1,2,4,8$ correspond to
\textquotedblleft sweet spots" within the class $L(q,1)$; in fact, for $%
q=1,2,4,8$ an \textit{enhancement} of the electric-magnetic ($U$-)duality
group $\mathcal{G}$ (\ref{GG})-(\ref{GG2}), as well as of its $\mathbf{R}_{4}
$ module (\ref{R42}) and of the corresponding BPS orbit $\mathcal{O}_{BPS}$ (%
\ref{HBPS2}), takes place (for further detail, see e.g. the discussion on
triality in \cite{Marrani-Group32}, as well as \cite{bfgm} for \textit{%
enhanced} BPS orbits). It should be noticed that the dimension of the scalar
manifold as well as of the BPS \textquotedblleft large" charge orbit is
invariant under such an enhancement, because the same amount of (compact)
generators is added at the numerator and at the denominator of the
corresponding cosets. Explicitly, one has :

\begin{itemize}
\item $q=1$ : $L(q,1)\leftrightarrow J_{3}^{\mathbb{R}}$,%
\begin{equation}
\begin{array}{cccc}
\mathcal{G}: & \left( SO\left( 3,2\right) \times SO(1,1)\right) _{0}\ltimes
\left( \mathbf{4}_{1}\times \mathbf{1}_{2}\right) & \longrightarrow & Sp(6,%
\mathbb{R}); \\
\mathbf{R}_{4}: & \mathbf{5}_{-1}\oplus \mathbf{4}_{0}\oplus \mathbf{5}_{1}
& \longrightarrow & \mathbf{14}^{\prime }\equiv \wedge _{0}^{3}; \\
\mathcal{O}_{BPS}: & \frac{\left( SO\left( 3,2\right) \times SO(1,1)\right)
_{0}\ltimes \left( \mathbf{4}_{1}\times \mathbf{1}_{2}\right) }{SO(3)\times
SO(2)} & \longrightarrow & \frac{Sp(6,\mathbb{R})}{SU(3)}.%
\end{array}%
\end{equation}

\item $q=2$ : $L(q,2)\leftrightarrow J_{3}^{\mathbb{C}}$,%
\begin{equation}
\begin{array}{cccc}
\mathcal{G}: & \left( SO\left( 4,2\right) \times U(1)\times SO(1,1)\right)
_{0,0}\ltimes \left( \mathbf{4}_{1,1}\times \mathbf{4}_{1,-1}^{\prime
}\times \mathbf{1}_{2,0}\right) & \longrightarrow & SU(3,3); \\
\mathbf{R}_{4}: & \mathbf{6}_{-1,0}\oplus \mathbf{4}_{0,1}\oplus \mathbf{4}%
_{0,-1}^{\prime }\oplus \mathbf{6}_{1,0} & \longrightarrow & \mathbf{20}%
\equiv \wedge ^{3}; \\
\mathcal{O}_{BPS}: & \frac{\left( SO\left( 4,2\right) \times U(1)\times
SO(1,1)\right) _{0,0}\ltimes \left( \mathbf{4}_{1,1}\times \mathbf{4}%
_{1,-1}^{\prime }\times \mathbf{1}_{2,0}\right) }{SO(4)\times SO(2)\times
U(1)} & \longrightarrow & \frac{SU(3,3)}{SU(3)\times SU(3)}.%
\end{array}%
\end{equation}

\item $q=4$ : $L(q,4)\leftrightarrow J_{3}^{\mathbb{H}}$,%
\begin{equation}
\begin{array}{cccc}
\mathcal{G}: & \left( SO\left( 6,2\right) \times USp(2)\times SO(1,1)\right)
_{0}\ltimes \left( \left( \mathbf{8}_{s},\mathbf{2}\right) _{1}\times \left(
\mathbf{1,1}\right) _{2}\right) & \longrightarrow & SO^{\ast }(12); \\
\mathbf{R}_{4}: & \left( \mathbf{8}_{v},\mathbf{1}\right) _{-1}\oplus \left(
\mathbf{8}_{c},\mathbf{2}\right) _{0}\oplus \left( \mathbf{8}_{v},\mathbf{1}%
\right) _{1} & \longrightarrow & \mathbf{32}^{(\prime )}; \\
\mathcal{O}_{BPS}: & \frac{\left( SO\left( 6,2\right) \times USp(2)\times
SO(1,1)\right) _{0}\ltimes \left( \left( \mathbf{8}_{s},\mathbf{2}\right)
_{1}\times \left( \mathbf{1,1}\right) _{2}\right) }{SO(6)\times SO(2)\times
USp(2)} & \longrightarrow & \frac{SO^{\ast }(12)}{SU(6)}.%
\end{array}%
\end{equation}

\item $q=8$ : $L(q,8)\leftrightarrow J_{3}^{\mathbb{O}}$,%
\begin{equation}
\begin{array}{cccc}
\mathcal{G}: & \left( SO\left( 10,2\right) \times SO(1,1)\right) _{0}\ltimes
\left( \mathbf{32}_{1}\times \mathbf{1}_{2}\right) & \longrightarrow &
E_{7(-25)}; \\
\mathbf{R}_{4}: & \mathbf{12}_{-1}\oplus \mathbf{32}_{0}\oplus \mathbf{12}%
_{1} & \longrightarrow & \mathbf{56}; \\
\mathcal{O}_{BPS}: & \frac{\left( SO\left( 10,2\right) \times SO(1,1)\right)
_{0}\ltimes \left( \mathbf{32}_{1}\times \mathbf{1}_{2}\right) }{%
SO(10)\times SO(2)} & \longrightarrow & \frac{E_{7(-25)}}{E_{6(-78)}}.%
\end{array}%
\end{equation}
\end{itemize}

Finally, we remark that, while for $0\leqslant q\leqslant 8$ the $L(q,1)$
models are sub-models obtainable by suitable truncation of the exceptional
\textquotedblleft magic" Maxwell-Einstein $\mathcal{N}=2$, $D=4$
supergravity model $L(8,1)$ based on $J_{3}^{\mathbb{O}}$, for $q\geqslant 9$
such models provide an \textit{infinite} (countable) sequence of homogeneous
\textit{non-symmetric} models which are \textit{not} truncations of any
symmetric model.

\section{Basics on Euclidean Clifford algebras\label{clifmat}}

We recall the definition of a Clifford algebra and the basic results on
(matrix) representations of such algebras. We then discuss $\Gamma $%
-matrices and Clifford sets of $\Gamma $-matrices, which define such
representations. We recall the Pauli matrices and introduce quadratic forms
defined by symmetric $\Gamma $-matrices that are tensor products of Pauli
matrices. These quadratic forms are building blocks in the definition of the
cubic forms in the $L(q,P,\dot{P})$-models.

\subsection{Euclidean Clifford algebras\label{eca}}

The \textit{Euclidean} \textit{Clifford algebra} $Cl(n,0)$ is the quotient
of the tensor algebra on $V={\mathbb{R}}^{n}$ by the relations $%
q_{n}(v)=v\otimes v$ for all $v\in V$, where $q_{n}(v)$ is the Euclidean
quadratic form (\ref{q_n(x)}) :
\begin{equation}
Cl(n,0)\,\equiv \,Cl(V,q_{n})\,:=\,(\oplus _{k=0}^{\infty }V^{\otimes
k})/<q_{n}(v)-v\otimes v>~.
\end{equation}%
We identify $V$ with the corresponding subspace in $Cl(n,0)$ and we write $%
xy $ for the product of $x$ and $y\in Cl(n,0)$. If $e_{1},\ldots ,e_{n}$ is
an orthonormal basis of $V$ and we write $v=x^{1}e_{1}+\ldots +x^{n}e_{n}$,
then in $Cl(n,0)$ we have $q_{n}(v)=v^{2}$, so:
\begin{equation}
\left( x^{1}\right) ^{2}\,+\,\ldots \,+\,\left( x^{n}\right)
^{2}\,=\,(x^{1}e_{1}+\ldots +x^{n}e_{n})^{2},
\end{equation}%
and expanding the right hand side we see that in $Cl(n,0)$ we have (here $%
I,J=1,...,n$)
\begin{equation}
e_{I}^{2}\,=\,1,\qquad e_{I}e_{J}+e_{J}e_{I}\,=\,0\quad (I\neq J)~.
\label{clifgen}
\end{equation}%
The Euclidean Clifford algebra $Cl(n,0)$ is a vector space of dimension $%
2^{n}$ with basis the $e_{I_{1}}e_{I_{2}}\ldots e_{I_{k}}$ with $%
I_{1}<I_{2}<\ldots <I_{k}$ and $k=1,2,\ldots ,n$.

\subsection{Representations of $Cl(n,0)$ and Clifford sets\label{clnrep}}

The structure of the Euclidean Clifford algebra $Cl(n,0)$ is well-known, we
just recall here the `Bott periodicity' : denoting by $M_{16}(\mathbb{R})=%
\mathbb{R}(16)$ the algebra of real $16\times 16$ matrices (of dimension $%
16\times 16=2^{8}$), it holds that
\begin{equation}
Cl(n+8,0)\,\cong \,Cl(n,0)\otimes M_{16}(\mathbb{R}).  \label{Bott}
\end{equation}%
For $n=1,\ldots ,8$, there are isomorphisms of algebras:
\begin{equation}
Cl(1,0)\cong \mathbb{R\times R},\quad Cl(2,0)\,\cong \,M_{2}(\mathbb{R}%
),\quad Cl(3,0)\,\cong \,M_{2}(\mathbb{C}),\quad Cl(4,0)\,\cong \,M_{2}(%
\mathbb{H}),  \label{iso1}
\end{equation}%
where $\mathbb{H}$ is the division algebra of quaternions,
\begin{equation}
Cl(5,0)\cong \,M_{2}(\mathbb{H})\times M_{2}(\mathbb{H}),\qquad
Cl(6,0)\,\cong \,M_{4}(\mathbb{H}),  \label{iso2}
\end{equation}%
\begin{equation}
\qquad Cl(7,0)\,\cong \,M_{8}(\mathbb{C}),\qquad Cl(8,0)\,\cong \,M_{16}(%
\mathbb{R})~.  \label{iso3}
\end{equation}%
Bott periodicity then implies that
\begin{equation}
Cl(9,0)\cong M_{16}(\mathbb{R})\times M_{16}(\mathbb{R}),\qquad
Cl(10,0)\cong M_{2}(\mathbb{R})\otimes M_{16}(\mathbb{R})\cong M_{32}(%
\mathbb{R})~.  \label{pre-hom1}
\end{equation}%
Since $\mathbb{C}$ and $\mathbb{H}$ can be identified with subalgebras of $%
M_{2}(\mathbb{R})$ and $M_{4}(\mathbb{R})$ respectively, we obtain
non-trivial homomorphisms (which are injective unless $n=1\mod 4$, and in
that case they are projections on one of the two factors):
\begin{equation}
Cl(1,0)\rightarrow \mathbb{R},\quad Cl(2,0)\rightarrow M_{2}(\mathbb{R}%
),\quad Cl(3,0)\,\rightarrow M_{4}(\mathbb{R}),\quad
Cl(4,0),\;Cl(5,0)\,\rightarrow \,M_{8}(\mathbb{R})~,  \label{hom1}
\end{equation}%
and similarly
\begin{equation}
Cl(6,0),\;Cl(7,0),\;Cl(8,0),\;Cl(9,0)\rightarrow M_{16}(\mathbb{R})~.
\label{hom2}
\end{equation}%
Notice that these are homomorphisms $Cl(q+1,0)\rightarrow M_{\mathcal{D}%
_{q+1}}(\mathbb{R})$ with $\mathcal{D}_{q+1}$ as in (\ref{table D}).

Under the homomorphisms (\ref{hom1}) and (\ref{hom2}), the elements $%
e_{I}\in Cl(n,0)$ map to so-called $\Gamma $-matrices $\Gamma _{I}$ which
satisfy the Clifford relations ($I,J=1,...,n$):
\begin{gather}
\Gamma _{I}^{2}\,=\,\mathbb{I},\qquad \Gamma _{I}\Gamma _{J}+\Gamma
_{J}\Gamma _{I}\,=\,0\quad (I\neq J);  \notag \\
\Updownarrow  \notag \\
\Gamma _{I}\Gamma _{J}+\Gamma _{J}\Gamma _{I}=2\delta _{IJ}\mathbb{I}.
\label{clifm}
\end{gather}%
A set of $n$ matrices $\{\Gamma _{1},\ldots ,\Gamma _{n}\}$ of size $m$ will
be called a \textit{Clifford set} (of cardinality $n$ and size $m$). In (\ref%
{Gamm}) we will define specific $\Gamma$-matrices.

Conversely, given a Clifford set $\{\Gamma _{1},\ldots ,\Gamma _{n}\}\subset
M_{m}(\mathbb{R})$ the definition of the Clifford algebra shows that there
is a unique homomorphism $Cl(n,0)\rightarrow M_{m}(\mathbb{R})$ sending $%
e_{I}\mapsto \Gamma _{I}$. Since the matrix algebras $M_{m}(\mathbb{K})$,
with $\mathbb{K}=\mathbb{R},\mathbb{C},\mathbb{H}$ are simple (so any
homomorphism $M_{m}(\mathbb{K})\rightarrow M_{r}(\mathbb{R})$ is either
injective or it is the zero homomorphism), it follows from the isomorphisms (%
\ref{iso1}) (\ref{iso2}) and (\ref{iso3}) that a Clifford set with matrices
of size $1,2,4,8,16$
can have cardinality at most $1$, $2$, $3$, $5$, $9$, respectively.

Furthermore, if $\{\Gamma _{1},\ldots ,\Gamma _{n}\}$ is a Clifford set, one
easily checks that
\begin{equation}  \label{clifquadrel}
\left( \sum_{I=1}^{n}x^{I}\Gamma _{I}\right) ^{2}\,=\,\left(
\sum_{I=1}^{n}\left( x^{I}\right) ^{2}\right) {\mathbb{I}}_{2^{g}}~.
\end{equation}

\subsection{The product decomposition for $n\equiv 1\mod 4$\label{cliffprod}}

We recall the product decomposition of the Euclidean Clifford algebras $%
Cl(n,0)$ in case $n\equiv 1\mod 4$. The key point are the following elements
in such a Clifford algebra, where we use the notation from Sec.\ \ref{eca}:
\begin{equation}
c\,:=\,e_{1}e_{2}\ldots e_{n},\qquad c_{+}\,:=\,(1+c)/2,\quad
c_{-}\,:=\,(1-c)/2\qquad (\in Cl(n,0))~.
\end{equation}%
Since $e_{I}^{2}=1$ and $e_{I}e_{J}=-e_{J}e_{I}$ if $I\neq J$, one easily
verifies that
\begin{equation}
c^{2}\,=\,e_{1}e_{2}\ldots e_{n}e_{1}e_{2}\ldots
e_{n}\,=\,(-1)^{n-1}e_{1}e_{2}e_{3}\ldots e_{n-1}e_{n}^2e_1e_{2}e_{3}\ldots
e_{n-1}\,=\,(-1)^{n(n-1)/2}~,
\end{equation}%
and that for any $I$ we have
\begin{equation}
ce_{I}\,=\,e_{1}e_{2}\ldots e_{n}c\,=\,(-1)^{n-1}e_{I}e_{1}e_{2}\ldots
e_{n}\,=\,(-1)^{n-1}e_{I}c~.
\end{equation}%
In particular, if $n\equiv 1\mod 2$ then we have $ce_{I}=e_{I}c$ which
implies that $cx=xc$ for all $x\in Cl(n)$ since the $e_{I}$ generate the
Clifford algebra. Notice that then also the $c_{\pm }$ are central elements,
$xc_{\pm }=c_{\pm }x$ for all $x\in Cl(n,0)$. If moreover $n\equiv 1\mod 4$
then we also have that $c^{2}=1$.

Now we assume that $n\equiv 1\mod 4$. Then we find that
\begin{equation}
c_{\pm}^2\,=(1\pm c)(1\pm c)/4\,=\,(1\pm 2c\,+\,c^2)/4\,=\,(1\pm
c)/2\,=\,c_\pm,\qquad
\end{equation}
and recall that the $c_\pm$ are central:
\begin{equation}
c_+c_-\,=\,c_-c_+\,=\,(c^2-1)/4\,=\,0~,
\end{equation}
so the $c_\pm$ are central idempotents in $Cl(n,0)$. This causes the algebra
to split into two components as follows. For $x\in Cl(n,0)$ let
\begin{equation}
x\,=\,x(1+c)/2\,+\,x(1-c)/2\,=\,x_+\,+\,x_-,
\end{equation}
and thus the vector space $Cl(n,0)$ splits as
\begin{equation}
Cl(n,0)\,=\,\,Cl(n,0)_+\,\oplus\,Cl(n,0)_-,\qquad
Cl(n,0)_\pm\,:=\,\{xc_\pm\,=\,c_\pm x:x\in Cl(n,0)\}~,
\end{equation}
the sum is direct since if $z=xc_+=yc_-$ then $zc_+=yc_-c_+=0$ and $%
z_-=xc_+c_-=0$ hence $z=zc_++zc_-=0$. The subspaces $Cl(n,0)_\pm$ are
actually subalgebras: if $u,v\in Cl(n,0)_\pm$ then also $uv\in Cl(n,0)_\pm$
since we can write $u=xc_\pm$, $v=yc_\pm$ and thus
\begin{equation}
uv\,=\,(xc_\pm)(yc_\pm)\,=\,xyc_\pm^2\,=\,xyc_\pm\quad\in Cl(n,0)_\pm~.
\end{equation}

In particular, given a matrix representation (an algebra homomorphism) $%
\phi:Cl(n,0)\rightarrow M_m(\mathbb{R})$ defined by $\phi(e_I)=\Gamma_I$,
the element $c$ maps to $\Gamma_1\Gamma_2\ldots \Gamma_n$ and $c^2=1$ maps
to ${\mathbb{I}}_m$. If the representation is irreducible, then by Schur's
lemma any element, like $\phi(c)$, that commutes with all $\phi(x)$, $x\in
Cl(n,0)$, is a scalar multiple of the identity and since $c^2=1$ it follows
that $\phi(c)=\pm {\mathbb{I}}_m$. In case $c={\mathbb{I}}_m$ one finds $%
\phi(c_+)={\mathbb{I}}_m$ and $\phi(c_-)=0$, so that $Cl(n,0)_-=\ker(\phi)$,
but in case $c=-{\mathbb{I}}_m$ one finds $\phi(c_-)={\mathbb{I}}_m$ and $%
\phi(c_+)=0$, so that $Cl(n,0)_+=\ker(\phi)$.

Notice also that changing the sign of all $\Gamma_I$, $I=1,\ldots,n$, one
obtains again a representation $\phi_-:Cl(n,0)\rightarrow M_m(\mathbb{R})$
(with $\phi_-(e_I)=-\Gamma_I$) but now, since $n$ is odd, we have $%
\phi_-(c)=-\phi(c)$. So, as is well-known, the representations $\phi$ and $%
\phi_-$ are not equivalent. More generally, changing the sign of an odd
number of the $\Gamma_I$ defines a representation which is not equivalent to
$\phi$ since it changes the sign of $\phi(c)$.

\subsection{Quadratic forms and $\Gamma$-matrices}

Recall the definition of the four $2\times 2$ (Pauli) matrices:%
\begin{equation}
\begin{array}{lll}
\gamma _{00}\,:=\,{\mathbb{I}}_{2}\,=\,\left(
\begin{array}{rc}
1 & 0 \\
0 & 1%
\end{array}%
\right) , & ~ & \gamma _{10}\,:=\,\sigma _{1}\,=\,\left(
\begin{array}{rc}
0 & 1 \\
1 & 0%
\end{array}%
\right) , \\
~ & ~ & ~ \\
\gamma _{01}\,=\,\sigma _{3}\,=\,\left(
\begin{array}{rc}
1 & 0 \\
0 & -1%
\end{array}%
\right) , & ~ & \gamma _{11}\,=\,i\sigma _{2}\,=\,\left(
\begin{array}{rc}
0 & 1 \\
-1 & 0%
\end{array}%
\right) ~.%
\end{array}
\label{gammas}
\end{equation}%
The notation is chosen such that (we recall that $i,j,k,l=0,1$)%
\begin{equation}
\gamma _{ij}\gamma _{kl}=\pm \gamma _{(i+k)(j+l)},
\end{equation}%
where the indices are summed modulo $2$. Notice that
\begin{equation}
\gamma _{ij}^{2}=(-1)^{ij}{\mathbb{I}}_{2},\qquad \gamma _{ij}\gamma
_{kl}\,=\,(-1)^{il+jk}\gamma _{kl}\gamma _{ij},\qquad {}^{T}\gamma
_{ij}=(-1)^{ij}\gamma _{ij}~.
\end{equation}

Recall that the tensor product of the square matrices $M=(M_{ij})$ and $%
N=(N_{kl})$, of size $m\times m$ and $n\times n$ respectively, is the matrix
of size $nm\times nm$ given by the (block) matrix
\begin{equation}
M\otimes N\,:=\,\left(
\begin{array}{ccc}
MN_{11} & \ldots & MN_{1n} \\
\vdots & \vdots & \vdots \\
MN_{n1} & \ldots & MN_{nn}%
\end{array}%
\right) ~.
\end{equation}

By a $\Gamma $-matrix (of size $2^{g}\times 2^{g}$, and characteristic $%
[{}_{j_{1}\ldots j_{g}}^{i_{1}\ldots i_{g}}]$), in this paper we intend the
following tensor product :%
\begin{equation}
\Gamma \equiv \Gamma \lbrack {}_{j_{1}\ldots j_{g}}^{i_{1}\ldots
i_{g}}]:=\gamma _{i_{1}j_{1}}\otimes \ldots \otimes \gamma _{i_{g}j_{g}}.
\label{Gamm}
\end{equation}%
Such a $\Gamma $-matrix is symmetric \textit{iff} the sum of the products $%
i_{k}j_{k}$ is zero modulo $2$ :
\begin{eqnarray}
^{T}\Gamma {} &=&^{T}(\gamma _{i_{1}j_{1}}\otimes \ldots \otimes \gamma
_{i_{g}j_{g}})\,=\,({}^{T}\gamma _{i_{1}j_{1}})\otimes \ldots \otimes
({}^{T}\gamma _{i_{g}j_{g}})\,  \notag \\
&=&\,(-1)^{\sum_{a=1}^{g}i_{a}j_{a}}(\gamma _{i_{1}j_{1}}\otimes \ldots
\otimes \gamma _{i_{g}j_{g}})=\,(-1)^{\sum_{a=1}^{g}i_{a}j_{a}}\Gamma .
\label{Gamm2}
\end{eqnarray}%
The quadratic form (with characteristic $[{}_{j_{1}\ldots
j_{g}}^{i_{1}\ldots i_{g}}]$) in $y^{1},\ldots ,y^{2^{g}}$ associated to the
$\Gamma $-matrix (\ref{Gamm}) is defined as
\begin{equation}
Q[{}_{j_{1}\ldots j_{g}}^{i_{1}\ldots i_{g}}](y)\,:=\,{}^{T}y(\gamma
_{i_{1}j_{1}}\otimes \ldots \otimes \gamma _{i_{g}j_{g}})y~.
\end{equation}%
If $\sum_{a=1}^{g}i_{a}j_{a}=0$ modulo $2$, the parity of the characteristic
is said to be even, and then (\ref{Gamm2}) shows that the $\Gamma $-matrix
is symmetric. The (anti)symmetric $\Gamma$ matrices of size $2^g$ are a
basis of the vector space of (skew)symmetric matrices of size $2^g$.

\subsection{Examples of the quadratic forms}

In case $g=1$ there are $3$ symmetric $\Gamma $-matrices, $\gamma
_{00},\gamma _{01},\gamma _{10}$ and the corresponding quadratic forms are
respectively:
\begin{equation}
Q[{}_{0}^{0}]\,:=\left( \,y^{1}\right) ^{2}+\left( \,y^{2}\right)
^{2},\qquad Q[{}_{1}^{0}]\,:=\,\left( \,y^{1}\right) ^{2}-\left(
\,y^{2}\right) ^{2},\qquad Q[{}_{0}^{1}]\,:=\,2y^{1}y^{2}~.  \label{quadg1}
\end{equation}%
For $g=2$ there are $10$ symmetric (and $6$ antisymmetric) $\Gamma $
matrices, here are some examples of the corresponding quadratic forms with $%
\Gamma $-matrices $\sigma _{1}\otimes \sigma _{1}$ and $\sigma _{1}\otimes
\sigma _{3}$ respectively:
\begin{equation}
Q[{}_{00}^{11}]\,:=\,2(y^{1}y^{4}\,+\,y^{2}y^{3}),\qquad
Q[{}_{10}^{01}]\,:=\,2(y^{1}y^{2}-y^{3}y^{4})~.
\end{equation}%
In case $g=3$ there are $36$ symmetric and $28$ antisymmetric $\Gamma $%
-matrices and for example the quadratic form corresponding to $\gamma
_{10}\otimes \gamma _{11}\otimes \gamma _{11}$ is
\begin{equation}
Q[{}_{110}^{111}]\,:=\,2(y^{1}y^{8}+y^{2}y^{7}-y^{3}y^{6}-y^{4}y^{5})~.
\end{equation}

\subsection{Clifford sets of $\Gamma$-matrices \label{CliffGamm}}

One easily verifies:
\begin{eqnarray}
\Gamma \lbrack {}_{j_{1}\ldots j_{g}}^{i_{1}\ldots i_{g}}]\Gamma \lbrack
{}_{l_{1}\ldots l_{g}}^{k_{1}\ldots k_{g}}] &=&(\gamma _{i_{1}j_{1}}\otimes
\ldots \otimes \gamma _{i_{g}j_{g}})(\gamma _{k_{1}l_{1}}\otimes \ldots
\otimes \gamma _{k_{g}l_{g}})\,  \notag \\
&=&\,(-1)^{\sum_{a=1}^{g}i_{a}l_{a}+j_{a}k_{a}}(\gamma _{k_{1}l_{1}}\otimes
\ldots \otimes \gamma _{k_{g}l_{g}})(\gamma _{i_{1}j_{1}}\otimes \ldots
\otimes \gamma _{i_{g}j_{g}})  \notag \\
&=&\,(-1)^{\sum_{a=1}^{g}i_{a}l_{a}+j_{a}k_{a}}\Gamma \lbrack
{}_{l_{1}\ldots l_{g}}^{k_{1}\ldots k_{g}}]\Gamma \lbrack {}_{j_{1}\ldots
j_{g}}^{i_{1}\ldots i_{g}}];  \label{prod} \\
\left( \Gamma \lbrack {}_{j_{1}\ldots j_{g}}^{i_{1}\ldots i_{g}}]\right)
^{2} &=&(\gamma _{i_{1}j_{1}}\otimes \ldots \otimes \gamma
_{i_{g}j_{g}})^{2}=(-1)^{\sum_{a=1}^{g}i_{a}j_{a}}{\mathbb{I}}_{2^{g}}.
\label{Gamsq}
\end{eqnarray}
Note that (\ref{Gamm2}) and (\ref{Gamsq}) show that if a $\Gamma$-matrix $%
\Gamma[{}^i_j]$ is symmetric then $\Gamma[{}^i_j]^2={\mathbb{I}}_{2^g}$.
Moreover, from (\ref{prod}) it follows that two distinct symmetric $\Gamma$%
-matrices $\Gamma[{}^i_j]$ and $\Gamma[{}^k_l]$ anti-commute iff the
characteristic $[{}^{i+k}_{j+l}]$ is odd, that is $%
\sum_{a=1}^g(i_a+k_a)(j_a+l_a)\equiv 1\mod 2$, (use that the symmetry
implies $\sum_{a=1}^g i_aj_a\equiv \sum_{a=1}^g k_al_a\equiv 0\mod 2$).

Thus a set of $\Gamma$-matrices $\{\Gamma[{}^{i^1}_{j^1}],\ldots,\Gamma[%
{}^{i^n}_{j^n}]\}$ is a Clifford set iff all characteristics $%
[{}^{i^1}_{j^1}],\ldots,[{}^{i^n}_{j^n}]$ are even and the sum of any two
distinct characteristics is odd.

\subsection{Quadratic identities between quadratic forms\label{GammQI}}

Whereas the quadratic forms $Q[{}_{j}^{i}]$ (with even characteristics of
length $g$) are a basis of the vector space of all quadratic forms in $%
m=2^{g}$ variables, their squares, homogeneous polynomials of degree four,
are not linear independent. The most basic example, known as Jacobi's
identity, is:
\begin{equation}
-Q[{}_{0}^{0}]^{2}\,+\,Q[{}_{1}^{0}]^{2}\,+\,Q[{}_{0}^{1}]^{2}\,=\,0~,
\end{equation}%
which is easily verified since, by definition,
\begin{equation}
Q[{}_{0}^{0}]\,:=\,\left( \,y^{1}\right) ^{2}+\left( \,y^{2}\right)
^{2},\qquad Q[{}_{1}^{0}]\,:=\,\left( \,y^{1}\right) ^{2}-\left(
\,y^{2}\right) ^{2},\qquad Q[{}_{1}^{0}]\,:=\,2y^{1}y^{2}~,
\end{equation}%
and thus we indeed have the identity
\begin{equation}
\left(\left( \,y^{1}\right) ^{2}+\left( \,y^{2}\right) ^{2}\right)^{2}\,=\,
\left(\left(\,y^{1}\right) ^{2}-\left( \,y^{2}\right) ^{2}\right)^{2}\,+\,
\left(2y^{1}y^{2}\right)^{2}~.
\end{equation}%
For $g=2,3,4$ there are similar identities between $4,6,10$ respectively
such quadrics, see (\ref{GammQI2}), (\ref{lorL410}), (\ref{GammQI4})
respectively. A remarkable fact, and crucial for the invertibility results
in this paper, is that in all these four cases the the Gamma-matrices of the
quadrics are ${\mathbb{I}}_{m}$ and the remaining $\Gamma _{I}$ form a
(maximal) Clifford set. Actually, these identities are classical identities
between theta functions. The case $g=1$ was known to Jacobi, and the other
cases were already known to Max Noether, see \cite[p.332, p.334]{Noether}.

Unfortunately, for $g>4$ it seems that the squares of the quadrics defined
by ${\mathbb{I}}_{m}$ and a maximal Clifford set are \textit{no longer}
linearly dependent.

\subsection{Heisenberg groups}

The $\Gamma $-matrices generate a finite (non-Abelian) subgroup of $GL(2^{g},%
\mathbb{C})$, which is called a Heisenberg group (cfr.\ e.g. \cite{CG}) or a
Clifford group. The subgroup has order $2\cdot 2^{2g}$ and each element is
of the form $\pm \Gamma $ where $\Gamma $ is one of the $2^{2g}$ $\Gamma $%
-matrices. See \cite[ App. A]{CG} for the quadratic forms $%
Q_{m}=Q[{}_{\epsilon ^{\prime }}^{\epsilon }]$. The quadratic relations
among the $Q_{m}$'s that we discussed can also be found in A.3 of \cite{CG},
where they are discussed in the context of Hopf maps.

\section{BPS entropy and attractors in $L\left( q,P\right) $ models \label%
{invertgamma}}

In the present paper we focus on BPS systems related to models $L(q,P,\dot{P}%
)$ of homogeneous very special geometry \cite{dWVP}: we will thus only
consider the case of a Lorentzian quadratic form in the $x^{i}$ and a
related Clifford set of $\Gamma $-matrices. In particular (see Sec.\ \ref%
{HNS}), we consider $V$ (cfr.\ (\ref{Vbold})) to have Lorentzian (mostly
plus) signature and%
\begin{equation}
\dim V=q+2,
\end{equation}%
such that (cfr. (\ref{exprr}))%
\begin{equation}
N=\dim V+1+\mathcal{D}_{\dim V-1}\cdot \left( P+\dot{P}\right) .
\label{exprr2}
\end{equation}%
For later convenience, in order to highlight the Lorentzian (mostly plus)
signature, we also shift the labeling of the $q+2$ $I$-indices from $%
1,...,q+2$ to $0,1,...,q+1$; moreover, the matrix $\Gamma _{0}$ will be
nothing but the identity matrix of size $m=\mathcal{D}_{q+1}\cdot \left( P+%
\dot{P}\right) $.

Physically, the $L(q,P,\dot{P})$ models of homogeneous very special geometry
determine the scalar manifolds (i.e., the target spaces of scalar fields) in
ungauged $\mathcal{N}=2$ Maxwell-Einstein supergravity theory coupled to
vector multiplets in three, four or five Lorentzian space-time dimensions
(the corresponding spaces are quaternionic, K\"{a}hler and real,
respectively).

Since the closed form expression of the Bekenstein-Hawking entropy (\ref%
{S-BPS}) as well as of the attractor values of scalar fields (\ref%
{BPS-crit-points-2}) of extremal BPS black holes are already known for
symmetric spaces (see e.g. \cite{FGimonK}), in the present paper we will
focus on homogeneous \textit{non-symmetric} spaces.

In the present section, we will start from a Clifford set, and we will
define a certain cubic form; then, we will show that under certain
conditions its gradient map is invertible; consequently, the corresponding
BPS system (\ref{BPS}) can be explicitly solved by (\ref{sol}), thus
allowing for a novel, closed form expression of the Bekenstein-Hawking
entropy (\ref{S-BPS}) as well as of the attractor values of scalar fields (%
\ref{BPS-crit-points-2}) of extremal BPS black holes, respectively given by (%
\ref{SSd}) and (\ref{zH-1})-(\ref{zH-3}) below.

After such a general treatment, we will consider some classes of models,
namely :

\begin{itemize}
\item $L(q,1)$ with $q=1,...,8$ (see Sec.\ \ref{explmodels});

\item $L(q,2)$ with $q=1,2,3$ (see Sec.\ \ref{L420});

\item $L(q,P)$, $q=1,2,3$, $P\geqslant 3$ (see Sec.\ \ref{LqP0}), with
explicit emphasis given to the models $L(1,P)$ with $P\geqslant 2$ given in
Sec.\ \ref{L1P0};

\item $L(4,1,1)$ (see Sec.\ \ref{L411}); in the present paper, this is the
unique model having a non-vanishing $\dot{P}$ (namely, with $P=\dot{P}=1$).
\end{itemize}

In particular, the models $L(1,2)$ and $L(1,3)$ will be analyzed in full
detail in Secs.\ \ref{L120} and \ref{L130}, respectively, and their analysis
will be explicitly generalized to the class of models $L(1,P)$ with $%
P\geqslant 2$ in Sec.\ \ref{L1P0}. In this respect, such Secs.\ extend the
treatment given\footnote{%
It would be interesting to investigate the geometric aspects of the examples
of non-homogeneous very special geometry discussed in Sec.\ 4 of \cite%
{Shmakova} (cfr. Refs. therein, as well); we leave this task for further
future work.} in Sec.\ 4 of \cite{Shmakova}, by providing explicit
expressions for the BPS black hole entropy as well as for the BPS attractors
in such an infinite class of non-symmetric (homogeneous) models of $\mathcal{%
N}=2$, $D=4$ supergravity with cubic prepotential.

It is also here worth anticipating that similar results can be obtained also
starting with quadratic forms with different signatures; for instance, we
will briefly deal with a ten-dimensional quadratic form with Kleinian
signature (i.e., with $\epsilon _{1}=...=\epsilon _{5}=1=-\epsilon
_{6}=...=-\epsilon _{10}$) in Sec.\ \ref{55}.

\subsection{The cubic form\label{defpol}}

Let $\{\Gamma _{1},\ldots ,\Gamma _{q+1}\}$ be a Clifford set of symmetric $%
\Gamma $-matrices of size $m\times m$, with 
$m={\mathcal{D}}_{q+1}$ as in (\ref{table D}) and Section \ref{CliffGamm}.
Let $\Gamma _{0}:={\mathbb{I}}_{m}$.

We define a Lorentzian quadratic form $q$ in $q+2$ variables $x^{0},\ldots
,x^{q+1}$ and $q+2$ quadratic forms $Q_{I}$, $I=0,\ldots ,q+1$, in $m={%
\mathcal{D}}_{q+1}$ variables $y^{1},\ldots ,y^{m}$: (cf.\ Eq.\ (\ref{qqqq})
above):
\begin{eqnarray}
q(x)\, &:&=\,-\left( x^{0}\right) ^{2}\,+\,\left( x^{1}\right)
^{2}\,+\,\ldots \,+\,\left( x^{q+1}\right) ^{2};  \label{qqqq-Lor} \\
Q_{I}(y)\, &:&=\,{}^{T}y\Gamma _{I}y~.  \label{Q_I}
\end{eqnarray}%
Using the these quadratic forms, we define the cubic form in $1+(q+2)+m$
variables $s$, $x^{0},\ldots ,x^{q+1}$, $y^{1},\ldots ,y^{m}$ :
\begin{equation}
\mathcal{V}\,\left( s,x,y\right) :=\,sq(x)\,+\,x^{0}Q_{0}(y)\,+\,\ldots
\,+\,x^{q+1}Q_{q+1}(y)~.  \label{cubb}
\end{equation}

\subsection{The invertibility condition \label{main}}

Given the Clifford set $\{\Gamma _{1},\ldots ,\Gamma _{q+1}\}$, in order to
invert the gradient map of ${\mathcal{V}}$, we need the existence of further
symmetric $m\times m$ matrices $\Omega _{1},\ldots ,\Omega _{r}$, which
anti-commute with the matrices in the Clifford set:
\begin{equation}
\Gamma _{I}\Omega _{K}\,=\,-\Omega _{K}\Gamma _{I}\qquad (I=1,\ldots
,q+1,\quad K=1,\ldots ,r)~.  \label{GOcom}
\end{equation}%
Moreover, if we denote the associated quadratic forms defined by these $%
\Omega _{K}$ by
\begin{equation}
R_{{K}}(y):={}^{T}y\Omega _{{K}}y~,\qquad K\,=\,1,\ldots ,r,  \label{R_k}
\end{equation}%
then the following Lorentzian quadratic identity should hold:
\begin{equation}
-Q_{0}\left( y\right) \,^{2}+\,\sum_{I=1}^{q+1}Q_{I}(y)^{2}\;+\;\sum_{{K}%
=1}^{r}R_{{K}}(y)^{2}\,=\,0~.  \label{lorrel}
\end{equation}

If auxiliary matrices $\Omega _{K}$ with all these properties exist, then
the gradient map (with $I=0,...,q+1$, and $\alpha =1,...,m$)
\begin{eqnarray}
\, &&\nabla _{\mathcal{V}}:\;\mathbb{R}^{q+3+m}\,\longrightarrow \,\mathbb{R}%
^{q+3+m}~; \\
&&\nabla _{\mathcal{V}}\,=\,^{T}(\mathcal{V}_{s},\ldots ,\mathcal{V}%
_{I},\ldots ,\mathcal{V}_{\alpha },\ldots )~, \\
&&\mathcal{V}_{s}\,:=\,\frac{\partial \mathcal{V}}{\partial s},\qquad
\mathcal{V}_{I}\,:=\,\frac{\partial \mathcal{V}}{\partial x^{I}},\qquad
\mathcal{V}_{\alpha }\,:=\,\frac{\partial \mathcal{V}}{\partial y^{\alpha }}%
~,
\end{eqnarray}%
is \textit{invertible}, with (birational) inverse given by polynomials of
degree $2$ if $r=0$ and of degree $4$ if $r>0$. An explicit expression of
the birational inverse map of the gradient map $\nabla _{\mathcal{V}}$ will
be given in Sec.\ \ref{explin}. This results in a closed form expression of
the solution to the BPS system (\ref{BPS}), given in Sec.\ \ref{Sol-BPS}. In
turn, in Sec.\ \ref{BPS-S-gen} this will allow for a closed form expression
of Bekenstein-Hawking entropy (\ref{S-BPS}) as well as of the attractor
values of scalar fields (\ref{BPS-crit-points-2}) of extremal BPS black
holes in the homogeneous non-symmetric very special geometry characterizing
the corresponding model of ungauged $\mathcal{N}=2$ Maxwell-Einstein
supergravity theory coupled to vector multiplets in four space-time
dimensions.

\subsection{Remarks}

Notice that only in the case that we do have a Lorentzian identity $%
-Q_{0}(y)^{2}+\sum_{I=1}^{q+1}Q_{I}(y)^{2}=0$, there is no need for the
extra $\Omega _{K}$'s (and one can then take $r=0$).

As it will be seen in Sec.\ \ref{complete}, this is a quite special case,
corresponding to some symmetric very special spaces, and to the very rich
geometry related to simple cubic Jordan algebras \cite{PR1, PR2}. In the
present investigation, since we want to solve the BPS system (\ref{BPS}) and
study the expression of the BPS black hole entropy (\ref{S-BPS}) in cases
not treated in literature, we will be interested in some classes of
homogeneous non-symmetric spaces which all have $r>0$.

It seems rather restrictive (and mysterious) to request the existence of the
(Lorentzian) quadratic identity (\ref{lorrel}), but it is crucial for us in
order to show the existence of a (birational) inverse map of the gradient
map of the corresponding cubic form $\mathcal{V}$, and thus to provide a
closed form expression of the solution to the associated BPS system (\ref%
{BPS}).

It is here worth remarking that we will not impose any condition on products
involving only $\Omega $-matrices; in particular, for $P\geqslant 3$, we
will consider symmetric $m\times m$ matrices $\Omega _{j}$'s that are
\textit{not} invertible (i.e. whose rank is less than $m$), so they cannot
be $\Gamma $-matrices (see Sec.\ \ref{LqP0}).

We conclude with some remarks on the inverse of the gradient map $\nabla _{%
\mathcal{V}}$. This gradient map is given by the partial derivatives of ${%
\mathcal{V}}$ which are homogeneous polynomials of degree two in $%
\xi:=(s,x,y)$. Therefore $\nabla{\mathcal{V}}(\xi)=\nabla{\mathcal{V}}(-\xi)$
and this implies that the inverse image of the image of any non-zero point
contains at least two points which differ by a sign.

The inverse map $\nabla {\mathcal{V}}^{-1}$ will be given by homogeneous
polynomials of degree four in general, and thus the composition $\nabla {%
\mathcal{V}}^{-1}\circ \nabla {\mathcal{V}}$ has coordinate functions that
are homogeneous of degree eight. Therefore it cannot be the identity map
(since the identity map has coordinate functions that are homogeneous of
degree one), but one has:
\begin{equation}
\Big((\nabla {\mathcal{V}})^{-1}\circ \nabla {\mathcal{V}}\Big)(\xi
)\,=\,f(\xi )\xi ,\qquad f(\xi )\,=\,4q(\xi )^{2}{\mathcal{V}}(\xi )~,
\end{equation}%
(cfr.\ (\ref{comp})), with $f(\xi )$ homogeneous of degree $2\cdot 2+3=7$.
In particular, for the points $\xi $ with $f(\xi )=0$ the inverse of the
gradient map does not provide useful information.

All this should not be surprising, in fact in the simple cubic Jordan
algebra models the gradient map is given by $M\mapsto M^\sharp$ where $%
M^\sharp$ is the adjoint of a $3\times 3$ matrix and one has $%
M^\sharp=(-M)^\sharp$; the inverse of the adjoint map is then $%
M^\sharp\mapsto (M^\sharp)^\sharp$ which is $M$, up to a homogeneous
polynomial of degree three which is the determinant of $M$: $%
(M^\sharp)^\sharp=\det(M)M$ (and $\det(M)$ is basically the cubic form $%
\mathcal{V}$ of the corresponding $L(q,1)$ model), see also Sec.\ \ref{L110}.

\subsection{Explicit inversion of $\protect\nabla _{\mathcal{V}}$\label%
{explin}}

When $r>0$, the inverse of the gradient map $\nabla _{\mathcal{V}}$ is given
as a composition of two maps, namely :
\begin{equation}
{\mathbb{R}}_{(s,x,y)}^{q+3+m}\,\overset{\nabla _{\mathcal{V}}}{%
\longrightarrow }\,{\mathbb{R}}_{z}^{q+3+m}\,\overset{\mathbf{\alpha }}{%
\longrightarrow }\,{\mathbb{R}}_{(t,u,v,w)}^{q+3+m+r}\,\overset{\mathbf{\mu }%
}{\longrightarrow }\,{\mathbb{R}}_{(s,x,y)}^{q+3+m}~,~\text{such~that}~~%
\mathbf{\mu }\circ \mathbf{\alpha }\circ \nabla _{\mathcal{V}}\,=\,4q(x)^{2}%
\mathcal{V}(s,x,y)\mathbb{I}_{q+3+m}.  \label{comp}
\end{equation}

The map $\mathbf{\alpha }$, which has $q+3+m+r$ components that are
homogeneous polynomials of degree $2$ in the variables $z_{1},\ldots
,z_{q+3+m}$, is given by
\begin{equation}
\mathbf{\alpha }(z_{1},\ldots
,z_{q+3+m})\,:=\,^{T}(z_{1}^{2},\,z_{1}z_{2},\ldots
,\,z_{1}z_{q+3+m},\,R_{1}(z),\ldots ,R_{r}(z))~,  \label{alpha}
\end{equation}%
where the $r$ quadratic forms $R_{K}(z)$'s depend only on the last $m$
variables $z_{q+3+1},\ldots ,z_{q+3+m}$:
\begin{equation}
R_{K}(z)\,:=\,R_{K}(z_{q+3+1},\ldots ,z_{q+3+m})~.  \label{RK}
\end{equation}%
The composition $\mathbf{\alpha }\circ \nabla _{\mathcal{V}}$ will be
explicitly computed in Sec.\ \ref{mapa} below, and it is given by
\begin{equation}
\mathbf{\alpha }\circ \nabla _{\mathcal{V}}\left( s,x,y\right) :=\mathbf{%
\alpha }(\nabla _{\mathcal{V}}(s,x,y))\,=\,q(x)~^{T}(\nabla _{\mathcal{V}%
}(s,x,y),\,-4R_{1}(y),\ldots ,\,-4R_{r}(y))~.  \label{alphaphi}
\end{equation}

Next, the map $\mathbf{\mu }$, which has $q+3+m$ components that are
homogeneous polynomials of degree $2$ in the variables $t,u_{0},\ldots
,u_{q+1},v_{1},\ldots ,v_{m},w_{1}\ldots ,w_{r}$, is given by
\begin{equation}
\mathbf{\mu }(t,u,v,w):\,=\,\left( \renewcommand{\arraystretch}{1.3}%
\begin{array}{rcl}
q(u) & + & \frac{1}{16}\sum_{K=1}^{r}w_{K}^{2} \\
-2tu_{0} & + & \frac{1}{2}Q_{0}(v) \\
2tu_{1} & + & \frac{1}{2}Q_{1}(v) \\
\vdots &  &  \\
2tu_{q+1} & + & \frac{1}{2}Q_{q+1}(v) \\
\left( \sum_{I=0}^{q+1}u_{I}\Gamma _{I}\right) v & - & \frac{1}{4}\left(
\sum_{K=1}^{r}w_{K}\Omega _{K}\right) v \\
&  &
\end{array}%
\right) ~,  \label{mu}
\end{equation}%
where the definition of the last $m$ components involves all the $q+2$
symmetric $\Gamma $-matrices $\left\{ \Gamma _{I}\right\} $ as well as the $%
r $ auxiliary symmetric matrices $\left\{ \Omega _{K}\right\} $, required in
the invertibility condition enounced in Sec.\ \ref{main}; all these matrices
have size $m\times m$.

Recalling the relabelling (\ref{csii}), we will verify in Sec.\ \ref%
{checkinvmap} that the composition of maps $\mathbf{\mu \circ \alpha }\circ
\nabla _{\mathcal{V}}$ is given by (cfr. (\ref{cubb}))
\begin{gather}
\left(
\begin{array}{c}
\left( \mathbf{\mu \circ \alpha }\right) ^{s}(\nabla _{\mathcal{V}}\left(
s,x,y\right) ) \\
\left( \mathbf{\mu \circ \alpha }\right) ^{I}(\nabla _{\mathcal{V}}\left(
s,x,y\right) ) \\
\left( \mathbf{\mu \circ \alpha }\right) ^{\alpha }(\nabla _{\mathcal{V}%
}\left( s,x,y\right) )%
\end{array}%
\right) =4q^{2}\left( x\right) \mathcal{V}\left( s,x,y\right) \left(
\begin{array}{c}
s \\
x^{I} \\
y^{\alpha }%
\end{array}%
\right)  \label{jj} \\
\Updownarrow  \notag \\
\mathbf{\mu \circ \alpha \circ }\nabla _{\mathcal{V}}\left( \xi \right)
=4q^{2}\left( x\right) \mathcal{V}\left( \xi \right) \mathbb{\xi }.
\label{jj-2}
\end{gather}%
As we will see below, in \textit{complete} models (see Sec.\ \ref{complete})
$r=0$ by definition, and one can omit the map $\mathbf{\alpha }$ (because it
becomes proportional to the identity map), and then (\ref{jj-2}) reduces to%
\begin{equation}
\mathbf{\mu }\circ \nabla _{\mathcal{V}}\left( \xi \right) =4\mathcal{V}%
\left( \xi \right) \mathbb{\xi }.  \label{complete!}
\end{equation}

The identity (\ref{jj-2}) implies that the composed map%
\begin{equation}
{\mathbb{R}}_{z}^{q+3+m}\,\,\overset{\mathbf{\mu \circ \alpha }}{%
\longrightarrow }\,{\mathbb{R}}_{(s,x,y)}^{q+3+m}  \label{mmm}
\end{equation}%
can be regarded as the inverse map of the gradient map $\nabla _{\mathcal{V}%
} $. In order to determine the general form of the map $\mathbf{\mu \circ
\alpha }$ (\ref{mmm}), the map $\mathbf{\mu }$ (\ref{mu}) must be evaluated
on the image of the map $\mathbf{\alpha }$ (\ref{alpha}) :%
\begin{equation}
\left( \mathbf{\mu \circ \alpha }\right) \left( z_{1},\ldots
,z_{q+3+m}\right) :=\mathbf{\mu }\left( \mathbf{\alpha }\left( z_{1},\ldots
,z_{q+3+m}\right) \right) .
\end{equation}%
The replacement of the variables $t,u_{0},\ldots ,u_{q+1},v_{1},\ldots
,v_{m},w_{1}\ldots ,w_{r}$ with the corresponding components (homogeneous
polynomials in the variables $z_{1},\ldots ,z_{q+3+m}$) in the image of $%
\mathbf{\alpha }$ (\ref{alpha}) reads as follows ($I=0,1,...,q+1$, $\alpha
=1,...,m$, $K=1,...,r$):
\begin{equation}
\left\{
\begin{array}{l}
t\rightarrow z_{1}^{2}; \\
u_{I}\rightarrow z_{1}z_{I+2}; \\
v_{\alpha }\rightarrow z_{1}z_{q+3+\alpha }; \\
w_{K}\rightarrow R_{K}\left( z\right) \overset{\text{(\ref{RK})}}{=}%
R_{K}(z_{q+3+1},\ldots ,z_{q+3+m}).%
\end{array}%
\right.
\end{equation}%
By defining the $m\times 1$ vector%
\begin{equation}
\hat{z}:=~^{T}\left( z_{q+4},...,z_{q+3+m}\right) ,  \label{z-hat}
\end{equation}%
one can easily compute :%
\begin{equation}
\left( \mathbf{\mu \circ \alpha }\right) \left( z_{1},\ldots
,z_{q+3+m}\right) :=\,\left( \renewcommand{\arraystretch}{1.3}%
\begin{array}{rcl}
q(u_{I}\rightarrow z_{1}z_{I+2}) & + & \frac{1}{16}\sum_{K=1}^{r}R_{K}^{2}%
\left( \hat{z}\right) \\
-2z_{1}^{3}z_{2} & + & \frac{1}{2}Q_{0}(v\rightarrow z_{1}\hat{z}) \\
2z_{1}^{3}z_{3} & + & \frac{1}{2}Q_{1}(v\rightarrow z_{1}\hat{z}) \\
\vdots &  &  \\
2z_{1}^{3}z_{q+3} & + & \frac{1}{2}Q_{q+1}(v\rightarrow z_{1}\hat{z}) \\
z_{1}^{2}\left( \sum_{I=0}^{q+1}z_{I+2}\Gamma _{I}\right) \hat{z} & - &
\frac{1}{4}z_{1}\left( \sum_{K=1}^{r}R_{K}\left( \hat{z}\right) \Omega
_{K}\right) \hat{z} \\
&  &
\end{array}%
\right) ,  \label{mu-alpha}
\end{equation}%
where the last line of the r.h.s. contains the product of the $m\times m$
matrices $\Gamma _{I}$ and $\Omega _{K}$ with the $m\times 1$ vector $\hat{z}
$ (\ref{z-hat}). From the definitions (\ref{qqqq-Lor}) and (\ref{R_k}), one
computes%
\begin{equation}
\begin{array}{l}
q(u_{I}\rightarrow z_{1}z_{I+2})=z_{1}^{2}\left(
-z_{2}^{2}+z_{3}^{2}+...+z_{q+3}^{2}\right) ; \\
Q_{0}(v\rightarrow z_{1}\hat{z})=z_{1}^{2}~^{T}\hat{z}\Gamma _{0}\hat{z}%
=z_{1}^{2}~^{T}\hat{z}\mathbb{I}_{m}\hat{z}=z_{1}^{2}\left(
z_{q+4}^{2}+...+z_{q+3+m}^{2}\right) ; \\
Q_{1}(v\rightarrow z_{1}\hat{z})=z_{1}^{2}~^{T}\hat{z}\Gamma _{1}\hat{z}; \\
\vdots \\
Q_{q+1}(v\rightarrow z_{1}\hat{z})=z_{1}^{2}~^{T}\hat{z}\Gamma _{q+1}\hat{z},%
\end{array}%
\end{equation}%
and therefore (\ref{mu-alpha}) can be further elaborated as follows :
\begin{eqnarray}
\left( \mathbf{\mu \circ \alpha }\right) \left( z_{1},\ldots
,z_{q+3+m}\right) &:&=\,\left( \renewcommand{\arraystretch}{1.3}%
\begin{array}{rcl}
z_{1}^{2}\left( -z_{2}^{2}+z_{3}^{2}+...+z_{q+3}^{2}\right) & + & \frac{1}{16%
}\sum_{K=1}^{r}R_{K}^{2}\left( \hat{z}\right) \\
-2z_{1}^{3}z_{2} & + & \frac{1}{2}z_{1}^{2}\left(
z_{q+4}^{2}+...+z_{q+3+m}^{2}\right) \\
2z_{1}^{3}z_{3} & + & \frac{1}{2}z_{1}^{2}~^{T}\hat{z}\Gamma _{1}\hat{z} \\
\vdots &  &  \\
2z_{1}^{3}z_{q+3} & + & \frac{1}{2}z_{1}^{2}~^{T}\hat{z}\Gamma _{q+1}\hat{z}
\\
z_{1}^{2}\left( \sum_{I=0}^{q+1}z_{I+2}\Gamma _{I}\right) \hat{z} & - &
\frac{1}{4}z_{1}\left( \sum_{K=1}^{r}R_{K}\left( \hat{z}\right) \Omega
_{K}\right) \hat{z} \\
&  &
\end{array}%
\right) ,  \notag \\
&&  \label{mu-alpha-2}
\end{eqnarray}%
where (by recalling (\ref{R_k}) and (\ref{RK}))%
\begin{equation}
R_{K}\left( \hat{z}\right) =~^{T}\hat{z}\Omega _{{K}}\hat{z}.
\end{equation}%
Each of the $q+3+m$ components of this composed map is given by an
homogeneous polynomial of degree 4 in the $q+3+m$ variables $z_{1},\ldots
,z_{q+3+m}$. As it is evident, the explicit form of such polynomials depends
on $q+2$ (symmetric) $\Gamma $-matrices $\Gamma _{I}$ (such that $\Gamma
_{0}=\mathbb{I}_{m}$) as well as on the $r$ symmetric auxiliary matrices $%
\Omega _{K}$ and the corresponding quadratic forms $R_{K}$ defined in (\ref%
{R_k}) and in (\ref{RK}). Note that both the $\Gamma _{I}$'s and the $\Omega
_{K}$'s, as well as the corresponding quadratic forms $Q_{I}$'s (\ref{Q_I})
and $R_{K}$'s (\ref{R_k}), occur in the invertibility condition enounced in
Sec.\ \ref{main} (cfr. Eqs. (\ref{GOcom}) and (\ref{lorrel}), respectively),
which is assumed to hold throughout the treatment of this Section, as well
as of the subsequent Secs. (\ref{Sol-BPS})-(\ref{complete}) and in the whole
Sec.\ \ref{checkinvmap}.

\subsection{\label{Sol-BPS}Solution of the BPS system}

From (\ref{jj})-(\ref{jj-2}), by replacing $\nabla _{\mathcal{V}}$ with $%
3\partial _{p}\Delta =3~^{T}\left( \Delta _{s},\Delta _{I},\Delta _{\alpha
}\right) $, one obtains%
\begin{equation}
\mathcal{V}\xi =\frac{9}{4}\frac{1}{\Delta _{s}^{2}}\left( \mathbf{\mu \circ
\alpha }\right) \left( \partial _{p}\Delta \right) .  \label{rr-bis}
\end{equation}%
By defining%
\begin{eqnarray}
\nabla _{\mathcal{V}}\left( \xi \right) \cdot \xi &:&=\frac{\partial
\mathcal{V}}{\partial s}s+\frac{\partial \mathcal{V}}{\partial x^{J}}x^{J}+%
\frac{\partial \mathcal{V}}{\partial y^{\beta }}y^{\beta };  \label{jjj1} \\
\left( \partial _{p}\Delta \right) \cdot \xi &:&=\Delta _{s}s+\Delta
_{J}x^{J}+\Delta _{\beta }y^{\beta };  \label{jjj2} \\
\left( \partial _{p}\Delta \right) \cdot \left( \mathbf{\mu \circ \alpha }%
\right) \left( \partial _{p}\Delta \right) &:&=\Delta _{s}\left( \mathbf{\mu
\circ \alpha }\right) ^{s}\left( \partial _{p}\Delta \right)  \notag \\
&&+\Delta _{J}\left( \mathbf{\mu \circ \alpha }\right) ^{J}\left( \partial
_{p}\Delta \right) +\Delta _{\beta }\left( \mathbf{\mu \circ \alpha }\right)
^{\beta }\left( \partial _{p}\Delta \right) ,  \label{jjj3}
\end{eqnarray}%
and recalling the Euler formula (\ref{Euler}), one can replace $\nabla _{%
\mathcal{V}}\left( \xi \right) $ with $3\partial _{p}\Delta $ and use (\ref%
{rr-bis}) in order to obtain%
\begin{equation}
3\mathcal{V}=\nabla _{\mathcal{V}}\left( \xi \right) \cdot \xi =3\left(
\partial _{p}\Delta \right) \cdot \xi =\frac{27}{4}\frac{1}{\mathcal{V}%
\Delta _{s}^{2}}\left( \partial _{p}\Delta \right) \cdot \left( \mathbf{\mu
\circ \alpha }\right) \left( \partial _{p}\Delta \right) ,  \label{jjj4}
\end{equation}%
leading to%
\begin{gather}
\mathcal{V}^{2}=\frac{9}{4}\frac{1}{\Delta _{s}^{2}}\left( \partial
_{p}\Delta \right) \cdot \left( \mathbf{\mu \circ \alpha }\right) \left(
\partial _{p}\Delta \right) \\
\Updownarrow  \notag \\
\left\vert \mathcal{V}\right\vert =\frac{3}{2\left\vert \Delta
_{s}\right\vert }\sqrt{\left( \partial _{p}\Delta \right) \cdot \left(
\mathbf{\mu \circ \alpha }\right) \left( \partial _{p}\Delta \right) },
\label{rrr-2-bis}
\end{gather}%
which is well defined for
\begin{equation}
\left( \partial _{p}\Delta \right) \cdot \left( \mathbf{\mu \circ \alpha }%
\right) \left( \partial _{p}\Delta \right) >0.
\end{equation}%
Then, from (\ref{rr-bis}) and (\ref{rrr-2-bis}), it follows that%
\begin{equation}
\xi =\pm \frac{9}{4}\frac{1}{\left\vert \mathcal{V}\right\vert \Delta
_{s}^{2}}\left( \mathbf{\mu \circ \alpha }\right) \left( \partial _{p}\Delta
\right) =\pm \frac{3}{2\left\vert \Delta _{s}\right\vert }\frac{\left(
\mathbf{\mu \circ \alpha }\right) \left( \partial _{p}\Delta \right) }{\sqrt{%
\left( \partial _{p}\Delta \right) \cdot \left( \mathbf{\mu \circ \alpha }%
\right) \left( \partial _{p}\Delta \right) }}  \label{solsol}
\end{equation}%
is the general solution of the BPS system (\ref{BPS}). However, we must also
recall the condition (\ref{d<0-2}), which in this case reads%
\begin{equation}
\left( \partial _{p}\Delta \right) \cdot \xi >0\Leftrightarrow \mathcal{V}%
\left( \xi \right) >0\Leftrightarrow \pm \frac{3}{2\left\vert \Delta
_{s}\right\vert }\sqrt{\left( \partial _{p}\Delta \right) \cdot \left(
\mathbf{\mu \circ \alpha }\right) \left( \partial _{p}\Delta \right) }>0,
\end{equation}%
which thus implies that only the branch \textquotedblleft $+$" of (\ref%
{solsol}) is consistent.

Summarising, \textit{at least} in those homogeneous $d$-spaces \cite%
{dWVP,dWVVP} in which the invertibility condition enounced in Sec.\ \ref%
{main} is satisfied, there exists a quartic homogeneous polynomial map ($%
i=1,...,q+3+m$, where - cfr. below (\ref{exprr2}) - $m=\left( P+\dot{P}%
\right) \mathcal{D}_{q+1}$)%
\begin{eqnarray}
\mathbf{\mu \circ \alpha } &:&\left( \mathbb{R}^{q+3+m}\right) ^{\ast
}\rightarrow \mathbb{R}^{q+3+m};  \label{pre-symmm} \\
\left( \mathbf{\mu \circ \alpha }\right) ^{i}(z) &:&=\left( \mathbf{\mu
\circ \alpha }\right) ^{ijklm}z_{j}z_{k}z_{l}z_{m},  \label{pre-symmm2}
\end{eqnarray}%
namely (cfr. the index splitting\footnote{%
The first value of the index $i$ has been denoted with $s$ in the splitting (%
\ref{csii}), whereas it has been denoted with $1$ in the result (\ref%
{mu-alpha-2}). Throughout the following treatment, we will assume $%
z_{1}\equiv z_{s}$.} (\ref{csii}) as well as the result (\ref{mu-alpha-2})) :%
\begin{eqnarray}
i &=&s:\left( \mathbf{\mu \circ \alpha }\right)
^{sjklm}z_{j}z_{k}z_{l}z_{m}:=z_{s}^{2}\left(
-z_{2}^{2}+z_{3}^{2}+...+z_{q+3}^{2}\right) +\frac{1}{16}%
\sum_{K=1}^{r}R_{K}^{2}\left( \hat{z}\right) ; \\
i &=&I:\left( \mathbf{\mu \circ \alpha }\right)
^{Ijklm}z_{j}z_{k}z_{l}z_{m}:=\delta _{s}^{(j|}\delta _{s}^{k}\left( \mathbf{%
\mu \circ \alpha }\right) ^{I|lm)}z_{j}z_{k}z_{l}z_{m}=z_{s}^{2}\left(
\mathbf{\mu \circ \alpha }\right) ^{I~lm}z_{l}z_{m}; \\
i &=&\alpha :\left( \mathbf{\mu \circ \alpha }\right) ^{\alpha
jklm}z_{j}z_{k}z_{l}z_{m}:=\delta _{s}^{(j|}\left( \mathbf{\mu \circ \alpha }%
\right) ^{\alpha |klm)}z_{j}z_{k}z_{l}z_{m}=z_{s}\left( \mathbf{\mu \circ
\alpha }\right) ^{\alpha ~klm}z_{k}z_{l}z_{m},
\end{eqnarray}%
where%
\begin{eqnarray}
\left( \mathbf{\mu \circ \alpha }\right) ^{I~lm}z_{l}z_{m} &:&=2z_{s}\eta
_{II}z_{I+2}+\frac{1}{2}~^{T}\hat{z}\Gamma _{I}\hat{z}=\left\{
\begin{array}{l}
I=0:\left( \mathbf{\mu \circ \alpha }\right) ^{0lm}z_{l}z_{m}:=-2z_{s}z_{2}+%
\frac{1}{2}\sum_{\alpha =1}^{m}z_{q+3+\alpha }^{2}; \\
\\
I=1,...,q+1:\left( \mathbf{\mu \circ \alpha }\right)
^{I~lm}z_{l}z_{m}:=2z_{s}z_{I+2}+\frac{1}{2}~^{T}\hat{z}\Gamma _{I}\hat{z};%
\end{array}%
\right.  \notag \\
&& \\
\left( \mathbf{\mu \circ \alpha }\right) ^{\alpha ~klm}z_{k}z_{l}z_{m}
&:&=z_{s}\left( \sum_{I=0}^{q+1}z_{I+2}\Gamma _{I}\right) \hat{z}-\frac{1}{4}%
\left( \sum_{K=1}^{r}R_{K}\left( \hat{z}\right) \Omega _{K}\right) \hat{z},
\end{eqnarray}%
where $\eta _{II}$ (no sum on the repeated index $I$) denotes the
non-vanishing (diagonal) components of the $\left( q+2\right) $-dimensional
(mostly plus) Lorentzian metric $\eta _{IJ}$ introduced in (\ref{d_ijk}).

Therefore, one can compute the quintic homogeneous polynomial (\ref{jjj3})
in $\partial _{p^{i}}\Delta \equiv \Delta _{i}$ to read%
\begin{eqnarray}
\left( \partial _{p}\Delta \right) \cdot \left( \mathbf{\mu \circ \alpha }%
\right) \left( \partial _{p}\Delta \right) &:&=\left( \mathbf{\mu \circ
\alpha }\right) ^{ijklm}\Delta _{i}\Delta _{j}\Delta _{k}\Delta _{l}\Delta
_{m}=  \notag \\
&=&\Delta _{s}\left( \mathbf{\mu \circ \alpha }\right) ^{sjklm}\Delta
_{j}\Delta _{k}\Delta _{l}\Delta _{m}+\Delta _{I}\left( \mathbf{\mu \circ
\alpha }\right) ^{Ijklm}\Delta _{j}\Delta _{k}\Delta _{l}\Delta _{m}+\Delta
_{\alpha }\left( \mathbf{\mu \circ \alpha }\right) ^{\alpha jklm}\Delta
_{j}\Delta _{k}\Delta _{l}\Delta _{m}  \notag \\
&=&\Delta _{s}\left[ \left( \mathbf{\mu \circ \alpha }\right) ^{sjklm}\Delta
_{j}\Delta _{k}+\Delta _{s}\left( \mathbf{\mu \circ \alpha }\right)
^{Ilm}\Delta _{I}+\left( \mathbf{\mu \circ \alpha }\right) ^{\alpha
klm}\Delta _{\alpha }\Delta _{k}\right] \Delta _{l}\Delta _{m}  \notag \\
&=&\Delta _{s}^{3}\left( -\Delta _{2}^{2}+\Delta _{3}^{2}+...+\Delta
_{q+3}^{2}\right) +\frac{\Delta _{s}}{16}\sum_{K=1}^{r}R_{K}^{2}\left( \hat{%
\Delta}\right)  \notag \\
&&+\Delta _{s}^{2}\sum_{I=0}^{q+1}\Delta _{I}\left( 2\Delta _{s}\eta
_{II}\Delta _{I+2}+\frac{1}{2}~^{T}\hat{\Delta}\Gamma _{I}\hat{\Delta}\right)
\notag \\
&&+\Delta _{s}^{2}\left( \sum_{I=0}^{q+1}\Delta _{I+2}~^{T}\hat{\Delta}%
\Gamma _{I}\hat{\Delta}\right) -\frac{\Delta _{s}}{4}\sum_{K=1}^{r}R_{K}%
\left( \hat{\Delta}\right) ~^{T}\hat{\Delta}\Omega _{K}\hat{\Delta}  \notag
\\
&=&\Delta _{s}^{3}\left( -\Delta _{2}^{2}+\Delta _{3}^{2}+...+\Delta
_{q+3}^{2}\right) +\frac{\Delta _{s}}{16}\sum_{K=1}^{r}R_{K}^{2}\left( \hat{%
\Delta}\right)  \notag \\
&&+2\Delta _{s}^{3}\sum_{I,J=0}^{q+1}\Delta _{I}\eta _{IJ}\Delta _{J+2}+%
\frac{\Delta _{s}^{2}}{2}\sum_{I=0}^{q+1}\Delta _{I}~^{T}\hat{\Delta}\Gamma
_{I}\hat{\Delta}  \notag \\
&&+\Delta _{s}^{2}\sum_{I=0}^{q+1}\Delta _{I+2}~^{T}\hat{\Delta}\Gamma _{I}%
\hat{\Delta}-\frac{\Delta _{s}}{4}\sum_{K=1}^{r}R_{K}^{2}\left( \hat{\Delta}%
\right)  \notag \\
&=&\Delta _{s}^{3}\left[ -\Delta _{2}\left( 2\Delta _{0}+\Delta _{2}\right)
+\Delta _{3}\left( 2\Delta _{1}+\Delta _{3}\right) +...+\Delta _{q+3}\left(
2\Delta _{q+1}+\Delta _{q+3}\right) \right]  \notag \\
&&+\Delta _{s}^{2}\left( \frac{\Delta _{0}}{2}+\Delta _{2}\right) \left(
\Delta _{q+4}^{2}+...+\Delta _{q+3+m}^{2}\right)  \notag \\
&&-\frac{3}{16}\Delta _{s}\sum_{K=1}^{r}R_{K}^{2}\left( \hat{\Delta}\right)
+\Delta _{s}^{2}\sum_{I=1}^{q+1}\left( \frac{\Delta _{I}}{2}+\Delta
_{I+2}\right) Q_{I}\left( \hat{\Delta}\right)  \notag \\
&=&\Delta _{s}^{3}\sum_{I=0}^{q+1}\eta _{II}\Delta _{I+2}\left( 2\Delta
_{I}+\Delta _{I+2}\right) +\Delta _{s}^{2}\left( \frac{\Delta _{0}}{2}%
+\Delta _{2}\right) \sum_{\alpha =1}^{m}\hat{\Delta}_{\alpha }^{2}  \notag \\
&&-\frac{3}{16}\Delta _{s}\sum_{K=1}^{r}R_{K}^{2}\left( \hat{\Delta}\right)
+\Delta _{s}^{2}\sum_{I=1}^{q+1}\left( \frac{\Delta _{I}}{2}+\Delta
_{I+2}\right) Q_{I}\left( \hat{\Delta}\right) ,  \label{ress}
\end{eqnarray}%
where in the last steps we have explicited all sums, and%
\begin{eqnarray}
\hat{\Delta} &:&=~^{T}\left( \hat{\Delta}_{1},...,\hat{\Delta}_{m}\right)
\equiv ~^{T}\left( \Delta _{q+4},...,\Delta _{q+3+m}\right) ; \\
R_{K}\left( \hat{\Delta}\right) &:&=~^{T}\hat{\Delta}\Omega _{K}\hat{\Delta}%
=\sum_{\alpha ,\beta =1}^{m}\left( \Omega _{K}\right) _{\alpha \beta }\hat{%
\Delta}_{\alpha }\hat{\Delta}_{\beta }; \\
Q_{I}\left( \hat{\Delta}\right) &:&=~~^{T}\hat{\Delta}\Gamma _{I}\hat{\Delta}%
=\sum_{\alpha ,\beta =1}^{m}\left( \Gamma _{I}\right) _{\alpha \beta }\hat{%
\Delta}_{\alpha }\hat{\Delta}_{\beta }.
\end{eqnarray}%
We should remark that the following identifications of labels have been
understood throughout:%
\begin{equation}
\begin{array}{cccccccc}
z_{1}, & z_{2}, & z_{3}, & ..., & z_{q+3}, & z_{q+4}, & ... & z_{q+3+m}; \\
\downarrow & \downarrow & \downarrow &  & \downarrow & \downarrow &  &
\downarrow \\
z_{s}, & z_{0}, & z_{1}, & ..., & z_{q+1}, & \hat{z}_{1}, &  & \hat{z}_{m},%
\end{array}%
\end{equation}%
where $z_{0},z_{1},z_{2},...,z_{q+1}=z_{I}$, and $\hat{z}_{1},...,\hat{z}%
_{m}=\hat{z}_{\alpha }\equiv z_{\alpha }$.

Consequently, (\ref{ress}) yields that if%
\begin{gather}
\left( \partial _{p}\Delta \right) \cdot \left( \mathbf{\mu \circ \alpha }%
\right) (\partial _{p}\Delta )>0 \\
\Updownarrow  \notag \\
\Delta _{s}\sum_{I=0}^{q+1}\eta _{II}\Delta _{I+2}\left( 2\Delta _{I}+\Delta
_{I+2}\right) +\left( \frac{\Delta _{0}}{2}+\Delta _{2}\right) \sum_{\alpha
=1}^{m}\hat{\Delta}_{\alpha }^{2}-\frac{3}{16\Delta _{s}}%
\sum_{K=1}^{r}R_{K}^{2}\left( \hat{\Delta}\right) +\sum_{I=1}^{q+1}\left(
\frac{\Delta _{I}}{2}+\Delta _{I+2}\right) Q_{I}\left( \hat{\Delta}\right)
>0,  \notag \\
\end{gather}%
the solution to the system (\ref{BPS}) reads, in vector notation,%
\begin{equation}
\xi =\frac{3}{2\left\vert \Delta _{s}\right\vert }\frac{\left( \mathbf{\mu
\circ \alpha }\right) \left( \partial _{p}\Delta \right) }{\sqrt{\left(
\partial _{p}\Delta \right) \cdot \left( \mathbf{\mu \circ \alpha }\right)
\left( \partial _{p}\Delta \right) }},  \label{sol}
\end{equation}%
or, more explicitly (recall (\ref{csii}))%
\begin{eqnarray}
s &=&\frac{3}{2\left\vert \Delta _{s}\right\vert }\frac{\left[ \Delta
_{s}^{2}\left( -\Delta _{2}^{2}+\Delta _{3}^{2}+...+\Delta _{q+3}^{2}\right)
+\frac{1}{16}\sum_{K=1}^{r}R_{K}^{2}\left( \hat{\Delta}\right) \right] }{%
\sqrt{\left( \partial _{p}\Delta \right) \cdot \left( \mathbf{\mu \circ
\alpha }\right) \left( \partial _{p}\Delta \right) }};  \label{pre-sol} \\
&&  \notag \\
x^{I} &=&\frac{3}{2}\frac{\left\vert \Delta _{s}\right\vert \left( 2\Delta
_{1}\eta _{II}\Delta _{I+2}+\frac{1}{2}~^{T}\hat{\Delta}\Gamma _{I}\hat{%
\Delta}\right) }{\sqrt{\left( \partial _{p}\Delta \right) \cdot \left(
\mathbf{\mu \circ \alpha }\right) \left( \partial _{p}\Delta \right) }};
\label{pre-sol-2} \\
&&  \notag \\
y^{\alpha } &=&\frac{3}{2}\text{sgn}\left( \Delta _{s}\right) \frac{\left[
\Delta _{s}\sum_{I=0}^{q+1}\Delta _{I+2}\sum_{\beta =1}^{m}\left( \Gamma
_{I}\right) _{\alpha \beta }z_{q+3+\beta }-\frac{1}{4}\sum_{K=1}^{r}R_{K}%
\left( \hat{z}\right) \sum_{\beta =1}^{m}\left( \Omega _{K}\right) _{\alpha
\beta }z_{q+3+\beta }\right] }{\sqrt{\left( \partial _{p}\Delta \right)
\cdot \left( \mathbf{\mu \circ \alpha }\right) \left( \partial _{p}\Delta
\right) }},  \label{pre-sol-3}
\end{eqnarray}%
where $\Delta _{i}\equiv \partial _{p^{i}}\Delta $ has been defined in (\ref%
{rec0}) (see also (\ref{Delta}) and (\ref{Delta!})), and all sums have been
made explicit in the numerators; the quantity $\left( \partial _{p}\Delta
\right) \cdot \left( \mathbf{\mu \circ \alpha }\right) \left( \partial
_{p}\Delta \right) $ in the square root in the denominator is given by (\ref%
{ress}).\bigskip

In order to prove that (\ref{sol}) is a solution to the BPS system (\ref{BPS}%
), we compute%
\begin{equation}
\xi \cdot \partial _{p}\Delta =\frac{3}{2}\frac{\sqrt{\left( \partial
_{p}\Delta \right) \cdot \left( \mathbf{\mu \circ \alpha }\right) \left(
\partial _{p}\Delta \right) }}{\left\vert \Delta _{S}\right\vert }=\mathcal{V%
}\left( \xi \right) .  \label{jjj}
\end{equation}%
Thus, by recalling the Euler formula%
\begin{equation}
\mathcal{V}\left( \xi \right) =\frac{1}{3}\nabla _{\mathcal{V}}\left( \xi
\right) \cdot \xi ,
\end{equation}%
one obtains that%
\begin{gather}
0=\xi \cdot \left( \partial _{p}\Delta -\frac{1}{3}\nabla _{\mathcal{V}%
}\left( \xi \right) \right) \quad \forall \xi ,\;\forall \partial _{p}\Delta
; \\
\Updownarrow  \notag \\
\partial _{p}\Delta =\frac{1}{3}\nabla _{\mathcal{V}}\left( \xi \right) =%
\frac{1}{3!}d_{ijk}x^{j}x^{k},
\end{gather}%
which is the BPS system (\ref{BPS}) itself.\bigskip

Thus, in all models explicitly treated below, after the checking that the
invertibility condition enounced in Sec.\ \ref{main} holds true, the crucial
data to be known are the symmetric $m\times m$ $\Gamma $-matrices and the
symmetric, auxiliary $m\times m$ matrices $\Omega _{K}$ such that (\ref%
{GOcom}) and (\ref{lorrel}) both hold true.

\subsubsection{The symmetric case\label{symmetric}}

It should be remarked that those homogeneous $d$-spaces \cite{dWVP,dWVVP} in
which the invertibility condition enounced in Sec.\ \ref{main} is satisfied%
\textbf{\ }include the noteworthy class of homogeneous \textit{symmetric} $d$%
-spaces (cfr. Sec.\ \ref{symmd}), in which it holds that%
\begin{equation}
\left( \mathbf{\mu \circ \alpha }\right) ^{i}\left( \partial _{p}\Delta
\right) =\left( \mathbf{\mu \circ \alpha }\right) ^{ijklm}\Delta _{j}\Delta
_{k}\Delta _{l}\Delta _{m}=2\Delta _{S}^{2}d^{ijk}\Delta _{j}\Delta _{k},
\label{symmm}
\end{equation}%
where $d^{ijk}$ is defined in (\ref{d^ijk})-(\ref{jjjj}), and in the case of
symmetric $d$-spaces it is a constant (numerical) tensor (i.e., it does not
depend on any scalar fields' degree of freedom). (\ref{symmm}) implies that%
\begin{gather}
\left( \partial _{p}\Delta \right) \cdot \left( \mathbf{\mu \circ \alpha }%
\right) \left( \partial _{p}\Delta \right) =\left( \mathbf{\mu \circ \alpha }%
\right) ^{ijklm}\Delta _{i}\Delta _{j}\Delta _{k}\Delta _{l}\Delta
_{m}=2\Delta _{S}^{2}d^{ijk}\Delta _{i}\Delta _{j}\Delta _{k};
\label{symmm2} \\
\Downarrow  \notag \\
\left( \partial _{p}\Delta \right) \cdot \left( \mathbf{\mu \circ \alpha }%
\right) \left( \partial _{p}\Delta \right) >0\quad \Leftrightarrow \quad
d^{ijk}\Delta _{i}\Delta _{j}\Delta _{k}>0.
\end{gather}%
Indeed, by plugging (\ref{symmm2}) into (\ref{sol}), one obtains%
\begin{equation}
\xi ^{i}=\frac{3}{2\left\vert \Delta _{s}\right\vert }\frac{\left( \mathbf{%
\mu \circ \alpha }\right) ^{i}\left( \partial _{p}\Delta \right) }{\sqrt{%
\left( \partial _{p}\Delta \right) \cdot \left( \mathbf{\mu \circ \alpha }%
\right) \left( \partial _{p}\Delta \right) }}=\frac{3}{\sqrt{2}}\frac{%
d^{ijk}\Delta _{j}\Delta _{k}}{\sqrt{d^{mnp}\Delta _{m}\Delta _{n}\Delta _{p}%
}},  \label{expl-attr}
\end{equation}%
thus matching (\ref{BPS-symm-2}) (recalling (\ref{csii})).

\subsection{\label{BPS-S-gen}BPS black hole entropy and attractors}

Let us recall that $\xi $ enters the expression (\ref{S-BPS}) of the black
hole entropy only through the quantity (\ref{expr1}), and (\ref{jjj}) holds
true. Remarkably, by virtue of (\ref{expr1}), the solutions of the BPS
system (\ref{BPS}) enter the expression of the black hole entropy only
through the square of the quantity (\ref{jjj}) :%
\begin{equation}
\left( \xi \cdot \partial _{p}\Delta \right) ^{2}=\mathcal{V}^{2}(\xi )=%
\frac{9}{4}\frac{\left( \partial _{p}\Delta \right) \cdot \left( \mathbf{\mu
\circ \alpha }\right) \left( \partial _{p}\Delta \right) }{\Delta _{s}^{2}}.
\label{qq-bis}
\end{equation}%
Then, by recalling (\ref{S-BPS}), the Bekenstein-Hawking entropy of static,
spherically symmetric, BPS extremal dyonic black holes in the model under
consideration of $\mathcal{N}=2$, $D=4$ Maxwell-Einstein supergravity has
the following expression :%
\begin{equation}
\frac{S}{\pi }=\frac{1}{3\left\vert p^{0}\right\vert }\sqrt{3\frac{\left(
\partial _{p}\Delta \right) \cdot \left( \mathbf{\mu \circ \alpha }\right)
\left( \partial _{p}\Delta \right) }{\Delta _{s}^{2}}-9\left[ p^{0}\left(
p\cdot q\right) -2I_{3}(p)\right] ^{2}}.  \label{SSd}
\end{equation}%
In both formul\ae\ (\ref{qq-bis}) and (\ref{SSd}) the scalar product $\left(
\partial _{p}\Delta \right) \cdot \left( \mathbf{\mu \circ \alpha }\right)
\left( \partial _{p}\Delta \right) $ is given by (\ref{ress}). Note that, as
it must be, $S$ (\ref{SSd}) is a homogeneous positive function of degree $2$
in the black hole charges. The consistency conditions for (\ref{qq-bis}) and
the BPS black hole entropy (\ref{SSd}) to hold formally read as follows :%
\begin{gather}
\left\{
\begin{array}{l}
\left( \partial _{p}\Delta \right) \cdot \left( \mathbf{\mu \circ \alpha }%
\right) \left( \partial _{p}\Delta \right) >0; \\
\\
\left( \partial _{p}\Delta \right) \cdot \left( \mathbf{\mu \circ \alpha }%
\right) \left( \partial _{p}\Delta \right) -3\left[ p^{0}\left( p\cdot
q\right) -2I_{3}(p)\right] ^{2}\Delta _{s}^{2}>0;%
\end{array}%
\right. \\
\Downarrow  \notag \\
\left( \partial _{p}\Delta \right) \cdot \left( \mathbf{\mu \circ \alpha }%
\right) \left( \partial _{p}\Delta \right) >3\left[ p^{0}\left( p\cdot
q\right) -2I_{3}(p)\right] ^{2}\Delta _{s}^{2},
\end{gather}%
where again the scalar product $\left( \partial _{p}\Delta \right) \cdot
\left( \mathbf{\mu \circ \alpha }\right) \left( \partial _{p}\Delta \right) $
is given by (\ref{ress}).

By exploiting the results (\ref{solsol}) and (\ref{qq-bis}) and defining%
\begin{eqnarray}
\mathcal{Q} &=&~^{T}\left( p^{0},p^{i},q_{0},q_{i}\right) ; \\
z_{H}(\mathcal{Q}) &\equiv &\left\{ z_{H}^{i}\left( \mathcal{Q}\right)
\right\} _{i}:=~^{T}\left( z_{H}^{s}(\mathcal{Q}),z_{H}^{I}(\mathcal{Q}%
),z_{H}^{\alpha }(\mathcal{Q})\right) ; \\
p &\equiv &\left\{ p^{i}\right\} _{i}:=~^{T}\left( p^{s},p^{I},p^{\alpha
}\right) ,
\end{eqnarray}%
the expression of BPS attractor points (\ref{BPS-crit-points-2}) is given,
in vector notation, by%
\begin{equation}
z_{H}(\mathcal{Q})=\frac{3}{2}\frac{\left( \mathbf{\mu \circ \alpha }\right)
\left( \partial _{p}\Delta \right) }{\left( \partial _{p}\Delta \right)
\cdot \left( \mathbf{\mu \circ \alpha }\right) \left( \partial _{p}\Delta
\right) }\left[ \frac{p^{0}\left( p\cdot q\right) -2I_{3}(p)}{p^{0}}-\mathbf{%
i}\frac{3}{2}\frac{S}{\pi }\right] +\frac{p}{p^{0}},  \label{zH}
\end{equation}%
or, more explicitly :%
\begin{eqnarray}
z_{H}^{s}(\mathcal{Q}) &=&\frac{3}{2}\frac{\left[ \Delta _{s}^{2}\left(
-\Delta _{2}^{2}+\Delta _{3}^{2}+...+\Delta _{q+3}^{2}\right) +\frac{1}{16}%
\sum_{K=1}^{r}R_{K}^{2}\left( \hat{\Delta}\right) \right] }{\left( \partial
_{p}\Delta \right) \cdot \left( \mathbf{\mu \circ \alpha }\right) \left(
\partial _{p}\Delta \right) }\left[ \frac{p^{0}\left( p\cdot q\right)
-2I_{3}(p)}{p^{0}}-\mathbf{i}\frac{3}{2}\frac{S}{\pi }\right] +\frac{p^{s}}{%
p^{0}};  \notag \\
&&  \label{zH-1} \\
z_{H}^{I}(\mathcal{Q}) &=&\frac{3}{2}\frac{\Delta _{s}^{2}\left( 2\Delta
_{s}\eta _{II}\Delta _{I+2}+\frac{1}{2}~^{T}\hat{z}\Gamma _{I}\hat{z}\right)
}{\left( \partial _{p}\Delta \right) \cdot \left( \mathbf{\mu \circ \alpha }%
\right) \left( \partial _{p}\Delta \right) }\left[ \frac{p^{0}\left( p\cdot
q\right) -2I_{3}(p)}{p^{0}}-\mathbf{i}\frac{3}{2}\frac{S}{\pi }\right] +%
\frac{p^{I}}{p^{0}};  \notag \\
&&  \label{zH-2} \\
z_{H}^{\alpha }(\mathcal{Q}) &=&\frac{3}{2}\frac{\left[ \Delta
_{s}\sum_{I=0}^{q+1}\Delta _{I+2}\sum_{\beta =1}^{m}\left( \Gamma
_{I}\right) _{\alpha \beta }\Delta _{q+3+\beta }-\frac{1}{4}%
\sum_{K=1}^{r}R_{K}\left( \hat{\Delta}\right) \sum_{\beta =1}^{m}\left(
\Omega _{K}\right) _{\alpha \beta }\Delta _{q+3+\beta }\right] }{\left(
\partial _{p}\Delta \right) \cdot \left( \mathbf{\mu \circ \alpha }\right)
\left( \partial _{p}\Delta \right) }\cdot  \notag \\
&&\cdot \left[ \frac{p^{0}\left( p\cdot q\right) -2I_{3}(p)}{p^{0}}-\mathbf{i%
}\frac{3}{2}\frac{S}{\pi }\right] +\frac{p^{\alpha }}{p^{0}},  \label{zH-3}
\end{eqnarray}%
where all sums are explicitly indicated in the numerator of (\ref{zH-3}),
and $\left( \partial _{p}\Delta \right) \cdot \left( \mathbf{\mu \circ
\alpha }\right) \left( \partial _{p}\Delta \right) $ and $S$ are
respectively given by (\ref{ress}) and (\ref{SSd}).

Again, by considering the general formula (\ref{SSd}) in homogeneous
symmetric models in which the results discussed in Sec.\ \ref{symmetric} as
well as the adjoint identity (\ref{adj-id}) hold true, it can be checked
that the entropy acquires the simple form given by (\ref{S-BPS-symm}) (see
discussion in Sec.\ \ref{symmd}).

\subsection{Complete models: $r=0$\label{complete}}

The classification, completed in \cite{dWVP} (see also \cite{dWVVP}), shows
that any $L(q,1,0)\equiv L(q,1),$ $q\geqslant -1$ model of homogeneous very
special geometry is defined by $\Gamma _{0}={\mathbb{I}}_{m}$ and a Clifford
set $\Gamma $-matrices, of size $m\times m$, namely by $\{\Gamma _{1},\ldots
,\Gamma _{q+1}\}$, where $m=\mathcal{D}_{q+1}$ (given e.g. in Table 1 of
\cite{dWVVP} and in (\ref{table D})). For fixed $q$, such matrices are
unique (up to a choice of basis in $\mathbb{R}^{m}$). We will choose in
Sec.\ \ref{L810} a Clifford set $\left\{ \Gamma _{I}\right\} _{I=1,...,9}$
of symmetric $\Gamma $-matrices. The associated quadratic forms $\left\{
Q_{I}\right\} _{I\neq 0}$'s have the additional pleasant property that
putting the last $m/2$ coordinate $y^{\alpha }$'s equal to zero, some
quadrics vanish identically, whereas the non-vanishing ones are the quadrics
associated to an $L(q^{\prime },1)$ model with $q^{\prime }<8$.

A \textit{complete} model is defined to be a model in which $r=0$ in the
invertibility condition enounced in Sec.\ \ref{main}. Then, (\ref{lorrel})
implies that the associated quadratic forms satisfy a Lorentzian quadratic
relation%
\begin{equation}
-Q_{0}^{2}\left( y\right) +Q_{1}^{2}\left( y\right) +\ldots
+Q_{q+1}^{2}\left( y\right) =0\Leftrightarrow q\left( Q(y)\right) =0,
\label{lorrel-r=0}
\end{equation}%
where (\ref{qqqq}) has been recalled. Defining the associated cubic form $%
\mathcal{V}$ as in (\ref{cubb}), the main result then states that the
gradient map $\nabla _{\mathcal{V}}$ will be invertible, with the
(birational) inverse map being polynomial of degree $2$.

Indeed, since for $r=0$ all coordinate functions of the map $\mathbf{\alpha }
$ are multiples of $z_{1}$ and $\mathbf{\mu }$ is homogeneous, so, as
mentioned above, one can redefine $\mathbf{\alpha }$ to be the identity map.
As a consequence, $\mathbf{\mu }$, which is given by quadratic polynomials,
is the birational inverse of $\nabla _{\mathcal{V}}$; cfr. (\ref{complete!}%
), which 
can be regarded as a consequence of the so-called \textquotedblleft adjoint
identity" (\ref{adj-id}) of cubic Jordan algebras.

The only\footnote{%
Actually, (\ref{adj-id}) holds also for $J_{3}=\mathbb{R}$ (corresponding to
the $T^{3}$ model of $\mathcal{N}=2$, $D=4$ supergravity) as well as for
semi-simple cubic Jordan algebras (named \textquotedblleft spin factors") $%
\mathbb{R}\oplus \mathbf{\Gamma }_{a,b}$ \cite{JVNW}. Therefore, strictly
speaking, also these models should be complete.} complete models known to us
have $q=1,2,4,8$ and $m=2q$, so $N=3q+3=6,9,15,27$ respectively; such models
have been discussed at the end of Sec.\ \ref{Lq1}, and they will be further
discussed below. In these models, which correspond to the \textquotedblleft
magic" class of symmetric $d$-manifolds \cite{GST1, GST2, GST3}, the cubic
forms are the well known norm forms on simple cubic Jordan algebras $J_{3}^{%
\mathbb{A}}$, for $\mathbb{A}=\mathbb{R},\mathbb{C},\mathbb{H},\mathbb{O}$
\cite{Russo}. These complete models correspond to the models $L(q,1)$ \cite%
{dWVP, dWVVP} with $q=1,2,4,8=\dim _{\mathbb{R}}\mathbb{A}$ for $\mathbb{A}=%
\mathbb{R},\mathbb{C},\mathbb{H},\mathbb{O}$ respectively, provided that the
$\left( q+2\right) $-dimensional vector has a (mostly plus) Lorentzian
signature $(1_{-},\left( q+1\right) _{+})$, which can always be arranged.

As we will see in Sec.\ \ref{55}, also quadratic forms in $q+2=4,6,10$
dimensions of Kleinian signatures $(2_{+},2_{-})$, $(3_{+},3_{-})$ and $%
(5_{+},5_{-})$ can be considered: they are associated to simple cubic Jordan
algebras over split composition algebras $J_{3}^{\mathbb{A}_{s}}$, for $%
\mathbb{A}=\mathbb{C}_{s},\mathbb{H}_{s},\mathbb{O}_{s}$. Moreover, they
correspond to non-supersymmetric Maxwell-Einstein theories (for $\mathbb{C}%
_{s}$ and $\mathbb{H}_{s}$), as well as to maximal supergravity (in the case
of $\mathbb{O}_{s}$); cfr. \cite{Marrani-Romano-1,Marrani-Romano-2}.

\section{Verifying the inverse map\label{checkinvmap}}

In this section we prove that, if the condition enounced in Sec.\ \ref{main}
is satisfied, then the formulas in Sec.\ \ref{explin} indeed provide the
(birational) inverse map of the gradient map of the cubic form ${\mathcal{V}}
$ under consideration.

\subsection{Factorization of $R_{K}$'s\label{propR}}

First of all, we derive a useful property of the quadratic polynomials $%
R_{K} $'s, defined by the extra symmetric matrices $\Omega _{K}$, with $%
1\leqslant {K}\leqslant r$ (with $r\geqslant 0$), of size $m$.

The last $m=2^{g}$ coordinate functions of $\nabla _{\mathcal{V}}$ are
denoted by ($y=~^{T}(y^{1},\ldots ,y^{m})$)
\begin{equation}
\nabla _{y}\mathcal{V}\,:=\,^{T}(\frac{\partial \mathcal{V}}{\partial y^{1}}%
,\ldots ,\frac{\partial \mathcal{V}}{\partial y^{m}})=\,2{}^{T}y\left(
\sum_{I=0}^{q+1}x^{I}\Gamma _{I}\right) .  \label{nablayN}
\end{equation}

We show that, upon substituting $\nabla _{y}\mathcal{V}$ into $R_{K}$ (\ref%
{R_k}), the following factorization holds :
\begin{equation}
R_{K}(\nabla _{y}\mathcal{V})\,=\,-4q(x)R_{K}(y).  \label{factorr}
\end{equation}%
In fact, by definition, one has:
\begin{equation}
R_{K}(\nabla _{y}\mathcal{V})\,=\,4{}^{T}y\left(
\sum_{I=0}^{q+1}x^{I}~\Gamma _{I}\right) \cdot \Omega _{K}\cdot \left(
\sum_{J=0}^{q+1}x^{J}\Gamma _{J}\right) y\,=\,4{}^{T}y\left(
\sum_{I,J=0}^{q+1}x^{I}x^{J}(\Gamma _{I}\Omega _{K}\Gamma _{J})\right) y~.
\label{onee}
\end{equation}

Recall that $\Gamma _{0}={\mathbb{I}}_{m}$ and that from (\ref{GOcom}) we
have $\Gamma _{I}\Omega _{K}=-\Omega _{K}\Gamma _{I}$ for $I=1,\ldots ,q+1$.
Therefore, for $I>0$ it holds that
\begin{equation}
\Gamma _{0}\Omega _{K}\Gamma _{0}\,=\,\Omega _{K},\qquad \Gamma _{0}\Omega
_{K}\Gamma _{I}\,=\,-\Gamma _{I}\Omega _{K}\,=\,-\Gamma _{I}\Omega
_{K}\Gamma _{0}~.
\end{equation}%
If $I,J>0$ and $I\neq J$, we have $\Gamma _{I}\Gamma _{J}=-\Gamma _{J}\Gamma
_{I}$ since we have a Clifford set, and thus
\begin{equation}
\Gamma _{I}\Omega _{K}\Gamma _{J}\,=\,-\Gamma _{I}\Gamma _{J}\Omega
_{K}\,=\,\Gamma _{J}\Gamma _{I}\Omega _{K}\,=\,-\Gamma _{J}\Omega _{K}\Gamma
_{I}~.
\end{equation}%
Furthermore, when $I=J>0$ we have $\Gamma _{I}\Gamma _{J}=\Gamma _{I}^{2}={%
\mathbb{I}}_{m}$, and thus
\begin{equation}
\Gamma _{I}\Omega _{K}\Gamma _{J}\,=\,-\Gamma _{I}\Gamma _{J}\Omega
_{K}\,=\,-\Omega _{K}~.
\end{equation}%
Therefore, all terms with $I\neq J$ in (\ref{onee}) cancel, and we are left
with
\begin{eqnarray}
R_{K}(\nabla _{y}\mathcal{V})\, &=&\,4{}^{T}y\left( \left( x^{0}\right)
^{2}\,\Omega _{K}-\,\sum_{I=1}^{q+1}(x^{I})^{2}\Omega _{K}\right) y  \notag
\\
&=&4\left( \left( x^{0}\right) ^{2}-\left( \left( x^{1}\right) ^{2}+\ldots
\,\left( x^{q+1}\right) ^{2}\right) \right) {}^{T}y\Omega
_{K}y=-4q(x)R_{K}(y)~,
\end{eqnarray}%
which proves (\ref{factorr}).

\subsection{The map $\mathbf{\protect\alpha \circ }\protect\nabla _{\mathcal{%
V}}$\label{mapa}}

For $\xi =(s,x,y)\in \mathbb{R}^{q+3+m}$ (cfr. (\ref{csii})), we now verify
that the image $(t,u,v,w)=\mathbf{\alpha }(\nabla _{\mathcal{V}}(\xi )) $ in
$\mathbb{R}^{q+3+m+r}$ is given by (\ref{alphaphi}).

The first $q+3+m$ components of $\mathbf{\alpha }(z)$ are the $z_{1}z_{a}$%
's, with $a=1,\ldots ,q+3+m$. Since the first component of $\nabla _{%
\mathcal{V}}$ is $\partial {\mathcal{V}}/\partial s=q(x)$, we see that the
first $q+3+m$ components of $\mathbf{\alpha }(\nabla _{\mathcal{V}}(\xi )):=%
\mathbf{\alpha \circ }\nabla _{\mathcal{V}}\left( \xi \right) $ are $%
q(x)\nabla _{\mathcal{V}}(\xi )$.

The last $r$ components of $\mathbf{\alpha }\left( z\right) $ are the $R_{K}$%
's, evaluated on the last $m$ variables. Hence, the last $r$ components of $%
\mathbf{\alpha \circ }\nabla _{\mathcal{V}}\left( \xi \right) $ are the $%
R_{K}$'s evaluated on $m$-vector $\nabla _{y}{\mathcal{V}}$. From (\ref%
{factorr}) we see that these components are $-4q(x)R_{K}(y)$, $K=1,...,r$.

Thus, all components of $\mathbf{\alpha \circ }\nabla _{\mathcal{V}}\left(
\xi \right) $ are stated as in (\ref{alphaphi}). For later convenience, we
write these components explicitly, with $I=1,\ldots ,q+1$, $\alpha =1,\ldots
,m$ and $K=1,\ldots ,r$ :
\begin{eqnarray}
t\, &=&\,q(x)^{2};  \label{ress-1} \\
u\,_{0} &=&\,q(x)\left( -2x^{0}s\,+\,Q_{0}(y)\right) ; \\
u\,_{I} &=&\,q(x)\left( 2x^{I}s\,+\,Q_{I}(y)\right) ; \\
v\,_{\alpha } &=&\,2q(x)\left( \sum_{I=0}^{q+1}x^{I}\Gamma _{I}\right)
_{\alpha \beta }y^{\beta }; \\
w\,_{K} &=&-4\,q(x)R_{K}(y)~.  \label{ress-4}
\end{eqnarray}

\subsection{The map $\mathbf{\protect\mu \circ \protect\alpha \circ }\protect%
\nabla _{\mathcal{V}}$\label{mueva}}

We now aim at proving (\ref{jj}), which in particular implies that the
composition $\mathbf{\mu }\circ \mathbf{\alpha }$ is the birational inverse
of the gradient map $\nabla _{\mathcal{V}}$.

To do so, we first compute $\mathbf{\mu }(\mathbf{\alpha }(\nabla _{\mathcal{%
V}}(\xi )))$ with $\xi =~^{T}(s,x,y)$ (cfr. (\ref{csii})). We already
computed $(\mathbf{\alpha }(\nabla _{\mathcal{V}}(\xi ))$ in (\ref{alphaphi}%
) and (\ref{ress-1})-(\ref{ress-4}), so it remains to evaluate the map $%
\mathbf{\mu }$, defined in (\ref{mu}), on $^{T}(t,u,v,w)=\mathbf{\alpha }%
(\nabla _{\mathcal{V}}(\xi ))$.

From (\ref{mu}), the first component of $\mathbf{\mu }$ is $q(u)\,+\,\frac{1%
}{16}\sum_{K=1}^{r}w_{K}^{2}$, and it holds that%
\begin{equation}
\frac{1}{16}\sum_{K=1}^{r}w_{K}^{2}=\frac{1}{16}%
\sum_{K=1}^{r}(-4q(x)R_{K}(y))^{2}\,=\,q(x)^{2}\sum_{K=1}^{r}R_{K}(y)^{2}.
\label{ress-5}
\end{equation}%
Next, from (\ref{mu}) the second component of $\mathbf{\mu }$ is $%
-2tu_{0}\,+\,\frac{1}{2}Q_{0}(v)$, and it holds that
\begin{equation}
2tu_{0}\,=\,-2q(x)^{3}(-2sx^{0}+Q_{0}(y))\,=\,2q(x)^{3}(2sx^{0}-Q_{0}(y))~.
\end{equation}%
Therefore, the next $J$ components of $\mathbf{\mu }$ ($J=1,...,q+1$) are,
from (\ref{mu}), $2tu_{J}+\frac{1}{2}Q_{J}(v)$, and it holds that
\begin{equation}
2tu_{J}\,=\,2q(x)^{3}(2sx^{J}+Q_{J}(y)).  \label{ress-6}
\end{equation}

Therefore, from (\ref{mu}), (\ref{ress-1})-(\ref{ress-4}) and (\ref{ress-5}%
)-(\ref{ress-6}), one obtains%
\begin{eqnarray}
(\mathbf{\mu }\circ \mathbf{\alpha }\circ \nabla _{\mathcal{V}})(s,x,y)\,
&=&\,q(x)^{2}\left( {\renewcommand{\arraystretch}{1.3}%
\begin{array}{l}
q\left(
-2sx^{0}\,+\,Q_{0}(y),2sx^{1}+Q_{1}(y),...,2sx^{q+1}\,+\,Q_{q+1}(y)\right)
+\sum_{K=1}^{r}R_{K}(y)^{2} \\
2q(x)\left( 2sx^{0}-Q_{0}(y)\right) +\,\frac{1}{2}Q_{0}(\nabla _{y}{\mathcal{%
V}}) \\
2q(x)(2sx^{1}+Q_{1}(y))\,+\,\frac{1}{2}Q_{1}(\nabla _{y}{\mathcal{V}}) \\
\quad \vdots \\
2q(x)(2sx^{q+1}+Q_{q+1}(y))\,+\,\frac{1}{2}Q_{q+1}(\nabla _{y}{\mathcal{V}})
\\
\left( \sum_{I=0}^{q+1}{\mathcal{V}}_{I}\Gamma _{I}\right) \nabla _{y}{%
\mathcal{V}}\,+\,(\sum_{K=1}^{r}R_{K}(y)\Omega _{K})\nabla _{y}{\mathcal{V}}%
\end{array}%
}\right) .  \notag \\
&&  \label{maf}
\end{eqnarray}

To verify (\ref{jj}), the treatment can be split in three parts,
respectively concerning the first, $s$-component, the middle $q+2$
components and the last $m$ components of $\mathbf{\mu }(\mathbf{\alpha }%
(\nabla _{\mathcal{V}}(s,x,y)))\equiv $ $\left( \mathbf{\mu \circ \alpha
\circ }\nabla _{\mathcal{V}}\right) \left( s,x,y\right) \equiv \mathbf{\mu
\circ \alpha \circ }\nabla _{\mathcal{V}}\left( \xi \right) $:

\begin{enumerate}
\item For what concerns the first, $s$- component of the map $\left( \mathbf{%
\mu \circ \alpha \circ }\nabla _{\mathcal{V}}\right) \left( s,x,y\right) $,
one needs to show that
\begin{equation}
4{\mathcal{V}}s\,=\,q\left(
-2sx^{0}\,+\,Q_{0}(y),2sx^{1}+Q_{1}(y),...,2sx^{q+1}\,+\,Q_{q+1}(y)\right)
+\,\sum_{K=1}^{r}R_{K}(y)^{2}.  \label{one}
\end{equation}%
We start and consider {\renewcommand{\arraystretch}{1.5}}%
\begin{eqnarray}
&&q(-2sx^{0}+Q_{0}(y),2sx^{1}+Q_{1}(y),\ldots
,2sx^{q+1}+Q_{q+1}(y))=-(-2sx^{0}+Q_{0}(y))^{2}\,+\,%
\sum_{I=1}^{q+1}(2sx^{I}+Q_{I}(y))^{2}  \notag \\
&=&4s^{2}q(x)\,+\,4s\sum_{I=0}^{q+1}x^{I}Q_{I}(y)\,-\,Q_{0}(y)^{2}\,+\,%
\sum_{I=1}^{q+1}Q_{I}(y)^{2}=4\mathcal{V}s\,-\,Q_{0}(y)^{2}\,+\,%
\sum_{I=1}^{q+1}Q_{I}(y)^{2}~.  \label{one-2}
\end{eqnarray}%
{\ }
Using the identity (\ref{lorrel}), one finds
\begin{equation}
\sum_{K=1}^{r}R_{K}(y)^{2}\,=\,Q_{0}(y)^{2}-\sum_{I=1}^{q+1}Q_{I}(y)^{2}~,
\end{equation}%
and thus (\ref{one-2}) proves (\ref{one}).

\item For the next $q+2$ components
of the map $\left( \mathbf{\mu \circ \alpha \circ }\nabla _{\mathcal{V}%
}\right) \left( s,x,y\right) $, one needs to prove that (cfr.\ (\ref{maf})):%
\begin{equation}
\left\{
\begin{array}{l}
4{\mathcal{V}}x^{0}=2q(x)(2sx^{0}-Q_{0}(y))\,+\,\frac{1}{2}Q_{0}(\nabla _{y}{%
\mathcal{V}}), \\
\\
4{\mathcal{V}}x^{I}=2q(x)(2sx^{I}+Q_{I}(y))\,+\,\frac{1}{2}Q_{I}(\nabla _{y}{%
\mathcal{V}}),\quad I=1,\ldots ,q+1~.%
\end{array}%
\right.  \label{mume}
\end{equation}%
We claim that {\renewcommand{\arraystretch}{1.3}
\begin{equation}
\frac{1}{2}Q_{I}(\nabla _{y}\mathcal{V})\,=\,\left\{
\begin{array}{rl}
2q(x)Q_{0}(y)\,+\,4x^{0}\sum_{J=0}^{q+1}x^{J}Q_{J}(y) & \quad \mbox{if}\quad
I=0; \\
-2q(x)Q_{I}(y)\,+\,4x^{I}\sum_{J=0}^{q+1}x^{J}Q_{J}(y) & \quad \mbox{if}%
\quad I=1,...,q+1~.%
\end{array}%
\right.  \label{claim2}
\end{equation}%
}
Indeed, it holds that
\begin{equation}
\frac{1}{2}Q_{I}(\nabla _{y}\mathcal{V})\,=\,\frac{1}{2}Q_{I}\left(
2{}^{T}y\left( \sum_{J=0}^{q+1}x^{J}\Gamma _{J}\right) \right)
=\,2{}^{T}y\left( \sum_{J=0}^{q+1}x^{J}\Gamma _{J}\right) \Gamma _{I}\left(
\sum_{J^{\prime }=0}^{q+1}x^{J^{\prime }}\Gamma _{J^{\prime }}\right) y~.
\label{gm}
\end{equation}%
If $I=0$, then $\Gamma _{0}={\mathbb{I}}_{m}$ and thus (\ref{gm}), combined
with (\ref{clifquadrel}), leads to%
\begin{eqnarray}
\frac{1}{2}Q_{0}(\nabla _{y}\mathcal{V}) &=&2{}^{T}y\left(
\sum_{J=0}^{q+1}x^{J}\Gamma _{J}\right) \Gamma _{0}\left( \sum_{J^{\prime
}=0}^{q+1}x^{J^{\prime }}\Gamma _{J^{\prime }}\right) y  \notag \\
&=&2{}^{T}y\left( x^{0}{\mathbb{I}}_{m}+\sum_{J=1}^{q+1}x^{J}\Gamma
_{J}\right) \left( 2x^{0}{\mathbb{I}}\,_{m}-x^{0}{\mathbb{I}}%
_{m}+\sum_{J^{\prime }=1}^{q+1}x^{J^{\prime }}\Gamma _{J^{\prime }}\right) y
\notag \\
&=&4x^{0}{}^{T}y\left( \sum_{J=0}^{q+1}x^{J}\Gamma _{J}\right) y\,+\,2\left(
-\left( x^{0}\right) ^{2}+\left( x^{1}\right) ^{2}+\ldots +\left(
x^{q+1}\right) ^{2}\right) ^{T}y{\mathbb{I}}_{m}y  \notag \\
&=&4x^{0}\sum_{J=0}^{q+1}x^{J}Q_{J}(y)+2q(x)Q_{0}(y).
\end{eqnarray}

On the other hand, if $I=1,...,q+1$ then $\Gamma _{I}$ and $\Gamma _{J}$
commute only for $J=0$ or $J=I$, and they anti-commute otherwise;
consequently, by moving $\Gamma _{I}$ to the right, it follows that%
\begin{eqnarray}
\frac{1}{2}Q_{I}(\nabla _{y}\mathcal{V}) &=&2{}^{T}y\left(
\sum_{J=0}^{q+1}x^{J}\Gamma _{J}\right) \Gamma _{I}\left( \sum_{J^{\prime
}=0}^{q+1}x^{J^{\prime }}\Gamma _{J^{\prime }}\right) y  \notag \\
&=&2{}^{T}y\left( x^{0}{\mathbb{I}}_{m}+\sum_{J=1}^{q+1}x^{J}\Gamma
_{J}\right) \left( x^{0}{\mathbb{I}}_{m}-\left( \sum_{J^{\prime
}=1}^{q+1}x^{J^{\prime }}\Gamma _{J^{\prime }}\right) \,+\,2x^{I}\Gamma
_{I}\right) \Gamma _{I}y  \notag \\
&=&2\left[ \left( x^{0}\right) ^{2}-(\left( x^{1}\right) ^{2}+\ldots +\left(
x^{q+1}\right) ^{2})\right] ~^{T}y\Gamma _{I}y\,+\,4x^{I}~^{T}y\left(
\sum_{J=0}^{q+1}x^{J}\Gamma _{J}\right) \Gamma _{I}^{2}y  \notag \\
&=&-2q(x)Q_{I}(y)\,+\,4x^{I}\sum_{J=0}^{q+1}x^{J}Q_{J}(y)~.
\end{eqnarray}%
Hence, the claim (\ref{claim2}) is proven. Now, we show that the equalities
in (\ref{mume}) follow from the claim (\ref{claim2}). In fact, by recalling (%
\ref{cubb}) for $I=0$ one obtains%
\begin{eqnarray}
&&2q(x)(2sx^{0}-Q_{0}(y))\,+\,\frac{1}{2}Q_{0}(\nabla _{y}{\mathcal{V}})
\notag \\
&=&4sq(x)x^{0}-\,2q(x)Q_{0}(y)+\,2q(x)Q_{0}(y)\,+\,4x^{0}%
\sum_{J=0}^{q+1}x^{J}Q_{J}(y)=4x^{0}{\mathcal{V}}.
\end{eqnarray}%
Analogously, for $I=1,...,q+1$ it holds that%
\begin{eqnarray}
&&2q(x)(2x^{I}+Q_{I}(y))\,+\,\frac{1}{2}Q_{I}(\nabla _{y}{\mathcal{V}})
\notag \\
&=&4sq(x)x^{I}+\,2q(x)Q_{I}(y)\,-2q(x)Q_{I}(y)\,+\,4x^{I}%
\sum_{J=0}^{q+1}x^{J}Q_{J}(y)=4x^{I}{\mathcal{V}}.
\end{eqnarray}%
This concludes the verification for the $q+2$ components under consideration.

\item Finally, for the last $m$ components of $\left( \mathbf{\mu \circ
\alpha \circ }\nabla _{\mathcal{V}}\right) \left( s,x,y\right) $ one must
check that
\begin{equation}
4{\mathcal{V}}y\,=\,\left( \sum_{I=0}^{q+1}{\mathcal{V}}_{I}\Gamma
_{I}\,+\,\sum_{K=1}^{r}R_{K}(y)\Omega _{K}\right) \nabla _{y}{\mathcal{V}}~.
\label{mulast}
\end{equation}%
First of all, we substitute ${\mathcal{V}}_{0}=-2sx^{0}+Q_{0}(y)$ and ${%
\mathcal{V}}_{I}=2sx^{I}+Q_{I}(y)$ for $I=1,\ldots ,q+1$, in the r.h.s. of (%
\ref{mulast}), obtaining
\begin{equation}
\sum_{I=0}^{q+1}{\mathcal{V}}_{I}\Gamma _{I}\,\,=\,2s(-x^{0}\Gamma
_{0}+x^{1}\Gamma _{1}+\ldots +x^{q+1}\Gamma
_{q+1})\,+\,\sum_{I=0}^{q+1}Q_{I}(y)\Gamma _{I}\,.
\end{equation}%
Thus, since ${\mathcal{V}}=sq(x)+\sum_{I=0}^{q+1}x^{I}Q_{I}\left( y\right) $%
, the formula (\ref{mulast}) follows if we verify the following two
identities:
\begin{eqnarray}
4sq(x)y &=&2s(-x^{0}\Gamma _{0}+x^{1}\Gamma _{1}+\ldots +x^{q+1}\Gamma
_{q+1})\nabla _{y}{\mathcal{V}};  \label{mul1} \\
&&  \notag \\
4\left( \sum_{I=0}^{q+1}x^{I}Q_{I}\left( y\right) \right) y &=&\left(
\sum_{I=0}^{q+1}Q_{I}(y)\Gamma _{I}\,+\,\sum_{K=1}^{r}R_{K}(y)\Omega
_{K}\right) \nabla _{y}{\mathcal{V}}.  \label{mul2}
\end{eqnarray}%
We recall that $\nabla _{y}{\mathcal{V}}=2(\sum_{I=0}^{q+1}x^{I}\Gamma
_{I})y $. The first identity (\ref{mul1}) is easy to verify by substituting
this and using (\ref{clifquadrel}); in fact, its r.h.s. can be elaborated as
follows :
\begin{eqnarray}
&&4s\left( -x^{0}\Gamma _{0}+x^{1}\Gamma _{1}+\ldots +x^{q+1}\Gamma
_{q+1}\right) \left( x^{0}\Gamma _{0}+x^{1}\Gamma _{1}+\ldots +x^{q+1}\Gamma
_{q+1}\right) y  \notag \\
&=&4s\left( -(x^{0})^{2}+(x^{1})^{2}+\ldots +(x^{q+1})^{2}\right) y=4sq(x)y~.
\end{eqnarray}%
In order to prove the second identity (\ref{mul2}), we observe that, since $%
\Gamma _{I}^{2}={\mathbb{I}}_{m}$, the first term in its r.h.s. can be
elaborated as
\begin{equation}
\left( \sum_{J=0}^{q+1}Q_{J}(y)\Gamma _{J}\right) \left(
2\sum_{I=0}^{q+1}x^{I}\Gamma _{I}\right) y\,=\,2\left(
\sum_{I=0}^{q+1}x^{I}Q_{I}(y)\right) y\,+\,2\sum_{I=0}^{q+1}x^{I}\sum_{I\neq
J=0}^{q+1}Q_{J}(y)\Gamma _{J}\Gamma _{I}y~.  \label{onet}
\end{equation}%
The second term in the r.h.s. of (\ref{mul2}) reads
\begin{equation}
\left( \sum_{K=1}^{r}R_{K}(y)\Omega _{K}\right) \left(
2\sum_{I=0}^{q+1}x^{I}\Gamma _{I}\right) y\,=\,2\sum_{I=0}^{q+1}x^{I}\left(
\sum_{K=1}^{r}R_{K}(y)\Omega _{K}\right) \Gamma _{I}.  \label{onett}
\end{equation}%
So, the second identity (\ref{mul2}) follows from (\ref{onet}) and (\ref%
{onett}) if we can show that
\begin{equation}
\left( \sum_{I=0}^{q+1}x^{I}Q_{I}\left( y\right) \right)
y\,=\,\sum_{I=0}^{q+1}x^{I}\left( \left( \sum_{I\neq
J=0}^{q+1}Q_{J}(y)\Gamma _{J}\right) \,+\,\left(
\sum_{K=1}^{r}R_{K}(y)\Omega _{K}\right) \right) \Gamma _{I}y~,
\end{equation}%
or, comparing the (matrix) coefficients of the $x^{I}$'s, equivalently, for
all $I=0,1,\ldots ,q+1$:
\begin{equation}
Q_{I}(y)\,=\,\left( \left( \sum_{I\neq J=0}^{q+1}Q_{J}(y)\Gamma _{J}\right)
+\,\left( \sum_{K=1}^{r}R_{K}(y)\Omega _{K}\right) \right) \Gamma _{I}~.
\label{muq}
\end{equation}%
To verify (\ref{muq}), we use the identity (\ref{lorrel}), which we write as
\begin{equation}
\mathbf{F}(y)\,:=\,-\,Q_{0}(y)^{2}\,+\,Q_{1}(y)^{2}\,+\,\ldots
\,+\,Q_{q+1}(y)^{2}\,+\,R_{1}(y)^{2}\,+\,\ldots
\,+\,R_{r}(y)^{2}\,=\,0~,\qquad
\end{equation}%
with, as above, $Q_{I}(y)\,=\,{}^{T}y\Gamma _{I}y$, $R_{K}(y):={}^{T}y\Omega
_{K}y$; note that $\mathbf{F}(y)$ is identically zero as a polynomial in $%
y=(y^{1},\ldots ,y^{m})$. Therefore all partial derivatives of $\mathbf{F}$
w.r.t.\ the $y^{\alpha }$ are also identically zero (as cubics in $y$).
Notice that
\begin{eqnarray}
0\, &=&\,\nabla _{y}\mathbf{F}\,\left( y\right)  \notag \\
&=&\,2\left( -Q_{0}(y)\Gamma _{0}\,+\,Q_{1}(y)\Gamma _{1}\,+\,\ldots
\,+\,Q_{q+1}(y)\Gamma _{q+1}\,+\,R_{1}(y)\Omega _{1}\,+\,\ldots
\,+\,R_{r}(y)\Omega _{r}\right) y~.  \notag \\
&&  \label{nabf}
\end{eqnarray}%
Multiplying (\ref{nabf}) from the left by one of the $\Gamma _{I}$'s ($%
I=1,\ldots ,q+1$) and using $\Gamma _{I}\Gamma _{J}=-\Gamma _{J}\Gamma _{I}$
if $J\neq 0,I$ and $\Gamma _{I}\Omega _{K}=-\Omega _{K}\Gamma _{I}$, we get:
\begin{eqnarray}
0 &=&\Gamma _{I}\nabla _{y}\mathbf{F}\left( y\right)  \notag \\
&=&2\left( -Q_{0}(y)\Gamma _{0}\,-\,Q_{1}(y)\Gamma _{1}\,-\,\ldots
\,+\,Q_{I}(y)\Gamma _{I}\,-\,\ldots \,-\,Q_{q+1}(y)\Gamma _{q+1}\right.
\notag \\
&&\,\left. -\,R_{1}(y)\Omega _{1}\,-\,\ldots \,-\,R_{r}(y)\Omega _{r}\right)
\Gamma _{I}y~.  \label{line3}
\end{eqnarray}%
Thus, using $\Gamma _{I}^{2}={\mathbb{I}}_{m}$, we obtain
\begin{equation}
Q_{I}(y)y\,=\,\left( \left( \sum_{I\neq J=0}^{q+1}Q_{J}(y)\Gamma _{J}\right)
+\,\left( \sum_{K=1}^{r}R_{K}(y)\Omega _{K}\right) \right) \Gamma _{I}~,
\end{equation}%
and therefore we have verified (\ref{muq}) for $I=1,..,q+1$. For $I=0$, it
holds that $\Gamma _{0}={\mathbb{I}}_{m}$ and (\ref{nabf}) implies
\begin{equation}
Q_{0}(y)y\,=\,\left( \left( \sum_{J=1}^{q+1}Q_{J}(y)\Gamma _{J}\right)
+\,\left( \sum_{K=1}^{r}R_{K}(y)\Omega _{K}\right) \right) \Gamma _{0}~.
\end{equation}%
Consequently, we have verified (\ref{muq}) for all $I=0,1,...,q+1$, and
therefore we proved the second identity (\ref{mul2}), as well. Hence, we
proved (\ref{mulast}). This concludes the verification that $\mathbf{\mu }%
\circ \mathbf{\alpha }$ provides a birational inverse of $\nabla _{\mathcal{V%
}}$.
\end{enumerate}

The results in Sec.\ \ref{checkinvmap}, holding when the condition enounced
in Sec.\ \ref{main} is satisfied, provide the proof of the invertibility of
the gradient map $\nabla _{\mathcal{V}}$ by giving its explicit birational
inverse map, as discussed in Sec.\ \ref{explin}.

\section{Examples, I : $L(q,1)$, $q=1,...,8$ \label{explmodels}}

Whenever $P=1,2$, the various $L(q,P)$ models for which we can prove
invertibility, namely those with $q=1,..,8$ and $P=1$ as well as those with $%
q=1,2,3$ and $P=2$, are conveniently presented below as suitable \textit{%
linear sections} (also named \textit{descendants}\footnote{%
But not necessarily \textit{submodels}; cfr. Sec.\ \ref{Des-Sub}.}) of the
complete model $L(8,1)$. Note that the invertibility of these descendants is
\textit{a priori} by no means guaranteed by the invertibility of $L(8,1)$
(which is basically the $J_{3}^{\mathbb{O}}$ and thus is well-known to be
invertible, cfr. Sec.\ \ref{L810}); however, the general treatment given
above does allow us to verify invertibility of such descendants.
Furthermore, the models $L(q,P)$ with $q=1,2,3$ and $P\geqslant 3$ will then
be treated with a different method in Sec.\ \ref{LqP0}.

For the $L(q,1)$ models (with $1\leqslant q\leqslant 8$) that are not
complete, so for $q=3,5,6,7$, the invertibility condition enounced in Sec.\ %
\ref{main} requires that a Clifford set of $q+1$ symmetric $\Gamma $%
-matrices $\{\Gamma _{1},\ldots ,\Gamma _{q+1}\}$ is contained in a larger
set $\{\Gamma _{0}:=\mathbb{I}_{m},\Gamma _{1},\ldots ,\Gamma _{q+1},\Omega
_{1},\ldots ,\Omega _{r}\}$ of symmetric matrices with certain properties.
It turns out that the $\Omega _{K}$'s (with $K=1,...,r$) can again be taken
to be $\Gamma $-matrices and that $\{\Gamma _{1},\ldots ,\Gamma
_{q+1},\Omega _{1},\ldots ,\Omega _{r}\}$ is in fact a Clifford set (of
square matrices of size $m={\mathcal{D}}_{q+1}$) for a complete model of
type $L(q^{\prime },1)$ with $q^{\prime }>q$. More precisely :

\begin{itemize}
\item $L(3,1)\subset L(4,1)$ (for which $m=2^{3}=8$) and the corresponding
Clifford sets are $\{\Gamma _{1},\ldots ,\Gamma _{4}\}\subset \{\Gamma
_{1},\ldots ,\Gamma _{5}\}$, so $L(3,1)$ is a submodel of the complete model
$L(4,1)$;

\item $L(5,1)$, $L(6,1)$ and $L(7,1)$ are submodels of the complete model $%
L(8,1)$, because they all have Clifford sets of $16\times 16$ $\Gamma $%
-matrices which are contained in the Clifford set (still of $16\times 16$ $%
\Gamma $-matrices) of the model $L(8,1)$ : for each of such submodels, the
missing $\Gamma $-matrices w.r.t. $L(8,1)$ can be identified with the needed
$\Omega $-matrices.
\end{itemize}

So, given a complete model with Clifford set $\{\Gamma _{1}^{\prime },\ldots
,\Gamma _{M}^{\prime }\}$, for any proper subset $\mathbb{S}\subset
\{1,\ldots ,M\}$, one obtains a \textit{descendant} model (possibly with $%
\dot{P}>0$), associated to a cubic form $\mathcal{V}_{\mathbb{S}}$ defined
by the $\Gamma $-matrices $\Gamma _{I}^{\prime }$, $I\in \mathbb{S}$, and
with $\Omega _{K}$'s defined as the $\Gamma _{K}^{\prime }$'s with $K\in
\mathbb{\breve{S}}$, where $\mathbb{\breve{S}}$ is the complementary of $%
\mathbb{S}$ in $\{1,\ldots ,M\}$. The explicit inverse of the gradient map $%
\nabla _{\mathcal{V}_{\mathbb{S}}}$ is thus obtained through the procedure
treated in Secs.\ \ref{main}-\ref{complete}.

From the treatment of Sec.\ \ref{invertgamma}, the solution of the BPS
system of the \textit{descendant} model under consideration is then given by
(\ref{pre-sol})-(\ref{pre-sol-3}), and (\ref{SSd}) and (\ref{zH-1})-(\ref%
{zH-3}) yield the corresponding expression of the BPS black hole entropy and
of the BPS attractors, respectively. We refer to Sec.\ \ref{L120}, \ref{L130}
for examples of lower dimensional models $L(q,P)$ (namely, $L(1,2)$ and $%
L(1,3)$) which we explicitly work out in detail (see also the generalization
to $L(1,P)$ models with $P\geqslant 2$ in Sec.\ \ref{L1P0}).

\subsection{$L(8,1)\equiv J_{3}^{\mathbb{O}}$ \label{L810}}

This model is the `largest' among the known complete models (see Sec.\ \ref%
{complete}): it has $q=8,P=1$, and $\mathcal{D}_{9}=16$ \cite{dWVP}; the
number of variables is%
\begin{equation}
\left( 1+q+2+P\cdot \mathcal{D}_{q+1}\right) _{q=8,P=1}=1+10+16=27,
\end{equation}%
and therefore $I=0,1,...,9$, and $\alpha =1,...,16$. The model corresponds
to a symmetric space, and it is related to the exceptional cubic Jordan
algebra\footnote{%
In fact, one could in principle write down an explicit $3\times 3$ Hermitian
matrix $M$ with octonionic components such that ${\mathcal{V}}=-\mathcal{N}%
\left( M\right) $, where $\mathcal{N}$ is the cubic norm (generalizing the
determinant) of $M$, but we refrain from doing so; see \cite{Krut}.} $J_{3}^{%
\mathbb{O}}$, whose reduced structure group is the minimally non-compact
real form $E_{6(-26)}$ of $E_{6}$.

In this case, the Bekenstein-Hawking entropy and the attractor values of
scalar fields of BPS extremal black holes are explicitly known; see e.g.
\cite{FGimonK}. Thus, we will not consider the solution of the BPS system in
this model, but rather we will give some treatment useful to discuss some
descendants from $L(8,1)$ itself (cfr. Sec.\ \ref{subL810}).

We choose the $q+2=10$ $\Gamma $-matrices $\Gamma _{0}={\mathbb{I}}%
_{16},\Gamma _{1},\ldots ,\Gamma _{9}$ (of size $16\times 16$) such that $%
\{\Gamma _{1},\ldots ,\Gamma _{9}\}$ is a Clifford set and such that the
corresponding quadrics are:
\begin{eqnarray}
&&%
\begin{array}{lll}
Q_{0}\equiv Q[{}_{0000}^{0000}] & := & \left( y^{1}\right) ^{2}+\,\ldots
\,+\,\left( y^{8}\right) ^{2}\,+\,\left( y^{9}\right) ^{2}\,+\,\ldots
\,+\,\left( y^{16}\right) ^{2}; \\
Q_{1}\equiv Q[{}_{0001}^{0000}] & := & \left\{
\begin{array}{l}
\left( y^{1}\right) ^{2}-\left( y^{2}\right) ^{2}\,+\,\ldots \,+\,\left(
y^{7}\right) ^{2}\,-\,\left( y^{8}\right) ^{2} \\
+\left( y^{9}\right) ^{2}-\left( y^{10}\right) ^{2}\,+\,\ldots \,+\left(
y^{15}\right) ^{2}-\left( y^{16}\right) ^{2};%
\end{array}%
\right. \\
Q_{2}\equiv Q[{}_{1010}^{0001}] & := & 2(y^{1}y^{2}-y^{3}y^{4}\,+%
\,y^{5}y^{6}-y^{7}y^{8}\,-\,y^{9}y^{10}+y^{11}y^{12}\,-%
\,y^{13}y^{14}+y^{15}y^{16}); \\
Q_{3}\equiv Q[{}_{0000}^{0011}] & := & 2(y^{1}y^{4}+y^{2}y^{3}\,+%
\,y^{5}y^{8}+y^{6}y^{7}\,+\,y^{9}y^{12}+y^{10}y^{11}\,+%
\,y^{13}y^{16}+y^{14}y^{15}); \\
Q_{4}\equiv Q[{}_{0101}^{0101}] & := &
2(y^{1}y^{6}-y^{2}y^{5}+y^{3}y^{8}-y^{4}y^{7}\,+%
\,y^{9}y^{14}-y^{10}y^{13}+y^{11}y^{16}-y^{12}y^{15}); \\
Q_{5}\equiv Q[{}_{0110}^{0111}] & := &
2(y^{1}y^{8}+y^{2}y^{7}-y^{3}y^{6}-y^{4}y^{5}\,+%
\,y^{9}y^{16}+y^{10}y^{15}-y^{11}y^{14}-y^{12}y^{13}); \\
Q_{6}\equiv Q[{}_{0010}^{1001}] & := &
2(y^{1}y^{10}+y^{2}y^{9}-y^{3}y^{12}-y^{4}y^{11}+y^{5}y^{14}+y^{6}y^{13}-y^{7}y^{16}-y^{8}y^{15});
\\
Q_{7}\equiv Q[{}_{1101}^{1011}] & := &
2(y^{1}y^{12}-y^{2}y^{11}+y^{3}y^{10}-y^{4}y^{9}-y^{5}y^{16}+y^{6}y^{15}-y^{7}y^{14}+y^{8}y^{13});
\\
Q_{8}\equiv Q[{}_{1110}^{1101}] & := &
2(y^{1}y^{14}+y^{2}y^{13}-y^{3}y^{16}-y^{4}y^{15}-y^{5}y^{10}-y^{6}y^{9}+y^{7}y^{12}+y^{8}y^{11});
\\
Q_{9}\equiv Q[{}_{1001}^{1111}] & := &
2(y^{1}y^{16}-y^{2}y^{15}+y^{3}y^{14}-y^{4}y^{13}+y^{5}y^{12}-y^{6}y^{11}+y^{7}y^{10}-y^{8}y^{9}).%
\end{array}
\notag \\
&&  \label{this3}
\end{eqnarray}%
The cubic form of this model is
\begin{equation}
{\mathcal{V}}_{L(8,1)}\,:=\,sq(x)\,+\,\sum_{I=0}^{9}x^{I}Q_{I}(y),
\label{v8}
\end{equation}%
where
\begin{equation}
\quad q(x)\,:=\,-\left( x^{0}\right) ^{2}\,+q_{9}(x)=-\left( x^{0}\right)
^{2}\,+\,\left( x^{1}\right) ^{2}+\left( x^{2}\right) ^{2}\,+\,\ldots
\,+\,\left( x^{9}\right) ^{2}~.  \label{q8}
\end{equation}

Since the quadratic forms $Q_{0},\ldots ,Q_{9}$ satisfy the Lorentzian
quadratic relation
\begin{equation}
-Q_{0}(y)^{2}\,+\,Q_{1}(y)^{2}\,+\,\ldots \,+\,Q_{9}(y)^{2}\,=\,0,
\label{GammQI4}
\end{equation}%
the $L(8,1)$ model satisfies the invertibility condition of Sec.\ \ref{main}
with $r=0$ (so no extra $\Omega _{K}$ are needed) : in fact, as mentioned
above, $L(8,1)$ is a complete model. Thus, the gradient map $\nabla _{%
\mathcal{V}_{L(8,1)}}$ of $\mathcal{V}_{L(8,1)}$ is invertible and the
inverse map $\mathbf{\mu }$ is given by homogeneous polynomials of degree
two. It holds that
\begin{equation}
{\mathbb{R}}_{\xi }^{27}\,\overset{\nabla _{\mathcal{V}_{L(8,1)}}}{%
\longrightarrow }\,{\mathbb{R}}_{z}^{27}\,\,\overset{\mathbf{\mu }}{%
\longrightarrow }\,{\mathbb{R}}_{\xi }^{27}~,
\end{equation}%
with%
\begin{equation}
\mathbf{\mu }\circ \nabla _{\mathcal{V}_{L(8,1)}}\left( \xi \right) =4%
\mathcal{V}_{L(8,1)}\left( \xi \right) \mathbb{\xi }.
\end{equation}%
The gradient map
\begin{equation*}
\nabla _{\mathcal{V}_{L(8,1)}}\,:\,\,{\mathbb{R}}^{27}\,\longrightarrow \,{%
\mathbb{R}}^{27},\qquad \xi \longmapsto \,^{T}\left( \mathcal{V}_{s}\left(
\xi \right) ,\mathcal{V}_{I}\left( \xi \right) ,\mathcal{V}_{\alpha }\left(
\xi \right) \right)
\end{equation*}%
can be identified with the adjoint map
\begin{equation}
J_{3}^{\mathbb{O}}\,\longrightarrow \,J_{3}^{\mathbb{O}},\qquad
M\,\longmapsto \,M^{\sharp }
\end{equation}%
here $M^{\sharp }$ is the adjoint matrix of $M\in J_{3}^{\mathbb{O}}$. This
adjoint map $\#$ is (birationally) invertible, with inverse given by the map
$M^{\sharp }\mapsto (M^{\sharp })^{\sharp }$, which is thus essentially the
map $\mathbf{\mu }$.

\subsection{Some submodels of $L(8,1)$\label{subL810}}

\subsubsection{$L(7,1)$}

$L(7,1)$ is a submodel of $L(8,1)$, and it corresponds to a homogeneous
non-symmetric (sub)manifold (of the symmetric scalar manifold pertaining to $%
L(8,1)$). Indeed, by setting $x^{9}=0$ in the $L(8,1)$ model treated in
Sec.\ \ref{L810}, one has $1+(10-1)+16=26$ variables. The $\Gamma $-matrices
involved are $\Gamma _{0}={\mathbb{I}}_{16}$ and the Clifford set, now with
only $8$ matrices, namely $\{\Gamma _{1},\ldots ,\Gamma _{8}\}$, still of
size ${\mathcal{D}}_{9}={\mathcal{D}}_{8}=16$. The $x^{9}=0$ restriction of
the Lorentzian quadratic form $q(x)$ (\ref{q8}) is the corresponding
Lorentzian form in $x^{0},\ldots ,x^{8}$ :%
\begin{equation}
q(x)\,:=\,-\left( x^{0}\right) ^{2}\,+q_{8}(x)=-\left( x^{0}\right)
^{2}\,+\,\left( x^{1}\right) ^{2}+\left( x^{2}\right) ^{2}\,+\,\ldots
\,+\,\left( x^{8}\right) ^{2}~.  \label{q7}
\end{equation}%
Therefore, ${\mathcal{V}}_{L(8,1,0)}$ restricts to the cubic form ${\mathcal{%
V}}_{L(7,1)}$ of the $L(7,1)$ model : $L(7,1)$ can be regarded as $L(8,1)$
with one linear constraint:
\begin{eqnarray}
L(7,1)\, &=&\,\left. L(8,1)\right\vert _{x^{9}=0}; \\
{\mathcal{V}}_{L(7,1)}\, &=&\,\left. {\mathcal{V}}_{L(8,1)}\right\vert
_{x^{9}=0}~.
\end{eqnarray}%
\texttt{\ }

As discussed in \cite{Russo}, it should be remarked that the invertibility
of the gradient map of ${\mathcal{V}}_{L(8,1)}$ would, in general, not
guarantee the invertibility of its restriction to a linear subspace of ${%
\mathbb{R}}^{27}$. However, the invertibility condition enounced in Sec.\ %
\ref{main} is satisfied, once we define
\begin{equation}
\Omega _{1}\,:=\,\Gamma _{9},\qquad \text{so}\quad
R_{1}(y)\,=\,Q_{9}(y),\quad r:=1~.
\end{equation}%
In fact, $\Omega _{1}$ anti-commutes with all matrices of the Clifford set $%
\{\Gamma _{1},\ldots ,\Gamma _{8}\}$, because it is originally part of the
Clifford set $\{\Gamma _{1},\ldots ,\Gamma _{9}\}$ of $L(8,1)$. The
quadratic form $R_{1}(y)$ defined by $\Omega _{1}=\Gamma _{9}$ is of course $%
R_{1}(y)=Q_{9}(y)$ of (\ref{this3}), so that the required Lorentzian
identity (\ref{lorrel}) holds true: it is nothing but (\ref{GammQI4}).

Since $r=1$, the inverse of the gradient map $\nabla _{{\mathcal{V}}%
_{L(7,1)}}$ is given as a composition of two maps, namely (cfr. (\ref{csii}%
)):%
\begin{equation}
{\mathbb{R}}_{\xi }^{1+9+16}\,\overset{\nabla _{{\mathcal{V}}_{L(7,1)}}}{%
\longrightarrow }\,{\mathbb{R}}_{z}^{1+9+16}\,\overset{\mathbf{\alpha }}{%
\longrightarrow }\,{\mathbb{R}}_{(t,u,v,w)}^{1+9+16+1}\,\overset{\mathbf{\mu
}}{\longrightarrow }\,{\mathbb{R}}_{\xi }^{1+9+16}~.
\end{equation}%
The map $\mathbf{\alpha }$, as in (\ref{alpha}), is given by:
\begin{equation}
\mathbf{\alpha }(z_{1},\ldots
,z_{26})\,:=\,^{T}(z_{1}^{2},\,z_{1}z_{2},\ldots ,\,z_{1}z_{26},\,R_{1}(z))~,
\end{equation}%
where the quadratic form $R_{1}(z)$ is obtained by substituting $y^{\alpha
}:=z_{\alpha +10}$ in $Q_{9}$:
\begin{equation}
R_{1}(z)\,:=%
\,2(z_{11}z_{26}-z_{12}z_{25}+z_{13}z_{24}-z_{14}z_{23}+z_{15}z_{22}-z_{16}z_{21}+z_{17}z_{20}-z_{18}z_{19})~.
\end{equation}%
The map $\mathbf{\mu }$ (\ref{mu}) has $1+9+16$ components which are
homogeneous polynomials of degree $2$ in the $27$ variables $t,u_{0},\ldots
,u_{8},v_{1},\ldots ,v_{16},w$. Since%
\begin{equation}
\mathbf{\mu }\circ \mathbf{\alpha }\circ \nabla _{{\mathcal{V}}%
_{L(7,1)}}\,(\xi )\,=\,4q(x)^{2}{\mathcal{V}}_{L(7,1)}(\xi )\mathbb{\xi },
\end{equation}%
where $q(x)$ is given by (\ref{q7}), the (birational) inverse of the
gradient map $\nabla _{\mathcal{V}_{L(7,1)}}$ is the map $\mathbf{\mu \circ
\alpha }$, which is a homogeneous polynomial map of degree four.

\subsubsection{$L(6,1)$}

A similar treatment can be given for $L(6,1)$, corresponding to an
homogeneous non-symmetric (sub)manifold (of the symmetric scalar manifold
pertaining to $L(8,1)$) : $L(6,1)$ can be regarded as $L(8,1)$ with two
linear constraints:
\begin{eqnarray}
L(6,1)\, &=&\,\left. L(8,1)\right\vert _{x^{8}=x^{9}=0}; \\
{\mathcal{V}}_{L(6,1)}\, &=&\,\left. {\mathcal{V}}_{L(8,1)}\right\vert
_{x^{8}=x^{9}=0}~,
\end{eqnarray}%
Therefore, from the invertibility condition enounced in Sec.\ \ref{main},
the model $L(6,1)$ is invertible, because one can take
\begin{equation}
\Omega _{1}=\Gamma _{9},\quad \Omega _{2}=\Gamma _{8},\quad \text{so}\quad
R_{1}(y)=Q_{9}(y),\quad R_{2}(y)=Q_{8}(y),\quad r=2.
\end{equation}%
The $x^{8}=x^{9}=0$ restriction of the Lorentzian quadratic form $q(x)$ (\ref%
{q8}) is the corresponding Lorentzian form in $x^{0},\ldots ,x^{7}$ :%
\begin{equation}
q(x)\,:=\,-\left( x^{0}\right) ^{2}\,+q_{7}(x)=-\left( x^{0}\right)
^{2}\,+\,\left( x^{1}\right) ^{2}+\left( x^{2}\right) ^{2}\,+\,\ldots
\,+\,\left( x^{7}\right) ^{2}~.  \label{q6}
\end{equation}

Since $r=2$, the inverse of the gradient map $\nabla _{{\mathcal{V}}%
_{L(6,1)}}$ is given as a composition of two maps, namely (cfr. (\ref{csii}%
)):%
\begin{equation}
{\mathbb{R}}_{\xi }^{1+8+16}\,\overset{\nabla _{{\mathcal{V}}_{L(6,1)}}}{%
\longrightarrow }\,{\mathbb{R}}_{z}^{1+8+16}\,\overset{\mathbf{\alpha }}{%
\longrightarrow }\,{\mathbb{R}}_{(t,u,v,w)}^{1+8+16+2}\,\overset{\mathbf{\mu
}}{\longrightarrow }\,{\mathbb{R}}_{\xi }^{1+8+16}~.
\end{equation}%
The map $\mathbf{\alpha }$, as in (\ref{alpha}), is given by:
\begin{equation}
\mathbf{\alpha }(z_{1},\ldots
,z_{25})\,:=\,^{T}(z_{1}^{2},\,z_{1}z_{2},\ldots
,\,z_{1}z_{25},\,R_{1}(z),\,R_{2}(z))~,
\end{equation}%
where the quadratic forms $R_{K}(z)$'s ($K=1,2$) depend only on the last $16$
variables $z_{10},\ldots ,z_{25}$ and they are obtained by substituting $%
y^{\alpha }:=z_{\alpha +9}$ in $Q_{9}$ and $Q_{8}$ respectively:
\begin{eqnarray}
R_{1}(z):=2(z_{10}z_{25}-z_{11}z_{24}+z_{12}z_{23}-z_{13}z_{22}+z_{14}z_{21}-z_{15}z_{20}+z_{16}z_{19}-z_{17}z_{18}); &&
\\
R_{2}(z):=2(z_{10}z_{23}+z_{11}z_{22}-z_{12}z_{25}-z_{13}z_{24}-z_{14}z_{19}-z_{15}z_{18}+z_{16}z_{21}+z_{17}z_{20}). &&
\end{eqnarray}%
The map $\mathbf{\mu }$ (\ref{mu}) has $1+8+16$ components which are
homogeneous polynomials of degree $2$ in the $27$ variables $t,u_{0},\ldots
,u_{7},v_{1},\ldots ,v_{16},w_{1},w_{2}$. Since
\begin{equation}
\mathbf{\mu }\circ \mathbf{\alpha }\circ \nabla _{{\mathcal{V}}%
_{L(6,1)}}\,(\xi )\,=\,4q(x)^{2}{\mathcal{V}}_{L(6,1)}(\xi )\mathbb{\xi },
\end{equation}%
where $q(x)$ is given by (\ref{q6}), the (birational) inverse of the
gradient map $\nabla _{\mathcal{V}_{L(6,1)}}$ is the map $\mathbf{\mu \circ
\alpha }$, which is a homogeneous polynomial map of degree four.

\subsubsection{$L(5,1)$\label{L51}}

One can take one step further (since the $\Gamma $-matrices still have size $%
16={\mathcal{D}}_{9}={\mathcal{D}}_{8}={\mathcal{D}}_{7}={\mathcal{D}}_{6}$%
), and consider $L(5,1)$, corresponding to a homogeneous non-symmetric
(sub)manifold (of the symmetric scalar manifold pertaining to $L(8,1)$) : $%
L(5,1)$ can be regarded as $L(8,1)$ with three linear constraints:
\begin{eqnarray}
L(5,1)\, &=&\,\left. L(8,1)\right\vert _{x^{7}=x^{8}=x^{9}=0}; \\
{\mathcal{V}}_{L(5,1)}\, &=&\,\left. {\mathcal{V}}_{L(8,1)}\right\vert
_{x^{7}=x^{8}=x^{9}=0}~.
\end{eqnarray}%
One can thus take
\begin{equation}
\Omega _{1}=\Gamma _{9},\quad \Omega _{2}=\Gamma _{8},\quad \Omega
_{3}=\Gamma _{7}\quad \text{so}\quad R_{1}(y)=Q_{9}(y),\quad
R_{2}(y)=Q_{8}(y),\quad R_{3}(y)\,=\,Q_{7}(y),
\end{equation}%
and $r=3$, in order to satisfy the invertibility condition of Sec.\ \ref%
{main}. The $x^{7}=x^{8}=x^{9}=0$ restriction of the Lorentzian quadratic
form $q(x)$ (\ref{q8}) is the corresponding Lorentzian form in $x^{0},\ldots
,x^{6}$ :%
\begin{equation}
q(x)\,:=\,-\left( x^{0}\right) ^{2}\,+q_{6}(x)=-\left( x^{0}\right)
^{2}\,+\,\left( x^{1}\right) ^{2}+\left( x^{2}\right) ^{2}\,+\,\ldots
\,+\,\left( x^{6}\right) ^{2}~.  \label{q5}
\end{equation}

Since $r=3$, the inverse of the gradient map $\nabla _{{\mathcal{V}}%
_{L(5,1)}}$ is given as a composition of two maps, namely (cfr. (\ref{csii}%
)):%
\begin{equation}
{\mathbb{R}}_{\xi }^{1+7+16}\,\overset{\nabla _{{\mathcal{V}}_{L(5,1)}}}{%
\longrightarrow }\,{\mathbb{R}}_{z}^{1+7+16}\,\overset{\mathbf{\alpha }}{%
\longrightarrow }\,{\mathbb{R}}_{(t,u,v,w)}^{1+7+16+3}\,\overset{\mathbf{\mu
}}{\longrightarrow }\,{\mathbb{R}}_{\xi }^{1+7+16}~.
\end{equation}%
The map $\mathbf{\alpha }$, as in (\ref{alpha}), is given by:
\begin{equation}
\mathbf{\alpha }(z_{1},\ldots
,z_{24})\,:=\,^{T}(z_{1}^{2},\,z_{1}z_{2},\ldots
,\,z_{1}z_{24},\,R_{1}(z),R_{2}(z),R_{3}(z))~,
\end{equation}%
where the quadratic forms $R_{K}(z)$'s ($K=1,2,3$) depend only on the last $%
16$ variables $z_{9},\ldots ,z_{24}$ and they are obtained by substituting $%
y^{\alpha }:=z_{\alpha +8}$ in $Q_{9},Q_{8},Q_{7}$ respectively:
\begin{eqnarray}
R_{1}(z):=2(z_{9}z_{24}-z_{10}z_{23}+z_{11}z_{22}-z_{12}z_{21}+z_{13}z_{20}-z_{14}z_{19}+z_{15}z_{18}-z_{16}z_{17}); &&
\\
R_{2}(z):=2(z_{9}z_{22}+z_{10}z_{21}-z_{11}z_{24}-z_{12}z_{23}-z_{13}z_{18}-z_{14}z_{17}+z_{15}z_{20}+z_{16}z_{19}); &&
\\
R_{3}(z):=2(z_{9}z_{20}-z_{10}z_{19}+z_{11}z_{18}-z_{12}z_{17}-z_{13}z_{24}+z_{14}z_{23}-z_{15}z_{22}+z_{16}z_{21}). &&
\end{eqnarray}%
The map $\mathbf{\mu }$ (\ref{mu}) has $1+7+16$ components which are
homogeneous polynomials of degree $2$ in the $27$ variables $t,u_{0},\ldots
,u_{6},v_{1},\ldots ,v_{16},w_{1},w_{2},w_{3}$. Since
\begin{equation}
\mathbf{\mu }\circ \mathbf{\alpha }\circ \nabla _{{\mathcal{V}}%
_{L(5,1)}}\,(\xi )\,=\,4q(x)^{2}{\mathcal{V}}_{L(5,1)}(\xi )\mathbb{\xi },
\end{equation}%
where $q(x)$ is given by (\ref{q5}), the (birational) inverse of the
gradient map $\nabla _{\mathcal{V}_{L(5,1)}}$ is the map $\mathbf{\mu \circ
\alpha }$, which is a homogeneous polynomial map of degree four.

\subsubsection{$L(4,1)\equiv J_{3}^{\mathbb{H}}$\label{L410JH}}

The model $L(4,1)$ corresponds to a symmetric space, and it is related to
the simple cubic Jordan algebra over the quaternions, $J_{3}^{\mathbb{H}}$.
In this case, the Bekenstein-Hawking entropy and the attractor values of
scalar fields of extremal BPS black holes are explicitly known; see e.g.
\cite{FGimonK}. In the present treatment, we will highlight its relation to
the complete `parent' model $L(8,1)$.

It should be remarked that going one step further from $L(5,1)$ and imposing
$x^{6}=x^{7}=x^{8}=x^{9}=0$ does not yield to the $L(4,1)$ model, since the $%
\Gamma $-matrices in such a model have size ${\mathcal{D}}_{5}=8$, and not
size $16$ as the ones of $L(5,1)$. Fortunately, our choice of $\Gamma $%
-matrices in (\ref{this3}) is such that taking the upper left $8\times 8$
block of $\Gamma _{0}={\mathbb{I}}_{16},\Gamma _{1},\ldots ,\Gamma _{5}$,
one obtains ${\mathbb{I}}_{8}$ and the $5$ other $\Gamma $-matrices of size $%
8$ that are easily verified to be again a Clifford set; taking similar
blocks of $\Gamma _{6},\ldots ,\Gamma _{9}$ one finds the zero matrix of
size $8$. In other words, the restriction of the quadrics $Q_{i}(y)$ from (%
\ref{this3}) to the subspace $y^{9}=\ldots =y^{16}=0$ yields to the quadrics
of $L(4,1)$. To be explicit:
\begin{eqnarray}
&&{\renewcommand{\arraystretch}{1.9}}{\
\begin{array}{lll}
\bar{Q}_{0}\equiv (Q_{0})_{|y^{9}=\ldots =y^{16}=0}=Q[{}_{000}^{000}] & := &
\left( y^{1}\right) ^{2}+\left( y^{2}\right) ^{2}\,+\,\ldots \,+\,\left(
y^{7}\right) ^{2}\,+\,\left( y^{8}\right) ^{2}; \\
\bar{Q}_{1}\equiv (Q_{1})_{|y^{9}=\ldots =y^{16}=0}=Q[{}_{001}^{000}] & := &
\left( y^{1}\right) ^{2}-\left( y^{2}\right) ^{2}\,+\,\ldots \,+\,\left(
y^{7}\right) ^{2}\,-\,\left( y^{8}\right) ^{2}; \\
\bar{Q}_{2}\equiv (Q_{2})_{|y^{9}=\ldots =y^{16}=0}=Q[{}_{010}^{001}] & := &
2(y^{1}y^{2}-y^{3}y^{4}\,+\,y^{5}y^{6}-y^{7}y^{8}); \\
\bar{Q}_{3}\equiv (Q_{3})_{|y^{9}=\ldots =y^{16}=0}=Q[{}_{000}^{011}] & := &
2(y^{1}y^{4}+y^{2}y^{3}\,+\,y^{5}y^{8}+y^{6}y^{7}); \\
\bar{Q}_{4}\equiv (Q_{4})_{|y^{9}=\ldots =y^{16}=0}=Q[{}_{101}^{101}] & := &
2(y^{1}y^{6}-y^{2}y^{5}+y^{3}y^{8}-y^{4}y^{7}); \\
\bar{Q}_{5}\equiv (Q_{5})_{|y^{9}=\ldots =y^{16}=0}=Q[{}_{110}^{111}] & := &
2(y^{1}y^{8}+y^{2}y^{7}-y^{3}y^{6}-y^{4}y^{5}); \\
(Q_{6})_{|y^{9}=\ldots =y^{16}=0}=0; &  &  \\
(Q_{7})_{|y^{9}=\ldots =y^{16}=0}=0; &  &  \\
(Q_{8})_{|y^{9}=\ldots =y^{16}=0}=0; &  &  \\
(Q_{9})_{|y^{9}=\ldots =y^{16}=0}=0~. &  &
\end{array}%
}  \notag \\
&&{\label{th410}}
\end{eqnarray}

The observant reader will have noticed that, in order to obtain the
characteristics in (\ref{th410}), we omitted the first column from the
characteristics in the first six quadrics in (\ref{this3}). It is easy to
check that the sum of any two distinct non-zero characteristics is odd,
hence the $\Gamma $-matrices of size $8$ corresponding to the last $5$
quadrics compose a Clifford set.

Next, we observe that if we put $y^{9}=\ldots =y^{16}=0$ in the Lorentzian
identity (\ref{GammQI4}) between $Q_{0}(y),\ldots ,Q_{9}(y)$, we obtain the
following Lorentzian identity between the `restricted' $Q_{0},\ldots ,Q_{5}$%
, denoted by $\bar{Q}_{0}(\bar{y}),\ldots ,\bar{Q}_{5}(\bar{y})$ (with $\bar{%
y}=(y^{1},\ldots ,y^{8})$) :
\begin{equation}
-\bar{Q}_{0}(\bar{y})^{2}\,+\,\bar{Q}_{1}(\bar{y})^{2}\,+\,\ldots \,+\,\bar{Q%
}_{5}(\bar{y})^{2}\,=\,0~.  \label{lorL410}
\end{equation}%
This implies that $L(4,1)$ is a complete model, that is $r=0$ in Sec.\ \ref%
{main}, since no extra $8\times 8$ matrices $\Omega _{K}$'s are needed.

To summarize, the complete model $L(4,1)$ can be regarded as the complete
model $L(8,1)$ with twelve linear constraints:
\begin{eqnarray}
L(4,1)\, &=&\,\left. L(8,1)\right\vert _{x^{6}=\ldots =x^{9}=0,~y^{9}=\ldots
=y^{16}=0};  \label{j1} \\
{\mathcal{V}}_{L(4,1)}\, &=&\,\left. {\mathcal{V}}_{L(8,1)}\right\vert
_{x^{6}=\ldots =x^{9}=0,~y^{9}=\ldots =y^{16}=0}~.  \label{j2}
\end{eqnarray}

Since the model is complete, so $r=0$, the map $\mathbf{\alpha }$ is not
needed and the inverse of the gradient map is now simply the map $\mathbf{%
\mu }$ from (\ref{mu}):
\begin{equation}
{\mathbb{R}}_{\xi }^{15}\,\overset{\nabla _{\mathcal{V}_{L(4,1)}}}{%
\longrightarrow }\,{\mathbb{R}}_{z}^{15}\,\,\overset{\mathbf{\mu }}{%
\longrightarrow }\,{\mathbb{R}}_{\xi }^{15}~,
\end{equation}%
with
\begin{equation}
\mathbf{\mu }\circ \nabla _{\mathcal{V}_{L(4,1)}}\left( \xi \right) =4%
\mathcal{V}_{L(4,1)}\left( \xi \right) \mathbb{\xi }.
\end{equation}%
The gradient map
\begin{equation*}
\nabla _{\mathcal{V}_{L(4,1)}}\,:\,\,{\mathbb{R}}^{15}\,\longrightarrow \,{%
\mathbb{R}}^{15},\qquad \xi \longmapsto \,^{T}\left( \mathcal{V}_{s}\left(
\xi \right) ,\mathcal{V}_{I}\left( \xi \right) ,\mathcal{V}_{\alpha }\left(
\xi \right) \right)
\end{equation*}%
can be identified with the adjoint map
\begin{equation}
J_{3}^{\mathbb{H}}\,\longrightarrow \,J_{3}^{\mathbb{H}},\qquad
M\,\longmapsto \,M^{\sharp }
\end{equation}%
here $M^{\sharp }$ is the adjoint matrix of $M\in J_{3}^{\mathbb{H}}$. This
adjoint map $\#$ is (birationally) invertible, with inverse given by the map
$M^{\sharp }\mapsto (M^{\sharp })^{\sharp }$, which is thus essentially the
map $\mathbf{\mu }$.

\paragraph{The cubic norm of $J_{3}^{\mathbb{H}}$.}

It is worth making more explicit the relation between the complete model $%
L(4,1)$ and the Euclidean simple cubic Jordan algebra $J_{3}^{\mathbb{H}}$.
Let $\mathbf{i}$, $\mathbf{j}$ and $\mathbf{k}$ denote the imaginary units
of $\mathbb{H}$, with standard multiplication rules of $\mathbb{H}$:%
\begin{equation}
\left.
\begin{array}{l}
\mathbf{ij}=-\mathbf{ji}=\mathbf{k}; \\
\mathbf{jk}=-\mathbf{kj}=\mathbf{i}; \\
\mathbf{ki}=-\mathbf{ik}=\mathbf{j},%
\end{array}%
\right\} \Longleftrightarrow \mathbf{i}^{2}=\mathbf{j}^{2}=\mathbf{k}^{2}=%
\mathbf{ijk}=-1.
\end{equation}%
Then, one can define the following matrix $\mathcal{M}$ belonging to $J_{3}^{%
\mathbb{H}}$:
\begin{equation}
J_{3}^{\mathbb{H}}\,\ni \,\mathcal{M}=\mathcal{M}^{\mathbb{H}}:=\left(
\begin{array}{lll}
a & \mathbf{z} & \overline{\mathbf{y}} \\
\overline{\mathbf{z}} & b & \mathbf{x} \\
{\mathbf{y}} & \overline{\mathbf{x}} & c%
\end{array}%
\right) ,  \label{matL41}
\end{equation}%
with%
\begin{equation}
\begin{array}{rcrrcr}
{a} & := & x^{0}-x^{1}\in \mathbb{R},\qquad & \mathbf{x} & := & y^{2}-%
\mathbf{i}y^{4}+\mathbf{j}y^{6}-\mathbf{k}y^{8}\in \mathbb{H}; \\
{b} & := & x^{0}+x^{1}\in \mathbb{R},\qquad & \mathbf{y} & := & y^{1}-%
\mathbf{i}y^{3}-\mathbf{j}y^{5}-\mathbf{k}y^{7}\in \mathbb{H}; \\
{c} & := & s\in \mathbb{R},\qquad & \mathbf{z} & := & -x^{2}-\mathbf{i}x^{3}+%
\mathbf{j}x^{4}-\mathbf{k}x^{5}\in \mathbb{H}.%
\end{array}%
\end{equation}%
The cubic norm ${\mathcal{N}}(\mathcal{M})$ of $\mathcal{M}$ is defined as
\begin{equation}
{\mathcal{N}}(\mathcal{M})\,:=\,abc\,-\,a\mathbf{x}\overline{\mathbf{x}}%
\,-\,b\mathbf{y}\overline{\mathbf{y}}\,-\,c\mathbf{z}\overline{\mathbf{z}}%
\,+\,(\mathbf{x}\mathbf{y})\mathbf{z}+\overline{\mathbf{z}}(\overline{%
\mathbf{y}}\overline{\mathbf{x}})~.
\end{equation}%
With this choice of coefficients, one can check
\begin{equation}
{\mathcal{V}}_{L(4,1)}\,=\,-{\mathcal{N}}(\mathcal{M})~.  \label{cubsL41}
\end{equation}

\subsubsection{$L(3,1)$\label{L310}}

This model, which also has size ${\mathcal{D}}_{4}=8$, is found by setting $%
x^{5}=0$ in $L(4,1)$ and taking $\Omega _{1}$ to be the $\Gamma $-matrix of $%
\bar{Q}_{5}$, so $r=1$:
\begin{eqnarray}
L(3,1)\, &=&\,\left. L(8,1)\right\vert _{x^{5}=\ldots =x^{9}=0,~y^{9}=\ldots
=y^{16}=0}=\left. L(4,1)\right\vert _{x^{5}=0}; \\
{\mathcal{V}}_{L(3,1)}\, &=&\,\left. {\mathcal{V}}_{L(8,1)}\right\vert
_{x^{5}=\ldots =x^{9}=0,~y^{9}=\ldots =y^{16}=0}~=\left. {\mathcal{V}}%
_{L(4,1)}\right\vert _{x^{5}=0}.
\end{eqnarray}%
Thus, $L(3,1)$ can be regarded as the complete model $L(8,1)$ with thirteen
linear constraints, or equivalently as the complete model $L(4,1)$ with one
linear constraint. In fact, $L(3,1)$ corresponds to a homogeneous
non-symmetric (sub)manifold (of the symmetric scalar manifolds pertaining to
$L(8,1)$ and $L(4,1)$).

In order to satisfy the invertibility condition of Sec.\ \ref{main} one can
thus take
\begin{equation}
\Omega _{1}=\bar{\Gamma}_{5}\quad \text{so}\quad R_{1}(y)=\bar{Q}%
_{5}(y),\quad r=1.
\end{equation}%
The $x^{5}=x^{6}=x^{7}=x^{8}=x^{9}=0$ restriction of the Lorentzian
quadratic form $q(x)$ (\ref{q8}) is the corresponding Lorentzian form in $%
x^{0},\ldots ,x^{4}$ :%
\begin{equation}
q(x)\,:=\,-\left( x^{0}\right) ^{2}\,+q_{4}(x)=-\left( x^{0}\right)
^{2}\,+\,\left( x^{1}\right) ^{2}+\left( x^{2}\right) ^{2}\,+\,\left(
x^{3}\right) ^{2}\,+\,\left( x^{4}\right) ^{2}~.  \label{q3}
\end{equation}%
Since $r=1$, the inverse of the gradient map $\nabla _{{\mathcal{V}}%
_{L(3,1)}}$ is given as a composition of two maps, namely (cfr. (\ref{csii}%
)):%
\begin{equation}
{\mathbb{R}}_{\xi }^{1+5+8}\,\overset{\nabla _{{\mathcal{V}}_{L(7,1)}}}{%
\longrightarrow }\,{\mathbb{R}}_{z}^{1+5+8}\,\overset{\mathbf{\alpha }}{%
\longrightarrow }\,{\mathbb{R}}_{(t,u,v,w)}^{1+5+8+1}\,\overset{\mathbf{\mu }%
}{\longrightarrow }\,{\mathbb{R}}_{\xi }^{1+5+8}~.
\end{equation}%
The map $\mathbf{\alpha }$, as in (\ref{alpha}), is given by:
\begin{equation}
\mathbf{\alpha }(z_{1},\ldots
,z_{14})\,:=\,^{T}(z_{1}^{2},\,z_{1}z_{2},\ldots ,\,z_{1}z_{14},\,R_{1}(z))~,
\end{equation}%
where the quadratic form $R_{1}(z)$ is defined by $\bar{Q}_{5}$ and depends
only on the last $8$ variables $z_{7},\ldots ,z_{14}$, obtained by
substituting $y^{\alpha }=z_{\alpha +6}$ in $\bar{Q}_{5}(y)$ :
\begin{equation}
R_{1}(z)\,:=\,2(z_{7}z_{14}+z_{8}z_{13}-z_{9}z_{12}-z_{10}z_{11})~.
\end{equation}%
The map $\mathbf{\mu }$ (\ref{mu}) has $1+5+8$ components which are
homogeneous polynomials of degree $2$ in the $15$ variables $t,u_{0},\ldots
,u_{4},v_{1},\ldots ,v_{8},w$. Since
\begin{equation}
\mathbf{\mu }\circ \mathbf{\alpha }\circ \nabla _{{\mathcal{V}}%
_{L(3,1)}}\,(\xi )=\,4q(x)^{2}{\mathcal{V}}_{L(3,1)}(\xi )\mathbb{\xi },
\end{equation}%
where $q(x)$ is given by (\ref{q3}), the (birational) inverse of the
gradient map $\nabla _{\mathcal{V}_{L(3,1)}}$ is the map $\mathbf{\mu \circ
\alpha }$, which is a homogeneous polynomial map of degree four.

\subsubsection{$L(2,1)\equiv J_{3}^{\mathbb{C}}$\label{L210}}

The model $L(2,1)$ corresponds to a symmetric space, and it is related to
the simple cubic Jordan algebra over the complex numbers, $J_{3}^{\mathbb{C}%
} $. In this case, the Bekenstein-Hawking entropy and the attractor values
of scalar fields of extremal BPS black holes are explicitly known; see e.g.
\cite{FGimonK}. In the present treatment, we will highlight its relation to
the complete `parent' models $L(8,1)$ and $L(4,1)$.

This model is complete : has $q=2,P=1$, and $\mathcal{D}_{3}=4$ \cite{dWVP};
thus, the number of variables is%
\begin{equation}
\left( 1+q+2+P\cdot D_{q+1}\right) _{q=2,P=1}=1+4+4=9.
\end{equation}%
This complete model can be regarded as a linearly constrained $L(4,1)$
model, and thus as a linearly constrained $L(8,1)$ model, as well. We
substitute $(y^{1},y^{2},y^{3},y^{4})=(y^{1},y^{2},0,0)$ in the quadratic
forms from the $L(4,1)$ model in (\ref{th410}), then one obtains the
following quadratic forms with associated $\Gamma $-matrices:
\begin{equation}
\begin{array}{llllll}
\Gamma _{0} & = & {\mathbb{I}}_{4}, & Q_{0}\equiv Q[{}_{00}^{00}] & = &
\left( y^{1}\right) ^{2}+\left( y^{2}\right) ^{2}\,+\left( y^{3}\right)
^{2}+\left( y^{4}\right) ^{2}; \\
\Gamma _{1} & = & \sigma _{3}\otimes {\mathbb{I}}_{2}, & Q_{1}\equiv
Q[{}_{01}^{00}] & = & \left( y^{1}\right) ^{2}-\left( y^{2}\right)
^{2}\,+\left( y^{3}\right) ^{2}-\left( y^{4}\right) ^{2}; \\
\Gamma _{2} & = & \sigma _{1}\otimes \sigma _{3},\qquad & Q_{2}\equiv
Q[{}_{10}^{01}] & = & 2(y^{1}y^{2}-y^{3}y^{4}); \\
\Gamma _{3} & = & \sigma _{1}\otimes \sigma _{1}, & Q_{3}\equiv
Q[{}_{00}^{11}] & = & 2(y^{1}y^{4}+y^{2}y^{3})~;%
\end{array}%
\end{equation}%
notice that%
\begin{equation}
(Q_{4})_{|y^{5}=\ldots =y^{16}=0}=(Q_{5})_{|y^{5}=\ldots =y^{16}=0}=0.
\end{equation}

Thus, $L(2,1)$ can be regarded as $L(8,1)$ with $6+12=18$ linear
constraints, or equivalently as $L(4,1)$ with $2+4=6$ linear constraints :
\begin{eqnarray}
L(2,1)\, &=&\left. \,L(8,1)\right\vert _{x^{4}=\ldots =x^{9}=0,~y^{5}=\ldots
=y^{16}=0}=\left. L(4,1)\right\vert _{x^{4}=x^{5}=0,~y^{5}=...=y^{8}=0}; \\
{\mathcal{V}}_{L(2,1)}\, &=&\,\left. {\mathcal{V}}_{L(8,1)}\right\vert
_{x^{4}=\ldots =x^{9}=0,~y^{5}=\ldots =y^{16}=0}~=\left. {\mathcal{V}}%
_{L(4,1)}\right\vert _{x^{4}=x^{5}=0,~y^{5}=\ldots =y^{8}=0}.
\end{eqnarray}%
The set $\{\Gamma _{1},\Gamma _{2},\Gamma _{3}\}$ is a Clifford set, and the
associated quadrics satisfy the Lorentzian quadratic relation%
\begin{equation}
-Q_{0}^{2}+Q_{2}^{1}+Q_{2}^{2}+Q_{3}^{2}=0.  \label{GammQI2}
\end{equation}%
Thus, the cubic form reads
\begin{equation}
{\mathcal{V}_{L(2,1)}}(s,x^{0},\ldots ,x^{3},y^{1},\ldots
,y^{4})\,:=\,sq(x)+\sum_{I=0}^{3}x^{I}Q_{I}(y)~,
\end{equation}%
where the $x^{4}=x^{5}=x^{6}=x^{7}=x^{8}=x^{9}=0$ restriction of the
Lorentzian quadratic form $q(x)$ (\ref{q8}) is the corresponding Lorentzian
form in $x^{0},\ldots ,x^{3}$ :
\begin{equation}
q(x)\,:=\,-\left( x^{0}\right) ^{2}\,+q_{3}(x)=-\left( x^{0}\right)
^{2}+\left( x^{1}\right) ^{2}+\left( x^{2}\right) ^{2}+\left( x^{3}\right)
^{2}.  \label{q2}
\end{equation}%
The invertibility condition enounced in Sec.\ \ref{main} yields that the
gradient map $\nabla _{\mathcal{V}}$ is invertible, with $r=0$. Thus, the
map $\mathbf{\alpha }$ can be identified with the identity map and the
inverse map of the gradient map $\nabla _{\mathcal{V}_{L(2,1)}}$ is simply $%
\mathbf{\mu }$, which is a polynomial map of degree $2$ :
\begin{equation}
{\mathbb{R}}_{\xi }^{9}\,\overset{\nabla _{\mathcal{V}_{L(2,1)}}}{%
\longrightarrow }\,{\mathbb{R}}_{z}^{9}\,\,\overset{\mathbf{\mu }}{%
\longrightarrow }\,{\mathbb{R}}_{\xi }^{9}~;
\end{equation}%
with
\begin{equation}
\mathbf{\mu }\circ \nabla _{\mathcal{V}_{L(2,1)}}\left( \xi \right) =4%
\mathcal{V}_{L(2,1)}\left( \xi \right) \mathbb{\xi };
\end{equation}%
The gradient map
\begin{equation*}
\nabla _{\mathcal{V}_{L(2,1)}}\,:\,\,{\mathbb{R}}^{9}\,\longrightarrow \,{%
\mathbb{R}}^{9},\qquad \xi \longmapsto \,^{T}\left( \mathcal{V}_{s}\left(
\xi \right) ,\mathcal{V}_{I}\left( \xi \right) ,\mathcal{V}_{\alpha }\left(
\xi \right) \right)
\end{equation*}%
can be identified with the adjoint map
\begin{equation}
J_{3}^{\mathbb{C}}\,\longrightarrow \,J_{3}^{\mathbb{C}},\qquad
M\,\longmapsto \,M^{\sharp }
\end{equation}%
here $M^{\sharp }$ is the adjoint matrix of $M\in J_{3}^{\mathbb{C}}$. This
adjoint map $\#$ is (birationally) invertible, with inverse given by the map
$M^{\sharp }\mapsto (M^{\sharp })^{\sharp }$, which is thus essentially the
map $\mathbf{\mu }$.

\paragraph{The determinant of $J_{3}^{\mathbb{C}}$.}

Again, it is worth making more explicit the relation between the complete
model $L(2,1)$ and the Euclidean simple cubic Jordan algebra $J_{3}^{\mathbb{%
C}}$. If $\mathbf{i}$ denotes the imaginary unit of $\mathbb{C}$, then to $%
\xi =~^{T}(s,x^{0},x^{1},x^{2},x^{3},y^{1},y^{2},y^{3},y^{4})$ (cfr. (\ref%
{csii})) we define the $3\times 3$ complex Hermitian matrix
\begin{equation}
J_{3}^{\mathbb{C}}\,\ni \,{M}\,=\,M^{\mathbb{C}}\,:=\,\left(
\begin{array}{ccc}
x^{0}-x^{1} & -x^{2}-\mathbf{i}x^{3} & y^{1}+\mathbf{i}y^{3} \\
-x^{2}+\mathbf{i}x^{3} & x^{0}+x^{1} & y^{2}-\mathbf{i}y^{4} \\
y^{1}-\mathbf{i}y^{3} & y^{2}+\mathbf{i}y^{4} & s%
\end{array}%
\right) ,  \label{matL21}
\end{equation}%
Notice that, consistently, this matrix is the restriction of $\mathcal{M}^{%
\mathbb{H}}\,$\ (\ref{matL41}) to ${x^{4}=x^{5}=0}$ and $y{^{5}=\ldots
=y^{8}=0}$. Then we find

\begin{equation}
{\mathcal{V}_{L(2,1)}}=\,-\det M~.
\end{equation}

\subsubsection{$L(1,1)\equiv J_{3}^{\mathbb{R}}$\label{L110}}

This model corresponds to a symmetric space \cite{dWVP}, and it is related
to the simple cubic Jordan algebra over the reals, $J_{3}^{\mathbb{R}}$,
namely the algebra of symmetric real $3\times 3$ matrices. In this case, the
Bekenstein-Hawking entropy and the attractor values of scalar fields of
extremal black holes are explicitly known; see e.g. \cite{FGimonK}. In the
present treatment, we will highlight its relation to the complete `parent'
models $L(8,1)$, $L(4,1)$ and $L(2,1)$.

From (\ref{table D}), it holds that $\mathcal{D}_{2}=2$, and there are $%
1+3+2=6$ variables, namely $s,x^{0},x^{1},x^{2},y^{1},y^{2}$. The symmetric $%
2\times 2$ $\Gamma $-matrices are $\Gamma _{0}={\mathbb{I}}_{2},\Gamma
_{1}=\sigma _{3},\Gamma _{2}=\sigma _{1}$, and thus $\{\Gamma _{1},\Gamma
_{2}\}$ is a Clifford set. The corresponding quadratic forms $%
Q_{I}={}^{T}y\Gamma _{I}y$ from (\ref{quadg1}) read
\begin{equation}
Q_{0}\,\equiv \,Q[{}_{0}^{0}],\qquad Q_{1}\,\equiv \,Q[{}_{1}^{0}],\qquad
Q_{2}\,\equiv \,Q[{}_{0}^{1}]~,
\end{equation}%
and again they can be obtained by suitably restricting the $y$-variables
from the quadratic forms of $L(8,1)$ (or, equivalently, from $L(4,1)$ or
from $L(2,1)$):
\begin{equation}
\begin{array}{lll}
(Q_{0})_{|y^{3}=\ldots =y^{16}=0}\equiv Q[{}_{0}^{0}] & := & \left(
y^{1}\right) ^{2}+\left( y^{2}\right) ^{2}; \\
(Q_{1})_{|y^{3}=\ldots =y^{16}=0}\equiv Q[{}_{1}^{0}] & := & \left(
y^{1}\right) ^{2}-\left( y^{2}\right) ^{2}; \\
(Q_{2})_{|y^{3}=\ldots =y^{16}=0}=Q[{}_{0}^{1}] & := & 2y^{1}y^{2}; \\
(Q_{3})_{|y^{3}=\ldots =y^{16}=0}=0, &  &
\end{array}
\label{th110}
\end{equation}%
thus implying%
\begin{eqnarray}
L(1,1)\, &=&\,\left. L(8,1)\right\vert _{x^{3}=\ldots =x^{9}=0,~y^{3}=\ldots
=y^{16}=0}  \notag \\
&=&\left. L(4,1)\right\vert _{x^{3}=x^{4}=x^{5}=0,~y^{3}=...=y^{8}=0}  \notag
\\
&=&\left. L(2,1)\right\vert _{x^{3}=0,~y^{3}=y^{4}=0}; \\
{\mathcal{V}}_{L(1,1)}\, &=&\,\left. {\mathcal{V}}_{L(8,1)}\right\vert
_{x^{3}=\ldots =x^{9}=0,~y^{3}=\ldots =y^{16}=0}~  \notag \\
&=&\left. {\mathcal{V}}_{L(4,1)}\right\vert
_{x^{3}=x^{4}=x^{5}=0,~y^{3}=\ldots =y^{8}=0}=\left. {\mathcal{V}}%
_{L(2,1)}\right\vert _{x^{3}=0,~y^{3}=y^{4}=0}.
\end{eqnarray}

The following Lorentzian quadratic identity holds :
\begin{equation}
-Q_{0}^{2}\,+\,Q_{1}^{2}\,+\,Q_{2}^{2}\,=\,0~,\qquad \text{in~fact}\quad
(\left( \,y^{1}\right) ^{2}+\left( \,y^{2}\right) ^{2})^{2}\,=\,(\left(
\,y^{1}\right) ^{2}-\left( \,y^{2}\right) ^{2})^{2}\,+\,(2y^{1}y^{2})^{2}.
\label{qidg1}
\end{equation}%
The cubic form $\mathcal{V}$ on ${\mathbb{R}}^{6}$ is thus given by:
\begin{equation}
\mathcal{V}_{L(1,1)}\,=\,sq(x)\,+\,\sum_{I=0}^{2}x^{I}Q_{I}(y),
\end{equation}%
where the $x^{3}=x^{4}=x^{5}=x^{6}=x^{7}=x^{8}=x^{9}=0$ restriction of the
Lorentzian quadratic form $q(x)$ (\ref{q8}) is the corresponding Lorentzian
form in $x^{0},x^{1},x^{2}$ :
\begin{equation}
q(x)\,:=\,-\left( x^{0}\right) ^{2}\,+q_{2}(x)=-\left( x^{0}\right)
^{2}+\left( x^{1}\right) ^{2}+\left( x^{2}\right) ^{2}.  \label{q1}
\end{equation}

(\ref{qidg1}) implies that the invertibility condition enounced in Sec.\ \ref%
{main} is satisfied with $r=0$, so $L(1,1)$ is a complete model. The inverse
of the gradient map
\begin{equation}
\nabla _{\mathcal{V}_{L(1,1)}}:\,{\mathbb{R}}_{s,x,y}^{6}\,\longrightarrow \,%
{\mathbb{R}}_{t,u,v}^{6},
\end{equation}%
is then simply the map $\mathbf{\mu }$ in (\ref{mu}), whose components are
homogeneous polynomials of degree $2$:
\begin{eqnarray}
\mathbf{\mu } &:&{\mathbb{R}}_{t,u,v}^{6}\equiv {\mathbb{R}}%
^{6}\,\longrightarrow \,{\mathbb{R}}_{s,x,y}^{6}\equiv {\mathbb{R}}^{6};
\notag \\
\mathbf{\mu }(t,u,v)\, &=&\,\left(
\begin{array}{c}
q(u) \\
-2tu_{0}+\,\frac{1}{2}Q_{0}(v) \\
2tu_{1}\,+\,\frac{1}{2}Q_{1}(v) \\
2tu_{2}\,+\,\frac{1}{2}Q_{2}(v) \\
\left( \sum_{I=0}^{2}u_{I}\Gamma _{I}\right) v \\
\end{array}%
\right) \;=\;\left(
\begin{array}{c}
-u_{0}^{2}+u_{1}^{2}+u_{2}^{2} \\
-2tu_{0}+\frac{1}{2}(v_{1}^{2}+v_{2}^{2}) \\
2tu_{1}+\frac{1}{2}(v_{1}^{2}-v_{2}^{2}) \\
2tu_{2}+v_{1}v_{2} \\
u_{0}v_{1}+u_{1}v_{1}+u_{2}v_{2} \\
u_{0}v_{2}-u_{1}v_{2}+u_{2}v_{1}%
\end{array}%
\right) ~.  \label{muL110}
\end{eqnarray}%
The following identity, for any $\xi =~^{T}(s,x,y)\in \mathbb{R}^{6}$ (cfr. (%
\ref{csii})), holds :
\begin{equation}
\mathbf{\mu }(\nabla _{\mathcal{V}_{L(1,1)}}(\xi ))\,=\,4{\mathcal{V}}%
_{L(1,1)}(\xi )\xi ~.
\end{equation}

\paragraph{The determinant of $J_{3}^{\mathbb{R}}$, and an alternative
construction of the map $\mathbf{\protect\mu }$.}

An alternative construction of the inverse map $\mathbf{\mu }$ can be given,
by exploiting the relation to $J_{3}^{\mathbb{R}}$. In fact, using the
restriction of the matrix $M^{\mathbb{C}}$ (\ref{matL21}) to $x^{3}=0$ and $%
y^{3}=y^{4}=0$, we get
\begin{equation}
\mathcal{V}_{L(1,1)}\,=\,-\det \mathfrak{M}~,\qquad \mathfrak{M}\,:=\,\left(
\begin{array}{ccc}
x^{0}-x^{1} & -x^{2} & y^{1} \\
-x^{2} & x^{0}+x^{1} & y^{2} \\
y^{1} & y^{2} & s%
\end{array}%
\right) \;\in \,J_{3}^{\mathbb{R}}~.
\end{equation}%
Thus, the derivatives of $\mathcal{V}_{L(1,1)}$ are linear combinations of
the $\left( 2\times 2\right) $-minors of $\mathfrak{M}$. By denoting with $%
M_{ab}$ the determinant of the submatrix of $\mathfrak{M}$ obtained by
deleting the $a$-th row and $b$-th column, one obtains, with $%
(t,u^{0},u^{1},u^{2},v^{1},v^{2})^{T}=\nabla _{\mathcal{V}_{L(1,1)}}$:
\begin{equation}
\begin{array}{llllrrcr}
t= & {\partial _{s}}\mathcal{V}_{L(1,1)} & = & -M_{33}, & u_{2}= & {\partial
_{x^{2}}}\mathcal{V}_{L(1,1)} & = & -2M_{12}, \\
u_{0}= & {\partial _{x^{0}}}\mathcal{V}_{L(1,1)} & = & -M_{11}-M_{22}, &
\qquad v_{1}= & {\partial _{y^{1}}}\mathcal{V}_{L(1,1)} & = & -2M_{13}, \\
u_{1}= & {\partial _{x^{1}}}\mathcal{V}_{L(1,1)} & = & M_{11}-M_{22}, &
v_{2}= & {\partial _{y^{2}}}\mathcal{V}_{L(1,1)} & = & 2M_{23}. \\
&  &  &  &  &  &  &
\end{array}%
\end{equation}%
Therefore, $\nabla _{\mathcal{V}_{L(1,1)}}$ determines the coefficients of
the adjoint matrix $\mathfrak{M}^{\sharp }$:
\begin{equation}
2\mathfrak{M}^{\sharp }\,:=\,2\left(
\begin{array}{ccc}
M_{11} & -M_{12} & M_{13} \\
-M_{12} & M_{22} & -M_{23} \\
M_{13} & -M_{23} & M_{33}%
\end{array}%
\right) \,=\,\left(
\begin{array}{ccc}
-u_{0}+u_{1} & u_{2} & -v_{1} \\
u_{2} & -u_{0}-u_{1} & -v_{2} \\
-v_{1} & -v_{2} & -2t%
\end{array}%
\right) ~.
\end{equation}%
The inverse of $\nabla _{\mathcal{V}_{L(1,1)}}$ is then basically given by%
\footnote{%
This also results from the adjoint identity (\ref{adj-id}) for $J_{3}^{%
\mathbb{R}}$, and as such this method actually works for all four complete
models $L(q,1)$ with $q=1,2,4,8$.}%
\begin{equation}
\mathfrak{M}^{\sharp }\mapsto (\mathfrak{M}^{\sharp })^{\sharp }=\left( \det
\mathfrak{M}\right) \mathfrak{M}=-{\mathcal{V}}_{L(1,1)}(\xi )\mathfrak{M},
\end{equation}%
since $\mathfrak{M}$ determines $\xi $. Explicitly, one finds
\begin{equation}
(2\mathfrak{M}^{\sharp })^{\sharp }\,=\,\left(
\begin{array}{ccc}
2tu_{0}+2tu_{1}-v_{2}^{2} & 2tu_{2}+v_{1}v_{2} &
-u_{0}v_{1}-u_{1}v_{1}+u_{2}v_{2} \\
2tu_{2}+v_{1}v_{2} & 2tu_{0}-2tu_{1}-v_{1}^{2} &
-u_{0}v_{2}+u_{1}v_{2}-u_{2}v_{1} \\
-u_{0}v_{1}-u_{1}v_{1}-u_{2}v_{2} & -u_{0}v_{2}+u_{1}v_{2}-u_{2}v_{1} &
u_{0}^{2}-u_{1}^{2}-u_{2}^{2}%
\end{array}%
\right) ~.
\end{equation}%
Next, since $(2\mathfrak{M}^{\sharp })^{\sharp }=4(\mathfrak{M}^{\sharp
})^{\sharp }=-4{\mathcal{V}}_{L(1,1)}(\xi )\mathfrak{M}$ one recovers the
inverse $\mathbf{\mu }$ of $\nabla _{\mathcal{V}_{L(1,1)}}$ given in (\ref%
{muL110}).

\section{Examples, II : $L(q,2)$, $q=1,2,3$\label{L420}}

In Sec.\ \ref{subL810} we showed that%
\begin{equation}
L(q,1)\subset L(8,1),~1\leqslant q\leqslant 7,
\end{equation}%
with the `inclusion' realized mostly by simply setting the last $x^{I}$%
-variables equal to zero, as well as by (occasionally) setting the `last
half' of the $y^{\alpha }$-variables to zero. In the present Section, we
will consider the inversion of the gradient map $\nabla _{\mathcal{V}}$ in
models $L(q,2)$, with $q=1,2,3$; in these models, it holds that (cfr. Table
1 of \cite{dWVP})%
\begin{equation}
m\equiv \mathcal{D}_{q+1}\cdot 2=2^{q}\cdot 2=2^{q+1}.
\end{equation}

\subsection{Block decompositions\label{blocks}}

We define the $P$-block diagonal form $M^{(P)}$ of a matrix $M$ of size $%
m\times m$ as the matrix of size $mP\times mP$ defined as
\begin{equation}
M^{(P)}\,:=\,M\otimes {\mathbb{I}}_{P}\,=\,\left(
\begin{array}{ccc}
M &  &  \\
\vdots & \ddots & \vdots \\
&  & M%
\end{array}%
\right) ~.
\end{equation}%
For a vector $y\in {\mathbb{R}}^{mP}$, we define vectors $y^{(k)}\in {%
\mathbb{R}}^{m}$ by taking the vector with components $(k-1)P+1,\ldots ,kP$
so that
\begin{equation}
y=~^{T}(y^{(1)},\ldots ,y^{(P)})\in (\underset{P\text{ times}}{\underbrace{{%
\mathbb{R}}^{m}\oplus \ldots \oplus {\mathbb{R}}^{m}}}\,=\,{\mathbb{R}}%
^{mP})~.  \label{blocky}
\end{equation}%
Let now $\Gamma _{0}={\mathbb{I}}_{m}$ and let $\{\Gamma _{1},\ldots ,\Gamma
_{q+1}\}$ be a Clifford set of $m\times m$ matrices. The symmetric matrices $%
\Gamma _{I}^{(P)}$ are, in general, not $\Gamma $-matrices, since $mP$ is
not a power of $2$. They still define quadratic forms on ${\mathbb{R}}^{mP}$%
, which we denote by $Q_{I}^{(P)}$:
\begin{equation}
Q_{I}^{(P)}(y)\,:=\,Q_{I}(y^{(1)})\,+\,Q_{I}(y^{(2)})\,+\,\ldots
\,+\,Q_{I}(y^{(P)})~.  \label{QP}
\end{equation}

\subsection{$P=2$\label{P=2}}

Given $\Gamma _{0}={\mathbb{I}}_{m}$ and a Clifford set of $m\times m$
matrices $\{\Gamma _{1},\ldots ,\Gamma _{q+1}\}$, the matrices $\Gamma
_{I}^{(2)}$'s are again $\Gamma $-matrices of size $2m=2^{g+1}$, with $%
\Gamma _{0}^{(2)}={\mathbb{I}}_{2m}$. Moreover, the set $\{\Gamma
_{1}^{(2)},\ldots ,\Gamma _{q+1}^{(2)}\}$ is again a Clifford set, now of $%
2m\times 2m$ matrices, since distinct matrices still anti-commute and they
square to ${\mathbb{I}}_{2m}$. In the subsequent treatment, we will show that%
\begin{equation}
L(q,2)\subset L\left( 2^{q},1\right) ,~q=1,2,3,\qquad\mbox{that is:}\quad
\left\{
\begin{array}{c}
L(1,2)\,\subset \,L(2,1), \\
L(2,2)\,\subset \,L(4,1), \\
L(3,2)\,\subset \,L(8,1),%
\end{array}%
\right.
\end{equation}%
where $\subset $ denotes a descendant relation. In each case, the quadratic
forms $Q_{I}^{(2)}$, $I=0,\ldots ,q+1$, are all in the set of the $q^{\prime
}+2$ quadratic forms defining the $L(q^{\prime },1)$ model. The remaining $%
r=q^{\prime }-q$ forms of the $L(q^{\prime },1)$ model (where $r=1,2,5$
respectively) can be used as the $R_{K}$'s in the invertibility condition of
Sec.\ \ref{main}. In particular, the Lorentzian relation between the
quadratic forms of the complete model implies the following Lorentzian
quadratic relation (\ref{lorrel}) among the $Q_{I}^{(2)}$'s and the $R_{K}$%
's:
\begin{equation}
-Q_{0}^{(2)}(y)^{2}\,+\,Q_{1}^{(2)}(y)^{2}\,+\,\ldots
\,+\,Q_{q+1}^{(2)}(y)^{2}\,+\,R_{1}(y)^{2}\,+\,\ldots
\,+\,R_{r}(y)^{2}\,=\,0~.  \label{P2}
\end{equation}%
Thus, the cubic form ${\mathcal{V}}={\mathcal{V}}_{L(q,2)}$ for $q=1,2,3$
has an invertible gradient map $\nabla _{\mathcal{V}_{L(q,2)}}$, and
correspondingly the BPS system (\ref{BPS}) can be explicitly solved.

Again, from the treatment of Sec.\ \ref{invertgamma}, the solution of the
BPS system of $L(q,2)$ for $q=1,2,3$ is then given by (\ref{pre-sol})-(\ref%
{pre-sol-3}), and (\ref{SSd}) and (\ref{zH-1})-(\ref{zH-3}) yield the
corresponding expression of the BPS black hole entropy and of the BPS
attractors, respectively. We refer to Sec.\ \ref{L120}, \ref{L130} for
examples of lower dimensional models $L(q,P)$ (namely, $L(1,2)$ and $L(1,3)$%
) which we explicitly work out in detail (see also the generalization to $%
L(1,P)$ models with $P\geqslant 2$ in Sec.\ \ref{L1P0}).

\subsection{$L(3,2)$\label{L320}}

The $L(3,2)$ model has%
\begin{equation}
\left( 1+q+2+P\cdot {\mathcal{D}}_{q+1}\right) _{q=3,P=2}\,=\,1+5+2\cdot 8=22
\end{equation}%
variables, denoted as $s,x^{0},\ldots ,x^{4},y^{1},\ldots ,y^{16}$. The
cubic form reads%
\begin{equation}
{\mathcal{V}}_{L(3,2)}\,=\,sq(x)\,+\,\sum_{I=0}^{4}x^{I}Q_{I}^{(2)},\qquad
q(x)\,:=\,-\left( x^{0}\right) ^{2}\,+\,\left( x^{1}\right) ^{2}+\left(
x^{2}\right) ^{2}\,+\,(x^{3})^{2}+\,\left( x^{4}\right) ^{2}~,  \label{V_3,2}
\end{equation}%
where the $Q_{I}^{(2)}$'s are given by (\ref{this2}) below.

Let us now show that $L(3,2)\subset L(8,1)$ : first of all, both models are
defined by $\Gamma $-matrices of size $16\times 16$, and let $\Gamma
_{0},\ldots ,\Gamma _{9}$ be those of $L(8,1)$ as in Sec.\ \ref{L810}. We
can write the five $\Gamma $-matrices (of size $8\times 8$) of the $L(3,1)$
model as $\bar{\Gamma}_{0}={\mathbb{I}}_{8}$ and the others $\{\bar{\Gamma}%
_{1},\ldots ,\bar{\Gamma}_{4}\}$ are a Clifford set. Obviously, $\Gamma _{0}=%
\bar{\Gamma}_{0}^{(2)}$. Next, we observe that $\Gamma _{1},\Gamma
_{3},\Gamma _{4},\Gamma _{5}$ (but not $\Gamma _{2}$) are of the form $%
\Gamma _{I}=\bar{\Gamma}_{J}^{(2)}$, for some $\Gamma $-matrices $\bar{\Gamma%
}_{J}$ of size $8\times 8$. Moreover, these four $\bar{\Gamma}_{J}$ form a
Clifford set (since the four $\Gamma _{I}$ are a Clifford set). Thus, we see
that $L(3,2)\subset L(8,1)$, by using
\begin{equation}
\bar{\Gamma}_{0}^{(2)}:=\Gamma _{0},\quad \bar{\Gamma}_{1}^{(2)}:=\Gamma
_{1},\quad \bar{\Gamma}_{2}^{(2)}:=\Gamma _{3},\quad \bar{\Gamma}%
_{3}^{(2)}:=\Gamma _{4},\quad \bar{\Gamma}_{4}^{(2)}:=\Gamma _{5}.
\end{equation}

Below, we list the ten quadratic forms of the $L(8,1)$ model in the same
order as in Sec.\ \ref{L810} but with the names adapted to the $L(3,2)$
model. Five of the forms are denoted by $Q_{I}^{(2)}$, $I=0,\ldots ,4$, and
these are in the $L(3,2)$ model, the remaining five, $R_{1},\ldots ,R_{5}$,
will be used to satisfy the invertibility condition\footnote{%
Notice that the $Q_{I}^{(2)}$ listed here are not the quadrics obtained from
the $Q_{I}$'s in Sec.\ \ref{L310}, since we have chosen a different Clifford
set for $L(3,1)$ in that section.} enounced in Sec.\ \ref{main} (with $r=5$%
); cfr. (\ref{this3}):%
\begin{eqnarray}
&&%
\begin{array}{lll}
Q_{0}^{(2)} & \equiv Q[{}_{0000}^{0000}]:= & \left( y^{1}\right)
^{2}+\,\ldots \,+\,\left( y^{8}\right) ^{2}\,+\,\left( y^{9}\right)
^{2}\,+\,\ldots \,+\,\left( y^{16}\right) ^{2}; \\
Q_{1}^{(2)} & \equiv Q[{}_{0001}^{0000}]:= & \left\{
\begin{array}{l}
\left( y^{1}\right) ^{2}-\left( y^{2}\right) ^{2}\,+\,\ldots \,+\,\left(
y^{7}\right) ^{2}\,-\,\left( y^{8}\right) ^{2} \\
+\left( y^{9}\right) ^{2}-\left( y^{10}\right) ^{2}\,+\,\ldots \,+\left(
y^{15}\right) ^{2}-\left( y^{16}\right) ^{2};%
\end{array}%
\right. \\
R_{1} & \equiv Q[{}_{1010}^{0001}]:= & 2(y^{1}y^{2}-y^{3}y^{4}\,+%
\,y^{5}y^{6}-y^{7}y^{8}\,-\,y^{9}y^{10}+y^{11}y^{12}\,-%
\,y^{13}y^{14}+y^{15}y^{16}); \\
Q_{2}^{(2)} & \equiv Q[{}_{0000}^{0011}]:= & 2(y^{1}y^{4}+y^{2}y^{3}\,+%
\,y^{5}y^{8}+y^{6}y^{7}\,+\,y^{9}y^{12}+y^{10}y^{11}\,+%
\,y^{13}y^{16}+y^{14}y^{15}); \\
Q_{3}^{(2)} & \equiv Q[{}_{0101}^{0101}]:= &
2(y^{1}y^{6}-y^{2}y^{5}+y^{3}y^{8}-y^{4}y^{7}\,+%
\,y^{9}y^{14}-y^{10}y^{13}+y^{11}y^{16}-y^{12}y^{15}); \\
Q_{4}^{(2)} & \equiv Q[{}_{0110}^{0111}]:= &
2(y^{1}y^{8}+y^{2}y^{7}-y^{3}y^{6}-y^{4}y^{5}\,+%
\,y^{6}y^{16}+y^{10}y^{15}-y^{11}y^{14}-y^{12}y^{13}); \\
R_{2} & \equiv Q[{}_{0010}^{1001}]:= &
2(y^{1}y^{10}+y^{2}y^{9}-y^{3}y^{12}-y^{4}y^{11}+y^{5}y^{14}+y^{6}y^{13}-y^{7}y^{16}-y^{8}y^{15});
\\
R_{3} & \equiv Q[{}_{1101}^{1011}]:= &
2(y^{1}y^{12}-y^{2}y^{11}+y^{3}y^{10}-y^{4}y^{9}-y^{5}y^{16}+y^{6}y^{15}-y^{7}y^{14}+y^{8}y^{13});
\\
R_{4} & \equiv Q[{}_{1110}^{1101}]:= &
2(y^{1}y^{14}+y^{2}y^{13}-y^{3}y^{16}-y^{4}y^{15}-y^{5}y^{10}-y^{6}y^{9}+y^{7}y^{12}+y^{8}y^{11});
\\
R_{5} & \equiv Q[{}_{1001}^{1111}]:= &
2(y^{1}y^{16}-y^{2}y^{15}+y^{3}y^{14}-y^{4}y^{13}+y^{5}y^{12}-y^{6}y^{11}+y^{7}y^{10}-y^{8}y^{9}).%
\end{array}
\notag \\
&&  \label{this2}
\end{eqnarray}

Thus, (notice the peculiar choice of the variables, we put $x^{2}=0$ but $%
x^{5}$ remains) it holds that
\begin{eqnarray}
L(3,2)\, &=&\,\left. L(8,1)\right\vert _{x^{2}=x^{6}=x^{7}=x^{8}=x^{9}=0}; \\
{\mathcal{V}}_{L(3,2)}\, &=&\,\left. {\mathcal{V}}_{L(8,1)}\right\vert
_{x^{2}=x^{6}=x^{7}=x^{8}=x^{9}=0}.
\end{eqnarray}%
Thus, up to a relabeling of the $x^{I}$'s, the model $L(3,2)$ can be
regarded as the $L(8,1)$ model with five linear constraints.

It is then clear that $L(3,2)$ is an invertible model : one may take the
\begin{equation}
\Omega _{1}=\Gamma _{2},\quad \Omega _{2}=\Gamma _{6},\quad \Omega
_{3}=\Gamma _{7},\quad \Omega _{4}=\Gamma _{8},\quad \Omega _{5}=\Gamma
_{9},\quad \text{so}\quad r=5,
\end{equation}%
and, since the set of $\Gamma $-matrices $\{\Gamma _{1},\ldots ,\Gamma
_{9}\} $ of $L(8,1)$ is a Clifford set and the Lorentzian identity (\ref%
{GammQI4}) holds, the invertibility condition enounced in Sec.\ \ref{main}
is satisfied.

Since $r\neq 0$, the inverse of the gradient map $\nabla _{{\mathcal{V}}%
_{L(3,2)}}$ is given as a composition of two maps, $\mathbf{\alpha }$ and $%
\mathbf{\mu }$ :%
\begin{equation}
{\mathbb{R}}_{\xi }^{1+5+16}\,\overset{\nabla _{{\mathcal{V}}_{L(3,2)}}}{%
\longrightarrow }\,{\mathbb{R}}_{z}^{1+5+16}\,\overset{\mathbf{\alpha }}{%
\longrightarrow }\,{\mathbb{R}}_{(t,u,v,w)}^{1+5+16+5}\,\overset{\mathbf{\mu
}}{\longrightarrow }\,{\mathbb{R}}_{\xi }^{1+5+16}~.
\end{equation}%
The map $\mathbf{\alpha }$, as in (\ref{alpha}), is given by
\begin{equation}
\mathbf{\alpha }(z_{1},\ldots
,z_{22})\,:=\,^{T}(z_{1}^{2},\,z_{1}z_{2},\ldots
,\,z_{1}z_{22},\,R_{1}(z),...,R_{5}(z)),
\end{equation}%
where the quadratic forms $R_{K}$'s ($K=1,...,5$) depend only on the last $%
2\cdot 8=16$ variables and they are obtained by substituting $y^{L}:=z_{L+6}$%
, $L=1,\ldots ,16$ in $Q_{2},Q_{6},Q_{7},Q_{8},Q_{9}$ given in (\ref{this3}%
), respectively:
\begin{eqnarray}
R_{1}(z):= &&Q_{2}(z_{6},\ldots ,z_{22}), \\
R_{2}(z):= &&Q_{6}(z_{6},\ldots ,z_{22}), \\
\vdots &&\vdots  \notag \\
R_{5}(z):= &&Q_{9}(z_{6},\ldots ,z_{22})~.
\end{eqnarray}%
The map $\mathbf{\mu }$ (\ref{mu}) has $1+5+16+5$ components which are
homogeneous polynomials of degree $2$ in the $27$ variables $t,u_{0},\ldots
,u_{4},v_{1},\ldots ,v_{16},w_{1},...,w_{5}$. Since
\begin{equation}
\mathbf{\mu }\circ \mathbf{\alpha }\circ \nabla _{{\mathcal{V}}%
_{L(3,2)}}\,(\xi )=\,4q(x)^{2}{\mathcal{V}}_{L(3,2)}(\xi )\mathbb{\xi },
\end{equation}%
where ${\mathcal{V}}_{L(3,2)}$ and the corresponding $q(x)$ are given by (%
\ref{V_3,2}), the (birational) inverse of the gradient map $\nabla _{%
\mathcal{V}_{L(3,2)}}$ is the map $\mathbf{\mu \circ \alpha }$, which is an
homogeneous polynomial map of degree four.

\subsection{$L(2,2)$\label{L220}}

The $L(2,2)$ model has $q=2,P=2$, and $\mathcal{D}_{3}=4$ \cite{dWVP}; thus,
the number of variables is%
\begin{equation}
\left( 1+q+2+P\cdot \mathcal{D}_{q+1}\right) _{q=2,P=2}=1+4+2\cdot 4=13.
\end{equation}%
Correspondingly, the cubic form reads as follows :
\begin{equation}
{\mathcal{V}}_{L(2,2)}\,=\,sq(x)\,+\,\sum_{I=0}^{3}x^{I}Q_{I}^{(2)},\qquad
q(x)\,:=\,-\left( x^{0}\right) ^{2}\,+\,\left( x^{1}\right) ^{2}+\left(
x^{2}\right) ^{2}\,+\,\left( x^{3}\right) ^{2}~,  \label{V_2,2}
\end{equation}%
where the $Q_{I}^{(2)}$'s are given by (\ref{this4}) below. Within the black
hole effective potential formalism, this model has been treated in \cite%
{DFT-Hom-07}.

Let us now show that $L(2,2)\subset L(4,1)$ : we write the four $\Gamma $%
-matrices (of size $4\times 4$) of the $L(2,1)$ model given in Sec.\ \ref%
{L210} as $\bar{\Gamma}_{0}={\mathbb{I}}_{4}$, and the others $\{\bar{\Gamma}%
_{1},\ldots ,\bar{\Gamma}_{3}\}$ are a Clifford set. Next, we observe that
if $\Gamma _{I}$, $I=0,\ldots ,5$ are the six $\Gamma $-matrices of the $%
L(4,1)$ model in (\ref{L410JH}), then we see that $L(2,2)\subset L(4,1)$,
since
\begin{equation}
\bar{\Gamma}_{0}^{(2)}:=\Gamma _{0},\quad \bar{\Gamma}_{1}^{(2)}:=\Gamma
_{1},\quad \bar{\Gamma}_{2}^{(2)}:=\Gamma _{2},\quad \bar{\Gamma}%
_{3}^{(2)}:=\Gamma _{3}.
\end{equation}

Below, we list the six quadratic forms of the $L(4,1)$ model in the same
order as in Sec.\ \ref{L410JH}, but with the names adapted to the $L(3,2)$
model. Four of the forms are denoted by $Q_{I}^{(2)}$, $I=0,\ldots ,3$, and
these are in the $L(2,2)$ model; the remaining two, $R_{1},R_{2}$, will be
used to satisfy the invertibility condition enounced in Sec.\ \ref{main}
(with $r=2$). Notice that both models are defined by $\Gamma $-matrices of
size $8\times 8$; let $\Gamma _{0},\ldots ,\Gamma _{5}$ be those of $L(4,1)$
as in Sec.\ \ref{L410JH}.

\begin{eqnarray}
&&%
\begin{array}{llllll}
\bar{\Gamma}_{0}^{(2)} & = & {\mathbb{I}}_{8}, & Q_{0}^{(2)}\equiv
Q[{}_{000}^{000}] & := & \left\{
\begin{array}{l}
\left( y^{1}\right) ^{2}+\left( y^{2}\right) ^{2}+\left( y^{3}\right)
^{2}+\left( y^{4}\right) ^{2} \\
+\,\left( y^{5}\right) ^{2}+\left( y^{6}\right) ^{2}+\,\left( y^{7}\right)
^{2}+\,\left( y^{8}\right) ^{2};%
\end{array}%
\right. \\
\bar{\Gamma}_{1}^{(2)} & = & \sigma _{3}\otimes {\mathbb{I}}_{2}\otimes {%
\mathbb{I}}_{2}, & Q_{1}^{(2)}\equiv Q[{}_{001}^{000}] & := & \left\{
\begin{array}{l}
\left( y^{1}\right) ^{2}-\left( y^{2}\right) ^{2}+\left( y^{3}\right)
^{2}-\left( y^{4}\right) ^{2} \\
+\,\left( y^{5}\right) ^{2}-\left( y^{6}\right) ^{2}+\,\left( y^{7}\right)
^{2}-\,\left( y^{8}\right) ^{2};%
\end{array}%
\right. \\
\bar{\Gamma}_{2}^{(2)} & = & \sigma _{1}\otimes \sigma _{3}\otimes {\mathbb{I%
}}_{2}, & Q_{2}^{(2)}\equiv Q[{}_{010}^{001}] & := & 2(y^{1}y^{2}-y^{3}y^{4}%
\,+\,y^{5}y^{6}-y^{7}y^{8}); \\
\bar{\Gamma}_{3}^{(2)} & = & \sigma _{1}\otimes \sigma _{1}\otimes {\mathbb{I%
}}_{2}, & Q_{3}^{(2)}\equiv Q[{}_{000}^{011}] & := & 2(y^{1}y^{4}+y^{2}y^{3}%
\,+\,y^{5}y^{8}+y^{6}y^{7}); \\
\Omega _{1} & = & \gamma _{11}\otimes {\mathbb{I}}_{2}\otimes \gamma _{11},
& R_{1}\equiv Q[{}_{101}^{101}] & := &
2(y^{1}y^{6}-y^{2}y^{5}+y^{3}y^{8}-y^{4}y^{7}); \\
\Omega _{2} & = & \sigma _{1}\otimes \gamma _{11}\otimes \gamma _{11}, &
R_{2}\equiv Q[{}_{110}^{111}] & := &
2(y^{1}y^{8}+y^{2}y^{7}-y^{3}y^{6}-y^{4}y^{5})~.%
\end{array}
\notag \\
&&  \label{this4}
\end{eqnarray}

Thus, it holds that
\begin{eqnarray}
L(2,2)\, &=&\,\left. L(4,1)\right\vert _{x^{4}=x^{5}=0}; \\
{\mathcal{V}}_{L(2,2)}\, &=&\,\left. {\mathcal{V}}_{L(4,1)}\right\vert
_{x^{4}=x^{5}=0}.
\end{eqnarray}%
Namely, the model $L(2,2)$ can be regarded as the $L(4,1)$ model with two
linear constraints.

It is then clear that the model $L(2,2)$ is invertible. Indeed, in order to
invert the gradient map $\nabla _{\mathcal{V}_{L(2,2)}}$, in the condition
of Sec.\ \ref{main}, one can take
\begin{equation}
\Omega _{1}\,:=\,\Gamma _{4},\qquad \Omega _{2}\,:=\,\Gamma _{5}\quad \text{%
so}\quad R_{1}(y)\,=\,Q_{4}(y),\quad R_{2}(y)\,=\,Q_{5}(y),\quad r=2.
\end{equation}%
Using that the set of $\Gamma $-matrices $\{\Gamma _{1},\ldots ,\Gamma
_{5}\} $ of $L(4,1)$ is a Clifford set and that the Lorentzian identity (\ref%
{lorL410}), which now reads
\begin{equation}
-Q_{0}^{(2)}(y)^{2}\,+\,Q_{1}^{(2)}(y)^{2}\,+\,Q_{2}^{(2)}(y)^{2}\,+%
\,Q_{3}^{(2)}(y)^{2}\,+\,R_{1}(y)^{2}\,+\,R_{2}(y)^{2}\,=\,0~,
\end{equation}%
holds, the condition of Sec.\ \ref{main} is satisfied.

Since $r=2$, the inverse of the gradient map $\nabla _{{\mathcal{V}}%
_{L(2,2)}}$ is given as a composition of two maps, $\mathbf{\alpha }$ and $%
\mathbf{\mu }$ (cfr. (\ref{csii})) :%
\begin{equation}
{\mathbb{R}}_{\xi }^{1+4+8}\,\overset{\nabla _{{\mathcal{V}}_{L(2,2)}}}{%
\longrightarrow }\,{\mathbb{R}}_{z}^{1+4+8}\,\overset{\mathbf{\alpha }}{%
\longrightarrow }\,{\mathbb{R}}_{(t,u,v,w)}^{1+4+8+2}\,\overset{\mathbf{\mu }%
}{\longrightarrow }\,{\mathbb{R}}_{\xi }^{1+4+8}~.
\end{equation}%
The map $\mathbf{\alpha }$, as in (\ref{alpha}), is given by
\begin{equation}
\mathbf{\alpha }(z_{1},\ldots
,z_{13})\,:=\,^{T}(z_{1}^{2},\,z_{1}z_{2},\ldots
,\,z_{1}z_{13},\,R_{1}(z),R_{2}(z)),
\end{equation}%
where the quadratic forms $R_{K}$'s ($K=1,2$) depend only on the last $%
2\cdot 4=8$ variables and they are obtained by substituting $y^{L}:=z_{L+5}$%
, $L=1,\ldots ,8$ in $\bar{Q}_{4}$ and $\bar{Q}_{5}$ given in (\ref{th410}),
respectively:%
\begin{eqnarray}
R_{1}(z) &=&2(z_{6}z_{11}-z_{7}z_{10}+z_{8}z_{13}-z_{9}z_{12}); \\
R_{2}(z) &=&2(z_{6}z_{13}+z_{7}z_{12}-z_{8}z_{11}-z_{9}z_{10}).
\end{eqnarray}%
The map $\mathbf{\mu }$ (\ref{mu}) has $1+4+8+2$ components that are
homogeneous polynomials of degree $2$ in the variables $t,u_{0},\ldots
,u_{3},v_{1},\ldots ,v_{8},w_{1},w_{2}$. Since
\begin{equation}
\mathbf{\mu }\circ \mathbf{\alpha }\circ \nabla _{{\mathcal{V}}%
_{L(2,2)}}\,(\xi )=\,4q(x)^{2}{\mathcal{V}}_{L(2,2)}(\xi )\mathbb{\xi },
\end{equation}%
where ${\mathcal{V}}_{L(2,2)}$ and the corresponding $q(x)$ are given by (%
\ref{V_2,2}), the (birational) inverse of the gradient map $\nabla _{%
\mathcal{V}_{L(2,2)}}$ is the map $\mathbf{\mu \circ \alpha }$, which is an
homogeneous polynomial map of degree four.

\subsection{$L(1,2)$\label{L120}}

The model $L(1,2)$ has $\mathcal{D}_{2}=2$ (cfr. (\ref{table D})). The total
dimension of the underlying vector space is then%
\begin{equation}
\left( 1+q+2+P\cdot \mathcal{D}_{q+1}\right) _{q=1,P=2}=1+3+2\cdot 2=8,
\end{equation}%
and the corresponding variables split as $s,\;x^{0},x^{1},x^{2},\;y^{1}%
\ldots ,y^{4}$; thus, in this model $I=0,1,2$ and $\alpha =1,...,4$. Despite
the treatment, within a different formalism, in \cite{H2-1} and in \cite%
{DFT-Hom-07}, to the best of our knowledge, the BPS black hole entropy and
attractors for such a model were not previously known in literature in terms
of the electric and magnetic black hole charges. It is here worth remarking
that the $L(1,2)$ model is one of the simplest models related to an
homogeneous \textit{non-symmetric} `special' manifold; therefore, in App. %
\ref{App-L(1,2)} we work out the computations in full detail.

\section{Examples, III : $L(q,P)$, $q=1,2,3$, $P\geqslant 3$\label{LqP0}}

In this section we consider the models $L(q,P)$ with $P\geqslant 3$ and $%
q=1,2,3$ : we will show that these models have an invertible gradient map%
\footnote{%
As to now, we don't know if this is the case also for other $q$'s $\geqslant
4$; see Sec.\ \ref{Conclusion}.}. While for the models with $P=1,2$
considered above the embedding as a linear section into a complete model $%
L(q^{\prime },1)$ (with $q^{\prime }=2,4,$ or $8$) was crucial in order to
establish invertibility, for the models with $P\geqslant 3$ treated below it
suffices that the invertibility condition of Sec.\ \ref{main} is satisfied
(for a fixed $q$) and $P=1$ and $2$; the invertibility for all other $%
P\geqslant 3$ then comes \textquotedblleft for free\textquotedblright .

\subsection{Block $(k,l)$-lifts}

In order to show that the gradient map of a model $L(q,P)$, with $q=1,2,3$
and $P\geqslant 3$ is invertible by the condition enounced in Sec.\ \ref%
{main}, we need to find symmetric matrices $\Omega _{K}$ satisfying the
anti-commutativity conditions (\ref{GOcom}) and the Lorentz identity (\ref%
{lorrel}).

To this aim, we will `lift' the $\Omega _{K}$'s, of size $2m$, from the $P=2$
models discussed in Sec.\ \ref{L420} to matrices $\Omega _{K}^{(kl)}$ of
size $mP$ for any $P\geqslant 3$. For $y=(y^{(1)},\ldots ,y^{(P)})\in {%
\mathbb{R}}^{mP}$ as in (\ref{blocky}) and $k,l$ with $1\leqslant
k<l\leqslant P$, we define the following vector in ${\mathbb{R}}^{2m}$:
\begin{equation}
y^{(kl)}\,:=\,^{T}(y^{(k)},y^{(l)})\qquad \in {\mathbb{R}}^{2m}.
\end{equation}%
Next, for a quadratic form $R$ in $2m$ variables, we define quadratic forms
in $mP$ variables, with $P\geqslant 2$, which actually depend only on $2m$
of them, namely on the $y$'s belonging to the blocks $k$ and $l$ (notice
that we suppress $P$ from the notation):
\begin{equation}
R^{(k,l)}(y)\,:=\,R(y^{(kl)})\qquad (y\in {\mathbb{R}}^{mP})~.  \label{Rkl}
\end{equation}

Given the matrices $\Omega _{K}$ of size $m\times m$ of the $L(q,2)$ model ($%
q=1,2,3$) discussed in Sec.\ \ref{L420}, let as before $R_{K}$ be the
quadratic form in $2m$ variables defined by $\Omega _{K}$ and let $\Omega
_{K}^{(kl)}$ be the symmetric matrix of size $mP$ defined by the quadratic
form $R_{K}^{(k,l)}$ in $mP$ variables, so that
\begin{equation}
R_{K}^{(k,l)}(y)\,=\,R_{K}(y^{(kl)})\,=\,{}^{T}y\Omega _{K}^{(kl)}y\qquad
(y\in \mathbb{R}^{mP})~.  \label{Omkl}
\end{equation}

\subsection{Invertibility\label{invP3}}

The $\Gamma $-matrices, of size $mP\times mP$, of the $L(q,P)$ ($P\geqslant
3 $) models are the $q+2$ matrices $\Gamma _{I}^{(P)}:=\Gamma _{I}\otimes {%
\mathbb{I}}_{P}$, $I=0,\ldots ,q+1$, where the $\Gamma _{I}$ are the $\Gamma
$-matrices of the $L(q,1)$ model.

Now, we claim that if we consider the $mP\times mP$ matrices $\Omega
_{K}^{(kl)}$, $1\leqslant k<l\leqslant P$ where the $\Omega _{K}$ are the
extra matrices in the $L(q,2)$ model, then the conditions (\ref{GOcom}) and (%
\ref{lorrel}) for the invertibility of the gradient map of these $L(q,P)$
models ($P\geqslant 3$) are all satisfied.

The anti-commutativity condition follows rather trivially from the fact that
in Sec.\ \ref{L420} we checked that the $\Gamma _{I}^{(2)}$, $I=1,\ldots
,q+1 $, anti-commute with all the $\Omega _{K}$'s of the $L(q,2)$ models we
have considered, and therefore also the $\Gamma $-matrices $\Gamma
_{I}^{(P)} $ of the $L(q,P)$ models (with $P\geqslant 3$) anti-commute with
all the $\Omega _{K}^{(kl)}$, so (\ref{GOcom}) is satisfied.

On the other hand, in Sec.\ \ref{verifeqn} below, we will check the
existence of the Lorentzian identity (\ref{lorrel}), namely we will verify
the following Lorentzian identity between quadratic forms in $mP$ variables
for all $P\geqslant 3$:
\begin{equation}
-Q_{0}^{(P)}(y)^{2}\,+\,Q_{1}^{(P)}(y)^{2}\,+\,\ldots
\,+\,Q_{q+1}^{(P)}(y)^{2}\,+\,\sum_{K=1}^{r}\,\sum_{1\leqslant k<l\leqslant
P}R_{K}^{(k,l)}(y)^{2}\,=\,0~.  \label{gqr}
\end{equation}

Thus, the invertibility condition of Sec.\ \ref{main} is satisfied and the
corresponding gradient map (and BPS system) can be inverted for any $q=1,2,3$
and $P\geqslant 3$.

Again, from the treatment of Sec.\ \ref{invertgamma}, the solution of the
BPS system of $L(q,P)$ for $q=1,2,3$ and $P\geqslant 3$ is then given by (%
\ref{pre-sol})-(\ref{pre-sol-3}), and (\ref{SSd}) and (\ref{zH-1})-(\ref%
{zH-3}) yield the corresponding expression of the BPS black hole entropy and
of the BPS attractors, respectively. We refer to Sec.\ \ref{L120}, \ref{L130}
for examples of lower dimensional models $L(q,P)$ (namely, $L(1,2)$ and $%
L(1,3)$) which we explicitly work out in detail (see also the generalization
to $L(1,P)$ models with $P\geqslant 2$ in Sec.\ \ref{L1P0}).

\subsection{Proof of the Lorentzian identity (\protect\ref{gqr}) \label%
{verifeqn}}

The definition (\ref{QP}) of $Q_{I}^{(P)}$'s allows us to rewrite the first
terms in the l.h.s. of (\ref{gqr}) as follows:
\begin{eqnarray}
&&-Q_{0}^{(P)}(y)^{2}\,+\,\sum_{I=1}^{q+1}Q_{I}^{(P)}(y)^{2}\,  \notag \\
&=&-\left( \sum_{k=1}^{P}Q_{0}(y^{(k)})\right) ^{2}+\sum_{I=1}^{q+1}\left(
\sum_{k=1}^{P}Q_{I}(y^{(k)})\right) ^{2}  \notag \\
&=&\,\sum_{k=1}^{P}\left(
-Q_{0}(y^{(k)})^{2}\,+\,\sum_{I=1}^{q+1}Q_{I}(y^{(k)})^{2}\right)
\,-\sum_{1\leqslant k<l\leqslant
P}S_{0,k,l}+\sum_{I=1}^{q+1}\,\sum_{1\leqslant k<l\leqslant P}S_{I,k,l},
\notag \\
&&  \label{QanyP}
\end{eqnarray}%
where the remaining cross terms are defined as follows (no sum on repeated
indices, $I=0,1,...,q+1$) :
\begin{equation}
S_{I,k,l}\,:=\,2Q_{I}(y^{(k)})Q_{I}(y^{(l)})~.  \label{SSS}
\end{equation}

\begin{itemize}
\item We do the cases $q=1,2$ first. The Lorentzian quadratic relations (\ref%
{qidg1}) resp. (\ref{GammQI2}) hold, thus for any $k$ we find that%
\begin{equation}
-Q_{0}(y^{(k)})^{2}\,+\,\sum_{I=1}^{q+1}Q_{I}(y^{(k)})^{2}=0,
\label{lorrelq12}
\end{equation}

which trivially implies
\begin{equation}
\sum_{k=1}^{P}\left(
-Q_{0}(y^{(k)})^{2}\,+\,\sum_{I=1}^{q+1}Q_{I}(y^{(k)})^{2}\right) \,=\,0~.
\end{equation}

Consequently, it remains to show that
\begin{equation}
-\sum_{1\leqslant k<l\leqslant
P}S_{0,k,l}+\sum_{I=1}^{q+1}\,\sum_{1\leqslant k<l\leqslant
P}S_{I,k,l}\,+\,\sum_{K=1}^{r}\sum_{1\leqslant k<l\leqslant
P}R_{K}^{(k,l)}(y)^{2}\,\,=\,0~.  \label{rema}
\end{equation}%
In the Lorentzian identity (\ref{P2}) for the $L(q,2)$ models, the vector $y$
is in fact $y^{(12)}$; so, replacing it by $y^{(kl)}$, one obtains the
identities, for any $1\leqslant k<l\leqslant P$:
\begin{equation}
-Q_{0}^{(2)}(y^{(kl)})^{2}\,+\,Q_{1}^{(2)}(y^{(kl)})^{2}\,+\,\ldots
\,+\,Q_{q+1}^{(2)}(y^{(kl)})^{2}\,+\,R_{1}(y^{(kl)})^{2}\,+\,\ldots
\,+\,R_{r}(y^{(kl)})^{2}\,=\,0~.  \label{Pkl}
\end{equation}%
Recalling the definition (\ref{QP}) of the quadratic forms $Q_{I}^{(2)}$,
one gets
\begin{equation}
Q_{I}^{(2)}(y^{(kl)})^{2}\,=\,\left(
Q_{I}(y^{(k)})\,+\,Q_{I}(y^{(l)})\right)
^{2}\,=\,Q_{I}(y^{(k)})^{2}\,+\,Q_{I}(y^{(l)})^{2}\,+%
\,2Q_{I}(y^{(k)})Q_{I}(y^{(l)})~.  \label{this}
\end{equation}%
By plugging (\ref{this}) into (\ref{Pkl}), two copies (one for $y^{(k)}$ and
one for $y^{(l)}$) of the identity (\ref{lorrelq12}) are obtained, and there
remains, for any pair $\left( k,l\right) $ with $1\leqslant k<l\leqslant P$,
the identity:%
\begin{equation}
\begin{array}{l}
2\left(
-Q_{0}(y^{(k)})Q_{0}(y^{(l)})\,+\,%
\sum_{I=1}^{q+1}Q_{I}(y^{(k)})Q_{I}(y^{(l)})\right) \\
+\,R_{1}(y^{(kl)})^{2}\,+\,\ldots \,+\,R_{r}(y^{(kl)})^{2}\,=\,0~.%
\end{array}
\label{reee}
\end{equation}%
By recalling the definition (\ref{SSS}) of $S_{I,k,l}$, we see that (\ref%
{reee}) is the identity
\begin{equation}
-S_{0,k,l}\,+\,\sum_{I=1}^{q+1}S_{I,k,l}+\,%
\sum_{K=1}^{r}R_{K}^{(k,l)}(y)^{2}\,\,=\,0.
\end{equation}%
Now we just sum over all $k,l$ and we obtain (\ref{rema}), which in turn
implies the identity (\ref{gqr}).

\item Let us now consider the case $q=3$. In this case we do not have a
Lorentzian relation (\ref{lorrelq12}) but the $L(3,1)$ model has an
auxiliary quadratic form $R$ (cfr.\ Sec.\ \ref{L310}). We first take a
closer look at the $L(3,2)$ model from Sec.\ \ref{L320}, which was shown to
be a descendant of $L(8,1)$. The main point of interest is the (auxiliary)
quadratic form $R_{1}$ in $2m=16$ variables $y^{1},\ldots ,y^{16}$, which is
$Q_{2}$ in the $L(8,1)$ model, and which can be written as
\begin{equation}
R_{1}(y)\,=\,R(y^{(1)})\,-\,R(y^{(2)}),\quad
R(z)=2(z_{1}z_{2}-z_{3}z_{4}+z_{5}z_{6}-z_{7}z_{8}),\quad
y=~^{T}(y^{(1)},y^{(2)})\in {\mathbb{R}}^{16}~.  \label{R32}
\end{equation}%
Next, we characterize the $L(3,1)$ model not as done in Sec.\ \ref{L310},
but rather as a linearly constrained $L(3,2)$ model:
\begin{equation}
L(3,1)\,=\,L(3,2)_{y^{9}=\ldots =y^{16}=0}~.
\end{equation}%
Within this approach, the five quadrics $Q_{0},\ldots ,Q_{4}$ from $L(3,1)$
are the restrictions of the quadrics of the same name from $L(3,2)$, which
again are $Q_{0},Q_{1},Q_{3},Q_{4},Q_{5}$ respectively of the $L(8,1)$
model, and the auxiliary quadric of $L(3,1)$ is the restriction of\footnote{%
Note that in the treatment of the $L(3,1)$ model in Sec.\ \ref{L310} we have
interchanged the roles of $Q_{2}$ and $Q_{5}$ from $L(8,1)$.} $Q_{2}$. With
these conventions, we have the Lorentzian relation
\begin{equation}
-Q_{0}(y)^{2}\,+\,\sum_{I=1}^{4}Q_{I}(y)^{2}\,+\,R\left( y\right) ^{2}\,=\,0
\label{lor31}
\end{equation}%
holding for the $L(3,1)$ model, as well as the Lorentzian relation for the $%
L(3,2)$ model:
\begin{equation}
-Q_{0}^{(2)}(y^{(12)})^{2}\,+\,\sum_{I=1}^{4}Q_{I}^{(2)}(y^{(12)})^{2}\,+%
\,R_{1}\left( y^{(12)}\right) \,+R_{2}\left( y^{(12)}\right) \,+\,\ldots
\,+\,R_{5}\left( y^{(12)}\right) \,=\,0~.  \label{lor32}
\end{equation}%
From (\ref{R32}), we obtain
\begin{equation}
R_{1}^{(k,l)}(y)^{2}\,=\,R_{1}(y^{(kl)})^{2}\,=%
\,R(y^{(k)})^{2}+R(y^{(l)})^{2}\,-\,S_{R,k,l},\qquad
S_{R,k,l}\,:=\,2R(y^{(k)})R(y^{(l)}).
\end{equation}%
Using this, (\ref{this}) and (\ref{lor31}), we obtain from (\ref{lor32}),
for any pair $(k,l)$ with $1\leqslant k<l\leqslant P$ the identity:
\begin{equation}
2\left(
-Q_{0}(y^{(k)})Q_{0}(y^{(l)})\,+\,\sum_{I=1}^{4}Q_{I}(y^{(k)})Q_{I}(y^{(l)})%
\,-\,R(y^{(k)})R(y^{(l)})\right) +\,R_{2}(y^{(kl)})^{2}\,+\,\ldots
\,+\,R_{5}(y^{(kl)})^{2}\,=\,0~,  \label{re3}
\end{equation}%
which can be rewritten as
\begin{equation}
-S_{0,k,l}\,+\,\sum_{I=1}^{4}S_{I,k,l}\,-\,S_{R,k,l}\,+\,%
\sum_{K=2}^{5}R_{K}^{(k,l)}(y)^{2}\,\,=\,0.  \label{relS}
\end{equation}%
Finally, using (\ref{QanyP}) and (\ref{lor31}), we obtain
\begin{equation}
-Q_{0}^{(P)}(y)^{2}\,+\,\sum_{I=1}^{4}Q_{I}^{(P)}(y)^{2}\,+\,%
\sum_{k<l}R_{1}^{(k,l)}(y)^{2}\,=\,\sum_{1\leqslant k<l\leqslant P}\left(
-S_{0,k,l}+\sum_{I=1}^{4}S_{I,k,l}\,-\,S_{R,k,l}\right) ~.  \label{QR1P}
\end{equation}%
Combining (\ref{relS}), summed over $k,l$, and (\ref{QR1P}) we conclude that
(\ref{gqr}) holds also for $q=3$.
\end{itemize}

\subsection{$L(3,3)$\label{L330}}

The $L(3,3)$ model has $q=3,P=3$, and $\mathcal{D}_{3}=8$ \cite{dWVP}; thus,
the number of variables is%
\begin{equation}
\left( 1+q+2+P\cdot \mathcal{D}_{q+1}\right) _{q=3,P=3}=1+5+24=30.
\end{equation}%
The $\Gamma ^{(3)}$-matrices, having size $24\times 24$, are \textit{not} $%
\Gamma $-matrices (differently from the $\Gamma ^{(2)}$-matrices). This
model has five $\Gamma $-matrices $\Gamma _{I}^{(3)}$ ($I=0,\ldots ,4$),
obtained from the five $\Gamma $-matrices of the $L(3,1)$ model and defining
the quadratic forms $Q_{I}^{(3)}$ as in (\ref{QP}):
\begin{equation}
Q_{I}^{(3)}(y^{1},\ldots ,y^{24})\,=\,Q_{I}(y^{1},\ldots
,y^{8})\,+\,Q_{I}(y^{9},\ldots ,y^{16})\,+\,Q_{I}(y^{17},\ldots ,y^{24})~.
\end{equation}%
For obvious reasons, we refrain from writing them down explicitly. Next,
there are the are $3\cdot 5=15$ matrices $\Omega _{K}^{(kl)}$ with $%
K=1,\ldots ,5$ and $1\leqslant k<l\leqslant 3$ which define the quadratic
forms $R_{K}^{(k,l)}$. Starting from the five quadratic forms $R_{K}$ in $16$
variables from the $L(3,2)$ model (given in (\ref{this2})), one finds these
quadratic forms $R_{K}^{(k,l)}$ in $24$ variables $y^{\alpha }$ (but
actually each on depends only on $16$ of these variables) as in (\ref{Rkl}):%
\begin{equation}
R_{K}^{(k,l)}(y)=R_{K}(y^{(kl)}),
\end{equation}%
so, for $K=1,\ldots ,5$ :
\begin{eqnarray}
R_{K}^{(1,2)}(y^{1},\ldots ,y^{24}) &=&R_{K}(y^{1},\ldots
,y^{8},y^{9},\ldots ,y^{16}); \\
R_{K}^{(1,3)}(y^{1},\ldots ,y^{24}) &=&R_{K}(y^{1},\ldots
,y^{8},y^{17},\ldots ,y^{24}); \\
R_{K}^{(2,3)}(y^{1},\ldots ,y^{24}) &=&R_{K}(y^{9},\ldots
,y^{16},y^{17},\ldots ,y^{24}).
\end{eqnarray}

The matrices (except for $\Gamma _{0}^{(3)}={\mathbb{I}}_{24}$) anti-commute
(cfr.\ Sec.\ \ref{invP3} for the general case) and it can be verified (see %
\ref{verifeqn} for the general case) that they satisfy the Lorenzian
quadratic relation given by (\ref{gqr}), with $P=3$:
\begin{equation}
-Q_{0}^{(3)}(y)^{2}\,+\,Q_{1}^{(3)}(y)^{2}\,+\,\ldots
\,+\,Q_{4}^{(3)}(y)^{2}\,+\,\sum_{K=1}^{5}\sum_{1\leqslant k<l\leqslant
3}R_{K}^{(k,l)}(y)^{2}\,=\,0~.
\end{equation}%
Therefore, the inverse of the gradient map $\nabla _{\mathcal{V}_{L(3,3)}}$
of the cubic form
\begin{equation}
{\mathcal{V}}\,_{L(3,3)}:=\,sq(x)\,+\,\sum_{I=0}^{4}x^{I}Q_{I}^{(3)}(y),~%
\text{with}\qquad q(x)\,=\,-\left( x^{0}\right) ^{2}+\left( x^{1}\right)
^{2}+\ldots +\left( x^{4}\right) ^{2}~,
\end{equation}%
can then be found following the procedure discussed in Sec.\ \ref{explin}.

\subsection{$L(2,3)$\label{L230}}

The $L(2,3)$ model has $q=2,P=3$, and $\mathcal{D}_{3}=4$ \cite{dWVP}; thus,
the number of variables is%
\begin{equation}
\left( 1+q+2+P\cdot \mathcal{D}_{q+1}\right) _{q=2,P=3}=1+4+12=17.
\end{equation}%
The $\Gamma ^{(3)}$-matrices have size $12\times 12$, thus, differently from
the $\Gamma ^{(2)}$-matrices, they are \textit{not} $\Gamma $-matrices;
besides the four $\Gamma _{I}^{(3)}$'s ($I=0,1,2,3$), there are the $3+3$
matrices $\Omega _{K}^{(kl)}$ with $K=1,2$ and $1\leqslant k<l\leqslant 3$,
which define the quadratic forms $R_{K}^{(k,l)}$. The quadratic forms
involved are
\begin{eqnarray}
&&%
\begin{array}{lll}
Q_{0}^{(3)} & = & \left\{
\begin{array}{l}
\left( y^{1}\right) ^{2}+\left( y^{2}\right) ^{2}+\left( y^{3}\right)
^{2}+\left( y^{4}\right) ^{2}+\left( y^{5}\right) ^{2}+\left( y^{6}\right)
^{2}+\,\left( y^{7}\right) ^{2} \\
+\,\left( y^{8}\right) ^{2}+\left( y^{9}\right) ^{2}+\left( y^{10}\right)
^{2}+\left( y^{11}\right) ^{2}+\left( y^{12}\right) ^{2};%
\end{array}%
\right. \\
Q_{1}^{(3)} & = & \left\{
\begin{array}{l}
\left( y^{1}\right) ^{2}-\left( y^{2}\right) ^{2}+\left( y^{3}\right)
^{2}-\left( y^{4}\right) ^{2}+\left( y^{5}\right) ^{2}-\left( y^{6}\right)
^{2}+\,\left( y^{7}\right) ^{2} \\
-\,\left( y^{8}\right) ^{2}+\left( y^{9}\right) ^{2}-\left( y^{10}\right)
^{2}+\left( y^{11}\right) ^{2}-\left( y^{12}\right) ^{2};%
\end{array}%
\right. \\
Q_{2}^{(3)} & = & 2(y^{1}y^{2}-y^{3}y^{4}\,+\,y^{5}y^{6}-y^{7}y^{8}\,+%
\,y^{9}y^{10}-y^{11}y^{12}); \\
Q_{3}^{(3)} & = & 2(y^{1}y^{4}+y^{2}y^{3}\,+\,y^{5}y^{8}+y^{6}y^{7}\,+%
\,y^{9}y^{12}+y^{10}y^{11}); \\
R_{1}^{(1,2)} & = & 2(y^{1}y^{6}-y^{2}y^{5}\,+\,y^{3}y^{8}-y^{4}y^{7}); \\
R_{1}^{(1,3)} & = & 2(y^{1}y^{10}-y^{2}y^{9}\,+\,y^{3}y^{12}-y^{4}y^{11});
\\
R_{1}^{(2,3)} & = & 2(y^{5}y^{10}-y^{6}y^{9}\,+\,y^{7}y^{12}-y^{8}y^{11});
\\
R_{2}^{(1,2)} & = & 2(y^{1}y^{8}+y^{2}y^{7}\,-\,y^{3}y^{6}-y^{4}y^{5}); \\
R_{2}^{(1,3)} & = & 2(y^{1}y^{12}+y^{2}y^{11}\,-\,y^{3}y^{10}+y^{4}y^{9});
\\
R_{2}^{(2,3)} & = & 2(y^{5}y^{12}+y^{6}y^{11}\,-\,y^{7}y^{10}+y^{8}y^{9})~.%
\end{array}
\notag \\
&&
\end{eqnarray}%
The matrices (except for $\Gamma _{0}^{(3)}={\mathbb{I}}_{12}$) anti-commute
(cfr.\ Sec.\ \ref{invP3} for the general case) and it can be verified (see %
\ref{verifeqn} for the general case) that they satisfy the Lorenzian
quadratic relation given by (\ref{gqr}), with $P=3$:
\begin{equation}
-Q_{0}^{(3)}(y)^{2}\,+\,Q_{1}^{(3)}(y)^{2}\,+\,Q_{2}^{(3)}(y)^{2}\,+%
\,Q_{3}^{(3)}(y)^{2}\,+\,\sum_{K=1}^{2}\sum_{1\leqslant k<l\leqslant
3}R_{K}^{(k,l)}(y)^{2}\,=\,0~.
\end{equation}%
Therefore, the inverse of the gradient map $\nabla _{\mathcal{V}_{L(2,3)}}$
of the cubic form
\begin{equation}
{\mathcal{V}}\,_{L(2,3)}:=\,sq(x)\,+\,\sum_{I=0}^{3}x^{I}Q_{I}^{(3)}(y),%
\text{ with}\qquad q(x)\,=\,-\left( x^{0}\right) ^{2}+\left( x^{1}\right)
^{2}+\left( x^{2}\right) ^{2}+\left( x^{3}\right) ^{2}~,
\end{equation}%
can then be found following the procedure discussed in Sec.\ \ref{explin}.

\subsection{$L(1,3)$\label{L130}}

This model has $q=1$, $P=3$, and $\mathcal{D}_{2}=2$ \cite{dWVP}; thus, the
number of variables is%
\begin{equation}
\left( 1+q+2+P\cdot \mathcal{D}_{q+1}\right) _{q=1,P=3}=1+3+6=10,
\end{equation}%
and the corresponding variables split as $s,\;x^{0},x^{1},x^{2},\;y^{1}%
\ldots ,y^{6}$; thus, in this model $I=0,1,2$ and $\alpha =1,...,6$. This is
one of the simplest models related to a homogeneous non-symmetric `special'
manifold, and for which, to the best of our knowledge, the BPS black hole
entropy and attractors are not known in literature. Therefore, In App.\ref%
{App-L(1,3)} we work out the computations of this model in full detail.

\subsection{$L(1,P)$\label{L1P0}}

Some models of type $L(1,P)$ have been treated explicitly: $L(1,1)$ in Sec.\ %
\ref{L110}, $L(1,2)$ in Sec.\ \ref{L120}, and $L(1,3)$ in Sec.\ \ref{L130}
(see also the discussion of $L(1,4)$ and $L(1,8)$ in Sec.\ \ref{L1PL81}
below). As far as we currently understand, for $P\geqslant 9$ such models
cannot be obtained as descendants of other complete models, so in Sec.\ \ref%
{Pgreq4} we will discuss the class of models $L(1,P)$ with $P\geqslant 2$,
following the approach of Sec.\ \ref{invP3}, and retrieving (for $P=2$ resp.
$3$) the explicit treatment of the models $L(1,2)$ and $L(1,3)$,
respectively given in Secs. \ref{L120} and \ref{L130}. Moreover, in Sec.\ %
\ref{geom-fact} we will also briefly present a geometric point of view on
the factorization of the inverse map of the gradient map, whose detailed
investigation goes beyond the scope of this paper.

\subsubsection{$P\geqslant 2$\label{Pgreq4}}

By definition, $\mathcal{D}_{2}=2$ \cite{dWVP}; thus, the number of
variables is%
\begin{equation}
\left( 1+q+2+P\cdot \mathcal{D}_{q+1}\right) _{q=1,P\geqslant 2}=1+3+2P=2P+4.
\end{equation}%
The cubic form $\mathcal{V}_{L(1,P)}$ reads
\begin{equation}
\mathcal{V\,}_{L(1,P)}=\,sq(x)\,+\,\sum_{I=0}^{2}x^{I}Q_{I}^{(P)}(y),\qquad
\text{with}\quad q(x)\,=\,-\left( x^{0}\right) ^{2}+\left( x^{1}\right)
^{2}+\left( x^{2}\right) ^{2}~,  \label{V-L13P}
\end{equation}%
where the $Q_{I}^{(P)}$ are obtained from the $Q_{I}$'s of the $L(1,1)$
model (cfr. (\ref{th110})):
\begin{equation}
Q_{0}^{(P)}\,=\,\sum_{i=1}^{2P}\left( y^{i}\right)
^{2};\;\;Q_{1}^{(P)}\,=\,\sum_{i=1}^{2P}(-1)^{i+1}\left( y^{i}\right)
^{2};\;\;Q_{2}^{(P)}\,=\,2(\sum_{j=1}^{P}y^{2j-1}y^{2j})~.
\end{equation}%
Moreover, we define the quadrics, based on the quadric $R$ of the $L(1,2)$
model (see (\ref{this5})), $R^{(k,l)}(y):=R(y^{(kl)})$ for every $(k,l)$
with $1\leqslant k<l\leqslant P$:
\begin{equation}
R^{(k,l)}\,=\,2(y^{k}y^{2l}\,-\,y^{k+1}y^{2l-1})~,
\end{equation}%
which depend only on the four variables $y^{k},y^{k+1},y^{2l-1},y^{2l}$.

As required by the invertibility condition of Sec.\ \ref{main} (cfr.\ (\ref%
{lorrel})), a Lorentzian identity of type (\ref{gqr}) holds true. 

Since $r=\binom{P}{2}$, the inverse of the gradient map $\nabla _{{\mathcal{V%
}}_{L(1,P)}}$ is given as a composition of two maps, namely (cfr. (\ref{csii}%
))
\begin{equation}
{\mathbb{R}}_{\xi }^{1+3+2P}\,\overset{\nabla _{{\mathcal{V}}_{L(1,P)}}}{%
\longrightarrow }\,{\mathbb{R}}_{z}^{1+3+2P}\,\overset{\mathbf{\alpha }}{%
\longrightarrow }\,{\mathbb{R}}_{(t,u,v,w)}^{1+3+2P+\binom{P}{2}}\,\overset{%
\mathbf{\mu }}{\longrightarrow }\,{\mathbb{R}}_{\xi }^{1+3+2P}~,
\end{equation}%
with%
\begin{equation}
\mathbf{\mu }\circ \mathbf{\alpha }\circ \nabla _{{\mathcal{V}}%
_{L(1,P)}}\,(\xi )\,=\,4q(x)^{2}{\mathcal{V}}_{L(1,P)}(\xi )\mathbb{\xi },
\end{equation}%
where $\xi =~^{T}(s,x,y)$, and ${\mathcal{V}}_{L(1,P)}$ and the
corresponding $q(x)$ are given by (\ref{V-L13P}). The map $\mathbf{\alpha }$
from (\ref{alpha}) is given by
\begin{equation}
\mathbf{\alpha }(z_{1},\ldots
,z_{2P+4})\,:=\,^{T}(z_{1}^{2},\,z_{1}z_{2},\ldots ,\,z_{1}z_{2P+4},\ldots
,\,R^{(k,l)}(z),\ldots ),
\end{equation}%
where each of the $\binom{P}{2}$ quadratic forms with $1\leq k<l\leq P$:%
\begin{equation}
R^{(k,l)}(z)=R(z_{4+2k-1},z_{4+2k},z_{4+2l-1},z_{4+2l})=2(z_{4+2k-1}z_{4+2l}-z_{4+2l-1}z_{4+2k})
\end{equation}%
depends only on four of the last $2P$ variables $z{5},\ldots, z_{2P+4}$.

Next, the map $\mathbf{\mu }$ from (\ref{mu}), which has $2P+4$ components
that are homogeneous polynomials of degree $2$ in the variables $%
t,u_{0},u_{1},u_{2},v_{1},\ldots ,v_{2P},\ldots w_{k,l}\ldots $ with $%
1\leqslant k<l\leqslant P$, is given by%
\begin{equation}
\mathbf{\mu }\,:\,{\mathbb{R}}_{t,u,v,w}^{1+3+2P+\binom{P}{2}%
}\,\longrightarrow \,{\mathbb{R}}_{\xi }^{2P+4},
\end{equation}%
\begin{equation}
\mathbf{\mu }(t,u,v,w)\,:=\,\left( \renewcommand{\arraystretch}{1.3}%
\begin{array}{rcl}
-u_{0}^{2}+u_{1}^{2}+u_{2}^{2} & + & \frac{1}{16}(w_{1,2}^{2}+\ldots
+w_{P-1,P}^{2}) \\
-2tu_{0} & + & \frac{1}{2}(v_{1}^{2}+v_{2}^{2}+\ldots
+v_{2P-1}^{2}+v_{2P}^{2}) \\
2tu_{1} & + & \frac{1}{16}(v_{1}^{2}-v_{2}^{2}+\ldots
+v_{2P-1}^{2}-v_{2P}^{2}) \\
2tu_{2} & + & v_{1}v_{2}+\ldots +v_{2P-1}v_{2P} \\
u_{0}v_{1}+u_{1}v_{1}+u_{2}v_{2} & - & \frac{1}{4}(w_{1,2}v_{4}+w_{1,3}v_{6}%
\ldots +w_{1,P}v_{2P})) \\
u_{0}v_{2}-u_{1}v_{2}+u_{2}v_{1} & - & \frac{1}{4}%
(-w_{1,2}v_{3}-w_{1,3}v_{5}\ldots -w_{1,P}v_{2P-1}) \\
& \vdots &  \\
u_{0}v_{2P-1}+u_{1}v_{2P-1}+u_{2}v_{2P} & - & \frac{1}{4}%
(-w_{1,P}v_{2}-w_{2,P}v_{4}\ldots -w_{P-1,P}v_{2P-2}) \\
u_{0}v_{2P}-u_{1}v_{2P}+u_{2}v_{2P-1} & - & \frac{1}{4}%
(w_{1,P}v_{1}+w_{2,P}v_{3}\ldots +w_{P-1,P}v_{2P-3})%
\end{array}%
\right) ~.
\end{equation}

The (birational) inverse map $(\nabla _{\mathcal{V}_{L(1,P)}})^{-1}=\mathbf{%
\mu \circ \alpha }$ of the gradient map $\nabla _{\mathcal{V}_{L(1,P)}}$
reads
\begin{equation}
\mathbf{\mu \circ \alpha }\,:\,{\mathbb{R}}_{z}^{2P+4}\,\overset{\mathbf{%
\alpha }}{\longrightarrow }\,{\mathbb{R}}_{(t,u,v,w)}^{1+3+2P+\binom{P}{2}}\,%
\overset{\mathbf{\mu }}{\longrightarrow }\,{\mathbb{R}}_{\xi }^{2P+4}~,
\end{equation}%
whose expression for arbitrary $P\geqslant 2$ is the following (cfr. (\ref%
{mu-alpha-2})):

\begin{equation}
(\mathbf{\mu \circ \alpha })\left( z\right) \,:=\left( {\renewcommand{%
\arraystretch}{1.3}%
\begin{array}{l}
z_{1}^{2}\left( -z_{2}^{2}+z_{3}^{2}+z_{4}^{2}\right) +\frac{1}{4}%
\sum_{g=2}^{P}\left( z_{2g+1}z_{2g+4}-z_{2g+2}z_{2g+3}\right) ^{2} \\
z_{1}^{2}\left[ -2z_{1}z_{2}+\frac{1}{2}\sum_{g=1}^{P}\left(
z_{2g+3}^{2}+z_{2g+4}^{2}\right) \right] \\
z_{1}^{2}\left[ 2z_{1}z_{3}+\frac{1}{2}\sum_{g=1}^{P}\left(
z_{2g+3}^{2}-z_{2g+4}^{2}\right) \right] \\
z_{1}^{2}\left( 2z_{1}z_{4}+\sum_{g=1}^{P}z_{2g+3}z_{2g+4}\right) \\
z_{1}^{2}\left( z_{2}z_{5}+z_{3}z_{5}+z_{4}z_{6}\right) -\frac{1}{2}%
z_{1}\sum_{g=2}^{P}(z_{5}z_{2g+4}-z_{6}z_{2g+3})z_{2g+4} \\
z_{1}^{2}\left( z_{2}z_{6}-z_{3}z_{6}+z_{4}z_{5}\right) +\frac{1}{2}%
z_{1}\sum_{g=2}^{P}(z_{5}z_{2g+4}-z_{6}z_{2g+3})z_{2g+3} \\
\vdots \\
z_{1}^{2}\left( z_{2}z_{2P+3}+z_{3}z_{2P+3}+z_{4}z_{2P+4}\right) -\frac{1}{2}%
z_{1}\sum_{g=2}^{P}(-z_{2g+1}z_{2P+4}+z_{2g+2}z_{2P+3})z_{2g+2} \\
z_{1}^{2}\left( z_{2}z_{2P+4}-z_{3}z_{2P+4}+z_{4}z_{2P+3}\right) +\frac{1}{2}%
z_{1}\sum_{g=2}^{P}(-z_{2g+1}z_{2P+4}+z_{2g+2}z_{2P+3})z_{2g+1}%
\end{array}%
}\right) ~.  \label{invL1P}
\end{equation}%
Note that for $P=2$ and $P=3$ one respectively retrieves the results (\ref%
{invL12}) and (\ref{invL31}).

Consequently, from the treatment of Sec.\ \ref{invertgamma}, the full
fledged expression of the solution (\ref{sol}) of the BPS system of $L(1,P)$
for arbitrary $P\geqslant 2$ is given by ($\Delta _{1}\equiv \Delta _{s}$,
and recall the definition (\ref{rec0})):%
\begin{eqnarray}
s &=&\frac{3}{2\left\vert \Delta _{s}\right\vert }\frac{\left[ \Delta
_{s}^{2}\left( -\Delta _{2}^{2}+\Delta _{3}^{2}+\Delta _{4}^{2}\right) +%
\frac{1}{4}\sum_{g=2}^{P}\left( \Delta _{2g+1}\Delta _{2g+4}-\Delta
_{2g+2}\Delta _{2g+3}\right) ^{2}\right] }{\sqrt{\left( \partial _{p}\Delta
\right) \cdot \left( \mathbf{\mu \circ \alpha }\right) \left( \partial
_{p}\Delta \right) }}; \\
&&  \notag \\
x^{0} &=&\frac{3}{2}\frac{\left\vert \Delta _{s}\right\vert \left[ -2\Delta
_{s}\Delta _{2}+\frac{1}{2}\sum_{g=1}^{P}\left( \Delta _{2g+3}^{2}+\Delta
_{2g+4}^{2}\right) \right] }{\sqrt{\left( \partial _{p}\Delta \right) \cdot
\left( \mathbf{\mu \circ \alpha }\right) \left( \partial _{p}\Delta \right) }%
}; \\
&&  \notag \\
x^{1} &=&\frac{3}{2}\frac{\left\vert \Delta _{s}\right\vert \left[ 2\Delta
_{s}\Delta _{3}+\frac{1}{2}\sum_{g=1}^{P}\left( \Delta _{2g+3}^{2}-\Delta
_{2g+4}^{2}\right) \right] }{\sqrt{\left( \partial _{p}\Delta \right) \cdot
\left( \mathbf{\mu \circ \alpha }\right) \left( \partial _{p}\Delta \right) }%
}; \\
&&  \notag \\
x^{2} &=&\frac{3}{2}\frac{\left\vert \Delta _{s}\right\vert \left( 2\Delta
_{s}\Delta _{4}+\sum_{g=1}^{P}\Delta _{2g+3}\Delta _{2g+4}\right) }{\sqrt{%
\left( \partial _{p}\Delta \right) \cdot \left( \mathbf{\mu \circ \alpha }%
\right) \left( \partial _{p}\Delta \right) }}; \\
&&  \notag \\
y^{1} &=&\frac{3\text{sgn}\left( \Delta _{s}\right) }{2}\frac{\left[ \Delta
_{s}\left( \Delta _{2}\Delta _{5}+\Delta _{3}\Delta _{5}+\Delta _{4}\Delta
_{6}\right) -\frac{1}{2}\sum_{g=2}^{P}(\Delta _{5}\Delta _{2g+4}-\Delta
_{6}\Delta _{2g+3})\Delta _{2g+4}\right] }{\sqrt{\left( \partial _{p}\Delta
\right) \cdot \left( \mathbf{\mu \circ \alpha }\right) \left( \partial
_{p}\Delta \right) }}; \\
&&  \notag \\
y^{2} &=&\frac{3\text{sgn}\left( \Delta _{s}\right) }{2}\frac{\left[ \Delta
_{s}\left( \Delta _{2}\Delta _{6}-\Delta _{3}\Delta _{6}+\Delta _{4}\Delta
_{5}\right) +\frac{1}{2}\sum_{g=2}^{P}(\Delta _{5}\Delta _{2g+4}-\Delta
_{6}\Delta _{2g+3})\Delta _{2g+3}\right] }{\sqrt{\left( \partial _{p}\Delta
\right) \cdot \left( \mathbf{\mu \circ \alpha }\right) \left( \partial
_{p}\Delta \right) }}; \\
&&\vdots  \notag \\
y^{2P-1} &=&\frac{3\text{sgn}\left( \Delta _{s}\right) }{2}\frac{\left[
\Delta _{s}\left( \Delta _{2}\Delta _{2P+3}+\Delta _{3}\Delta _{2P+3}+\Delta
_{4}\Delta _{2P+4}\right) -\frac{1}{2}\sum_{g=2}^{P}(-\Delta _{2g+1}\Delta
_{2P+4}+\Delta _{2g+2}\Delta _{2P+3})\Delta _{2g+2}\right] }{\sqrt{\left(
\partial _{p}\Delta \right) \cdot \left( \mathbf{\mu \circ \alpha }\right)
\left( \partial _{p}\Delta \right) }};  \notag \\
&& \\
y^{2P} &=&\frac{3\text{sgn}\left( \Delta _{s}\right) }{2}\frac{\left[ \Delta
_{s}\left( \Delta _{2}\Delta _{2P+4}-\Delta _{3}\Delta _{2P+4}+\Delta
_{4}\Delta _{2P+3}\right) +\frac{1}{2}\sum_{g=2}^{P}(-\Delta _{2g+1}\Delta
_{2P+4}+\Delta _{2g+2}\Delta _{2P+3})\Delta _{2g+1}\right] }{\sqrt{\left(
\partial _{p}\Delta \right) \cdot \left( \mathbf{\mu \circ \alpha }\right)
\left( \partial _{p}\Delta \right) }},  \notag \\
&&
\end{eqnarray}

where
\begin{eqnarray}
\left( \partial _{p}\Delta \right) \cdot \left( \mathbf{\mu \circ \alpha }%
\right) \left( \partial _{p}\Delta \right) &=&\Delta _{s}^{3}\left( -\Delta
_{2}^{2}+\Delta _{3}^{2}+\Delta _{4}^{2}\right) +\frac{1}{4}\Delta
_{s}\sum_{g=2}^{P}\left( \Delta _{2g+1}\Delta _{2g+4}-\Delta _{2g+2}\Delta
_{2g+3}\right) ^{2}  \notag \\
&&+\Delta _{s}^{2}\Delta _{2}\left[ -2\Delta _{s}\Delta _{2}+\frac{1}{2}%
\sum_{g=1}^{P}\left( \Delta _{2g+3}^{2}+\Delta _{2g+4}^{2}\right) \right]
\notag \\
&&+\Delta _{s}^{2}\Delta _{3}\left[ 2\Delta _{s}\Delta _{3}+\frac{1}{2}%
\sum_{g=1}^{P}\left( \Delta _{2g+3}^{2}-\Delta _{2g+4}^{2}\right) \right]
\notag \\
&&+\Delta _{s}^{2}\Delta _{4}\left( 2\Delta _{s}\Delta
_{4}+\sum_{g=1}^{P}\Delta _{2g+3}\Delta _{2g+4}\right)  \notag \\
&&+\Delta _{s}\Delta _{5}\left[ \Delta _{s}\left( \Delta _{2}\Delta
_{5}+\Delta _{3}\Delta _{5}+\Delta _{4}\Delta _{6}\right) -\frac{1}{2}%
\sum_{g=2}^{P}(\Delta _{5}\Delta _{2g+4}-\Delta _{6}\Delta _{2g+3})\Delta
_{2g+4}\right]  \notag \\
&&+\Delta _{s}\Delta _{6}\left[ \Delta _{s}\left( \Delta _{2}\Delta
_{6}-\Delta _{3}\Delta _{6}+\Delta _{4}\Delta _{5}\right) +\frac{1}{2}%
\sum_{g=2}^{P}(\Delta _{5}\Delta _{2g+4}-\Delta _{6}\Delta _{2g+3})\Delta
_{2g+3}\right]  \notag \\
&&+...  \notag \\
&&+\Delta _{s}\Delta _{2P+3}\left[ \Delta _{s}\left( \Delta _{2}\Delta
_{2P+3}+\Delta _{3}\Delta _{2P+3}+\Delta _{4}\Delta _{2P+4}\right) \right.
\notag \\
&&\left. -\frac{1}{2}\sum_{g=2}^{P}(-\Delta _{2g+1}\Delta _{2P+4}+\Delta
_{2g+2}\Delta _{2P+3})\Delta _{2g+2}\right]  \notag \\
&&+\Delta _{s}\Delta _{2P+4}\left[ \Delta _{s}\left( \Delta _{2}\Delta
_{2P+4}-\Delta _{3}\Delta _{2P+4}+\Delta _{4}\Delta _{2P+3}\right) \right.
\notag \\
&&\left. +\frac{1}{2}\sum_{g=2}^{P}(-\Delta _{2g+1}\Delta _{2P+4}+\Delta
_{2g+2}\Delta _{2P+3})\Delta _{2g+1}\right] .  \label{den1P}
\end{eqnarray}%
Then, (\ref{SSd}) and (\ref{zH-1})-(\ref{zH-3}) respectively yield the
corresponding full fledged expression of the BPS black hole entropy and of
the BPS attractors :%
\begin{equation}
\frac{S}{\pi }=\frac{1}{3\left\vert p^{0}\right\vert }\sqrt{3\frac{\left(
\partial _{p}\Delta \right) \cdot \left( \mathbf{\mu \circ \alpha }\right)
\left( \partial _{p}\Delta \right) }{\Delta _{s}^{2}}-9\left[ p^{0}\left(
p\cdot q\right) -2I_{3}(p)\right] ^{2}};
\end{equation}%
\begin{eqnarray}
z_{H}^{1}(\mathcal{Q}) &=&\frac{3}{2}\frac{\left[ \Delta _{s}^{2}\left(
-\Delta _{2}^{2}+\Delta _{3}^{2}+\Delta _{4}^{2}\right) +\frac{1}{4}%
\sum_{g=2}^{P}\left( \Delta _{2g+1}\Delta _{2g+4}-\Delta _{2g+2}\Delta
_{2g+3}\right) ^{2}\right] }{\left( \partial _{p}\Delta \right) \cdot \left(
\mathbf{\mu \circ \alpha }\right) \left( \partial _{p}\Delta \right) }\cdot
\notag \\
&&\cdot \left[ \frac{p^{0}\left( p\cdot q\right) -2I_{3}(p)}{p^{0}}-\mathbf{i%
}\frac{3}{2}\frac{S}{\pi }\right] +\frac{p^{1}}{p^{0}}; \\
&&  \notag \\
z_{H}^{2}(\mathcal{Q}) &=&\frac{3}{2}\frac{\Delta _{s}^{2}\left[ -2\Delta
_{s}\Delta _{2}+\frac{1}{2}\sum_{g=1}^{P}\left( \Delta _{2g+3}^{2}+\Delta
_{2g+4}^{2}\right) \right] }{\left( \partial _{p}\Delta \right) \cdot \left(
\mathbf{\mu \circ \alpha }\right) \left( \partial _{p}\Delta \right) }\cdot
\notag \\
&&\cdot \left[ \frac{p^{0}\left( p\cdot q\right) -2I_{3}(p)}{p^{0}}-\mathbf{i%
}\frac{3}{2}\frac{S}{\pi }\right] +\frac{p^{2}}{p^{0}}; \\
&&  \notag \\
z_{H}^{3}(\mathcal{Q}) &=&\frac{3}{2}\frac{\Delta _{s}^{2}\left[ 2\Delta
_{s}\Delta _{3}+\frac{1}{2}\sum_{g=1}^{P}\left( \Delta _{2g+3}^{2}-\Delta
_{2g+4}^{2}\right) \right] }{\left( \partial _{p}\Delta \right) \cdot \left(
\mathbf{\mu \circ \alpha }\right) \left( \partial _{p}\Delta \right) }\cdot
\notag \\
&&\cdot \left[ \frac{p^{0}\left( p\cdot q\right) -2I_{3}(p)}{p^{0}}-\mathbf{i%
}\frac{3}{2}\frac{S}{\pi }\right] +\frac{p^{3}}{p^{0}}; \\
&&  \notag \\
z_{H}^{4}(\mathcal{Q}) &=&\frac{3}{2}\frac{\Delta _{s}^{2}\left( 2\Delta
_{s}\Delta _{4}+\sum_{g=1}^{P}\Delta _{2g+3}\Delta _{2g+4}\right) }{\left(
\partial _{p}\Delta \right) \cdot \left( \mathbf{\mu \circ \alpha }\right)
\left( \partial _{p}\Delta \right) }\cdot  \notag \\
&&\cdot \left[ \frac{p^{0}\left( p\cdot q\right) -2I_{3}(p)}{p^{0}}-\mathbf{i%
}\frac{3}{2}\frac{S}{\pi }\right] +\frac{p^{4}}{p^{0}}; \\
&&  \notag \\
z_{H}^{5}(\mathcal{Q}) &=&\frac{3}{2}\frac{\Delta _{s}\left[ \Delta
_{s}\left( \Delta _{2}\Delta _{5}+\Delta _{3}\Delta _{5}+\Delta _{4}\Delta
_{6}\right) -\frac{1}{2}\sum_{g=2}^{P}(\Delta _{5}\Delta _{2g+4}-\Delta
_{6}\Delta _{2g+3})\Delta _{2g+4}\right] }{\left( \partial _{p}\Delta
\right) \cdot \left( \mathbf{\mu \circ \alpha }\right) \left( \partial
_{p}\Delta \right) }\cdot  \notag \\
&&\cdot \left[ \frac{p^{0}\left( p\cdot q\right) -2I_{3}(p)}{p^{0}}-\mathbf{i%
}\frac{3}{2}\frac{S}{\pi }\right] +\frac{p^{5}}{p^{0}}; \\
&&  \notag \\
z_{H}^{6}(\mathcal{Q}) &=&\frac{3}{2}\frac{\Delta _{s}\left[ \Delta
_{s}\left( \Delta _{2}\Delta _{6}-\Delta _{3}\Delta _{6}+\Delta _{4}\Delta
_{5}\right) +\frac{1}{2}\sum_{g=2}^{P}(\Delta _{5}\Delta _{2g+4}-\Delta
_{6}\Delta _{2g+3})z_{2g+3}\right] }{\left( \partial _{p}\Delta \right)
\cdot \left( \mathbf{\mu \circ \alpha }\right) \left( \partial _{p}\Delta
\right) }\cdot  \notag \\
&&\cdot \left[ \frac{p^{0}\left( p\cdot q\right) -2I_{3}(p)}{p^{0}}-\mathbf{i%
}\frac{3}{2}\frac{S}{\pi }\right] +\frac{p^{6}}{p^{0}}; \\
&&\vdots  \notag \\
z_{H}^{2P+3}(\mathcal{Q}) &=&\frac{3}{2}\frac{\Delta _{s}\left[ \Delta
_{s}\left( \Delta _{2}\Delta _{2P+3}+\Delta _{3}\Delta _{2P+3}+\Delta
_{4}\Delta _{2P+4}\right) -\frac{1}{2}\sum_{g=2}^{P}(-\Delta _{2g+1}\Delta
_{2P+4}+\Delta _{2g+2}\Delta _{2P+3})\Delta _{2g+2}\right] }{\left( \partial
_{p}\Delta \right) \cdot \left( \mathbf{\mu \circ \alpha }\right) \left(
\partial _{p}\Delta \right) }\cdot  \notag \\
&&\cdot \left[ \frac{p^{0}\left( p\cdot q\right) -2I_{3}(p)}{p^{0}}-\mathbf{i%
}\frac{3}{2}\frac{S}{\pi }\right] +\frac{p^{2P+3}}{p^{0}};  \notag \\
&& \\
z_{H}^{2P+4}(\mathcal{Q}) &=&\frac{3}{2}\frac{\Delta _{s}\left[ \Delta
_{s}\left( \Delta _{2}\Delta _{2P+4}-\Delta _{3}\Delta _{2P+4}+\Delta
_{4}\Delta _{2P+3}\right) +\frac{1}{2}\sum_{g=2}^{P}(-\Delta _{2g+1}\Delta
_{2P+4}+\Delta _{2g+2}\Delta _{2P+3})\Delta _{2g+1}\right] }{\left( \partial
_{p}\Delta \right) \cdot \left( \mathbf{\mu \circ \alpha }\right) \left(
\partial _{p}\Delta \right) }\cdot  \notag \\
&&\cdot \left[ \frac{p^{0}\left( p\cdot q\right) -2I_{3}(p)}{p^{0}}-\mathbf{i%
}\frac{3}{2}\frac{S}{\pi }\right] +\frac{p^{2P+4}}{p^{0}},  \notag \\
&&
\end{eqnarray}%
with $\left( \partial _{p}\Delta \right) \cdot \left( \mathbf{\mu \circ
\alpha }\right) \left( \partial _{p}\Delta \right) $ given by (\ref{den1P}),
and
\begin{eqnarray}
p\cdot q &=&p^{0}q_{0}+p^{1}q_{1}+....+p^{2P+4}q_{2P+4}; \\
I_{3}(p) &=&\mathcal{V}_{L(1,P)}\,\left( p^{1},...,p^{2P+4}\right) =p^{1}%
\left[ -\left( p^{2}\right) ^{2}\,+\,\left( p^{3}\right) ^{2}\,+\,\left(
p^{4}\right) ^{2}\right] \,  \notag \\
&&+\,p^{2}Q_{0}^{(P)}(p^{5},...,p^{2P+4})\,+%
\,p^{3}Q_{1}^{(P)}(p^{5},...,p^{2P+4})\,+%
\,p^{4}Q_{2}^{(P)}(p^{5},...,p^{2P+4}) \\
&=&p^{1}\left[ -\left( p^{2}\right) ^{2}\,+\,\left( p^{3}\right)
^{2}\,+\,\left( p^{4}\right) ^{2}\right] +p^{2}\sum_{i=1}^{2P}\left(
p^{i+4}\right) ^{2}  \notag \\
&&+\,p^{3}\,\sum_{i=1}^{2P}(-1)^{i+1}\left( p^{i+4}\right)
^{2}+2p^{4}\sum_{i=1}^{P}p^{2i+3}p^{2i+4}.
\end{eqnarray}%
Consistently, it should be remarked that for $P=2$ and $P=3$ the above formul%
\ae\ allow one to retrieve the explicit results obtained in Secs. \ref{L120}
and \ref{L130}, respectively.

It is here worth remarking that the above expressions provide, for the first
time to the best of our knowledge, the explicit form of the BPS black hole
entropy and attractors in an infinite class of (homogeneous) non-symmetric
models of $\mathcal{N}=2$, $D=4$ supergravity with cubic prepotential.

\subsubsection{On the geometry of $(\protect\nabla _{\mathcal{V}%
_{L(1,P)}})^{-1}$, $P\geqslant 1$\label{geom-fact}}

We will now describe briefly some properties of the factorization
\begin{equation}
(\nabla _{\mathcal{V}_{L(1,P)}})^{-1}=\mathbf{\mu \circ \alpha },~P\geqslant
1,
\end{equation}%
in order to have a different, geometric perspective on its nature and on the
arising of the quadratic forms $R_{k}$'s (or $R_{K}^{(k,l)}$'s) appearing in
the Lorentzian quadratic identities (\ref{lorrel}), (\ref{P2}), or (\ref{gqr}%
).

Since the map $\mathbf{\alpha }:\mathbb{R}^{2P+4}\rightarrow \mathbb{R}%
^{2P+4+\binom{P}{2}}$ is defined by homogenous polynomials of degree two,
the map $\mathbf{\alpha }$ has image $Z\subset \mathbb{R}^{2P+4+\binom{P}{2}%
} $ defined by homogeneous polynomials. Moreover, the dimension of $Z$, as
an algebraic variety, is $2P+4$, as it is easily seen by looking at the
expression of $\mathbf{\alpha }$. From the parametrization $\mathbf{\alpha }$
of $Z$, we can deduce that it is a \textit{cone} over a Grassmann variety $%
G(2,P+2)$. Moreover, the inverse of $\mathbf{\alpha }$ (as a birational map
from $\mathbb{R}^{2P+4}$ to $Z$) is the restriction to $Z$ of the projection
$\pi (t,u,v,w)=~^{T}(t,u,v)$. Indeed, the map $\mathbf{\alpha }:\mathbb{R}%
^{2P+4}\rightarrow Z$ is such that, letting $z=~^{T}(z_{1},\ldots ,z_{2P+4})$%
,
\begin{equation}
(\pi \circ \mathbf{\alpha })(z)=z_{1}z~.
\end{equation}

The map $\mathbf{\mu }:\mathbb{R}^{2P+4+\binom{P}{2}}\rightarrow \mathbb{R}%
^{2P+4}$ restricted to $Z$ induces a map $\mathbf{\mu }:Z\rightarrow \mathbb{%
R}^{2P+4}$. The inverse (as a birational map) is the map $\mathbf{\phi }:%
\mathbb{R}^{2P+4}\rightarrow Z\subset \mathbb{R}^{2P+4+\binom{P}{2}}$
defined by
\begin{equation}
\mathbf{\phi }(\xi ):=~^{T}(\nabla _{\mathcal{V}_{L(1,P)}}(\xi ),\ldots
,-4R^{(k,l)}(\xi ),\ldots )
\end{equation}%
because
\begin{equation}
(\mathbf{\mu }\circ \mathbf{\phi })(\xi )=4{\mathcal{V}_{L(1,P)}}(\xi )\xi ~.
\end{equation}%
Indeed,
\begin{equation}
q(x)\mathbf{\phi }(\xi )=\mathbf{\alpha }({\mathcal{V}_{L(1,P)}}(\xi ))
\end{equation}%
and
\begin{equation}
(\mathbf{\mu }\circ \mathbf{\alpha })({\mathcal{V}_{L(1,P)}}(\xi ))=4q(x)^{2}%
{\mathcal{V}_{L(1,P)}}(\xi )\xi ~.
\end{equation}%
\bigskip

A similar description of the images $Z$'s of the corresponding maps $\mathbf{%
\alpha }$'s and the definition of the corresponding $\mathbf{\phi }$'s can
actually be provided for \textit{all} the models $L(q,P)$ treated until now;
however, a thorough treatment goes well beyond the scope of this paper, and
we leave it for further future work.\bigskip

This geometric approach has many interesting applications for the study of
the geometry of the singular \textit{locus} of the cubic hypersurface ${%
\mathcal{V}_{L(q,P)}}$ in the associated projective space, which in many
cases is a notable algebraic variety (for example in the complete case, but
not only). In some cases, the quadratic homogeneous polynomials defining the
map $\mathbf{\phi }$ are a basis of all homogenous quadratic polynomials
vanishing on (some irreducible component of) the singular \textit{locus} of
the cubic. It may also happen that there are more quadratic equations
vanishing on the singular \textit{locus} than the partial derivatives of ${%
\mathcal{V}_{L(q,P)}}$ which define it, and this \textit{explains} why
correspondingly $r>0$. This geometric point of view and its applications to
Algebraic Geometry will be considered elsewhere.

\section{Non-uniqueness of $\Omega $'s\label{L1PL81}}

The models $L(1,P)$ for $P\geqslant 4$ can be handled with the general
results obtained in Sec.\ \ref{invP3}, as outlined in the previous Section.
However, those with $P\leqslant 8$ can also be embedded into the complete $%
L(8,1)$ model, and this provides an alternative way to determine the inverse
of their gradient maps.

In this respect, we should stress the fact that the extra $\Omega $%
-matrices, required for non-complete systems by the invertibility condition
enounced in Sec.\ \ref{main}, \textit{may not be unique}; even the number $r$
of such matrices is not determined uniquely by the model.

We illustrate this for $P=4$ and $8$, respectively in Secs.\ \ref{L14} and %
\ref{L18}.

\subsection{$L(1,4)$ $\subset L(4,1)$\label{L14}}

The $L(1,4)$ model has $q=1$, $P=4$, and $\mathcal{D}_{2}=2$ \cite{dWVP};
thus, the number of variables is%
\begin{equation}
\left( 1+q+2+P\cdot \mathcal{D}_{q+1}\right) _{q=1,P=4}=1+3+8=12.
\end{equation}%
The cubic form $\mathcal{V}_{L(1,4)}$ in the variables $%
s,x^{0},x^{1},x^{2},y^{1},\ldots ,y^{8}$ reads
\begin{equation}
\mathcal{V\,}_{L(1,4)}=\,sq(x)\,+\,\sum_{I=0}^{2}x^{I}Q_{I}^{(4)}(y),\qquad
\text{with}\quad q(x)\,=\,-\left( x^{0}\right) ^{2}+\left( x^{1}\right)
^{2}+\left( x^{2}\right) ^{2}~,  \label{V-L140}
\end{equation}%
where the $Q_{I}^{(4)}$'s are obtained from the $Q_{I}$'s in the $L(1,1)$
model (cfr. (\ref{th110})):
\begin{eqnarray}
Q_{0}^{(4)}\, &=&\,\left( y^{1}\right) ^{2}+\left( y^{2}\right)
^{2}\,+\left( y^{3}\right) ^{2}+\left( y^{4}\right) ^{2}\,+\,\left(
y^{5}\right) ^{2}+\left( y^{6}\right) ^{2}\,+\,\left( y^{7}\right)
^{2}+\left( y^{8}\right) ^{2}; \\
Q_{1}^{(4)}\, &=&\,\left( y^{1}\right) ^{2}-\left( y^{2}\right)
^{2}\,+\left( y^{3}\right) ^{2}-\left( y^{4}\right) ^{2}\,+\,\left(
y^{5}\right) ^{2}-\left( y^{6}\right) ^{2}\,+\,\left( y^{7}\right)
^{2}-\left( y^{8}\right) ^{2}; \\
Q_{2}^{(4)}\,
&=&\,2(y^{1}y^{2}\,+\,y^{3}y^{4}\,+\,y^{5}y^{6}\,+\,y^{7}y^{8})~.
\end{eqnarray}

In order to invert the gradient map of $\mathcal{V\,}_{L(1,4)}$, according
to the condition in Sec.\ \ref{main} we need extra quadrics $R_{K}$ defined
by matrices $\Omega _{K}$ satisfying a Lorentzian quadratic identity.

\begin{itemize}
\item One way to do so is to follow Sec.\ \ref{invP3} and thus to define the
quadrics, based on the quadric $R$ in the $L(1,2)$ model given in (\ref%
{this5}), $R^{(k,l)}(y):=R(y^{(kl)})$:
\begin{eqnarray}
R^{(1,2)}\,=\,2(y^{1}y^{4}\,-\,y^{2}y^{3}), &\qquad
&R^{(1,3)}\,=\,2(y^{1}y^{6}\,-\,y^{2}y^{5}),  \label{RklL41} \\
R^{(2,3)}\,=\,2(y^{3}y^{6}\,-\,y^{4}y^{5}), &\qquad
&R^{(1,4)}\,=\,2(y^{1}y^{8}\,-\,y^{2}y^{7}), \\
R^{(2,4)}\,=\,2(y^{3}y^{8}\,-\,y^{4}y^{7}), &\qquad
&R^{(3,4)}\,=\,2(y^{5}y^{8}\,-\,y^{6}y^{7})~.
\end{eqnarray}%
As shown in general in \ref{verifeqn}, a Lorentzian identity of type (\ref%
{gqr}) holds true, namely,
\begin{equation}
-Q_{0}^{(4)}(y)^{2}\,+\,Q_{1}^{(4)}(y)^{2}\,+\,Q_{2}^{(4)}(y)^{2}\,+\,%
\sum_{1\leqslant k<l\leqslant 4}R^{(k,l)}(y)^{2}~=0.  \label{lorL41A}
\end{equation}%
The $L(1,4)$ model is thus invertible with $r=6$.

\item Another approach is to take the $L(4,1)$ model discussed in Sec.\ \ref%
{L410JH} (itself a linearly constrained $L(8,1)$ model : cfr. (\ref{j1})-(%
\ref{j2})). The first three quadrics $\bar{Q}_{0},\bar{Q}_{1},\bar{Q}_{2}$
of this model, as listed in (\ref{th410}), are almost equal to the $%
Q_{i}^{(4)}$ above, and they become equal if we change the signs in some of
the variables\footnote{%
Notice that we did something similar in (\ref{L12sub}) in Sec.\ \ref{L120}
for the $L(1,2)$ model.}:
\begin{equation}
Q_{i}^{(4)}(y^{1},y^{2},y^{3},y^{4},y^{5},y^{6},y^{7},y^{8})\,=\,\bar{Q}%
_{i}(y^{1},y^{2},-y^{3},y^{4},y^{5},y^{6},-y^{7},y^{8})\qquad (i=0,1,2)~.
\end{equation}%
There are three more quadratic forms $\bar{Q}_{3},\bar{Q}_{4},\bar{Q}_{5}$
in the $L(4,1)$ model, and the Lorentzian identity (\ref{lorL410}) holds
between the six quadrics of the $L(4,1)$ model. In fact, if we define three
quadratic forms by
\begin{equation}
R_{K}(y^{1},y^{2},y^{3},y^{4},y^{5},y^{6},y^{7},y^{8})\,:=\,\bar{Q}%
_{K+2}(y^{1},y^{2},-y^{3},y^{4},y^{5},y^{6},-y^{7},y^{8})\qquad (K=1,2,3)~,
\label{RKL41}
\end{equation}%
then we obtain the Lorentzian identity
\begin{equation}
-Q_{0}^{(4)}(y)^{2}\,+\,Q_{1}^{(4)}(y)^{2}\,+\,Q_{2}^{(4)}(y)^{2}\,+%
\,R_{1}(y)^{2}\,+\,R_{2}(y)^{2}\,+\,R_{3}(y)^{2}\,=\,0~.  \label{lorL41B}
\end{equation}%
This shows that the descendant relation $L(1,4)$ $\subset L(4,1)$ holds,
and, again, the condition in Sec.\ \ref{main} is again satisfied, but now
with $r=3$.
\end{itemize}

\subsection{ $L(1,8)\subset L(8,1)$\label{L18}}

A similar treatment shows that $L(1,8)$ is a descendant of $L(8,1)$: $%
L(1,8)\subset L(8,1)$. Indeed, the first three of the ten quadrics in $%
L(8,1) $ are the $Q_{i}^{(8)}$, $i=0,1,2$, of the $L(1,8)$ model, upon
substituting%
\begin{equation}
y^{\alpha }\longrightarrow -y^{\alpha }~\text{for~}\alpha =3,7,11,15.
\label{subst}
\end{equation}

\begin{itemize}
\item In order to satisfy the invertibility condition enounced in Sec.\ \ref%
{main}, one can use the $R^{(k,l)}$, $1\leqslant k<l\leqslant 8$, as defined
in Sec.\ \ref{invP3}, and in this case one has $r=\binom{8}{2}=28$.

\item Alternatively, one can use the remaining seven quadrics $Q_{3},\ldots
,Q_{9}$ and after the substitution (\ref{subst}) one obtains seven quadrics $%
R_{K}$, $K=1,\ldots ,7$, which again imply the invertibility condition of
Sec.\ \ref{main} to be satisfied, now with $r=7$.
\end{itemize}

Clearly, also the models $L(1,P)$ with $P=5,6,7$ can be handled in a similar
way (setting the last $16-2P$ variables $y^{\alpha }$ equal to zero). Notice
that $L(1,5)$, $L(1,6)$, $L(1,7)$ and $L(1,8)$ are descendant, but \textit{%
not submodels}, of $L(8,1)$; see the discussion below (Sec.\ \ref{Des-Sub}).

More in general, it holds that%
\begin{equation}
L(1,P)\subset L(8,1),~1\leqslant P\leqslant 8.  \label{embb}
\end{equation}%
The key feature is that for $q+1\leqslant 9$ the $\Gamma $-matrices $\Gamma
_{1},\ldots ,\Gamma _{q+1}$ of $L(8,1)$ define a representation of the
Clifford algebra $Cl(q+1,0)$ : thus, there always exists $L(q,P,\dot{P}%
)\subset L(8,1)$, for appropriate $q$, $P$ and $\dot{P}$ with $(P+\dot{P}){%
\mathcal{D}}_{q+1}=16$ (setting some of the $y^{\alpha }$ variables equal to
zero allows one to lower $P,\dot{P}$).


\subsection{\label{Des-Sub}Descendant $\nrightarrow $ Submodel}

In the treatment given above, we have discussed various cases in which a
model $L(q,P,\dot{P})$ can be regarded as a model $L(q^{\prime },P^{\prime },%
\dot{P}^{\prime })$ with a larger number of variables (namely, $q+\left( P+%
\dot{P}\right) \cdot \mathcal{D}_{q+1}<q^{\prime }+\left( P^{\prime }+\dot{P}%
^{\prime }\right) \cdot \mathcal{D}_{q^{\prime }+1}$) with some linear
constraints (and possibly with some renaming of variables). $L(q,P,\dot{P})$
has thus been defined as a \textit{descendant} of $L(q^{\prime },P^{\prime },%
\dot{P}^{\prime })$, denoted by%
\begin{equation}
L(q,P,\dot{P})\subset L(q^{\prime },P^{\prime },\dot{P}^{\prime }),
\label{g}
\end{equation}%
and this relation has been instrumental in proving the invertibility of the
corresponding gradient map by using the invertibility condition enounced in
Sec.\ \ref{main}. Here, we want to point out that (\ref{g}) does \textit{not}
necessarily imply that $L(q,P,\dot{P})$ is a \textit{submodel} of $%
L(q^{\prime },P^{\prime },\dot{P}^{\prime })$ (while the converse is
trivially true).

Indeed, we have to recall that each model $L(q,P,\dot{P})$ corresponds to a
non-compact, Riemannian homogeneous (\textquotedblleft special") manifold,
or more rigorously to a triplet of quaternionic K\"{a}hler, special K\"{a}%
hler and real special manifolds, which can be coupled to Maxwell-Einstein
supergravity with $8$ supersymmetries respectively in 3,4 or 5 Lorentzian
space-time dimensions \cite{dWVP, dWVVP} :%
\begin{eqnarray}
L(q,P,\dot{P}) &\Longleftrightarrow &\left\{
\begin{array}{l}
\frac{G_{3}}{H_{3}}~\text{quat.~K\"{a}hler,} \\
\frac{G_{4}}{H_{4}}~\text{special~K\"{a}hler,} \\
\frac{G_{5}}{H_{5}}~\text{real~special,}%
\end{array}%
\right. \\
H_{i} &=&\text{mcs}\left( G_{i}\right) ,i=3,4,5\text{,} \\
\frac{G_{3}}{H_{3}} &\supsetneq &\frac{G_{4}}{H_{4}}\supsetneq \frac{G_{5}}{%
H_{5}}, \\
\text{with~}G_{3} &\supsetneq &G_{4}\supsetneq G_{5}~\text{and}%
~~H_{3}\supsetneq H_{4}\supsetneq H_{5},
\end{eqnarray}%
where `mcs' stands for maximal compact subgroup. Therefore, a necessary (but
generally not sufficient) condition for a model $L$ (associated to $\frac{%
G_{i}}{H_{i}}$, $i=3,4,5$) to be a \textit{submodel} of a model $L^{\prime }$
(associated to $\frac{G_{i}^{\prime }}{H_{i}^{\prime }}$, $i=3,4,5$) is that
the structure of the corresponding cosets is consistent with the immersions,
namely that%
\begin{equation}
G_{i}\subsetneq G_{i}^{\prime },~H_{i}\subsetneq H_{i}^{\prime }~\forall
i=3,4,5.
\end{equation}

A \textbf{counterexample} is provided by (\ref{embb}) with $P=5$ : in fact, $%
L(1,5)$ is a \textit{descendant} of $L(8,1)$, namely%
\begin{equation}
L(1,5)\subset L(8,1),
\end{equation}%
but $L(1,5)$ \textit{is not}\footnote{\textit{A fortiori}, the same holds
for the models $L(1,6)$, $L(1,7)$ and $L(1,8)$.}\textit{\ a submodel} of $%
L(8,1)$. In fact \cite{dWVVP,Cecotti}:%
\begin{eqnarray}
L(1,5) &\Longleftrightarrow &G_{5,L(1,5)}=\left( SO(2,1)\otimes SO(5)\otimes
SO(1,1)\right) _{0}\ltimes \left( \mathbf{2},\mathbf{5}\right) _{3/2}; \\
T_{3}(1,5,0) &\ni &\mathcal{T}:=\underset{SO(5)~\text{covariant}}{\left(
\begin{array}{ccc}
\mathbf{1} & \mathbf{1} & \mathbf{5} \\
\ast & \mathbf{1} & \mathbf{5} \\
\ast & \ast & \mathbf{1}%
\end{array}%
\right) },~~\dim _{\mathbb{R}}\left( T_{3}(1,5,0)\right) =14=\underset{%
SO(2,1)\otimes SO(5)\text{~covariant}}{\left( \mathbf{3},\mathbf{1}\right)
\oplus (\mathbf{1},\mathbf{1})\oplus (\mathbf{2},\mathbf{5})};  \notag \\
&& \\
L(8,1) &\Longleftrightarrow &G_{5,L(8,1)}=E_{6(-26)}; \\
J_{3}^{\mathbb{O}} &\ni &\mathcal{J}:=\underset{SO(8)~\text{covariant}}{%
\left(
\begin{array}{ccc}
\mathbf{1} & \mathbf{8}_{v} & \mathbf{8}_{s} \\
\ast & \mathbf{1} & \mathbf{8}_{c} \\
\ast & \ast & \mathbf{1}%
\end{array}%
\right) },~~\dim _{\mathbb{R}}\left( J_{3}^{\mathbb{O}}\right) =27=\underset{%
E_{6(-26)}\text{~covariant}}{\mathbf{27}},
\end{eqnarray}%
where $T_{3}(1,5,0)$ is the Hermitian part of a manifestly $SO(5)$-covariant
Vinberg's cubic T-algebra \cite{Vinberg}, and $J_{3}^{\mathbb{O}}$ denotes
the exceptional cubic Jordan algebra \cite{JVNW}. Note the Peirce
decomposition of $J_{3}^{\mathbb{O}}$ \cite{Peirce}
\begin{equation}
\begin{array}{l}
E_{6(-26)}\supsetneq SO(9,1)\otimes SO(1,1); \\
\mathbf{27}=\mathbf{1}_{-4}\oplus \mathbf{10}_{2}\oplus \mathbf{16}_{-1}.%
\end{array}%
\end{equation}%
Therefore, \textit{if} the $L(1,5)$ model is a \textit{submodel} $L(8,1)$
model, \textit{then} it should hold, among other things, that

\begin{description}
\item[\textbf{i)}]
\begin{gather}
G_{5,L(1,5)}\subsetneq G_{5,L(8,1)}; \\
\Updownarrow  \notag \\
\left( SO(2,1)\otimes SO(5)\otimes SO(1,1)\right) _{0}\ltimes \left( \mathbf{%
2},\mathbf{5}\right) _{3/2}\subsetneq SO(9,1)\otimes SO(1,1)\subsetneq
E_{6(-26)}.
\end{gather}

\item[\textbf{ii)}] as an $SO(9,1)$-representation (after Peirce
decomposition), $J_{3}^{\mathbb{O}}$ contains $T_{3}(1,5,0)$ (this latter as
an $\left( SO(2,1)\otimes SO(5)\right) $-representation).
\end{description}

By the results proven in App. \ref{App-Lie}, it follows that the model $%
L(1,5)$ is \textit{not} a submodel of $L(8,1)$, but rather only one of its
\textit{descendants}. It can thus be stated that the immersions giving rise
to $L(1,P)\subset L(8,1)$,$~5\leqslant P\leqslant 8$, are \textit{not}
consistent with the structure of (the coset spaces respectively
corresponding to) such models.


\section{Beyond $L(q,P)$ models : $L(4,1,1)$\label{L411}}

We will now consider the unique model of the present paper having a
non-vanishing $\dot{P}$ : the model $L(4,1,1)$.

As we have observed in Sec.\ \ref{L410JH}, the complete model $L(4,1)$ can
be obtained from the complete model $L(8,1)$ by setting $x^{6}=\ldots
=x^{9}=0$ and by taking the upper left $8\times 8$ block $\bar{\Gamma}_{I}$
of the first six $\Gamma $-matrices $\Gamma _{0},\ldots ,\Gamma _{5}$ (of
size $16\times 16$) of $L(8,1)$. A closer look at such six $\Gamma $%
-matrices reveals that ($I=0,1,3,4,5$) :
\begin{equation}
\Gamma _{I}\,=\,\bar{\Gamma}_{I}\otimes {\mathbb{I}}_{2}\,=\,\left(
\begin{array}{cc}
\bar{\Gamma}_{I} & 0 \\
0 & \bar{\Gamma}_{I}%
\end{array}%
\right) ,\qquad \Gamma _{2}\,=\,\bar{\Gamma}_{2}\otimes \sigma
_{3}\,=\,\left(
\begin{array}{cc}
\bar{\Gamma}_{2} & 0 \\
0 & -\bar{\Gamma}_{2}%
\end{array}%
\right) .
\end{equation}%
The representation $\psi :Cl(5,0)\rightarrow M_{16}(\mathbb{R})$ defined by $%
\psi (e_{I}):=\Gamma _{I}$, $I=1,\ldots ,5$, is thus reducible, and it is
the direct sum of the two non-equivalent irreducible $8$-dimensional
representations of $Cl(5,0)$, since the sign of only one $\bar{\Gamma}_{I}$
is changed in the second component (cfr.\ Sec.\ \ref{cliffprod}).

By recalling the discussion in Sec.\ \ref{HNS}, one can then conclude that $%
L(4,1,1)\subset L(8,1)$ : by using the first $1+5$ $\Gamma $-matrices of $%
L(8,1)$, one indeed obtains
\begin{equation}
L(4,1,1)\,=\,\left. L(8,1)\right\vert _{x^{6}=\ldots =x^{9}=0}~,
\end{equation}%
so $L(4,1,1)$ can be regarded as the $L(8,1)$ model with four linear
constraints. The $L(4,1,1)$ model has $q=4,P=\dot{P}=1$, and $\mathcal{D}%
_{5}=8$ \cite{dWVP}, so the number of variables is%
\begin{equation}
\left( 1+q+2+\left( P+\dot{P}\right) \cdot \mathcal{D}_{q+1}\right)
_{q=4,P=1,\dot{P}=1}=1+6+\left( 1+1\right) \cdot 8=23.
\end{equation}%
With the notation from (\ref{thL411}) below, the cubic form of this model is
\begin{equation}
{\mathcal{V}}_{L(4,1,1)}\,:=\,sq(x)\,+\,\sum_{I=0}^{5}x^{I}Q_{I}^{(1,1)}(y)%
\,=\,\left. {\mathcal{V}}_{L(8,1)}\right\vert _{x^{6}=\ldots =x^{9}=0},
\label{vL411}
\end{equation}%
where
\begin{equation}
q(x)=-\left( x^{0}\right) ^{2}\,+\,\left( x^{1}\right) ^{2}+\left(
x^{2}\right) ^{2}\,+\,\ldots \,+\,\left( x^{5}\right) ^{2}~.  \label{qL411}
\end{equation}

The remaining four $\Gamma $-matrices $\Gamma _{6},\ldots ,\Gamma _{9}$ of
the $L(8,1)$ model are used as the extra matrices $\Omega _{1},\ldots
,\Omega _{4}$ occurring in the invertibility condition of Sec.\ \ref{main}
(with $r=4$):
\begin{equation*}
\Omega _{K}\,=\,\Gamma _{5+K},\qquad \text{so}\quad
R_{K}(y)\,=\,Q_{5+K}(y),\quad (K=1,\ldots ,4).
\end{equation*}%
Therefore, the set of $\Gamma $-matrices $\{\Gamma _{1},\ldots ,\Gamma
_{9}\} $ of $L(8,1)$ is a Clifford set and the anti-commutativity conditions
in the condition of Sec.\ \ref{main} are satisfied. The Lorentzian quadratic
relation (\ref{GammQI4}), where we now change the names according to (\ref%
{thL411}), then shows that Lorentzian quadratic identity
\begin{equation}
-Q_{0}^{(1,1)}(y)^{2}\,+\,Q_{1}^{(1,1)}(y)^{2}\,+\,\ldots
\,+\,Q_{5}^{(1,1)}(y)^{2}\,+\,R_{1}(y)^{2}\,+\,\ldots
\,+\,R_{4}(y)^{2}\,=\,0~  \label{lorrelL411}
\end{equation}%
holds, as required by the aforementioned condition, which then implies that
the gradient map of $L(4,1,1)$ is invertible.

We now present the quadrics of the $L(8,1)$ system, as given in (\ref{this3}%
), with the appropriate renamings suitable for the $L(4,1,1)$ model:
\begin{eqnarray}
&&%
\begin{array}{lll}
Q_{0}^{(1,1)}\equiv Q[{}_{0000}^{0000}] & := & \left( y^{1}\right)
^{2}+\,\ldots \,+\,\left( y^{8}\right) ^{2}\,+\,\left( y^{9}\right)
^{2}\,+\,\ldots \,+\,\left( y^{16}\right) ^{2}; \\
Q_{1}^{(1,1)}\equiv Q[{}_{0001}^{0000}] & := & \left\{
\begin{array}{l}
\left( y^{1}\right) ^{2}-\left( y^{2}\right) ^{2}\,+\,\ldots \,+\,\left(
y^{7}\right) ^{2}\,-\,\left( y^{8}\right) ^{2} \\
+\left( y^{9}\right) ^{2}-\left( y^{10}\right) ^{2}\,+\,\ldots \,+\left(
y^{15}\right) ^{2}-\left( y^{16}\right) ^{2};%
\end{array}%
\right. \\
Q_{2}^{(1,1)}\equiv Q[{}_{1010}^{0001}] & := & 2(y^{1}y^{2}-y^{3}y^{4}\,+%
\,y^{5}y^{6}-y^{7}y^{8}\,-\,y^{9}y^{10}+y^{11}y^{12}\,-%
\,y^{13}y^{14}+y^{15}y^{16}); \\
Q_{3}^{(1,1)}\equiv Q[{}_{0000}^{0011}] & := & 2(y^{1}y^{4}+y^{2}y^{3}\,+%
\,y^{5}y^{8}+y^{6}y^{7}\,+\,y^{9}y^{12}+y^{10}y^{11}\,+%
\,y^{13}y^{16}+y^{14}y^{15}); \\
Q_{4}^{(1,1)}\equiv Q[{}_{0101}^{0101}] & := &
2(y^{1}y^{6}-y^{2}y^{5}+y^{3}y^{8}-y^{4}y^{7}\,+%
\,y^{9}y^{14}-y^{10}y^{13}+y^{11}y^{16}-y^{12}y^{15}); \\
Q_{5}^{(1,1)}\equiv Q[{}_{0110}^{0111}] & := &
2(y^{1}y^{8}+y^{2}y^{7}-y^{3}y^{6}-y^{4}y^{5}\,+%
\,y^{9}y^{16}+y^{10}y^{15}-y^{11}y^{14}-y^{12}y^{13}); \\
R_{1}\equiv Q[{}_{0010}^{1001}] & := &
2(y^{1}y^{10}+y^{2}y^{9}-y^{3}y^{12}-y^{4}y^{11}+y^{5}y^{14}+y^{6}y^{13}-y^{7}y^{16}-y^{8}y^{15});
\\
R_{2}\equiv Q[{}_{1101}^{1011}] & := &
2(y^{1}y^{12}-y^{2}y^{11}+y^{3}y^{10}-y^{4}y^{9}-y^{5}y^{16}+y^{6}y^{15}-y^{7}y^{14}+y^{8}y^{13});
\\
R_{3}\equiv Q[{}_{1110}^{1101}] & := &
2(y^{1}y^{14}+y^{2}y^{13}-y^{3}y^{16}-y^{4}y^{15}-y^{5}y^{10}-y^{6}y^{9}+y^{7}y^{12}+y^{8}y^{11});
\\
R_{4}\equiv Q[{}_{1001}^{1111}] & := &
2(y^{1}y^{16}-y^{2}y^{15}+y^{3}y^{14}-y^{4}y^{13}+y^{5}y^{12}-y^{6}y^{11}+y^{7}y^{10}-y^{8}y^{9}).%
\end{array}
\notag \\
&&  \label{thL411}
\end{eqnarray}%
Since $r=4$, the inverse of the gradient map $\nabla _{{\mathcal{V}}%
_{L(4,1,1)}}$ is given as a composition of two maps, namely (cfr. (\ref{csii}%
))
\begin{equation}
{\mathbb{R}}_{\xi }^{1+6+16}\,\overset{\nabla _{{\mathcal{V}}_{L(1,3)}}}{%
\longrightarrow }\,{\mathbb{R}}_{z}^{1+6+16}\,\overset{\mathbf{\alpha }}{%
\longrightarrow }\,{\mathbb{R}}_{(t,u,v,w)}^{1+6+16+4}\,\overset{\mathbf{\mu
}}{\longrightarrow }\,{\mathbb{R}}_{\xi }^{1+6+16}~,
\end{equation}%
where $\xi =(s,x,y)$, and ${\mathcal{V}}_{L(4,1,1)}$ and the corresponding $%
q(x)$ are given by (\ref{vL411}) and (\ref{qL411}), respectively. The map $%
\mathbf{\alpha }$ from (\ref{alpha}) is given by
\begin{equation}
\mathbf{\alpha }(z_{1},\ldots
,z_{23})\,:=\,^{T}(z_{1}^{2},\,z_{1}z_{2},\ldots
,\,z_{1}z_{23},\,R_{1}(z),R_{2}(z),R_{3}(z),R_{4}(z)),
\end{equation}%
where each of the three quadratic forms $R_{K}(z)$'s ($K=1,...,4$) depends
only on four of the last $\left( 1+1\right) \cdot 8=16$ variables :%
\begin{equation}
R_{K}(z)=R_{K}(z_{8},\ldots ,z_{23}).
\end{equation}%
The map $\mathbf{\mu }$ (\ref{mu}) has $1+3+16$ components that are
homogeneous polynomials of degree $2$ in the variables $%
t,u_{0},u_{1},u_{5},v_{1},\ldots ,v_{16},w_{1},...,w_{4}$. Since
\begin{equation}
\mathbf{\mu }\circ \mathbf{\alpha }\circ \nabla _{{\mathcal{V}}%
_{L(4,1,1)}}\,(\xi )\,=\,4q(x)^{2}{\mathcal{V}}_{L(4,1,1)}(\xi )\mathbb{\xi }%
,
\end{equation}%
the (birational) inverse map $(\nabla _{\mathcal{V}_{L(4,1,1)}})^{-1}=%
\mathbf{\mu \circ \alpha }$ of the gradient map $\nabla _{\mathcal{V}%
_{L(4,1,1)}}$ is a homogeneous polynomial map of degree four. From the
treatment of Sec.\ \ref{invertgamma}, the solution of the BPS system of $%
L(4,1,1)$ is then given by (\ref{pre-sol})-(\ref{pre-sol-3}), and (\ref{SSd}%
) and (\ref{zH-1})-(\ref{zH-3}) yield the corresponding expression of the
BPS black hole entropy and of the BPS attractors, respectively.

\section{Kleinian signatures and split algebras\label{55}}

As already mentioned at the end of Sec.\ \ref{complete}, one can also find
sets of symmetric $\Gamma $-matrices defining a Clifford algebra
representation of a quadratic form in $q+2=4,6,10$ dimensions with Kleinian
\textquotedblleft neutral" signatures $(2_{+},2_{-})$, $(3_{+},3_{-})$ and $%
(5_{+},5_{-})$ for $q=2,4$ and $8$ respectively : these Kleinian signatures
correspond to simple cubic Euclidean Jordan algebras over split composition
algebras $J_{3}^{\mathbb{A}_{s}}$, for $\mathbb{A}=\mathbb{C}_{s},\mathbb{H}%
_{s}$ and $\mathbb{O}_{s}$, respectively. These cases do not belong to the
homogeneous special manifolds classified by $L(q,P,\dot{P})$ : in fact, they
pertain to the so-called `magic' non-supersymmetric Maxwell-Einstein
theories, as well as to the maximal supergravity (in the case of split
octonions $\mathbb{O}_{s}$); cfr. \cite{Marrani-Romano-1,Marrani-Romano-2} :
they can be regarded as the `Kleinian counterparts' of the magic $\mathcal{N}%
=2$ supergravity theories discussed in Sec.\ \ref{clifmat} (recall (\ref%
{form-Eucl}) therein) \cite{Hasebe,Marrani-Romano-1,Marrani-Romano-2} :
\begin{eqnarray}
&&\quad
\begin{array}{c}
J_{3}^{\mathbb{C}_{s}} \\
\dim \mathbf{V}=1+4+4=9, \\
Cl\left( 1,2\right) , \\
\mathcal{N}=0,%
\end{array}%
\quad
\begin{array}{c}
J_{3}^{\mathbb{H}_{s}} \\
\dim \mathbf{V}=1+6+8=15, \\
Cl\left( 2,3\right) , \\
\mathcal{N}=0,%
\end{array}%
\quad
\begin{array}{c}
J_{3}^{\mathbb{O}_{s}} \\
\dim \mathbf{V}=1+10+16=27, \\
Cl\left( 4,5\right) , \\
\mathcal{N}=8.%
\end{array}
\notag \\
&&  \label{form-Klein}
\end{eqnarray}%
Notice that, as the magic $L(q,1)$ models ($q=1,2,4,8$) are related to
Euclidean Clifford algebras $Cl(q+1,0)$ (cfr. Sec.\ \ref{clifmat}), their
`Kleinian counterparts' models (existing for $q=2,4,8$) are related to
Clifford algebras $Cl\left( \frac{q}{2}+1,\frac{q}{2}\right) $ in $q+1$
dimensions with (mostly minus) signature $\left( \frac{q}{2}+1,\frac{q}{2}%
\right) $; in fact, the reality properties of the spinors are the same for
the $\left( q+1,0\right) $ and $\left( \frac{q}{2}+1,\frac{q}{2}\right) $
signatures in $q+1$ dimensions.

The extremal black hole entropy in maximal supergravity is explicitly known
(see e.g. \cite{FGimonK}), and by suitable truncations one obtains the same
quantity in the $J_{3}^{\mathbb{H}_{s}}$- and $J_{3}^{\mathbb{C}_{s}}$-
based theories.

For completeness's sake, we consider here the Kleinian model based on the
exceptional cubic Jordan algebra $J_{3}^{\mathbb{O}_{s}}$, which pertains to
maximal supergravity; in this case,
\begin{equation}
\mathcal{V}_{J_{3}^{\mathbb{O}_{s}}}\,=\,sq(x)\,+\,%
\sum_{I=1}^{10}x^{I}Q_{I}(y)\qquad \text{with}\quad q(x)\,=\,\left(
x^{1}\right) ^{2}\,+\ldots +\,\,\left( x^{5}\right) ^{2}\,-\,\left(
x^{6}\right) ^{2}\,-\,\ldots \,-\,\,\left( x^{10}\right) ^{2}.  \label{form1}
\end{equation}%
Thus, the quadratic form $q(x)$ has signature $(5,5)$, and the ten quadrics $%
Q_{I}$'s read
\begin{eqnarray}
&&%
\begin{array}{rcl}
Q_{1} & = &
2(y^{1}y^{6}-y^{2}y^{5}-y^{3}y^{8}+y^{4}y^{7}+y^{9}y^{14}-y^{10}y^{13}-y^{11}y^{16}+y^{12}y^{15});
\\
Q_{2} & = &
2(y^{1}y^{10}-y^{2}y^{9}+y^{3}y^{12}-y^{4}y^{11}-y^{5}y^{14}+y^{6}y^{13}-y^{7}y^{16}+y^{8}y^{15});
\\
Q_{3} & = &
2(y^{1}y^{11}-y^{2}y^{12}-y^{3}y^{9}+y^{4}y^{10}-y^{5}y^{15}+y^{6}y^{16}+y^{7}y^{13}-y^{8}y^{14});
\\
Q_{4} & = &
2(y^{1}y^{12}+y^{2}y^{11}-y^{3}y^{10}-y^{4}y^{9}-y^{5}y^{16}-y^{6}y^{15}+y^{7}y^{14}+y^{8}y^{13});
\\
Q_{5} & = &
2(y^{1}y^{14}-y^{2}y^{13}-y^{3}y^{16}+y^{4}y^{15}+y^{5}y^{10}-y^{6}y^{9}-y^{7}y^{12}+y^{8}y^{11});
\\
Q_{6} & = &
2(y^{1}y^{6}-y^{2}y^{5}-y^{3}y^{8}+y^{4}y^{7}-y^{9}y^{14}+y^{10}y^{13}+y^{11}y^{16}-y^{12}y^{15});
\\
Q_{7} & = &
2(y^{1}y^{10}-y^{2}y^{9}-y^{3}y^{12}+y^{4}y^{11}+y^{5}y^{14}-y^{6}y^{13}-y^{7}y^{16}+y^{8}y^{15});
\\
Q_{8} & = &
2(y^{1}y^{11}+y^{2}y^{12}-y^{3}y^{9}-y^{4}y^{10}-y^{5}y^{15}-y^{6}y^{16}+y^{7}y^{13}+y^{8}y^{14});
\\
Q_{9} & = &
2(y^{1}y^{12}-y^{2}y^{11}+y^{3}y^{10}-y^{4}y^{9}-y^{5}y^{16}+y^{6}y^{15}-y^{7}y^{14}+y^{8}y^{13});
\\
Q_{10} & = &
2(y^{1}y^{14}-y^{2}y^{13}-y^{3}y^{16}+y^{4}y^{15}-y^{5}y^{10}+y^{6}y^{9}+y^{7}y^{12}-y^{8}y^{11})~.%
\end{array}
\notag \\
&&  \label{form2}
\end{eqnarray}%
These quadrics satisfy the Kleinian quadratic identity\footnote{%
Notice that $Q_{I}^{2}-Q_{I+5}^{2}$ (for $I=1,2,3,4,5$) doesn't have that
many terms.}
\begin{equation}
q(Q_{1},\ldots ,Q_{10})\,=\,Q_{1}^{2}\,+\,\ldots
\,+\,Q_{5}^{2}\,-\,Q_{6}^{2}\,-\,\ldots \,-Q_{10}^{2}\,=\,0~,
\end{equation}%
and the matrices defining these quadratic forms satisfy the Clifford
relations (\ref{clifm}). The differences between the model $L(8,1)$ treated
in Sec.\ \ref{L810} and the Kleinian model of maximal supergravity discussed
here can be realized at a glance by comparing (\ref{v8})-(\ref{q8}) and (\ref%
{this3}) to (\ref{form1}) and (\ref{form2}), respectively.

The inverse of the gradient map $\nabla _{\mathcal{V}_{J_{3}^{\mathbb{O}%
_{s}}}}$ can then be computed to be a polynomial map of degree two, similar
to the one for the Lorenzian case discussed in Sec.\ \ref{L810}.

\section{Beyond the invertibility condition : $L(9,1)$\label{L910}}

Let us now consider the model $L(9,1)$.

The $q+2=11$ quadratic forms $Q_{0},\ldots ,Q_{10}$ are associated to $%
\Gamma _{0}={\mathbb{I}}_{32}$ and to a set of $32\times 32$ Clifford
matrices $\{\Gamma _{1},\ldots ,\Gamma _{10}\}$, generating the Euclidean
Clifford algebra\footnote{%
Recall the non-trivial homomorphism $Cl(10,0)\rightarrow M_{2}(\mathbb{R}%
)\otimes M_{16}(\mathbb{R})=M_{32}(\mathbb{R})$; cfr. (\ref{pre-hom1}).} $%
Cl(10,0)$. It is not hard to find such matrices : by denoting by $\Gamma
_{1}^{\prime },\ldots ,\Gamma _{9}^{\prime }$ the $16\times 16$ Clifford
matrices of $Cl(9,0)$ (which correspond to the quadrics $Q_{1},\ldots ,Q_{9}$
in the complete model $L(8,1)$ discussed in Sec.\ \ref{L810}), for the
Euclidean Clifford algebra $Cl(10,0)$, corresponding to the model $L(9,1)$,
it simply suffices to take the corresponding block matrices of size $%
32\times 32$, as follows:
\begin{equation}
\Gamma _{0}=\mathbb{I}_{32},~~~~\Gamma _{I}\,:=\,\left(
\begin{array}{cc}
\Gamma _{I}^{\prime } & 0 \\
0 & -\Gamma _{I}^{\prime }%
\end{array}%
\right) ~,\qquad (I=1,...,9)\qquad \Gamma _{10}\,:=\,{\mathbb{I}}%
_{16}\otimes \sigma _{1}\,=\,\left(
\begin{array}{cc}
0 & {\mathbb{I}}_{16} \\
{\mathbb{I}}_{16} & 0%
\end{array}%
\right) ~,
\end{equation}%
and one easily verifies that these matrices satisfy the Clifford relations (%
\ref{clifm}).

Notice that there is \textit{no} homomorphism $Cl(11,0)\rightarrow M_{32}(%
\mathbb{R})$, so the Clifford set of $\Gamma $-matrices $\{\Gamma _{1}:={%
\mathbb{I}}_{32},\Gamma _{2},\ldots ,\Gamma _{10}\}$ is \textit{maximal} for
the size $32\times 32$, namely one cannot add another $\Gamma $-matrix of
size $32\times 32$ and still have a Clifford set.

However, in striking contrast to the complete models related to $\Gamma $%
-matrices of size $m=2^{g}$ with $g=1,2,3,4$, discussed in Sec.\ \ref%
{complete} and respectively treated in Secs.\ \ref{L110}, \ref{L210}, \ref%
{L410JH} and \ref{L810}, \textit{there is no (Lorentzian) quadratic identity}
between the $11$ quadratic forms $Q_{0},\ldots ,Q_{10}$. Therefore, one
\textit{cannot} exploit the invertibility condition enounced in Sec.\ \ref%
{main} in order to determine the invertibility of the gradient map $\nabla _{%
\mathcal{V}_{L(9,1)}}$. Of course, such a condition provides a sufficient
but not necessary condition for invertibility, so the lack of a suitable
quadratic identity of quadrics does not necessarily imply the
non-invertibility of the gradient map of the corresponding cubic form.

At any rate, other approaches to prove invertibility or non-invertibility of
the gradient map $\nabla _{\mathcal{V}_{L(9,1)}}$ should be found, but they
are beyond the scope of the present investigation.


\section{\label{Conclusion}Final remarks and outlook}

We have considered the issue to obtain an explicit expression of the
attractor values of scalar fields as well as of the Bekenstein-Hawking
entropy, of static, asymptotically flat, dyonic, BPS extremal black holes in
ungauged $\mathcal{N}=2$ Maxwell-Einstein supergravity theories in four
space-time dimensions, coupled to non-linear sigma models of scalar fields
endowed with very special geometry; this class of theories encompasses all
four-dimensional $\mathcal{N}=2$ theories which can be obtained as an $S^{1}$%
-compactification of five-dimensional minimal supergravity theories. After
\cite{Shmakova}, this problem can be translated into the issue of solving
certain algebraic inhomogeneous systems of degree two, named \textit{BPS
systems}.

Within the so-called `very special' geometry (related to \textit{cubic}
holomorphic prepotentials), we have focused on \textit{homogeneous}
non-compact Riemannian spaces. For homogeneous \textit{symmetric} spaces,
which are related to (simple and semi-simple) cubic (Euclidean) Jordan
algebras, the solution to the BPS system is explicitly known, as is the
expression of the BPS entropy and attractors (cfr. \cite{FGimonK}, and Refs.
therein) : they can be formulated only in terms of a unique quartic invariant%
\footnote{%
In \cite{FGimonK}, the treatment of the present paper corresponds to the
manifestly `$G_{6}$-invariant' formalism discussed in Sec.\ 3.4 therein; see
also \cite{Magic-Wissanji}.} polynomial in the black hole electric-magnetic
charges. On the other hand, not much is known for the homogeneous \textit{%
non-symmetric }spaces; in fact, to the best of our knowledge, only \cite%
{DFT-Hom-07} and \cite{H2-1} briefly treated, within a different formalism,
the models $L(1,2)$ and $L(2,2)$. Therefore, in the present investigation we
have focussed on \textit{homogeneous} \textit{non-symmetric} very special
geometry, which has been classified, in terms of Euclidean Clifford
algebras, in \cite{dWVP}.

In Sec.\ \ref{main} we have formulated a (sufficient, but not necessary)
condition for the invertibility of the gradient map of the cubic form
defining the homogeneous non-symmetric very special geometry (and thus for
the resolution of the related BPS system) : this condition requires the
existence of a suitable Lorentzian quadratic identity involving the
quadratic forms defined by the symmetric $\Gamma $-matrices of the
corresponding Euclidean real Clifford algebra, as well as some other
quadratic forms defined by symmetric auxiliary matrices denoted by $\Omega
_{K}$. Subsequently, we have thus provided in Sec.\ \ref{explin} an explicit
expression for the (birational) inverse map of the gradient map of the
models for which the invertibility condition holds; the inverse map is a
homogeneous polynomial map of degree four. Then, in Sec.\ \ref{Sol-BPS}, we
have presented, within the assumption that the aforementioned condition
holds true, a procedure for the explicit solution of the related BPS system,
determining in Sec.\ \ref{BPS-S-gen} an explicit formula for the BPS
Bekenstein-Hawking entropy of extremal black holes, as well as for the
attractor values of the scalar fields in such a background. It is also here
worth remarking that the explicit solution of the BPS system is also
relevant for the solution of the attractor equations in asymptotically $AdS$%
, dyonic, extremal $\frac{1}{4}$-BPS black holes of $U(1)$ Fayet-Iliopoulos
gauged Maxwell-Einstein $\mathcal{N}=2$ supergravity in four space-time
dimensions \cite{Halmagyi1, Halmagyi2}.

Besides the general treatment given in Sec.\ \ref{invertgamma} (within the
validity of the invertibility condition enounced in Sec.\ \ref{main} and
then discussed in Sec.\ \ref{checkinvmap}) as well as in Sec.\ \ref{Pgreq4}
(for the models $L(1,P)$ with $P\geqslant 4$), we have explicitly considered
various homogeneous non-symmetric models, namely :

\begin{itemize}
\item $L(q,1)$, $q=1,2,...,8$ (Sec.\ \ref{explmodels}) and $9$ (Sec.\ \ref%
{L910});

\item $L(q,2)$, $q=1,2,3$ (Sec.\ \ref{L420});

\item $L(q,P)$, $q=1,2,3$, $P\geqslant 3$ (Sec.\ \ref{LqP0}), with explicit
emphasis given to the models $L(1,P)$ with $P\geqslant 2$ given in Sec.\ \ref%
{L1P0};

\item $L(4,1,1)$ (Sec.\ \ref{L411}).
\end{itemize}

In particular, the models $L(1,2)$ and $L(1,3)$ have been worked out in full
detail in Secs.\ \ref{L120} resp. \ref{L130}, and in Sec.\ \ref{L1P0} their
treatment has been generalized (in a $P$-dependent manner) to the infinite
class of $L(1,P)$ $P\geqslant 2$ non-symmetric models of $\mathcal{N}=2$, $%
D=4$ supergravity. In this respect, we have\ extended the treatment given in
Sec.\ 4 of \cite{Shmakova}, by providing, for the first time to the best of
our knowledge, the explicit form of the BPS black hole entropy and of the
BPS attractors in an infinite class of (homogeneous) non-symmetric models of
$\mathcal{N}=2$ supergravity with cubic prepotential.\bigskip

Still, many homogeneous non-symmetric models remain to be investigated for
what concerns the invertibility of the corresponding gradient map, and thus
the solution to the corresponding BPS system, aiming at obtaining explicit
expressions for the Bekenstein-Hawking entropy of extremal BPS black holes
as well as for the corresponding BPS attractor configurations of the scalar
fields. From the classification of \cite{dWVP} (see also \cite{dWVVP, dWVP2}%
), these models belong to the infinite series

\begin{itemize}
\item $L(q,1)$, $q\geqslant 10$;

\item $L(q,2)$, $q\geqslant 4$;

\item $L(q,P)$, $q\geqslant 4$, $P\geqslant 3$;

\item $L(-1,P)$, $P\geqslant 1$ (the so-called `non-Jordan symmetric
sequence' \cite{broken});

\item $L(4m,P,\dot{P})$, with\footnote{%
The models $L(0,P,\dot{P})$ have been treated, within a different formalism,
in \cite{H2-1} and \cite{DFT-Hom-07}.} $m\geqslant 0$, $P\geqslant 1$, and $%
\dot{P}\geqslant 1$ (excluding the model $L(4,1,1)$).
\end{itemize}

We leave the treatment of such classes to future work. It would also be
interesting to investigate the invertibility of the gradient map, and thus
the solution to the corresponding BPS system, of cubic forms associated to
noteworthy classes of non-homogeneous spaces.\bigskip

Also, we would like to recall that in Sec.\ \ref{geom-fact}, we have briefly
considered a geometric perspective on the factorized nature of the inverse
map of the gradient map of the cubic forms pertaining to the models $L(1,P)$
with $P\geqslant 1$. We conjecture that this holds essentially true for any $%
L(q,P)$ model, thus providing an explanation to $r>0$ in non-complete
models; in future works, it will be interesting to discuss this geometric
point of view in detail, as well as to study various subsequent applications
to Algebraic Geometry.

Within this research venue, it would be interesting to investigate the
geometric aspects of the examples of non-homogeneous very special geometry
discussed by Shmakova in Sec.\ 4 of \cite{Shmakova} (cfr. Refs. therein, as
well), as well as of the non-homogeneous 2-moduli cubic models in which
non-trivial involutory matrices determining multiple attractor solutions
exist \cite{multBPS1, multBPS2}; we leave these tasks for further future
work.

Moreover, some time ago, after the seminal paper of Moore \cite{Moore}, an
intriguing relationship was observed between Gauss's composition laws and
the arithmetics of BPS black holes in string theory and supergravity.
Remarkably, this relationship has been recently extended to include
Bhargava's higher composition laws, closely related to various classes of
black hole solutions appearing in string/M-theory \cite{Borsten-PhD, GKT,
Bhargava}. We leave to further future work the investigation of the
possibility of a further generalization and extension to BPS black holes in
theories with homogeneous non-symmetric scalar manifolds.

Finally, we remark that the stratification of the $U$-duality orbits
supporting extremal black hole attractors is well known for symmetric models
of $\mathcal{N}=2$, $D=4$ Maxwell-Einstein supergravity, since the seminal
paper \cite{FG1} (see e.g. \cite{bfgm,FG2}, and also \cite{Small-Orbits} for
a recollection of results). It woud be of utmost interest to extend the
known results to the infinite classes of homogeneous \textit{non-symmetric}
models, especially to those which cannot be obtained by suitable truncations
of symmetric models. Some advances along such a venue are discussed in \cite%
{Alek}, and we leave the investigation of this issue for future work.

\appendix

\section{\label{App-L(1,2)}Details on $L(1,2)$}

The cubic form $\mathcal{V}_{L(1,2)}$ is given by{%
\begin{equation}
\mathcal{V}_{L(1,2)}\,\left( s,x,y\right)
=\,sq(x)\,+\,x^{0}Q_{0}^{(2)}(y)\,+\,x^{1}Q_{1}^{(2)}(y)\,+%
\,x^{2}Q_{2}^{(2)}(y)~,  \label{V-L120}
\end{equation}%
}where
\begin{equation}
q(x)\,:=\,-\left( x^{0}\right) ^{2}\,+\left( x^{1}\right) ^{2}\,+\,\left(
x^{2}\right) ^{2}~,  \label{q-L120}
\end{equation}%
and the $Q_{I}^{(2)}$'s are given by (\ref{this5}) below.

To write this model explicitly and to show the invertibility of the gradient
map, we first show that $L(1,2)\subset L(2,1)$; to this aim, we start and
observe that both of the models are defined by $4\times 4$ $\Gamma $%
-matrices. Using the quadratic forms from Sec.\ \ref{L210} but substituting $%
y^{3}:=-y^{3}$, one finds that the first three of these forms are the $%
Q_{I}^{(2)}$ with the $Q_{I}$ from the $L(2,1)$ model and the remaining form
is denoted by $R\equiv R_{1}$:
\begin{equation}
{\renewcommand{\arraystretch}{1.3}%
\begin{array}{lllrll}
\Gamma _{0}^{(2)} & = & {\mathbb{I}}_{4}, & Q_{0}^{(2)}\equiv Q[{}_{00}^{00}]
& = & \left( y^{1}\right) ^{2}+\left( y^{2}\right) ^{2}\,+\left(
y^{3}\right) ^{2}+\left( y^{4}\right) ^{2}; \\
\Gamma _{1}^{(2)} & = & \sigma _{3}\otimes {\mathbb{I}}_{2}, &
Q_{1}^{(2)}\equiv Q[{}_{01}^{00}] & = & \left( y^{1}\right) ^{2}-\left(
y^{2}\right) ^{2}\,+\left( y^{3}\right) ^{2}-\left( y^{4}\right) ^{2}; \\
\Gamma _{2}^{(2)} & = & \sigma _{1}\otimes {\mathbb{I}}_{2},\qquad &
Q_{2}^{(2)}\equiv Q[{}_{10}^{01}] & = & 2(y^{1}y^{2}+y^{3}y^{4}); \\
\Omega _{1} & = & \sigma _{1}\otimes \sigma _{1}, & R\equiv Q[{}_{11}^{11}]
& = & 2(y^{1}y^{4}-y^{2}y^{3})~.%
\end{array}%
}  \label{this5}
\end{equation}%
Thus, it holds that:
\begin{eqnarray}
L(1,2)\, &=&\,\left. L(2,1)\right\vert _{x^{3}=0,~y^{3}:=-y^{3}};
\label{L12sub} \\
{\mathcal{V}}_{L(1,2)}\, &=&\,\left. {\mathcal{V}}_{L(2,1)}\right\vert
_{x^{3}=0,~y^{3}:=-y^{3}}.
\end{eqnarray}%
Namely, the model $L(1,2)$ can be regarded as the $L(2,1)$ model with one
linear constraint, upon the renaming $y^{3}:=-y^{3}$.

The model $L(1,2)$ is invertible since the the matrices $\Gamma
_{1}^{(2)},\Gamma _{2}^{(2)},\Omega _{1}$ anti-commute, the first two form a
Clifford set and finally the $Q_{I}^{(2)}$ 's and the $R(y)$ are related by
the following quadratic (Lorentzian) identity :
\begin{equation}
-Q_{0}^{(2)}(y)^{2}\,+\,Q_{1}^{(2)}(y)^{2}\,+\,Q_{2}^{(2)}(y)^{2}\,+%
\,R(y)^{2}\,=\,0~.
\end{equation}%
Therefore, the invertibility condition of Sec.\ \ref{main} is satisfied, and
$\nabla _{{\mathcal{V}}_{L(1,2)}}$ is invertible.

Since $r=1$, the inverse of the gradient map $\nabla _{{\mathcal{V}}%
_{L(1,2)}}$ is given as a composition of two maps $\mathbf{\alpha }$ and $%
\mathbf{\mu }$ (cfr. (\ref{csii})) :%
\begin{equation}
{\mathbb{R}}_{\xi }^{1+3+4}\,\overset{\nabla _{{\mathcal{V}}_{L(1,2)}}}{%
\longrightarrow }\,{\mathbb{R}}_{z}^{1+3+4}\,\overset{\mathbf{\alpha }}{%
\longrightarrow }\,{\mathbb{R}}_{(t,u,v,w)}^{1+3+4+1}\,\overset{\mathbf{\mu }%
}{\longrightarrow }\,{\mathbb{R}}_{\xi }^{1+3+4}~.
\end{equation}%
The map $\mathbf{\alpha }$, as in (\ref{alpha}), is given by
\begin{equation}
\mathbf{\alpha }(z_{1},\ldots
,z_{8})\,:=\,^{T}(z_{1}^{2},\,z_{1}z_{2},\ldots ,\,z_{1}z_{8},\,R(z)),
\end{equation}%
where the quadratic form $R$ depends only on the last $2\cdot 2=4$ variables
:%
\begin{equation}
R(z)=R(z_{5},\ldots ,z_{8})\,=\,2(z_{5}z_{8}-z_{6}z_{7}).
\end{equation}%
The map $\mathbf{\mu }$ (\ref{mu}) has $1+3+4=8$ components that are
homogeneous polynomials of degree $2$ in the variables $t,u_{0},\ldots
,u_{2},v_{1},\ldots ,v_{4},w$, and it is given by
\begin{equation}
\mathbf{\mu }(t,u,v,w)\,:=\,\left( {\renewcommand{\arraystretch}{1.3}%
\begin{array}{rcl}
-u_{0}^{2}+u_{1}^{2}+u_{2}^{2} & + & \mbox{$\frac{1}{16}$}w^{2} \\
-2tu_{0} & + & \mbox{$\frac{1}{2}$}(v_{1}^{2}+v_{2}^{2}+v_{3}^{2}+v_{4}^{2})
\\
2tu_{1} & + & \mbox{$\frac{1}{2}$}(v_{1}^{2}-v_{2}^{2}+v_{3}^{2}-v_{4}^{2})
\\
2tu_{2} & + & v_{1}v_{2}+v_{3}v_{4} \\
u_{0}v_{1}+u_{1}v_{1}+u_{2}v_{2} & - & \mbox{$\frac{1}{4}$}v_{4}w \\
u_{0}v_{2}-u_{1}v_{2}+u_{2}v_{1} & + & \mbox{$\frac{1}{4}$}v_{3}w \\
u_{0}v_{3}+u_{1}v_{3}+u_{2}v_{4} & + & \mbox{$\frac{1}{4}$}v_{2}w \\
u_{0}v_{4}-u_{1}v_{4}+u_{2}v_{3} & - & \mbox{$\frac{1}{4}$}v_{1}w%
\end{array}%
}\right) .~
\end{equation}%
Since
\begin{equation}
\mathbf{\mu }\circ \mathbf{\alpha }\circ \nabla _{{\mathcal{V}}%
_{L(1,2)}}\,(\xi )=\,4q(x)^{2}{\mathcal{V}}_{L(1,2)}(\xi )\mathbb{\xi },
\end{equation}%
where ${\mathcal{V}}_{L(1,2)}$ and the corresponding $q(x)$ are given by (%
\ref{V-L120}) and (\ref{q-L120}), the (birational) inverse of the gradient
map $\nabla _{\mathcal{V}_{L(1,2)}}$ is the map
\begin{equation}
\mathbf{\mu \circ \alpha }\,:\,{\mathbb{R}}_{z}^{1+3+4}\,\overset{\mathbf{%
\alpha }}{\longrightarrow }\,{\mathbb{R}}_{(t,u,v,w)}^{1+3+4+1}\,\overset{%
\mathbf{\mu }}{\longrightarrow }\,{\mathbb{R}}_{\xi }^{1+3+4}~,
\end{equation}%
given by the following homogeneous polynomials of degree four (cfr. (\ref%
{mu-alpha-2})) :%
\begin{equation}
(\mathbf{\mu \circ \alpha })\left( z\right) \,:=\,\left( {%
\renewcommand{\arraystretch}{1.3}%
\begin{array}{l}
z_{1}^{2}\left( -z_{2}^{2}+z_{3}^{2}+z_{4}^{2}\right) +\frac{1}{4}\left(
z_{5}z_{8}-z_{6}z_{7}\right) ^{2} \\
z_{1}^{2}\left[ -2z_{1}z_{2}+\frac{1}{2}\left(
z_{5}^{2}+z_{6}^{2}+z_{7}^{2}+z_{8}^{2}\right) \right] \\
z_{1}^{2}\left[ 2z_{1}z_{3}+\frac{1}{2}\left(
z_{5}^{2}-z_{6}^{2}+z_{7}^{2}-z_{8}^{2}\right) \right] \\
z_{1}^{2}\left( 2z_{1}z_{4}+z_{5}z_{6}+z_{7}z_{8}\right) \\
z_{1}\left[ z_{1}\left( z_{2}z_{5}+z_{3}z_{5}+z_{4}z_{6}\right) -\frac{1}{2}%
\left( z_{5}z_{8}-z_{6}z_{7}\right) z_{8}\right] \\
z_{1}\left[ z_{1}\left( z_{2}z_{6}-z_{3}z_{6}+z_{4}z_{5}\right) +\frac{1}{2}%
\left( z_{5}z_{8}-z_{6}z_{7}\right) z_{7}\right] \\
z_{1}\left[ z_{1}\left( z_{2}z_{7}+z_{3}z_{7}+z_{4}z_{8}\right) +\frac{1}{2}%
\left( z_{5}z_{8}-z_{6}z_{7}\right) z_{6}\right] \\
z_{1}\left[ z_{1}\left( z_{2}z_{8}-z_{3}z_{8}+z_{4}z_{7}\right) -\frac{1}{2}%
\left( z_{5}z_{8}-z_{6}z_{7}\right) z_{5}\right]%
\end{array}%
}\right) ~.  \label{invL12}
\end{equation}

Consequently, from the treatment of Sec.\ \ref{invertgamma}, the full
fledged expression of the solution (\ref{sol}) of the BPS system of $L(1,2)$
is given by ($\Delta _{1}\equiv \Delta _{s}$, and recall the definition (\ref%
{rec0})):%
\begin{eqnarray}
s &=&\frac{3}{2\left\vert \Delta _{s}\right\vert }\frac{\left[ \Delta
_{s}^{2}\left( -\Delta _{2}^{2}+\Delta _{3}^{2}+\Delta _{4}^{2}\right) +%
\frac{1}{4}\left( \Delta _{5}\Delta _{8}-\Delta _{6}\Delta _{7}\right) ^{2}%
\right] }{\sqrt{\left( \partial _{p}\Delta \right) \cdot \left( \mathbf{\mu
\circ \alpha }\right) \left( \partial _{p}\Delta \right) }}; \\
&&  \notag \\
x^{0} &=&\frac{3}{2}\frac{\left\vert \Delta _{s}\right\vert \left[ -2\Delta
_{s}\Delta _{2}+\frac{1}{2}\left( \Delta _{5}^{2}+\Delta _{6}^{2}+\Delta
_{7}^{2}+\Delta _{8}^{2}\right) \right] }{\sqrt{\left( \partial _{p}\Delta
\right) \cdot \left( \mathbf{\mu \circ \alpha }\right) \left( \partial
_{p}\Delta \right) }}; \\
&&  \notag \\
x^{1} &=&\frac{3}{2}\frac{\left\vert \Delta _{s}\right\vert \left[ 2\Delta
_{s}\Delta _{3}+\frac{1}{2}\left( \Delta _{5}^{2}-\Delta _{6}^{2}+\Delta
_{7}^{2}-\Delta _{8}^{2}\right) \right] }{\sqrt{\left( \partial _{p}\Delta
\right) \cdot \left( \mathbf{\mu \circ \alpha }\right) \left( \partial
_{p}\Delta \right) }}; \\
&&  \notag \\
x^{2} &=&\frac{3}{2}\frac{\left\vert \Delta _{s}\right\vert \left( 2\Delta
_{s}\Delta _{4}+\Delta _{5}\Delta _{6}+\Delta _{7}\Delta _{8}\right) }{\sqrt{%
\left( \partial _{p}\Delta \right) \cdot \left( \mathbf{\mu \circ \alpha }%
\right) \left( \partial _{p}\Delta \right) }}; \\
&&  \notag \\
y^{1} &=&\frac{3\text{sgn}\left( \Delta _{s}\right) }{2}\frac{\left[ \Delta
_{s}\left( \Delta _{2}\Delta _{5}+\Delta _{3}\Delta _{5}+\Delta _{4}\Delta
_{6}\right) -\frac{1}{2}\left( \Delta _{5}\Delta _{8}-\Delta _{6}\Delta
_{7}\right) \Delta _{8}\right] }{\sqrt{\left( \partial _{p}\Delta \right)
\cdot \left( \mathbf{\mu \circ \alpha }\right) \left( \partial _{p}\Delta
\right) }}; \\
&&  \notag \\
y^{2} &=&\frac{3\text{sgn}\left( \Delta _{s}\right) }{2}\frac{\left[ \Delta
_{s}\left( \Delta _{2}\Delta _{6}-\Delta _{3}\Delta _{6}+\Delta _{4}\Delta
_{5}\right) +\frac{1}{2}\left( \Delta _{5}\Delta _{8}-\Delta _{6}\Delta
_{7}\right) \Delta _{7}\right] }{\sqrt{\left( \partial _{p}\Delta \right)
\cdot \left( \mathbf{\mu \circ \alpha }\right) \left( \partial _{p}\Delta
\right) }}; \\
&&  \notag \\
y^{3} &=&\frac{3\text{sgn}\left( \Delta _{s}\right) }{2}\frac{\left[ \Delta
_{s}\left( \Delta _{2}\Delta _{7}+\Delta _{3}\Delta _{7}+\Delta _{4}\Delta
_{8}\right) +\frac{1}{2}\left( \Delta _{5}\Delta _{8}-\Delta _{6}\Delta
_{7}\right) \Delta _{6}\right] }{\sqrt{\left( \partial _{p}\Delta \right)
\cdot \left( \mathbf{\mu \circ \alpha }\right) \left( \partial _{p}\Delta
\right) }}; \\
&&  \notag \\
y^{4} &=&\frac{3\text{sgn}\left( \Delta _{s}\right) }{2}\frac{\left[ \Delta
_{s}\left( \Delta _{2}\Delta _{8}-\Delta _{3}\Delta _{8}+\Delta _{4}\Delta
_{7}\right) -\frac{1}{2}\left( \Delta _{5}\Delta _{8}-\Delta _{6}\Delta
_{7}\right) \Delta _{5}\right] }{\sqrt{\left( \partial _{p}\Delta \right)
\cdot \left( \mathbf{\mu \circ \alpha }\right) \left( \partial _{p}\Delta
\right) }},
\end{eqnarray}

where
\begin{eqnarray}
\left( \partial _{p}\Delta \right) \cdot \left( \mathbf{\mu \circ \alpha }%
\right) \left( \partial _{p}\Delta \right) &=&\Delta _{s}^{3}\left( -\Delta
_{2}^{2}+\Delta _{3}^{2}+\Delta _{4}^{2}\right) +\frac{1}{4}\Delta
_{s}\left( \Delta _{5}\Delta _{8}-\Delta _{6}\Delta _{7}\right) ^{2}  \notag
\\
&&+\Delta _{s}^{2}\Delta _{2}\left[ -2\Delta _{s}\Delta _{2}+\frac{1}{2}%
\left( \Delta _{5}^{2}+\Delta _{6}^{2}+\Delta _{7}^{2}+\Delta
_{8}^{2}\right) \right]  \notag \\
&&+\Delta _{s}^{2}\Delta _{3}\left[ 2\Delta _{s}\Delta _{3}+\frac{1}{2}%
\left( \Delta _{5}^{2}-\Delta _{6}^{2}+\Delta _{7}^{2}-\Delta
_{8}^{2}\right) \right]  \notag \\
&&+\Delta _{s}^{2}\Delta _{4}\left( 2\Delta _{s}\Delta _{4}+\Delta
_{5}\Delta _{6}+\Delta _{7}\Delta _{8}\right)  \notag \\
&&+\Delta _{s}\Delta _{5}\left[ \Delta _{s}\left( \Delta _{2}\Delta
_{5}+\Delta _{3}\Delta _{5}+\Delta _{4}\Delta _{6}\right) -\frac{1}{2}\left(
\Delta _{5}\Delta _{8}-\Delta _{6}\Delta _{7}\right) \Delta _{8}\right]
\notag \\
&&+\Delta _{s}\Delta _{6}\left[ \Delta _{s}\left( \Delta _{2}\Delta
_{6}-\Delta _{3}\Delta _{6}+\Delta _{4}\Delta _{5}\right) +\frac{1}{2}\left(
\Delta _{5}\Delta _{8}-\Delta _{6}\Delta _{7}\right) \Delta _{7}\right]
\notag \\
&&+\Delta _{s}\Delta _{7}\left[ \Delta _{s}\left( \Delta _{2}\Delta
_{7}+\Delta _{3}\Delta _{7}+\Delta _{4}\Delta _{8}\right) +\frac{1}{2}\left(
\Delta _{5}\Delta _{8}-\Delta _{6}\Delta _{7}\right) \Delta _{6}\right]
\notag \\
&&+\Delta _{s}\Delta _{8}\left[ \Delta _{s}\left( \Delta _{2}\Delta
_{8}-\Delta _{3}\Delta _{8}+\Delta _{4}\Delta _{7}\right) -\frac{1}{2}\left(
\Delta _{5}\Delta _{8}-\Delta _{6}\Delta _{7}\right) \Delta _{5}\right] .
\label{den1}
\end{eqnarray}%
Then, (\ref{SSd}) and (\ref{zH-1})-(\ref{zH-3}) respectively yield the
corresponding full fledged expression of the BPS black hole entropy and of
the BPS attractors :%
\begin{equation}
\frac{S}{\pi }=\frac{1}{3\left\vert p^{0}\right\vert }\sqrt{3\frac{\left(
\partial _{p}\Delta \right) \cdot \left( \mathbf{\mu \circ \alpha }\right)
\left( \partial _{p}\Delta \right) }{\Delta _{s}^{2}}-9\left[ p^{0}\left(
p\cdot q\right) -2I_{3}(p)\right] ^{2}};
\end{equation}%
\begin{eqnarray}
z_{H}^{1}(\mathcal{Q}) &=&\frac{3}{2}\frac{\left[ \Delta _{s}^{2}\left(
-\Delta _{2}^{2}+\Delta _{3}^{2}+\Delta _{4}^{2}\right) +\frac{1}{4}\left(
\Delta _{5}\Delta _{8}-\Delta _{6}\Delta _{7}\right) ^{2}\right] }{\left(
\partial _{p}\Delta \right) \cdot \left( \mathbf{\mu \circ \alpha }\right)
\left( \partial _{p}\Delta \right) }\left[ \frac{p^{0}\left( p\cdot q\right)
-2I_{3}(p)}{p^{0}}-\mathbf{i}\frac{3}{2}\frac{S}{\pi }\right] +\frac{p^{1}}{%
p^{0}}; \\
&&  \notag \\
z_{H}^{2}(\mathcal{Q}) &=&\frac{3}{2}\frac{\Delta _{s}^{2}\left[ -2\Delta
_{s}\Delta _{2}+\frac{1}{2}\left( \Delta _{5}^{2}+\Delta _{6}^{2}+\Delta
_{7}^{2}+\Delta _{8}^{2}\right) \right] }{\left( \partial _{p}\Delta \right)
\cdot \left( \mathbf{\mu \circ \alpha }\right) \left( \partial _{p}\Delta
\right) }\left[ \frac{p^{0}\left( p\cdot q\right) -2I_{3}(p)}{p^{0}}-\mathbf{%
i}\frac{3}{2}\frac{S}{\pi }\right] +\frac{p^{2}}{p^{0}}; \\
&&  \notag \\
z_{H}^{3}(\mathcal{Q}) &=&\frac{3}{2}\frac{\Delta _{s}^{2}\left[ 2\Delta
_{s}\Delta _{3}+\frac{1}{2}\left( \Delta _{5}^{2}-\Delta _{6}^{2}+\Delta
_{7}^{2}-\Delta _{8}^{2}\right) \right] }{\left( \partial _{p}\Delta \right)
\cdot \left( \mathbf{\mu \circ \alpha }\right) \left( \partial _{p}\Delta
\right) }\left[ \frac{p^{0}\left( p\cdot q\right) -2I_{3}(p)}{p^{0}}-\mathbf{%
i}\frac{3}{2}\frac{S}{\pi }\right] +\frac{p^{3}}{p^{0}}; \\
&&  \notag \\
z_{H}^{4}(\mathcal{Q}) &=&\frac{3}{2}\frac{\Delta _{s}^{2}\left( 2\Delta
_{s}\Delta _{4}+\Delta _{5}\Delta _{6}+\Delta _{7}\Delta _{8}\right) }{%
\left( \partial _{p}\Delta \right) \cdot \left( \mathbf{\mu \circ \alpha }%
\right) \left( \partial _{p}\Delta \right) }\left[ \frac{p^{0}\left( p\cdot
q\right) -2I_{3}(p)}{p^{0}}-\mathbf{i}\frac{3}{2}\frac{S}{\pi }\right] +%
\frac{p^{4}}{p^{0}}; \\
&&  \notag \\
z_{H}^{5}(\mathcal{Q}) &=&\frac{3}{2}\frac{\Delta _{s}\left[ \Delta
_{s}\left( \Delta _{2}\Delta _{5}+\Delta _{3}\Delta _{5}+\Delta _{4}\Delta
_{6}\right) -\frac{1}{2}\left( \Delta _{5}\Delta _{8}-\Delta _{6}\Delta
_{7}\right) \Delta _{8}\right] }{\left( \partial _{p}\Delta \right) \cdot
\left( \mathbf{\mu \circ \alpha }\right) \left( \partial _{p}\Delta \right) }%
\left[ \frac{p^{0}\left( p\cdot q\right) -2I_{3}(p)}{p^{0}}-\mathbf{i}\frac{3%
}{2}\frac{S}{\pi }\right] +\frac{p^{5}}{p^{0}};  \notag \\
&& \\
z_{H}^{6}(\mathcal{Q}) &=&\frac{3}{2}\frac{\Delta _{s}\left[ \Delta
_{s}\left( \Delta _{2}\Delta _{6}-\Delta _{3}\Delta _{6}+\Delta _{4}\Delta
_{5}\right) +\frac{1}{2}\left( \Delta _{5}\Delta _{8}-\Delta _{6}\Delta
_{7}\right) \Delta _{7}\right] }{\left( \partial _{p}\Delta \right) \cdot
\left( \mathbf{\mu \circ \alpha }\right) \left( \partial _{p}\Delta \right) }%
\left[ \frac{p^{0}\left( p\cdot q\right) -2I_{3}(p)}{p^{0}}-\mathbf{i}\frac{3%
}{2}\frac{S}{\pi }\right] +\frac{p^{6}}{p^{0}};  \notag \\
&& \\
z_{H}^{7}(\mathcal{Q}) &=&\frac{3}{2}\frac{\Delta _{s}\left[ \Delta
_{s}\left( \Delta _{2}\Delta _{7}+\Delta _{3}\Delta _{7}+\Delta _{4}\Delta
_{8}\right) +\frac{1}{2}\left( \Delta _{5}\Delta _{8}-\Delta _{6}\Delta
_{7}\right) \Delta _{6}\right] }{\left( \partial _{p}\Delta \right) \cdot
\left( \mathbf{\mu \circ \alpha }\right) \left( \partial _{p}\Delta \right) }%
\left[ \frac{p^{0}\left( p\cdot q\right) -2I_{3}(p)}{p^{0}}-\mathbf{i}\frac{3%
}{2}\frac{S}{\pi }\right] +\frac{p^{7}}{p^{0}};  \notag \\
&& \\
z_{H}^{8}(\mathcal{Q}) &=&\frac{3}{2}\frac{\Delta _{s}\left[ \Delta
_{s}\left( \Delta _{2}\Delta _{8}-\Delta _{3}\Delta _{8}+\Delta _{4}\Delta
_{7}\right) -\frac{1}{2}\left( \Delta _{5}\Delta _{8}-\Delta _{6}\Delta
_{7}\right) \Delta _{5}\right] }{\left( \partial _{p}\Delta \right) \cdot
\left( \mathbf{\mu \circ \alpha }\right) \left( \partial _{p}\Delta \right) }%
\left[ \frac{p^{0}\left( p\cdot q\right) -2I_{3}(p)}{p^{0}}-\mathbf{i}\frac{3%
}{2}\frac{S}{\pi }\right] +\frac{p^{8}}{p^{0}},  \notag \\
&&
\end{eqnarray}%
with $\left( \partial _{p}\Delta \right) \cdot \left( \mathbf{\mu \circ
\alpha }\right) \left( \partial _{p}\Delta \right) $ given by (\ref{den1}),
and
\begin{eqnarray}
p\cdot q &=&p^{0}q_{0}+p^{1}q_{1}+....+p^{8}q_{8}; \\
I_{3}(p) &=&\mathcal{V}_{L(1,2)}\,\left( p^{1},...,p^{8}\right) =p^{1}\left[
-\left( p^{2}\right) ^{2}\,+\,\left( p^{3}\right) ^{2}\,+\,\left(
p^{4}\right) ^{2}\right] \,  \notag \\
&&+\,p^{2}Q_{0}^{(2)}(p^{5},...,p^{8})\,+\,p^{3}Q_{1}^{(2)}(p^{5},...,p^{8})%
\,+\,p^{4}Q_{2}^{(2)}(p^{5},...,p^{8}) \\
&=&p^{1}\left[ -\left( p^{2}\right) ^{2}\,+\,\left( p^{3}\right)
^{2}\,+\,\left( p^{4}\right) ^{2}\right] +p^{2}\left[ \left( p^{5}\right)
^{2}+\left( p^{6}\right) ^{2}\,+\left( p^{7}\right) ^{2}+\left( p^{8}\right)
^{2}\right]  \notag \\
&&+\,p^{3}\left[ \left( p^{5}\right) ^{2}-\left( p^{6}\right) ^{2}\,+\left(
p^{7}\right) ^{2}-\left( p^{8}\right) ^{2}\right]
+2p^{4}(p^{5}p^{6}+p^{7}p^{8}).
\end{eqnarray}

\section{\label{App-L(1,3)}Details on $L(1,3)$}

The cubic form $\mathcal{V}_{L(1,3)}$ reads
\begin{equation}
\mathcal{V\,}_{L(1,3)}=\,sq(x)\,+\,\sum_{I=0}^{2}x^{I}Q_{I}^{(3)}(y),\qquad
\text{with}\quad q(x)\,=\,-\left( x^{0}\right) ^{2}+\left( x^{1}\right)
^{2}+\left( x^{2}\right) ^{2}~,  \label{V-L130}
\end{equation}%
where the $Q_{I}^{(3)}$ are obtained from the $Q_{I}$'s of the $L(1,1)$
model (cfr. (\ref{th110})):
\begin{eqnarray}
Q_{0}^{(3)}\, &=&\,\left( y^{1}\right) ^{2}+\left( y^{2}\right)
^{2}\,+\left( y^{3}\right) ^{2}+\left( y^{4}\right) ^{2}\,+\,\left(
y^{5}\right) ^{2}+\left( y^{6}\right) ^{2}; \\
Q_{1}^{(3)}\, &=&\,\left( y^{1}\right) ^{2}-\left( y^{2}\right)
^{2}\,+\left( y^{3}\right) ^{2}-\left( y^{4}\right) ^{2}\,+\,\left(
y^{5}\right) ^{2}-\left( y^{6}\right) ^{2}; \\
Q_{2}^{(3)}\, &=&\,2(y^{1}y^{2}\,+\,y^{3}y^{4}\,+\,y^{5}y^{6})~.
\end{eqnarray}%
Moreover, we define the quadrics, based on the quadric $R$ of the $L(1,2)$
model (see (\ref{this5})), $R^{(k,l)}(y):=R(y^{(kl)})$:
\begin{equation}
R^{(1,2)}\,=\,2(y^{1}y^{4}\,-\,y^{2}y^{3}),\qquad
R^{(1,3)}\,=\,2(y^{1}y^{6}\,-\,y^{2}y^{5}),\qquad
R^{(2,3)}\,=\,2(y^{3}y^{6}\,-\,y^{4}y^{5})~.
\end{equation}

As required by the invertibility condition of Sec.\ \ref{main} (cfr.\ (\ref%
{lorrel})), a Lorentzian identity of type (\ref{gqr}) holds true, namely,
\begin{equation}
-Q_{0}^{(3)}(y)^{2}\,+\,Q_{1}^{(3)}(y)^{2}\,+\,Q_{2}^{(3)}(y)^{2}\,+%
\,R^{(1,2)}(y)^{2}\,+\,R^{(1,3)}(y)^{2}\,+\,R^{(2,3)}(y)^{2}~=0.
\end{equation}%
Since $r=3$, the inverse of the gradient map $\nabla _{{\mathcal{V}}%
_{L(1,3)}}$ is given as a composition of two maps, namely (cfr. (\ref{csii}%
))
\begin{equation}
{\mathbb{R}}_{\xi }^{1+3+6}\,\overset{\nabla _{{\mathcal{V}}_{L(1,3)}}}{%
\longrightarrow }\,{\mathbb{R}}_{z}^{1+3+6}\,\overset{\mathbf{\alpha }}{%
\longrightarrow }\,{\mathbb{R}}_{(t,u,v,w)}^{1+3+6+3}\,\overset{\mathbf{\mu }%
}{\longrightarrow }\,{\mathbb{R}}_{\xi }^{1+3+6}~,
\end{equation}%
with%
\begin{equation}
\mathbf{\mu }\circ \mathbf{\alpha }\circ \nabla _{{\mathcal{V}}%
_{L(1,3)}}\,(\xi )\,=\,4q(x)^{2}{\mathcal{V}}_{L(1,3)}(\xi )\mathbb{\xi },
\end{equation}%
where $\xi =(s,x,y)$, and ${\mathcal{V}}_{L(1,3)}$ and the corresponding $%
q(x)$ are given by (\ref{V-L130}). The map $\mathbf{\alpha }$ from (\ref%
{alpha}) is given by
\begin{equation}
\mathbf{\alpha }(z_{1},\ldots
,z_{10})\,:=\,^{T}(z_{1}^{2},\,z_{1}z_{2},\ldots
,\,z_{1}z_{10},\,R^{(1,2)}(z),R^{(1,3)}(z),R^{(2,3)}(z)),
\end{equation}%
where each of the three quadratic forms $R^{(k,l)}(z)$'s depends only on
four of the last $3\cdot 2=6$ variables :%
\begin{equation}
R^{(1,2)}(z)=R(z_{5},\ldots ,z_{8}),\quad
R^{(1,3)}(z)=R(z_{5},z_{6},z_{9},z_{10}),\quad R^{(2,3)}(z)=R(z_{7},\ldots
,z_{10}).
\end{equation}%
The map $\mathbf{\mu }$ (\ref{mu}) has $1+3+6$ components which are
homogeneous polynomials of degree $2$ in the variables $%
t,u_{0},u_{1},u_{2},v_{1},\ldots ,v_{6},w_{1},w_{2},w_{3}$, and it is given
by%
\begin{equation}
\mathbf{\mu }\,:\,{\mathbb{R}}_{t,u,v,w}^{1+3+6+3}\,\longrightarrow \,{%
\mathbb{R}}_{\xi }^{1+3+6},
\end{equation}%
\begin{equation}
\mathbf{\mu }(t,u,v,w)\,:=\,\left( \renewcommand{\arraystretch}{1.3}%
\begin{array}{rcl}
-u_{0}^{2}+u_{1}^{2}+u_{2}^{2} & + & \frac{1}{16}%
(w_{1}^{2}+w_{2}^{2}+w_{3}^{2}) \\
-2tu_{0} & + & \frac{1}{2}%
(v_{1}^{2}+v_{2}^{2}+v_{3}^{2}+v_{4}^{2}+v_{5}^{2}+v_{6}^{2}) \\
2tu_{1} & + & \frac{1}{16}%
(v_{1}^{2}-v_{2}^{2}+v_{3}^{2}-v_{4}^{2}+v_{5}^{2}-v_{6}^{2}) \\
2tu_{2} & + & v_{1}v_{2}+v_{3}v_{4}+v_{5}v_{6} \\
u_{0}v_{1}+u_{1}v_{1}+u_{2}v_{2} & + & \frac{1}{4}(-v_{4}w_{1}-v_{6}w_{2})
\\
u_{0}v_{2}-u_{1}v_{2}+u_{2}v_{1} & + & \frac{1}{4}(v_{3}w_{1}+v_{5}w_{2}) \\
u_{0}v_{3}+u_{1}v_{3}+u_{2}v_{4} & + & \frac{1}{4}(v_{2}w_{1}-v_{6}w_{3}) \\
u_{0}v_{4}-u_{1}v_{4}+u_{2}v_{3} & + & \frac{1}{4}(-v_{1}w_{1}-v_{5}w_{3})
\\
u_{0}v_{5}+u_{1}v_{5}+u_{2}v_{6} & + & \frac{1}{4}(v_{2}w_{2}+v_{4}w_{3}) \\
u_{0}v_{6}-u_{1}v_{6}+u_{2}v_{5} & + & \frac{1}{4}(-v_{1}w_{2}-v_{3}w_{3})%
\end{array}%
\right) ~.
\end{equation}

Therefore, the (birational) inverse map $(\nabla _{\mathcal{V}%
_{L(1,3)}})^{-1}=\mathbf{\mu \circ \alpha }$ of the gradient map $\nabla _{%
\mathcal{V}_{L(1,3)}}$ reads (cfr. (\ref{mu-alpha-2}))
\begin{equation}
\mathbf{\mu \circ \alpha }\,:\,{\mathbb{R}}_{z}^{1+3+6}\,\overset{\mathbf{%
\alpha }}{\longrightarrow }\,{\mathbb{R}}_{(t,u,v,w)}^{1+3+6+3}\,\overset{%
\mathbf{\mu }}{\longrightarrow }\,{\mathbb{R}}_{\xi }^{1+3+6}~;
\end{equation}

\

\begin{eqnarray}
(\mathbf{\mu \circ \alpha })\left( z\right) \, &:&=\,\left( {%
\renewcommand{\arraystretch}{1.3}%
\begin{array}{l}
z_{1}^{2}\left( -z_{2}^{2}+z_{3}^{2}+z_{4}^{2}\right) +\frac{1}{4}\left[
\left( z_{5}z_{8}-z_{6}z_{7}\right) ^{2}+\left(
z_{7}z_{10}-z_{8}z_{9}\right) ^{2}\right] \\
z_{1}^{2}\left[ -2z_{1}z_{2}+\frac{1}{2}\left(
z_{5}^{2}+z_{6}^{2}+z_{7}^{2}+z_{8}^{2}+z_{9}^{2}+z_{10}^{2}\right) \right]
\\
z_{1}^{2}\left[ 2z_{1}z_{3}+\frac{1}{2}\left(
z_{5}^{2}-z_{6}^{2}+z_{7}^{2}-z_{8}^{2}+z_{9}^{2}-z_{10}^{2}\right) \right]
\\
z_{1}^{2}\left( 2z_{1}z_{4}+z_{5}z_{6}+z_{7}z_{8}+z_{9}z_{10}\right) \\
z_{1}^{2}\left( z_{2}z_{5}+z_{3}z_{5}+z_{4}z_{6}\right) +\frac{1}{2}%
z_{1}\left(
-z_{5}z_{8}^{2}-z_{5}z_{10}^{2}+z_{6}z_{7}z_{8}+z_{6}z_{9}z_{10}\right) \\
z_{1}^{2}\left( z_{2}z_{6}-z_{3}z_{6}+z_{4}z_{5}\right) +\frac{1}{2}%
z_{1}\left(
z_{5}z_{7}z_{8}+z_{5}z_{9}z_{10}-z_{6}z_{7}^{2}-z_{6}z_{9}^{2}\right) \\
z_{1}^{2}\left( z_{2}z_{7}+z_{3}z_{7}+z_{4}z_{8}\right) +\frac{1}{2}%
z_{1}\left(
z_{5}z_{6}z_{8}-z_{6}^{2}z_{7}-z_{7}z_{10}^{2}+z_{8}z_{9}z_{10}\right) \\
z_{1}^{2}\left( z_{2}z_{8}-z_{3}z_{8}+z_{4}z_{7}\right) +\frac{1}{2}%
z_{1}\left(
-z_{5}^{2}z_{8}+z_{5}z_{6}z_{7}+z_{7}z_{9}z_{10}-z_{8}z_{9}^{2}\right) \\
z_{1}^{2}\left( z_{2}z_{9}+z_{3}z_{9}+z_{4}z_{10}\right) +\frac{1}{2}%
z_{1}\left(
z_{5}z_{6}z_{10}-z_{6}^{2}z_{9}+z_{7}z_{8}z_{10}-z_{8}^{2}z_{9}\right) \\
z_{1}^{2}\left( z_{2}z_{10}-z_{3}z_{10}+z_{4}z_{9}\right) +\frac{1}{2}%
z_{1}\left(
-z_{5}^{2}z_{10}+z_{5}z_{6}z_{9}-z_{7}^{2}z_{10}+z_{7}z_{8}z_{9}\right)%
\end{array}%
}\right) ~.  \notag \\
&&  \label{invL31}
\end{eqnarray}%
Consequently, from the treatment of Sec.\ \ref{invertgamma}, the full
fledged expression of the solution (\ref{sol}) of the BPS system of $L(1,3)$
is given by ($\Delta _{1}\equiv \Delta _{s}$, and recall the definition (\ref%
{rec0})):%
\begin{eqnarray}
s &=&\frac{3}{2\left\vert \Delta _{s}\right\vert }\frac{\left\{ \Delta
_{s}^{2}\left( -\Delta _{2}^{2}+\Delta _{3}^{2}+\Delta _{4}^{2}\right) +%
\frac{1}{4}\left[ \left( \Delta _{5}\Delta _{8}-\Delta _{6}\Delta
_{7}\right) ^{2}+\left( \Delta _{7}\Delta _{10}-\Delta _{8}\Delta
_{9}\right) ^{2}\right] \right\} }{\sqrt{\left( \partial _{p}\Delta \right)
\cdot \left( \mathbf{\mu \circ \alpha }\right) \left( \partial _{p}\Delta
\right) }};  \notag \\
&& \\
x^{0} &=&\frac{3}{2}\frac{\left\vert \Delta _{s}\right\vert \left[ -2\Delta
_{s}\Delta _{2}+\frac{1}{2}\left( \Delta _{5}^{2}+\Delta _{6}^{2}+\Delta
_{7}^{2}+\Delta _{8}^{2}+\Delta _{9}^{2}+\Delta _{10}^{2}\right) \right] }{%
\sqrt{\left( \partial _{p}\Delta \right) \cdot \left( \mathbf{\mu \circ
\alpha }\right) \left( \partial _{p}\Delta \right) }}; \\
&&  \notag \\
x^{1} &=&\frac{3}{2}\frac{\left\vert \Delta _{s}\right\vert \left[ 2\Delta
_{s}\Delta _{3}+\frac{1}{2}\left( \Delta _{5}^{2}-\Delta _{6}^{2}+\Delta
_{7}^{2}-\Delta _{8}^{2}+\Delta _{9}^{2}-\Delta _{10}^{2}\right) \right] }{%
\sqrt{\left( \partial _{p}\Delta \right) \cdot \left( \mathbf{\mu \circ
\alpha }\right) \left( \partial _{p}\Delta \right) }}; \\
&&  \notag \\
x^{2} &=&\frac{3}{2}\frac{\left\vert \Delta _{s}\right\vert \left( 2\Delta
_{s}\Delta _{4}+\Delta _{5}\Delta _{6}+\Delta _{7}\Delta _{8}+\Delta
z_{9}\Delta _{10}\right) }{\sqrt{\left( \partial _{p}\Delta \right) \cdot
\left( \mathbf{\mu \circ \alpha }\right) \left( \partial _{p}\Delta \right) }%
}; \\
&&  \notag \\
y^{1} &=&\frac{3}{2}\frac{\left[ \left\vert \Delta _{s}\right\vert \left(
\Delta _{2}\Delta _{5}+\Delta _{3}\Delta _{5}+\Delta _{4}\Delta _{6}\right) +%
\frac{1}{2}\text{sgn}\left( \Delta _{s}\right) \left( -\Delta _{5}\Delta
_{8}^{2}-\Delta _{5}\Delta _{10}^{2}+\Delta _{6}\Delta _{7}\Delta
_{8}+\Delta _{6}\Delta _{9}\Delta _{10}\right) \right] }{\sqrt{\left(
\partial _{p}\Delta \right) \cdot \left( \mathbf{\mu \circ \alpha }\right)
\left( \partial _{p}\Delta \right) }}; \\
&&  \notag \\
y^{2} &=&\frac{3}{2}\frac{\left[ \left\vert \Delta _{s}\right\vert \left(
\Delta _{2}\Delta _{6}-\Delta _{3}\Delta _{6}+\Delta _{4}\Delta _{5}\right) +%
\frac{1}{2}\text{sgn}\left( \Delta _{s}\right) \left( \Delta _{5}\Delta
_{7}\Delta _{8}+\Delta _{5}\Delta _{9}\Delta _{10}-\Delta _{6}\Delta
_{7}^{2}-\Delta _{6}\Delta _{9}^{2}\right) \right] }{\sqrt{\left( \partial
_{p}\Delta \right) \cdot \left( \mathbf{\mu \circ \alpha }\right) \left(
\partial _{p}\Delta \right) }}; \\
&&  \notag \\
y^{3} &=&\frac{3}{2}\frac{\left[ \left\vert \Delta _{s}\right\vert \left(
\Delta _{2}\Delta _{7}+\Delta _{3}\Delta +\Delta _{4}\Delta _{8}\right) +%
\frac{1}{2}\text{sgn}\left( \Delta _{s}\right) \left( \Delta _{5}\Delta
_{6}\Delta _{8}-\Delta _{6}^{2}\Delta _{7}-\Delta _{7}\Delta
_{10}^{2}+\Delta _{8}\Delta _{9}\Delta _{10}\right) \right] }{\sqrt{\left(
\partial _{p}\Delta \right) \cdot \left( \mathbf{\mu \circ \alpha }\right)
\left( \partial _{p}\Delta \right) }}; \\
&&  \notag \\
y^{4} &=&\frac{3}{2}\frac{\left[ \left\vert \Delta _{s}\right\vert \left(
\Delta _{2}\Delta _{8}-\Delta _{3}\Delta _{8}+\Delta _{4}\Delta _{7}\right) +%
\frac{1}{2}\text{sgn}\left( \Delta _{s}\right) \left( -\Delta _{5}^{2}\Delta
_{8}+\Delta _{5}\Delta _{6}\Delta _{7}+\Delta _{7}\Delta _{9}\Delta
_{10}-\Delta _{8}\Delta _{9}^{2}\right) \right] }{\sqrt{\left( \partial
_{p}\Delta \right) \cdot \left( \mathbf{\mu \circ \alpha }\right) \left(
\partial _{p}\Delta \right) }}; \\
&&  \notag \\
y^{5} &=&\frac{3}{2}\frac{\left[ \left\vert \Delta _{s}\right\vert \left(
\Delta _{2}\Delta _{9}+\Delta _{3}\Delta _{9}+\Delta _{4}\Delta _{10}\right)
+\frac{1}{2}\text{sgn}\left( \Delta _{s}\right) \left( \Delta _{5}\Delta
_{6}\Delta _{10}-\Delta _{6}^{2}\Delta _{9}+\Delta _{7}\Delta _{8}\Delta
_{10}-\Delta _{8}^{2}\Delta _{9}\right) \right] }{\sqrt{\left( \partial
_{p}\Delta \right) \cdot \left( \mathbf{\mu \circ \alpha }\right) \left(
\partial _{p}\Delta \right) }}; \\
&&  \notag \\
y^{6} &=&\frac{3}{2}\frac{\left[ \left\vert \Delta _{s}\right\vert \left(
\Delta _{2}\Delta _{10}-\Delta _{3}\Delta _{10}+\Delta _{4}\Delta
_{9}\right) +\frac{1}{2}\text{sgn}\left( \Delta _{s}\right) \left( -\Delta
_{5}^{2}\Delta _{10}+\Delta _{5}\Delta _{6}\Delta _{9}-\Delta _{7}^{2}\Delta
_{10}+\Delta _{7}\Delta _{8}\Delta _{9}\right) \right] }{\sqrt{\left(
\partial _{p}\Delta \right) \cdot \left( \mathbf{\mu \circ \alpha }\right)
\left( \partial _{p}\Delta \right) }},  \notag \\
&&
\end{eqnarray}%
where
\begin{eqnarray}
\left( \partial _{p}\Delta \right) \cdot \left( \mathbf{\mu \circ \alpha }%
\right) \left( \partial _{p}\Delta \right) &=&\Delta _{s}^{3}\left( -\Delta
_{2}^{2}+\Delta _{3}^{2}+\Delta _{4}^{2}\right)  \notag \\
&&+\frac{1}{4}\Delta _{s}\left[ \left( \Delta _{5}\Delta _{8}-\Delta
_{6}\Delta _{7}\right) ^{2}+\left( \Delta _{7}\Delta _{10}-\Delta _{8}\Delta
_{9}\right) ^{2}\right]  \notag \\
&&+\Delta _{s}^{2}\Delta _{2}\left[ -2\Delta _{s}\Delta _{2}+\frac{1}{2}%
\left( \Delta _{5}^{2}+\Delta _{6}^{2}+\Delta _{7}^{2}+\Delta
_{8}^{2}+\Delta _{9}^{2}+\Delta _{10}^{2}\right) \right]  \notag \\
&&+\Delta _{s}^{2}\Delta _{3}\left[ 2\Delta _{s}\Delta _{3}+\frac{1}{2}%
\left( \Delta _{5}^{2}-\Delta _{6}^{2}+\Delta _{7}^{2}-\Delta
_{8}^{2}+\Delta _{9}^{2}-\Delta _{10}^{2}\right) \right]  \notag \\
&&+\Delta _{s}^{2}\Delta _{4}\left( 2\Delta _{s}\Delta _{4}+\Delta
_{5}\Delta _{6}+\Delta _{7}\Delta _{8}+\Delta z_{9}\Delta _{10}\right)
\notag \\
&&+\Delta _{s}^{2}\Delta _{5}\left( \Delta _{2}\Delta _{5}+\Delta _{3}\Delta
_{5}+\Delta _{4}\Delta _{6}\right)  \notag \\
&&+\frac{1}{2}\Delta _{s}\Delta _{5}\left( -\Delta _{5}\Delta
_{8}^{2}-\Delta _{5}\Delta _{10}^{2}+\Delta _{6}\Delta _{7}\Delta
_{8}+\Delta _{6}\Delta _{9}\Delta _{10}\right)  \notag \\
&&+\Delta _{s}^{2}\Delta _{6}\left( \Delta _{2}\Delta _{6}-\Delta _{3}\Delta
_{6}+\Delta _{4}\Delta _{5}\right)  \notag \\
&&+\frac{1}{2}\Delta _{s}\Delta _{6}\left( \Delta _{5}\Delta _{7}\Delta
_{8}+\Delta _{5}\Delta _{9}\Delta _{10}-\Delta _{6}\Delta _{7}^{2}-\Delta
_{6}\Delta _{9}^{2}\right)  \notag \\
&&+\Delta _{s}^{2}\Delta _{7}\left( \Delta _{2}\Delta _{7}+\Delta _{3}\Delta
+\Delta _{4}\Delta _{8}\right)  \notag \\
&&+\frac{1}{2}\Delta _{s}\Delta _{7}\left( \Delta _{5}\Delta _{6}\Delta
_{8}-\Delta _{6}^{2}\Delta _{7}-\Delta _{7}\Delta _{10}^{2}+\Delta
_{8}\Delta _{9}\Delta _{10}\right)  \notag \\
&&+\Delta _{s}^{2}\Delta _{8}\left( \Delta _{2}\Delta _{8}-\Delta _{3}\Delta
_{8}+\Delta _{4}\Delta _{7}\right)  \notag \\
&&+\frac{1}{2}\Delta _{s}\Delta _{8}\left( -\Delta _{5}^{2}\Delta
_{8}+\Delta _{5}\Delta _{6}\Delta _{7}+\Delta _{7}\Delta _{9}\Delta
_{10}-\Delta _{8}\Delta _{9}^{2}\right)  \notag \\
&&+\Delta _{s}^{2}\Delta _{9}\left( \Delta _{2}\Delta _{9}+\Delta _{3}\Delta
_{9}+\Delta _{4}\Delta _{10}\right)  \notag \\
&&+\frac{1}{2}\Delta _{s}\Delta _{9}\left( \Delta _{5}\Delta _{6}\Delta
_{10}-\Delta _{6}^{2}\Delta _{9}+\Delta _{7}\Delta _{8}\Delta _{10}-\Delta
_{8}^{2}\Delta _{9}\right)  \notag \\
&&+\Delta _{s}^{2}\Delta _{10}\left( \Delta _{2}\Delta _{10}-\Delta
_{3}\Delta _{10}+\Delta _{4}\Delta _{9}\right)  \notag \\
&&+\frac{1}{2}\Delta _{s}\Delta _{10}\left( -\Delta _{5}^{2}\Delta
_{10}+\Delta _{5}\Delta _{6}\Delta _{9}-\Delta _{7}^{2}\Delta _{10}+\Delta
_{7}\Delta _{8}\Delta _{9}\right) .  \label{den2}
\end{eqnarray}%
Then, (\ref{SSd}) and (\ref{zH-1})-(\ref{zH-3}) respectively yield the
corresponding full fledged expression of the BPS black hole entropy and of
the BPS attractors :%
\begin{equation}
\frac{S}{\pi }=\frac{1}{3\left\vert p^{0}\right\vert }\sqrt{3\frac{\left(
\partial _{p}\Delta \right) \cdot \left( \mathbf{\mu \circ \alpha }\right)
\left( \partial _{p}\Delta \right) }{\Delta _{s}^{2}}-9\left[ p^{0}\left(
p\cdot q\right) -2I_{3}(p)\right] ^{2}};
\end{equation}%
\begin{eqnarray}
z_{H}^{1}(\mathcal{Q}) &=&\frac{3}{2}\frac{\left\{ \Delta _{s}^{2}\left(
-\Delta _{2}^{2}+\Delta _{3}^{2}+\Delta _{4}^{2}\right) +\frac{1}{4}\left[
\left( \Delta _{5}\Delta _{8}-\Delta _{6}\Delta _{7}\right) ^{2}+\left(
\Delta _{7}\Delta _{10}-\Delta _{8}\Delta _{9}\right) ^{2}\right] \right\} }{%
\left( \partial _{p}\Delta \right) \cdot \left( \mathbf{\mu \circ \alpha }%
\right) \left( \partial _{p}\Delta \right) }\cdot  \notag \\
&&\cdot \left[ \frac{p^{0}\left( p\cdot q\right) -2I_{3}(p)}{p^{0}}-\mathbf{i%
}\frac{3}{2}\frac{S}{\pi }\right] +\frac{p^{1}}{p^{0}}; \\
z_{H}^{2}(\mathcal{Q}) &=&\frac{3}{2}\frac{\Delta _{s}^{2}\left[ -2\Delta
_{s}\Delta _{2}+\frac{1}{2}\left( \Delta _{5}^{2}+\Delta _{6}^{2}+\Delta
_{7}^{2}+\Delta _{8}^{2}+\Delta _{9}^{2}+\Delta _{10}^{2}\right) \right] }{%
\left( \partial _{p}\Delta \right) \cdot \left( \mathbf{\mu \circ \alpha }%
\right) \left( \partial _{p}\Delta \right) }\left[ \frac{p^{0}\left( p\cdot
q\right) -2I_{3}(p)}{p^{0}}-\mathbf{i}\frac{3}{2}\frac{S}{\pi }\right] +%
\frac{p^{2}}{p^{0}};  \notag \\
&& \\
z_{H}^{3}(\mathcal{Q}) &=&\frac{3}{2}\frac{\Delta _{s}^{2}\left[ 2\Delta
_{s}\Delta _{3}+\frac{1}{2}\left( \Delta _{5}^{2}-\Delta _{6}^{2}+\Delta
_{7}^{2}-\Delta _{8}^{2}+\Delta _{9}^{2}-\Delta _{10}^{2}\right) \right] }{%
\left( \partial _{p}\Delta \right) \cdot \left( \mathbf{\mu \circ \alpha }%
\right) \left( \partial _{p}\Delta \right) }\left[ \frac{p^{0}\left( p\cdot
q\right) -2I_{3}(p)}{p^{0}}-\mathbf{i}\frac{3}{2}\frac{S}{\pi }\right] +%
\frac{p^{3}}{p^{0}};  \notag \\
&& \\
z_{H}^{4}(\mathcal{Q}) &=&\frac{3}{2}\frac{\Delta _{s}^{2}\left( 2\Delta
_{s}\Delta _{4}+\Delta _{5}\Delta _{6}+\Delta _{7}\Delta _{8}+\Delta
z_{9}\Delta _{10}\right) }{\left( \partial _{p}\Delta \right) \cdot \left(
\mathbf{\mu \circ \alpha }\right) \left( \partial _{p}\Delta \right) }\left[
\frac{p^{0}\left( p\cdot q\right) -2I_{3}(p)}{p^{0}}-\mathbf{i}\frac{3}{2}%
\frac{S}{\pi }\right] +\frac{p^{4}}{p^{0}}; \\
z_{H}^{5}(\mathcal{Q}) &=&\frac{3}{2}\frac{\left[ \Delta _{s}^{2}\left(
\Delta _{2}\Delta _{5}+\Delta _{3}\Delta _{5}+\Delta _{4}\Delta _{6}\right) +%
\frac{1}{2}\Delta _{s}\left( -\Delta _{5}\Delta _{8}^{2}-\Delta _{5}\Delta
_{10}^{2}+\Delta _{6}\Delta _{7}\Delta _{8}+\Delta _{6}\Delta _{9}\Delta
_{10}\right) \right] }{\left( \partial _{p}\Delta \right) \cdot \left(
\mathbf{\mu \circ \alpha }\right) \left( \partial _{p}\Delta \right) }\cdot
\notag \\
&&\cdot \left[ \frac{p^{0}\left( p\cdot q\right) -2I_{3}(p)}{p^{0}}-\mathbf{i%
}\frac{3}{2}\frac{S}{\pi }\right] +\frac{p^{5}}{p^{0}}; \\
z_{H}^{6}(\mathcal{Q}) &=&\frac{3}{2}\frac{\left[ \Delta _{s}^{2}\left(
\Delta _{2}\Delta _{6}-\Delta _{3}\Delta _{6}+\Delta _{4}\Delta _{5}\right) +%
\frac{1}{2}\Delta _{s}\left( \Delta _{5}\Delta _{7}\Delta _{8}+\Delta
_{5}\Delta _{9}\Delta _{10}-\Delta _{6}\Delta _{7}^{2}-\Delta _{6}\Delta
_{9}^{2}\right) \right] }{\left( \partial _{p}\Delta \right) \cdot \left(
\mathbf{\mu \circ \alpha }\right) \left( \partial _{p}\Delta \right) }\cdot
\notag \\
&&\cdot \left[ \frac{p^{0}\left( p\cdot q\right) -2I_{3}(p)}{p^{0}}-\mathbf{i%
}\frac{3}{2}\frac{S}{\pi }\right] +\frac{p^{6}}{p^{0}}; \\
z_{H}^{7}(\mathcal{Q}) &=&\frac{3}{2}\frac{\left[ \Delta _{s}^{2}\left(
\Delta _{2}\Delta _{7}+\Delta _{3}\Delta +\Delta _{4}\Delta _{8}\right) +%
\frac{1}{2}\Delta _{s}\left( \Delta _{5}\Delta _{6}\Delta _{8}-\Delta
_{6}^{2}\Delta _{7}-\Delta _{7}\Delta _{10}^{2}+\Delta _{8}\Delta _{9}\Delta
_{10}\right) \right] }{\left( \partial _{p}\Delta \right) \cdot \left(
\mathbf{\mu \circ \alpha }\right) \left( \partial _{p}\Delta \right) }\cdot
\notag \\
&&\cdot \left[ \frac{p^{0}\left( p\cdot q\right) -2I_{3}(p)}{p^{0}}-\mathbf{i%
}\frac{3}{2}\frac{S}{\pi }\right] +\frac{p^{7}}{p^{0}}; \\
z_{H}^{8}(\mathcal{Q}) &=&\frac{3}{2}\frac{\left[ \Delta _{s}^{2}\left(
\Delta _{2}\Delta _{8}-\Delta _{3}\Delta _{8}+\Delta _{4}\Delta _{7}\right) +%
\frac{1}{2}\Delta _{s}\left( -\Delta _{5}^{2}\Delta _{8}+\Delta _{5}\Delta
_{6}\Delta _{7}+\Delta _{7}\Delta _{9}\Delta _{10}-\Delta _{8}\Delta
_{9}^{2}\right) \right] }{\left( \partial _{p}\Delta \right) \cdot \left(
\mathbf{\mu \circ \alpha }\right) \left( \partial _{p}\Delta \right) }\cdot
\notag \\
&&\cdot \left[ \frac{p^{0}\left( p\cdot q\right) -2I_{3}(p)}{p^{0}}-\mathbf{i%
}\frac{3}{2}\frac{S}{\pi }\right] +\frac{p^{8}}{p^{0}}; \\
z_{H}^{9}(\mathcal{Q}) &=&\frac{3}{2}\frac{\left[ \Delta _{s}^{2}\left(
\Delta _{2}\Delta _{9}+\Delta _{3}\Delta _{9}+\Delta _{4}\Delta _{10}\right)
+\frac{1}{2}\Delta _{s}\left( \Delta _{5}\Delta _{6}\Delta _{10}-\Delta
_{6}^{2}\Delta _{9}+\Delta _{7}\Delta _{8}\Delta _{10}-\Delta _{8}^{2}\Delta
_{9}\right) \right] }{\left( \partial _{p}\Delta \right) \cdot \left(
\mathbf{\mu \circ \alpha }\right) \left( \partial _{p}\Delta \right) }\cdot
\notag \\
&&\cdot \left[ \frac{p^{0}\left( p\cdot q\right) -2I_{3}(p)}{p^{0}}-\mathbf{i%
}\frac{3}{2}\frac{S}{\pi }\right] +\frac{p^{9}}{p^{0}}; \\
z_{H}^{10}(\mathcal{Q}) &=&\frac{3}{2}\frac{\left[ \Delta _{s}^{2}\left(
\Delta _{2}\Delta _{10}-\Delta _{3}\Delta _{10}+\Delta _{4}\Delta
_{9}\right) +\frac{1}{2}\Delta _{s}\left( -\Delta _{5}^{2}\Delta
_{10}+\Delta _{5}\Delta _{6}\Delta _{9}-\Delta _{7}^{2}\Delta _{10}+\Delta
_{7}\Delta _{8}\Delta _{9}\right) \right] }{\left( \partial _{p}\Delta
\right) \cdot \left( \mathbf{\mu \circ \alpha }\right) \left( \partial
_{p}\Delta \right) }\cdot  \notag \\
&&\cdot \left[ \frac{p^{0}\left( p\cdot q\right) -2I_{3}(p)}{p^{0}}-\mathbf{i%
}\frac{3}{2}\frac{S}{\pi }\right] +\frac{p^{10}}{p^{0}},
\end{eqnarray}%
with $\left( \partial _{p}\Delta \right) \cdot \left( \mathbf{\mu \circ
\alpha }\right) \left( \partial _{p}\Delta \right) $ given by (\ref{den2}),
and
\begin{eqnarray}
p\cdot q &=&p^{0}q_{0}+p^{1}q_{1}+....+p^{10}q_{10}; \\
I_{3}(p) &=&\mathcal{V}_{L(1,3)}\,\left( p^{1},...,p^{10}\right) =\,p^{1}%
\left[ -\left( p^{2}\right) ^{2}+\left( p^{3}\right) ^{2}+\left(
p^{4}\right) ^{2}\right] +p^{2}Q_{0}^{(3)}(p^{5},...,p^{10})\,  \notag \\
&&+\,p^{3}Q_{1}^{(3)}(p^{5},...,p^{10})\,+%
\,p^{4}Q_{2}^{(2)}(p^{5},...,p^{10})\, \\
&=&p^{1}\left[ -\left( p^{2}\right) ^{2}\,+\,\left( p^{3}\right)
^{2}\,+\,\left( p^{4}\right) ^{2}\right] +p^{2}\left[ \left( p^{5}\right)
^{2}+\left( p^{6}\right) ^{2}\,+\left( p^{7}\right) ^{2}+\left( p^{8}\right)
^{2}+\left( p^{9}\right) ^{2}+\left( p^{10}\right) ^{2}\right]  \notag \\
&&+\,p^{3}\left[ \left( p^{5}\right) ^{2}-\left( p^{6}\right) ^{2}\,+\left(
p^{7}\right) ^{2}-\left( p^{8}\right) ^{2}+\left( p^{9}\right) ^{2}-\left(
p^{10}\right) ^{2}\right] +2p^{4}(p^{5}p^{6}+p^{7}p^{8}+p^{9}p^{10}).
\end{eqnarray}

\section{\label{App-Lie}A calculation in Lie theory}

With no loss of generality, one can consider Lie algebras on $\mathbb{C}$,
and classify all inequivalent (up to $\mathfrak{d}_{5}$ inner automorphisms)
algebras $\mathfrak{a}_{1}\oplus \mathfrak{b}_{2}$ in $\mathfrak{d}_{5}$,
then considering also the branching of the $\mathbf{10}\oplus \mathbf{16}$
of $\mathfrak{d}_{5}$ in irreprs.\ of $\mathfrak{a}_{1}\oplus \mathfrak{b}%
_{2} $:

\begin{eqnarray}
&&%
\begin{array}{l}
\begin{array}{l}
I:\mathfrak{d}_{5}\rightarrow \underset{\text{symmetric~in~}I,II}{\mathfrak{a%
}_{1,I}\oplus \mathfrak{a}_{1,II}\oplus \mathfrak{a}_{3}}\rightarrow
\mathfrak{a}_{1,I}\oplus \mathfrak{a}_{1,II}\oplus \mathfrak{b}_{2} \\
\begin{array}{l}
\mathbf{45}=(\mathbf{3},\mathbf{1},\mathbf{1})\oplus (\mathbf{1},\mathbf{3},%
\mathbf{1})\oplus (\mathbf{1},\mathbf{1},\mathbf{15})\oplus (\mathbf{2},%
\mathbf{2},\mathbf{6}) \\
=(\mathbf{3},\mathbf{1},\mathbf{1})\oplus (\mathbf{1},\mathbf{3},\mathbf{1}%
)\oplus (\mathbf{1},\mathbf{1},\mathbf{10})\oplus (\mathbf{1},\mathbf{1},%
\mathbf{5})\oplus (\mathbf{2},\mathbf{2},\mathbf{5})\oplus (\mathbf{2},%
\mathbf{2},\mathbf{1});%
\end{array}
\\
\mathbf{10}=(\mathbf{2},\mathbf{2},\mathbf{1})\oplus (\mathbf{1},\mathbf{1},%
\mathbf{6})=(\mathbf{2},\mathbf{2},\mathbf{1})\oplus (\mathbf{1},\mathbf{1},%
\mathbf{5})\oplus (\mathbf{1},\mathbf{1},\mathbf{1}); \\
\mathbf{16}=(\mathbf{2},\mathbf{1},\mathbf{4})\oplus (\mathbf{1},\mathbf{2},%
\overline{\mathbf{4}})=(\mathbf{2},\mathbf{1},\mathbf{4})\oplus (\mathbf{1},%
\mathbf{2},\mathbf{4});%
\end{array}
\\
~ \\
\begin{array}{l}
II:\mathfrak{d}_{5}\rightarrow \mathfrak{d}_{4}\oplus T_{1}\rightarrow
\mathfrak{b}_{2}\oplus \mathfrak{a}_{1}\oplus T_{1} \\
\begin{array}{l}
\mathbf{45}=\mathbf{28}_{0}\oplus \mathbf{1}_{0}\oplus \mathbf{8}%
_{v,2}\oplus \mathbf{8}_{v,-2} \\
=(\mathbf{10},\mathbf{1})_{0}\oplus (\mathbf{1},\mathbf{3})_{0}\oplus (%
\mathbf{5},\mathbf{3})_{0}\oplus (\mathbf{1},\mathbf{1})_{0}\oplus (\mathbf{5%
},\mathbf{1})_{2}\oplus (\mathbf{1},\mathbf{3})_{2}\oplus (\mathbf{5},%
\mathbf{1})_{-2}\oplus (\mathbf{1},\mathbf{3})_{-2};%
\end{array}%
\end{array}
\\
~ \\
\begin{array}{l}
III:\mathfrak{d}_{5}\rightarrow \mathfrak{b}_{4}\rightarrow \mathfrak{d}%
_{4}\rightarrow \mathfrak{a}_{1}\oplus \mathfrak{b}_{2}; \\
\mathbf{45}=\mathbf{36}\oplus \mathbf{9}=\mathbf{28}\oplus \mathbf{8}%
_{v}\oplus \mathbf{8}_{v}\oplus \mathbf{1}=(\mathbf{3},\mathbf{1})\oplus (%
\mathbf{1},\mathbf{10})\oplus (\mathbf{3},\mathbf{5})\oplus (\mathbf{3},%
\mathbf{1})\oplus (\mathbf{1},\mathbf{5});%
\end{array}
\\
~ \\
\begin{array}{l}
IV:\mathfrak{d}_{5}\rightarrow \mathfrak{b}_{4}\rightarrow \underset{\text{%
symmetric~in~}I,II}{\mathfrak{a}_{1,I}\oplus \mathfrak{a}_{1,II}\oplus
\mathfrak{b}_{2}}; \\
\mathbf{45}=\mathbf{36}\oplus \mathbf{9}=\left( \mathbf{3,1,1}\right) \oplus
\left( \mathbf{1,3,1}\right) \oplus \left( \mathbf{1,1,10}\right) \oplus
\left( \mathbf{2,2,5}\right) \oplus \left( \mathbf{2,2,1}\right) \oplus
\left( \mathbf{1,1,5}\right) ; \\
\mathbf{10}=\mathbf{9}\oplus \mathbf{1}=\left( \mathbf{2,2,1}\right) \oplus
\left( \mathbf{1,1,5}\right) \oplus \left( \mathbf{1,1,1}\right) ; \\
\mathbf{16}=\mathbf{16}=\left( \mathbf{2,1,4}\right) \oplus \left( \mathbf{%
1,2,4}\right) ;%
\end{array}
\\
~ \\
\begin{array}{l}
V:\mathfrak{d}_{5}\rightarrow \mathfrak{b}_{4}\rightarrow \mathfrak{a}%
_{1}\oplus \mathfrak{a}_{3}\rightarrow \mathfrak{a}_{1}\oplus \mathfrak{b}%
_{2}; \\
\begin{array}{l}
\mathbf{45}=\mathbf{36}\oplus \mathbf{9}=\left( \mathbf{3,1}\right) \oplus
\left( \mathbf{1,15}\right) \oplus \left( \mathbf{3,6}\right) \oplus \left(
\mathbf{3,1}\right) \oplus \left( \mathbf{1,6}\right) \\
=\left( \mathbf{3,1}\right) \oplus \left( \mathbf{1,10}\right) \oplus \left(
\mathbf{1,5}\right) \oplus \left( \mathbf{3,5}\right) \oplus \left( \mathbf{%
3,1}\right) \oplus \left( \mathbf{3,1}\right) \oplus \left( \mathbf{1,5}%
\right) \oplus \left( \mathbf{1,1}\right) ;%
\end{array}%
\end{array}
\\
~ \\
\begin{array}{l}
VI:\mathfrak{d}_{5}\rightarrow \mathfrak{b}_{3}\oplus \mathfrak{a}%
_{I}\rightarrow \mathfrak{a}_{3}\oplus \mathfrak{a}_{1}\rightarrow \mathfrak{%
b}_{2}\oplus \mathfrak{a}_{1}; \\
\begin{array}{l}
\mathbf{45}=(\mathbf{21,1)}\oplus (\mathbf{1,3)}\oplus (\mathbf{7,3)}=(%
\mathbf{15,1)}\oplus (\mathbf{6,1)}\oplus (\mathbf{1,3)}\oplus (\mathbf{%
6,3)\oplus (\mathbf{6,1)}} \\
=(\mathbf{10,1)}\oplus (\mathbf{5,1)}\oplus (\mathbf{5,1)}\oplus (\mathbf{%
1,1)}\oplus (\mathbf{1,3)}\oplus (\mathbf{5,3)\oplus (1\mathbf{,3)}\oplus (5%
\mathbf{,1)}\oplus (1\mathbf{,1);}}%
\end{array}%
\end{array}
\\
~ \\
\begin{array}{l}
VII:\mathfrak{d}_{5}\rightarrow \mathfrak{b}_{3}\oplus \mathfrak{a}%
_{I}\rightarrow \mathfrak{b}_{2}\oplus \mathfrak{a}_{1}\oplus T_{1}; \\
\begin{array}{l}
\mathbf{45}=(\mathbf{21,1)}\oplus (\mathbf{1,3)}\oplus (\mathbf{7,3)} \\
=(\mathbf{10,1)}_{0}\oplus (\mathbf{1,1)}_{0}\oplus (\mathbf{5,1)}_{2}\oplus
(\mathbf{5,1)}_{-2}\oplus (\mathbf{1,3)}_{0}\oplus (\mathbf{5,3)}_{0}\oplus (%
\mathbf{1,3)}_{2}\oplus (\mathbf{1,3)}_{-2}\mathbf{\mathbf{;}}%
\end{array}%
\end{array}
\\
~ \\
\begin{array}{l}
VIII:\mathfrak{d}_{5}\rightarrow \underset{\text{symmetric~in~}I,II}{%
\mathfrak{b}_{2,I}\oplus \mathfrak{b}_{2,II}}\rightarrow \mathfrak{b}%
_{2,I}\oplus \mathfrak{a}_{1}\oplus \mathfrak{a}_{1}; \\
\mathbf{45}=\left( \mathbf{10,1}\right) \oplus \left( \mathbf{1,10}\right)
\oplus \left( \mathbf{5,5}\right) =\left( \mathbf{10,1,1}\right) \oplus
\left( \mathbf{1,3,1}\right) \oplus \left( \mathbf{1,1,3}\right) \oplus
\left( \mathbf{1,2,2}\right) \oplus \left( \mathbf{5,2,2}\right) \oplus
\left( \mathbf{5,1,1}\right) ; \\
\mathbf{10}=\left( \mathbf{5,1}\right) \oplus \left( \mathbf{1,5}\right)
=\left( \mathbf{5,1,1}\right) \oplus \left( \mathbf{1,2,2}\right) \oplus
\left( \mathbf{1,1,1}\right) ; \\
\mathbf{16}=\left( \mathbf{4,4}\right) =\left( \mathbf{4,2,1}\right) \oplus
\left( \mathbf{4,1,2}\right) ;%
\end{array}
\\
~ \\
\begin{array}{l}
IX:\mathfrak{d}_{5}\rightarrow \underset{\text{symmetric~in~}I,II}{\mathfrak{%
b}_{2,I}\oplus \mathfrak{b}_{2,II}}\rightarrow \mathfrak{b}_{2,I}\oplus
\mathfrak{a}_{1}\oplus T_{1}; \\
\mathbf{45}=\left( \mathbf{10,1}\right) \oplus \left( \mathbf{1,10}\right)
\oplus \left( \mathbf{5,5}\right) \\
=\left( \mathbf{10,1}\right) _{0}\oplus \left( \mathbf{1,3}\right)
_{0}\oplus \left( \mathbf{1,1}\right) _{0}\oplus \left( \mathbf{1,3}\right)
_{2}\oplus \left( \mathbf{1,3}\right) _{-2}\oplus \left( \mathbf{5,3}\right)
\oplus \left( \mathbf{5,1}\right) _{2}\oplus \left( \mathbf{5,1}\right)
_{-2};%
\end{array}
\\
~ \\
\begin{array}{l}
X:\mathfrak{d}_{5}\rightarrow \underset{\text{symmetric~in~}I,II}{\mathfrak{b%
}_{2,I}\oplus \mathfrak{b}_{2,II}}\rightarrow \mathfrak{b}_{2,I}\oplus
\mathfrak{a}_{1}; \\
\mathbf{45}=\left( \mathbf{10,1}\right) \oplus \left( \mathbf{1,10}\right)
\oplus \left( \mathbf{5,5}\right) =\left( \mathbf{10,1}\right) \oplus \left(
\mathbf{1,3}\right) \oplus \left( \mathbf{1,7}\right) \oplus \left( \mathbf{%
5,5}\right) .%
\end{array}%
\end{array}
\notag \\
&&
\end{eqnarray}

It thus follows that the cases\footnote{%
Some chains (such as $I$ and $IV$) of embeddings yield to the same $%
\mathfrak{b}_{2}\oplus \mathfrak{a}_{1}$ subalgebra of $\mathfrak{d}_{5}$.} $%
I$-$X$ do not satisfy \textit{at least} one of the aforementioned conditions
\textbf{i)} and \textbf{ii)} of Sec. \ref{Des-Sub} for $L(1,5)$ to be a
\textit{submodel} of $L(8,1)$. This proves that%
\begin{equation}
\begin{array}{l}
T_{3}(1,5,0)\nsubseteq J_{3}^{\mathbb{O}}; \\
\left( SO(2,1)\otimes SO(5)\otimes SO(1,1)\right) _{0}\ltimes \left( \mathbf{%
2},\mathbf{5}\right) _{3/2}\nsubseteq SO(9,1)\otimes SO(1,1)\subsetneq
E_{6(-26)},%
\end{array}%
\end{equation}%
and thus that the model $L(1,5)$ is \textit{not} a submodel of $L(8,1)$, but
rather only one of its \textit{descendants}.


\begin{thebibliography}{KLOPVP}
\bibitem[ABCDF]{N=2-Big} L. Andrianopoli, M. Bertolini, A. Ceresole, R.
D'Auria, S. Ferrara, $\mathcal{N}\mathit{=2}$\textit{\ supergravity and }$%
\mathcal{N}\mathit{=2}$\textit{\ superYang-Mills theory on general scalar
manifolds: Symplectic covariance, gaugings and the momentum map}, J. Geom.
Phys. \textbf{23} (1997) 111-189, \texttt{hep-th/9605032}.

\bibitem[ADF]{Andrianopoli} L. Andrianopoli, R. D'Auria, S. Ferrara, \textit{%
Flat symplectic bundles of }$\mathcal{N}$\textit{\ extended supergravities,
central charges and black hole entropy, }Lectures given at the APCTP Winter
School on Dualities of Gauge and String Theories, 17-28 February 1997, Seoul
and Sokcho, Korea, \texttt{hep-th/9707203 [hep-th]}.

\bibitem[ADF2]{uinvar} L.~Andrianopoli, R.~D'Auria, S.~Ferrara, \textit{%
U-invariants, black-hole entropy and fixed scalars}, Phys.\ Lett.\ \textbf{%
B403} (1997) 12, \texttt{hep-th/9703156}.

\bibitem[ADFL]{F-Flat-Gauging} L. Andrianopoli, R. D'Auria, S. Ferrara, M.A.
Lled\'{o}, \textit{Gauging of flat groups in four-dimensional supergravity},
JHEP \textbf{0207} (2002) 010, \texttt{hep-th/0203206}.

\bibitem[ADFM]{rev-Ferrara} L. Andrianopoli, R. D'Auria, S. Ferrara, M.
Trigiante, \textit{Extremal black holes in supergravity}, Lect. Notes Phys.
\textbf{737} (2008) 661, \texttt{hep-th/0611345}.

\bibitem[ADFT]{H2-1} L. Andrianopoli, R. D'Auria, S. Ferrara, M. Trigiante,
\textit{Black hole attractors in }$\mathcal{N}\mathit{=1}$\textit{\
supergravity}, JHEP \textbf{07} (2007) 019, \texttt{hep-th/0703178}.

\bibitem[Alb]{Peirce} A. A. Albert, \textit{A structure theory for Jordan
algebras}, Annals of Mathematics, Second Series, \textbf{48}: 546--567
(1947).

\bibitem[AM]{AM} S.~Ferrara, R.~Kallosh, A.~Strominger, $\mathcal{N}\mathit{%
=2}$\textit{\ extremal black holes}, Phys.\ Rev.\ \textbf{D52} (1995) 5412,
\texttt{hep-th/9508072}. A.~Strominger, \textit{Macroscopic Entropy of }$%
\mathcal{N}\mathit{=2}$\textit{\ Extremal Black Holes},\ Phys.\ Lett.\
\textbf{B383} (1996) 39, \texttt{hep-th/9602111}. S.~Ferrara, R.~Kallosh,
\textit{Supersymmetry and Attractors}, Phys.\ Rev.\ \textbf{D54} (1996)
1514, \texttt{hep-th/9602136}. S.~Ferrara, R.~Kallosh, \textit{Universality
of Supersymmetric Attractors}, Phys.\ Rev.\ \textbf{D54} (1996) 1525,
\texttt{hep-th/9603090}.

\bibitem[AMS]{Alek} D.~V.~Alekseevsky, A.~Marrani, A.~F.~Spiro, \textit{%
Special Vinberg Cones and the Entropy of BPS Black Holes in }$D=4$\textit{\
Supergravity}, \texttt{arXiv: 2107.06797 [hep-th]}.

\bibitem[Baez]{triality-refs} J. C. Baez, \textit{The Octonions}, Bull. Am.
Math. Soc. \textbf{39}, 145 (2002), \texttt{math/0105155 [math-ra]}.

\bibitem[BDFMR]{Small-Orbits} L. Borsten, M.J. Duff, S. Ferrara, A. Marrani,
W. Rubens, \textit{Small Orbits}, Phys. Rev. \textbf{D85} (2012) 086002,
\texttt{arXiv:1108.0424 [hep-th]}.

\bibitem[BDM]{Bhargava} L. Borsten, M.J. Duff, A. Marrani, \textit{Black
Holes and Higher Composition Laws}, \texttt{arXiv:2006.03574 [hep-th]}.

\bibitem[BFGM]{bfgm} S.~Bellucci, S.~Ferrara, M.~G\"{u}naydin, A.~Marrani,
\textit{Charge orbits of symmetric special geometries and attractors}, Int.\
J.\ Mod.\ Phys.\ \textbf{A21}, 5043 (2006), \texttt{hep-th/0606209}.

\bibitem[BFM]{bfm1} S.~Bellucci, S.~Ferrara, A.~Marrani, \textit{On some
properties of the attractor equations}, Phys.\ Lett.\ \textbf{B635} (2006)
172, \texttt{hep-th/0602161}.

\bibitem[BFMb]{book} S. Bellucci, S. Ferrara, A. Marrani : \textit{%
\textquotedblleft Supersymmetric mechanics. Vol. 2: The attractor mechanism
and space time singularities"}, Lect. Notes Phys. \textbf{701} (2006).

\bibitem[BFSY]{BFSY-D=5} S. Bellucci, S. Ferrara, A. Shcherbakov, A.
Yeranyan, \textit{Attractors and first order formalism in five dimensions
revisited}, Phys. Rev. \textbf{D83} (2011) 065003, \texttt{arXiv:1010.3516
[hep-th]}.

\bibitem[BH]{black} P.~Breitenlohner, D.~Maison and G.~W.~Gibbons, \textit{%
Four Dimensional Black Holes from Kaluza--klein theories}, Commun.\ Math.\
Phys.\ \textbf{120}, 295 (1988). R.~Kallosh, T.~Ort\'{\i}n, A.~W.~Peet,
\textit{Entropy and action of dilaton black holes},\ Phys.\ Rev.\ \textbf{D47%
}, 5400 (1993), \texttt{hep-th/9211015}. R.~Kallosh, \textit{Supersymmetric
black holes},\ Phys.\ Lett.\ \textbf{B282}, 80 (1992), \texttt{hep-th/9201029%
}.

\bibitem[BHst]{review} J. M. Maldacena, \textit{Black Holes in String Theory}%
, PhD Thesis, Princeton U. (1996), \texttt{hep-th/9607235}. A.~W.~Peet,
\textit{TASI lectures on black holes in string theory}, in : \textit{%
\textquotedblleft Theoretical Advanced Study Institute in Elementary
Particle Physics (TASI 99): Strings, Branes, and Gravity"}, 353-433 (1999),
\texttt{hep-th/0008241}. B.~Pioline, \textit{Lectures on on black holes,
topological strings and quantum attractors (2.0)}, Lect. Notes Phys. \textbf{%
755} (2008) 283. A.~Dabholkar, \textit{Black Hole Entropy And Attractors},
Class.\ Quant.\ Grav.\ \textbf{23} (2006) S957. S. Ferrara, K. Hayakawa, A.
Marrani, \textit{Lectures on Attractors and Black Holes}, Fortsch. Phys.
\textbf{56} (2008) 993, \texttt{arXiv:0805.2498 [hep-th]}. S. Bellucci, S.
Ferrara, R. Kallosh, A. Marrani, \textit{Extremal Black Hole and Flux Vacua
Attractors}, Lect. Notes Phys. \textbf{755} (2008) 115, \texttt{%
arXiv:0711.4547 [hep-th]}.

\bibitem[BMR1]{Raju-1} S. Bellucci, A. Marrani, R. Roychowdhury, \textit{On
Quantum Special K\"{a}hler Geometry}, Int. J. Mod. Phys. \textbf{A25} (2010)
1891-1935, \texttt{arXiv:0910.4249 [hep-th]}.

\bibitem[BMR2]{Raju-2} S. Bellucci, A. Marrani, R. Roychowdhury, \textit{%
Topics in Cubic Special Geometry}, J. Math. Phys. \textbf{52} (2011) 082302,
\texttt{arXiv:1011.0705 [hep-th]}.

\bibitem[Bor]{Borsten-PhD} L. Borsten, \textit{Aspects of M-Theory and
Quantum Information}. PhD thesis, Imperial College, 2010. \texttt{%
https://spiral.imperial.ac.uk:8443/handle/10044/1/6051}.

\bibitem[BPS]{BPS bound} E. B. Bogomol'nyi, \textit{Stability of Classical
Solutions}, Sov. J. Nucl. Phys. \textbf{24} (1976), 449; Yad. Fiz. \textbf{24%
} (1976), 861. M. K. Prasad, C. M. Sommerfield, \textit{Exact Classical
Solution for the 't Hooft Monopole and the Julia-Zee Dyon}, Phys. Rev. Lett.
\textbf{35} (12) 760 (1975).

\bibitem[CDF]{CDFp} A.~Ceresole, R.~D'Auria, S.~Ferrara, \textit{The
Symplectic Structure of }$\mathcal{N}\mathit{=2}$\textit{\ Supergravity and
its Central Extension}, Nucl.\ Phys.\ Proc.\ Suppl.\ \textbf{46} (1996) 67,
\texttt{hep-th/9509160}.

\bibitem[CdWM]{cdm} G.~Lopes Cardoso, B.~de Wit, T.~Mohaupt, \textit{%
Corrections to macroscopic supersymmetric black-hole entropy},\ Phys.\
Lett.\ \textbf{B451} (1999) 309, \texttt{hep-th/9812082}. G.~Lopes Cardoso,
B.~de Wit, T.~Mohaupt, \textit{Macroscopic entropy formulae and
non-holomorphic corrections for supersymmetric black holes},\ Nucl.\ Phys.\
\textbf{B567} (2000) 87, \texttt{hep-th/9906094}.

\bibitem[Cec]{Cecotti} S. Cecotti, \textit{Homogeneous Kahler Manifolds and
T Algebras in }$\mathcal{N}\mathit{=2}$\textit{\ Supergravity and
Superstrings}, Commun. Math. Phys. \textbf{124} (1989) 23-55.

\bibitem[CFGM]{d-geometries} A. Ceresole, S. Ferrara, A. Gnecchi, A.
Marrani, $\mathit{d}$\textit{-Geometries Revisited}, JHEP \textbf{02} (2013)
059, \texttt{arXiv:1210.5983 [hep-th]}{}.

\bibitem[CFM]{CFM} A. Ceresole, S. Ferrara, A. Marrani, \textit{4d/5d
Correspondence for the Black Hole Potential and its Critical Points}, Class.
Quant. Grav. \textbf{24} (2007) 5651-5666, \texttt{arXiv:0707.0964 [hep-th]}.

\bibitem[CG]{CG} B.L.\ Cerchiai, B.\ van Geemen, \textit{From qubits to E}$%
_{7}$, Journal of Mathematical Physics \textbf{51} (2010) 122203, \texttt{%
arXiv:1003.4255 [quant-ph]}.

\bibitem[CJ]{CJ-1} E. Cremmer, B. Julia, \textit{The }$\mathcal{N}=8$\textit{%
\ Supergravity Theory. 1. The Lagrangian}, Phys. Lett. \textbf{B80}, 48
(1978). E. Cremmer and B. Julia, \textit{The }$SO(8)$\textit{\ Supergravity}%
, Nucl. Phys. \textbf{B159}, 141 (1979).

\bibitem[Cor]{Cortes} V. Cort\'{e}s, \textit{Homogeneous special geometry},
Transf. Groups \textbf{1} (1996), 337-373.

\bibitem[Cve]{Cvetic} M. Cvetic, C. M. Hull, \textit{Black Holes and
U-Duality}, Nucl. Phys. \textbf{B480} (1996) 296, \texttt{hep-th/9606193}.
M. Cvetic, D. Youm, \textit{All the Static Spherically Symmetric Black Holes
of Heterotic String on a Six Torus}, Nucl. Phys. \textbf{B472} (1996) 249,
\texttt{hep-th/9512127}. M. Cvetic, A. A. Tseytlin, \textit{Solitonic
Strings and BPS Saturated Dyonic Black Holes}, Phys. Rev. \textbf{D53}
(1996) 5619, Phys. Rev. \textbf{D55} (1997) 3907 (erratum), \texttt{%
hep-th/9512031}.

\bibitem[CVP]{CVP} E. Cremmer, A. Van Proeyen, \textit{Classification of K%
\"{a}hler Manifolds in }$\mathcal{N}\mathit{=2}$\textit{\ Vector Multiplet
Supergravity Couplings}, Class. Quant. Grav. \textbf{2} (1985) 445.

\bibitem[D]{Duff} M.~J.~Duff, \textit{String triality, black-hole entropy
and Cayley's hyperdeterminant}, Phys. Rev. \textbf{D76} (2007) 025017,
\texttt{hep-th/0601134}. L. Borsten, M.J. Duff, A. Marrani, W. Rubens, On
the Black-Hole/Qubit Correspondence, Eur. Phys. J. Plus \textbf{126} (2011)
37, \texttt{arXiv:1101.3559 [hep-th]}. L. Borsten, M.J. Duff, P. Levay,
\textit{The black-hole/qubit correspondence: an up-to-date review}, Class.
Quant. Grav. \textbf{29} (2012) 224008, \texttt{arXiv:1206.3166 [hep-th]}.

\bibitem[DFT]{DFT-Hom-07} R. D'Auria, S. Ferrara, M. Trigiante, \textit{%
Critical points of the Black-Hole potential for homogeneous special
geometries}, JHEP \textbf{0703} (2007) 097, \texttt{hep-th/0701090}.

\bibitem[DHW]{Magic-Wissanji} K. Dasgupta, V. Hussin, A. Wissanji, \textit{%
Quaternionic K\"{a}hler Manifolds, Constrained Instantons and the Magic
Square. I}, Nucl. Phys. \textbf{B793} (2008) 34-82, \texttt{0708.1023
[hep-th]}.

\bibitem[dWVP1]{dWVP} B. de Wit, A. Van Proeyen, \textit{Special geometry,
cubic polynomials and homogeneous quaternionic spaces}, Commun. Math. Phys.
\textbf{149} (1992) 307-334, \texttt{hep-th/9112027}.

\bibitem[dWVP2]{broken} B. de Wit, A. Van Proeyen, \textit{Broken sigma
model isometries in very special geometry}, Phys. Lett. \textbf{B293} (1992)
94-99 \texttt{hep-th/9207091}.

\bibitem[dWVP3]{dWVP2} B. de Wit, A. Van Proeyen, \textit{Isometries of
special manifolds}, Proceedings of the Meeting on Quaternionic Structures in
Mathematics and Physics, Trieste (Italy), September 1994, \texttt{%
hep-th/9505097}.

\bibitem[dWVVP]{dWVVP} B. de Wit, F. Vanderseypen, A. Van Proeyen, \textit{%
Symmetry structure of special geometries}, Nucl. Phys. \textbf{B400} (1993)
463-524, \texttt{hep-th/9210068}.

\bibitem[Extr]{Extreme-BH} A.~Sen, \textit{Extremal black holes and
elementary string states},\ Mod.\ Phys.\ Lett.\ \textbf{A10} (1995) 2081,
\texttt{hep-th/9504147}. G.~T.~Horowitz, A.~Strominger, \textit{Counting
States of Near-Extremal Black Holes},\ Phys.\ Rev.\ Lett.\ \textbf{77}, 2368
(1996), \texttt{hep-th/9602051}.

\bibitem[FG1]{FG1} S. Ferrara, M. G\"{u}naydin, \textit{Orbits of
exceptional groups, duality and BPS states in string theory}, Int. J. Mod.
Phys. \textbf{A13}, 2075-2088 (1998), \texttt{hep-th/9708025}.

\bibitem[FG2]{FG2} S. Ferrara, M. G\"{u}naydin, \textit{Orbits and
Attractors for }$\mathcal{N}\mathit{=2}$\textit{\ Maxwell-Einstein
Supergravity Theories in Five Dimensions}, Nucl. Phys. \textbf{B759} 1-19
(2006), \texttt{hep-th/0606108}.

\bibitem[FGK]{fegika} S.~Ferrara, G.~W.~Gibbons, R.~Kallosh, \textit{Black
holes and critical points in moduli space}, Nucl.\ Phys.\ \textbf{B500}
(1997) 75, \texttt{hep-th/9702103}.

\bibitem[FGimK]{FGimonK} S. Ferrara, E. G. Gimon, R. Kallosh, \textit{Magic
supergravities, }$\mathcal{N}\mathit{=8}$\textit{\ and black hole composites}%
, Phys. Rev. \textbf{D74} (2006) 125018, \texttt{hep-th/0606211}.

\bibitem[FM]{FM-moduli} S. Ferrara, A. Marrani, \textit{On the Moduli Space
of non-BPS Attractors for }$\mathcal{N}=2$\textit{\ Symmetric Manifolds},
Phys. Lett. \textbf{B652} (2007) 111, \texttt{arXiv: 0706.1667 [hep-th]}.

\bibitem[Fre]{Freed} D. S. Freed, \textit{Special K\"{a}hler manifolds},
Commun. Math. Phys. \textbf{203} (1999) 31-52, \texttt{hep-th/9712042}.

\bibitem[GH]{Halmagyi2} A. Gnecchi, N. Halmagyi, \textit{Supersymmetric
black holes in }$AdS_{4}$\textit{\ from very special geometry}, JHEP \textbf{%
04} (2014) 173, \texttt{arXiv:1312.2766 [hep-th]}.

\bibitem[GKN]{GP1} M. G\"{u}naydin, K. Koepsell, H. Nicolai, \textit{%
Conformal and quasiconformal realizations of exceptional Lie groups},
Commun. Math. Phys. \textbf{221}, 57-76 (2001), \texttt{hep-th/0008063}.

\bibitem[GKT]{GKT} M. G\"{u}naydin, S. Kachru, A. Tripathy, Black holes and
Bhargava's invariant theory, J. Phys. \textbf{A53}, no.44, 444001 (2020),
\texttt{arXiv:1903.02323 [hep-th]}.

\bibitem[GMV]{gmv} M.~Gasperini, J.~Maharana, G.~Veneziano, \textit{From
trivial to nontrivial conformal string backgrounds via O(d,d) transformations%
}, Phys.\ Lett.\ \textbf{B272} (1991) 277. J.~Maharana, J.~H.~Schwarz,
\textit{Noncompact symmetries in string theory}, Nucl.\ Phys.\ \textbf{B390}
(1993) 3, \texttt{hep-th/9207016}

\bibitem[GP1]{GP2} M. G\"{u}naydin, O. Pavlyk, \textit{Generalized
spacetimes defined by cubic forms and the minimal unitary realizations of
their quasiconformal groups}, JHEP \textbf{08}, 101 (2005), \texttt{%
hep-th/0506010}.

\bibitem[GP2]{GP3} M. G\"{u}naydin, O. Pavlyk, \textit{Spectrum Generating
Conformal and Quasiconformal }$\mathit{U}$\textit{-Duality Groups}, \textit{%
Supergravity and Spherical Vectors}, JHEP \textbf{04}, 070 (2010), \texttt{%
arXiv:0901.1646 [hep-th]}.

\bibitem[GP3]{GP4} M. G\"{u}naydin, O. Pavlyk, \textit{Quasiconformal
Realizations of E}$_{6(6)}$\textit{, E}$_{7(7)}$\textit{, E}$_{8(8)}$\textit{%
\ and }$SO(n+3,m+3)$\textit{, }$N\geqslant 4$\textit{\ Supergravity and
Spherical Vectors}, Adv. Theor. Math. Phys. \textbf{13}, no.6, 1895-1940
(2009), \texttt{arXiv:0904.0784 [hep-th]}.

\bibitem[GR]{wald} B.~De Witt and C.~De Witt eds. : \textquotedblleft
\textit{Black Holes}\textquotedblright , Gordon and Breach, New York (1973).
S.~W.~Hawking, W.~Israel : \textquotedblleft \textit{General Relativity}%
\textquotedblright , Cambridge University Press, 1979. R.~M.~Wald :
\textquotedblleft \textit{General Relativity}\textquotedblright , University
of Chicago Press, 1984.

\bibitem[GST1]{GST1} M. G\"{u}naydin, G. Sierra, P.K. Townsend, \textit{%
Exceptional Supergravity Theories and the Magic Square}, Phys. Lett. \textbf{%
B133}, 72-76 (1983).

\bibitem[GST2]{GST2} M. G\"{u}naydin, G. Sierra, P.K. Townsend, \textit{The
Geometry of }$\mathcal{N}\mathit{=2}$\textit{\ Maxwell-Einstein Supergravity
and Jordan Algebras}, Nucl. Phys. \textbf{B242} (1984) 244-268.

\bibitem[GST3]{GST3} M. G\"{u}naydin, G. Sierra, P.K. Townsend, \textit{More
on d = 5 Maxwell Einstein Supergravity: Symmetric Spaces and Kinks}, Class.
Quant. Grav. \textbf{3}, 763 (1986).

\bibitem[Hal]{Halmagyi1} N. Halmagyi, \textit{BPS Black Hole Horizons in }$%
\mathcal{N}\mathit{=2}$\textit{\ Gauged Supergravity}, JHEP \textbf{1402}
(2014) 051, \texttt{arXiv:1308.1439 [hep-th]}.

\bibitem[Has]{Hasebe} K. Hasebe, \textit{The Split-Algebras and Non-compact
Hopf Maps}, J. Math. Phys. \textbf{51} (2010) 053524, \texttt{%
arXiv:0905.2792 [math-ph]}.

\bibitem[Haw]{Hawk} S. W. Hawking, \textit{Black hole explosions?}, Nature
\textbf{248}, 30--31 (1974).

\bibitem[HB]{entrop} S.~W.~Hawking, \textit{Gravitational radiation from
colliding black holes}, Phys.\ Rev.\ Lett.\ \textbf{26} (1971)
1344.~D.~Bekenstein, \textit{Black holes and entropy}, Phys.\ Rev.\ \textbf{%
D7} (1973) 2333.

\bibitem[HP]{ph} S.~W.~Hawking, R.~Penrose, \textit{The Singularities of
gravitational collapse and cosmology}, Proc.\ Roy.\ Soc.\ Lond.\ A \textbf{%
314} (1970) 529.

\bibitem[HT]{HT-1} C. Hull, P. K. Townsend, \textit{Unity of Superstring
Dualities}, Nucl. Phys. \textbf{B438}, 109 (1995), \texttt{hep-th/9410167}.

\bibitem[JVNW]{JVNW} P. Jordan, J. v. Neumann, E. Wigner, \textit{On an
Algebraic Generalization of the Quantum Mechanical Formalism}, Annals of
Mathematics, Second Series, Vol. \textbf{35}, No. 1 (Jan., 1934), pp. 29-64.

\bibitem[Kac]{Kac-80} V. G. Kac, \textit{Some Remarks on Nilpotent Orbits},
J. of Algebra \textbf{64}, 190-213 (1980).

\bibitem[KLOPVP]{CCC} R.~Kallosh, A.~D.~Linde, T.~Ort\'{\i}n, A.~W.~Peet,
A.~Van Proeyen, \textit{Supersymmetry as a cosmic censor}, Phys.\ Rev.\
\textbf{D46}, 5278 (1992), \texttt{hep-th/9205027}.

\bibitem[Kru]{Krut} S. Krutelevich, \textit{Jordan Algebras, Exceptional
Groups, and Bhargava Composition}, J. of Algebra \textbf{314}, Issue 2
(2007), 924-977, \texttt{math/0411104}.

\bibitem[KS]{KS} Y.\ Kim, F-O.\ Schreyer, \textit{An explicit matrix
factorization of cubic hypersurfaces of small dimension}, \texttt{%
arXiv:1905.09626 [math.AG]}.

\bibitem[LS]{Shahbazi} C.I. Lazaroiu, C.S. Shahbazi, \textit{%
Four-dimensional geometric supergravity and electromagnetic duality: a brief
guide for mathematicians}, \texttt{arXiv: 2006.16157 [math.DG]}.

\bibitem[Micro]{blackmicro} A.~Strominger, C.~Vafa, \textit{Microscopic
Origin of the Bekenstein-Hawking Entropy},\ Phys.\ Lett.\ \textbf{B379}
(1996) 99, \texttt{hep-th/9601029}. C.~G.~Callan, J.~M.~Maldacena, \textit{%
D-brane Approach to Black Hole Quantum Mechanics},\ Nucl.\ Phys.\ \textbf{%
B472}, 591 (1996), \texttt{hep-th/9602043}. R.~Dijkgraaf, E.~P.~Verlinde,
H.~L.~Verlinde, \textit{BPS spectrum of the five-brane and black-hole entropy%
},\ Nucl.\ Phys.\ \textbf{B486}, 77 (1997), \texttt{hep-th/9603126}.
D.~M.~Kaplan, D.~A.~Lowe, J.~M.~Maldacena, A.~Strominger, \textit{%
Microscopic entropy of }$\mathcal{N}\mathit{=2}$\textit{\ extremal black
holes},\ Phys.\ Rev.\ \textbf{D55}, 4898 (1997), \texttt{hep-th/9609204}.
J.~M.~Maldacena, $\mathcal{N}\mathit{=2}$\textit{\ extremal black holes and
intersecting branes},\ Phys.\ Lett.\ \textbf{B403}, 20 (1997), \texttt{%
hep-th/9611163}. J.~M.~Maldacena, A.~Strominger, E.~Witten, \textit{Black
hole entropy in M-theory},\ JHEP \textbf{9712} (1997) 002, \texttt{%
hep-th/9711053}.

\bibitem[MMT]{multBPS2} A. Marrani, T. Mandal, P. K. Tripathy, \textit{%
Supersymmetric Black Holes and Freudenthal Duality}, Int. J. Mod. Phys.
\textbf{A32} (2017) 19n20, 1750114, \texttt{arXiv:1703.08669 [hep-th]}.

\bibitem[Mo1]{Moore} G. W. Moore, \textit{Arithmetic and attractors},
\texttt{hep-th/9807087}.

\bibitem[Mo2]{moore} G.~W.~Moore, \textit{Strings and arithmetic}, in :
\textquotedblleft \textit{Les Houches School of Physics: Frontiers in Number
Theory, Physics and Geometry"}, 303-359, \texttt{hep-th/0401049}.

\bibitem[MPRR]{Marrani-Romano-1} A. Marrani, G. Pradisi, F. Riccioni, L.
Romano, \textit{Nonsupersymmetric magic theories and Ehlers truncations},
Int. J. Mod. Phys. \textbf{A32} (2017) 19n20, 1750120, \texttt{arXiv:
1701.03031 [hep-th]}.

\bibitem[MR]{Marrani-Romano-2} A. Marrani, L. Romano, \textit{Orbits in
nonsupersymmetric magic theorie}s, Int. J. Mod. Phys. \textbf{A34} (2019)
32, 1950190, \texttt{arXiv:1906.05830 [hep-th]}.

\bibitem[MT]{multBPS1} T. Mandal, P. K. Tripathy, \textit{On the Uniqueness
of Supersymmetric Attractors}, Phys. Lett. \textbf{B749} (2015) 221-225,
\texttt{arXiv:1506.06276 [hep-th]}.

\bibitem[MTR]{Marrani-Group32} A. Marrani, P. Truini, M. Rios, \textit{The
Magic of Being Exceptional}, J. Phys. Conf. Ser. \textbf{1194} (2019) 1,
012075, \texttt{arXiv:1811.11208 [hep-th]}.

\bibitem[nBPS]{non-BPS} K.~Goldstein, N.~Iizuka, R.~P.~Jena, S.~P.~Trivedi,
\textit{Non-supersymmetric attractors},\ Phys.\ Rev.\ \textbf{D72} (2005)
124021, \texttt{hep-th/0507096}. R.~Kallosh, \textit{New attractors},\ JHEP
\textbf{0512} (2005) 022, \texttt{hep-th/0510024}. P.~K.~Tripathy,
S.~P.~Trivedi, \textit{Non-supersymmetric attractors in string theory},\
JHEP \textbf{0603} (2006) 022, \texttt{hep-th/0511117}. R.~Kallosh,
N.~Sivanandam, M.~Soroush, \textit{The non-BPS black-hole attractor equation}%
,\ JHEP \textbf{0603} (2006) 060, \texttt{hep-th/0602005}. A.~Dabholkar,
A.~Sen, S.~P. Trivedi, \textit{Black hole microstates and attractor without
supersymmetry},\ JHEP \textbf{01} (2007) 096, \texttt{hep-th/0611143}.

\bibitem[Noe]{Noether} M.\ Noether, \textit{Zur Theorie der Thetafunctionen
van beliebig vielen Argumenten}, Math.\ Ann.\ \textbf{16} (1880) 270-344.

\bibitem[Ort]{ortino} R. Kallosh, T. Ort\'{\i}n, \textit{Charge quantization
of Axion-Dilaton black holes}, Phys. Rev. \textbf{D48} (1993) 742, \texttt{%
hep-th/9302109}. E. Bergshoeff, R. Kallosh, T. Ort\'{\i}n, \textit{%
Stationary Axion-Dilaton solutions and supersymmetry}, Nuc. Phys. \textbf{%
B478} (1996) 156, \texttt{hep-th 9605059}.

\bibitem[Ort2]{ortin} R.~R.~Khuri, T.~Ort\'{\i}n, \textit{A
Non-Supersymmetric Dyonic Extreme Reissner-Nordstr\"{o}m Black Hole}, Phys.\
Lett.\ \textbf{B373} (1996) 56, \texttt{hep-th/9512178}. T.~Ort\'{\i}n,
\textit{Extremality versus supersymmetry in stringy black holes}, Phys.\
Lett. \textbf{B422} (1998) 93, \texttt{hep-th/9612142}.

\bibitem[PR1]{PR1} L. Pirio, F. Russo, \textit{Extremal varieties
3-rationally connected by cubics, quadro-quadric Cremona transformations and
rank 3 Jordan algebras}, Journal f\" ur die reine und angewandte Mathematik
(Crelle's Journal) \textbf{716} (2016), 229--250., \texttt{arXiv:1109.3573
[math.AG]}.

\bibitem[PR2]{PR2} L. Pirio, F. Russo, \textit{Quadro-quadric cremona
transformations in low dimensions via the JC-correspondence}, Annales de
l'Institut Fourier, Volume \textbf{64} (2014) no. 1, p. 71-111, \texttt{%
arXiv:1204.0428 [math.AG]}.

\bibitem[Rus]{Russo} F. Russo, \textit{On the geometry of some special
projective varieties}, Lecture Notes of the UMI \textbf{18}, Springer (2016).

\bibitem[Sch]{sesch} J.~H.~Schwarz, A.~Sen, \textit{Duality symmetries of 4D
heterotic strings},\ Phys.\ Lett.\ \textbf{B312}, 105 (1993), \texttt{%
hep-th/9305185}. J.~H.~Schwarz, A.~Sen, \textit{Duality symmetric actions},\
Nucl.\ Phys.\ \textbf{B411}, 35 (1994), \texttt{hep-th/9304154}.

\bibitem[Sen]{Sen} A.~Sen, \textit{Black Hole Solutions In Heterotic String
Theory On A Torus},\ Nucl.\ Phys.\ \textbf{B440} (1995) 421, \texttt{%
hep-th/9411187}. A.~Sen, \textit{Quantization of dyon charge and electric
magnetic duality in string theory},\ Phys.\ Lett.\ \textbf{B303} (1993) 22,
\texttt{hep-th/9209016}.

\bibitem[Shm]{Shmakova} M. Shmakova, \textit{Calabi-Yau Black Holes}, Phys.
Rev. \textbf{D56} (1997) 540-544, \texttt{hep-th/9612076}.

\bibitem[SKG]{spegeo} S. Ferrara, A. Strominger, $\mathcal{N}\mathit{=2}$%
\textit{\ space-time supersymmetry and Calabi Yau Moduli Space}, in :
\textit{Proceedings of College Station Workshop \textquotedblleft Strings
`89"}, pag. 245, eds. Arnowitt \textit{et al.}, World Scientific 1989.
P.~Candelas, X.~de la Ossa, \textit{Moduli Space of CalabiYau Manifolds},\
Nucl.\ Phys.\ \textbf{B355}, 455 (1991). B.~de Wit, A.~Van Proeyen, \textit{%
Potentials And Symmetries Of General Gauged }$\mathcal{N}\mathit{=2}$\textit{%
\ Supergravity - Yang-Mills Models},\ Nucl.\ Phys.\ \textbf{B245}, 89
(1984). E.~Cremmer, C.~Kounnas, A.~Van Proeyen, J.~P.~Derendinger,
S.~Ferrara, B.~de Wit, L.~Girardello, \textit{Vector Multiplets Coupled To }$%
\mathcal{N}\mathit{=2}$\textit{\ Supergravity: Superhiggs Effect, Flat
Potentials And Geometric Structure},\ Nucl.\ Phys.\ \textbf{B250}, 385
(1985). B.~de Wit, P.~G.~Lauwers, A.~Van Proeyen, \textit{Lagrangians Of }$%
\mathcal{N}\mathit{=2}$\textit{\ Supergravity - Matter Systems},\ Nucl.\
Phys.\ \textbf{B255}, 569 (1985). L.~Castellani, R.~D'Auria, S.~Ferrara,
\textit{Special geometry without special coordinates},\ Class.\ Quant.\
Grav.\ \textbf{7}, 1767 (1990). L.~Castellani, R.~D'Auria, S.~Ferrara,
\textit{Special K\"{a}hler Geometry; an intrinsic Formulation from }$%
\mathcal{N}\mathit{=2}$\textit{\ Space-Time Supersymmetry},\ Phys.\ Lett.\
\textbf{B241}, 57 (1990).

\bibitem[Str]{schw} J.~H.~Schwarz, \textit{Lectures on superstring and M
theory dualities}, Nucl.\ Phys.\ Proc.\ Suppl.\ \textbf{55B}, 1 (1997),
\textit{hep-th/9607201}. M.~J.~Duff, \textit{M theory (the theory formerly
known as strings)}, Int.\ J.\ Mod.\ Phys.\ \textbf{A11} (1996) 5623, \texttt{%
hep-th/9608117}. A.~Sen, \textit{Unification of string dualities}, Nucl.\
Phys.\ Proc.\ Suppl.\ \textbf{58}, 5 (1997), \texttt{hep-th/9609176}.

\bibitem[Stro]{Strominger-SKG} A. Strominger, \textit{Special Geometry},
Commun. Math. Phys. \textbf{133} (1990) 163-180.

\bibitem[vGe]{vG} B.\ van Geemen, \textit{Schottky-Jung relations and vector
bundles on hyperelliptic curves}, Math.\ Ann.\ \textbf{281} (1988) 431-449.

\bibitem[Vin]{Vinberg} E. B. Vinberg, \textit{The Theory of Homogeneous
Convex Cones}, in : Transactions of the Moscow Math. Society for the year
1963, 340-403. Providence, RI : American Mathematical Society, 1965.

\bibitem[VSch]{vasch} J.~H.~Schwarz, \textit{M theory extensions of T duality%
},\ in : \textit{\textquotedblleft Frontiers in Quantum Field Theory, in
Honor of the 60th Birthday of Prof. K. Kikkawa"}, 3-14 (1996), \texttt{%
hep-th/9601077}. C.~Vafa, \textit{Evidence for F-Theory},\ Nucl. Phys.\
\textbf{B469}, 403 (1996), hep-th/9602022.

\bibitem[W]{witten} E.~Witten, \textit{String theory dynamics in various
dimensions}, Nucl.\ Phys.\ \textbf{B443} (1995) 85, \texttt{hep-th/9503124}.

\bibitem[Wa]{w} R.~M.~Wald, \textit{Black hole entropy in the Noether charge}%
, Phys.\ Rev.\ \textbf{D48} (1993) 3427, \texttt{gr-qc/9307038}.
\end{thebibliography}
\end{document}